\documentclass[preprint,authoryear,12pt]{elsarticle}

\usepackage{amsmath}

\usepackage{bm}

\usepackage{graphicx}
\usepackage[font= footnotesize]{caption}
\usepackage[font=footnotesize]{subcaption}

\usepackage[table]{xcolor}
\usepackage{booktabs}
\usepackage{tabu}
\usepackage{threeparttable}
\usepackage{multirow}
\usepackage{multicol}

\usepackage{xpatch}
\makeatletter
\chardef\TPT@@@asteriskcatcode=\catcode`*
\catcode`*=11
\xpatchcmd{\threeparttable}
  {\TPT@hookin{tabular}}
  {\TPT@hookin{tabular}\TPT@hookin{tabu}}
  {}{}
\catcode`*=\TPT@@@asteriskcatcode
\makeatother

\usepackage{array}
\newcolumntype{P}[1]{>{\centering\arraybackslash}p{#1}}

\usepackage{fullpage}
\usepackage{comment}

\usepackage[%
a4paper=true,%
breaklinks=true,%
colorlinks=true,%
pdfauthor={First Author et al.},%
pdftitle={Template for manuscripts in Advances in Space Research}%
]{hyperref}

\journal{Advances in Space Research}

\begin{document}

% -------------------------------------------------------------------------------------------------------------------------------------- 
%     FRONTMATTER
% -------------------------------------------------------------------------------------------------------------------------------------- 
\begin{frontmatter}


% Use the tnoteref command within \title and fnref within \author or \address for footnotes;
% use the corref command within \author for corresponding author footnotes;
% \author{Name\corref{cor1}\fnref{label2}}
% \address{Address\fnref{label3}}

\title{Medium Earth Orbit dynamical survey and its use in passive debris removal}

\author{Despoina K. Skoulidou\corref{cor}}
\ead{dskoulid@physics.auth.gr}
\address{Department of Physics, Aristotle University of Thessaloniki, 54124 Thessaloniki, Greece}
\cortext[cor]{Corresponding author}
\author{Aaron J. Rosengren}
\ead{ajrosengren@email.arizona.edu}
\address{Aerospace and Mechanical Engineering, University of Arizona, Tucson, AZ 85721, USA}
\address{Department of Physics, Aristotle University of Thessaloniki, 54124 Thessaloniki, Greece}
\author{Kleomenis Tsiganis}
\ead{tsiganis@auth.gr}
\author{George Voyatzis}
\ead{voyatzis@auth.gr}
\address{Department of Physics, Aristotle University of Thessaloniki, 54124 Thessaloniki, Greece}

% -------------------------------------------------------------------------------------------------------------------------------------- 
%     ABSTRACT
% -------------------------------------------------------------------------------------------------------------------------------------- 
\begin{abstract}

The Medium Earth Orbit (MEO) region hosts satellites for navigation, communication, and geodetic/space 
environmental science, among which are the Global Navigation Satellites Systems (GNSS). Safe and efficient removal of debris from 
MEO is problematic due to the high cost for maneuvers needed to directly reach the Earth (reentry orbits) and the relatively 
crowded GNSS neighborhood (graveyard orbits). Recent studies have highlighted the complicated secular dynamics in the MEO region, 
but also the possibility of exploiting these dynamics, for designing removal strategies. In this paper, we present our numerical 
exploration of the long-term dynamics in MEO, performed with the purpose of unveiling the set of reentry and graveyard solutions 
that could be reached with maneuvers of reasonable $\Delta V$ cost. We simulated the dynamics over 120-200 years for an extended grid of 
millions of fictitious MEO satellites that covered all inclinations from 0 to 90$^{\circ}$, using non-averaged equations of motion 
and a suitable dynamical model that accounted for the principal geopotential terms, 3rd-body perturbations and solar radiation 
pressure (SRP). We found a sizeable set of usable solutions with reentry times that exceed $\sim 40$~years, mainly around three 
specific inclination values: 46$^{\circ}$, 56$^{\circ}$, and 68$^{\circ}$; a result compatible with our understanding of MEO 
secular dynamics. For $\Delta V \leq 300$~m/s (i.e., achieved if you start from a typical GNSS orbit and target a 
disposal orbit with $e<0.3$), reentry times from GNSS altitudes exceed $\sim 
70~$years, while low-cost ($\Delta V \simeq 5-35~$m/s) graveyard orbits, stable for at lest 200~years, are found for 
eccentricities up to $e\approx 0.018$. This investigation was carried out in the framework of the EC-funded ``ReDSHIFT'' project. 

\end{abstract}


\end{frontmatter}

\parindent=0.5 cm

% -------------------------------------------------------------------------------------------------------------------------------------- 
%          INTRODUCTION         
% -------------------------------------------------------------------------------------------------------------------------------------- 
\section{Introduction}
\label{intro}

The Medium Earth Orbit (MEO) region of the near-Earth space environment is defined (with respect to orbital altitude, $h$) as the 
region higher than the Low Earth Orbit (LEO) protected region and lower than the Geosynchronous Earth Orbit (GEO) region, i.e.,\ 
$h=\left(2,000-35,786\right)$~km. However, in reality, the actual space used for operations is much more limited. Currently, one of the 
most populated places in the MEO region is occupied by Global Navigation Satellite Systems (GNSS), which are located at relatively high 
inclinations. An in-depth understanding of the long-term dynamics of the GNSS region is needed, given the importance of 
these systems for humanity. Similarly, the dynamics of the ``extended MEO'' region around GNSS altitudes -- encompassing eccentric orbits at all 
inclinations -- has to be understood, given the possibility of it becoming usable in the future. We refer the reader to 
\citet{rAfS18} for a quite complete and up-to-date description of GNSS secular dynamics and related issues. Here, we present 
the main characteristics of the MEO region and discuss some open issues, regarding end-of-life (EoL) satellite disposal.  \\

Numerous secular and semi-secular lunisolar resonances cross the circumterrestrial space. Their location and strength depend on 
the main orbital parameters, i.e.,\ the semi-major axis $a$, eccentricity $e$ and inclination $i$. These resonances induce a slow, 
large-amplitude variation in the eccentricity and/or inclination of an orbit. The orbital eccentricity being most relevant to the 
current discussion, as its increase leads to a decrease of perigee altitude. A visual inspection of the resonant effects can be 
obtained with the use of 2-D projections, typically referred to as {\it dynamical maps}, where variations 
of an orbital parameter (typically, $e$) are color-coded on a grid of initial conditions, and resonant lines are super-imposed 
\citep[as seen, e.g.,\ in ][] {gC62,sB01b,aR19}. Lunisolar resonances are known to overlap near the GNSS region, when mapped in the 
$(a,i)$ or $(e,i)$ plane \citep{aR15,jD16}, a property that adds complexity (chaos) in the dynamics. The long-term effect of resonances 
in the GNSS region have been studied, using both analytical and numerical methods, on the averaged equations of motion 
\citep[e.g., see ][]{aR15,lSgM15,aCcG16,jD16,iGetal16,aR17}.  \\

Mitigation of the space debris population and direct disposal of the non-operational satellites that are placed in the GNSS region 
is not an easy task, as unassisted (natural) reentry to Earth seems not to be possible, ever after century-long timescales. Hence the basic 
(passive) removal strategy would consist either in (a) assisting eccentricity build-up to reach a reentry solution within a 
reasonable time, or (b) moving to a long-term stable graveyard orbit. Both strategies would need to take into account the 
boundaries of the operation zones, the resonant dynamics in the neighborhood, and the need for low-cost maneuvers 
\citep{jR15,eA16,aR17,rAfS18}. In the eccentricity build-up scenario, usable disposal orbits should have low-to-moderate 
eccentricities, so to be reachable with low $\Delta V$; in this paper we set the limit to $300$~m/sec. Also the removal or waiting 
time (i.e.,\ the time spent by the disposed satellite on the reentry trajectory) should not be unrealistically long, nor the dwell 
times in the LEO and GEO protected regions. In the other scenario, a graveyard orbit -- even not necessarily strictly circular -- 
should be stable and not cross any of the neighboring operational zones for very long times; here we set the limit to 200 years. 
For an in-depth investigation of the long-term dynamics in the MEO region, several parameters have to be varied, including the 
initial epoch, secular orientation (i.e.,\ values of $\omega$ = argument of perigee and $\Omega$ = longitude of ascending node) 
and the assumed area-to-mass $A/m$ ratio of the debris. Here, we made the choice of extending our study over a dense grid in 
$(a,e)$ and for all inclinations between $0$ and $90^{\circ}$ (i.e.,\ an ``extended MEO'' region), hence necessarily limiting 
ourselves in the choice of initial secular orientations, epochs, and $A/m$ values.  \\

Our study is part of the EC-funded ``ReDSHIFT'' project\footnote{http://redshift-h2020.eu} \citep{ReDSHIFT18}. The main goal of 
this project is to introduce a holistic approach in the design of passive debris removal strategies. As such, it represents a 
combination of theoretical and experimental research activities, including astrodynamics, debris population evolution, legal 
aspects, advanced additive manufacturing (3D printing), and testing of components, with the scope of producing a small satellite 
that would be better ``designed for demise''. A significant part of the project comprises an in-depth investigation of the 
dynamics of the whole circumterrestrial space. A general overview of the dynamics over a coarse grid, covering LEO-to-GEO 
altitudes, was presented in \citet{aR19}, followed by publication of the results of higher-resolution simulations of the densely 
populated areas (LEO \citep{eA18a,eA18b}; MEO \citep{dS17,dS18}; GEO \citep{cC17,iG17}). In the latter, the possibility of using 
the resulting dynamical maps for locating ``natural highways'' for EoL disposal is discussed.  \\

The first goal of the present work is to provide an updated dynamical atlas of the MEO region around GNSS altitudes but extended 
over the whole eccentricity domain and all inclinations up to 90$^{\circ}$. To this purpose we integrated several million initial 
conditions, using a non-averaged symplectic propagator (called SWIFT-SAT). Apart from looking for natural reentry solution that 
can be reached with moderate $\Delta V$ over reasonably long timescales, the second goal is to extend our study to 
define and map the usable graveyard regions around the GNSS, a task that was previously shown to be complicated, at least for 
the Galileo constellation (see e.g., \citet{aR17}).  \\

The dynamical properties of the GNSS population are presented in Section \ref{subsec:21}, while the dynamical model and the 
grid of initial conditions used in our numerical simulations are defined in Section \ref{subsec:22}. In Section \ref{sec:3}, the 
main results of the numerical simulations are collected and presented in the form of a dynamical atlas. Our study on assisted 
disposal with $\Delta V$-maneuvers is presented in Section \ref{sec:maneuv}. Finally, the conclusions of this work are presented 
in Section \ref{sec:concl}.  \\

% -------------------------------------------------------------------------------------------------------------------------------------- 
%          PROBLEM FORMULATION
% -------------------------------------------------------------------------------------------------------------------------------------- 
\section{Problem formulation}
\label{sec:2}
\subsection{Medium Earth Orbit environment}
\label{subsec:21}

In MEO, some of the most populated groups of objects are the GNSS constellations, the Geosynchronous Transfer Orbits (GTOs) 
and the Molniyas. GTOs and Molniyas have eccentricities that range between $e\sim0.5$ and $e\sim0.8$, while for the GNSS 
$e\sim 0$. The inclinations vary, around $i\sim5^{\circ},~28^{\circ},~46^{\circ}$ (GTOs) and $i\sim64^{\circ}$ (Molniya), or 
between $i\sim55^{\circ}$ and $i\sim65^{\circ}$ for the GNSS. GTOs approach both LEO and GEO altitudes, as also do Molniyas. 
However, Molniyas and GNSS are placed near the 2:1 tesseral resonance. The GNSS consist of four constellations: GLONASS ($a\sim 
25510$~km, $i\sim64.8^{\circ}$); GPS ($a\sim 26561$~km, $i\sim55^{\circ}$); BEIDOU ($a\sim 27906$~km, $i\sim55^{\circ}$) and 
GALILEO ($a\sim 29601$~km, $i\sim56^{\circ}$).\footnote{GPS, GLONASS, GALILEO and BEIDOU are designed to be placed close to the 
$2:1$, $17:8$, $17:10$ and $17:9$ tesseral resonance, respectively.}  \\

\begin{figure}[htp!]
  \centering
    \begin{subfigure}[b]{0.4\textwidth}
      \includegraphics[width=\textwidth]{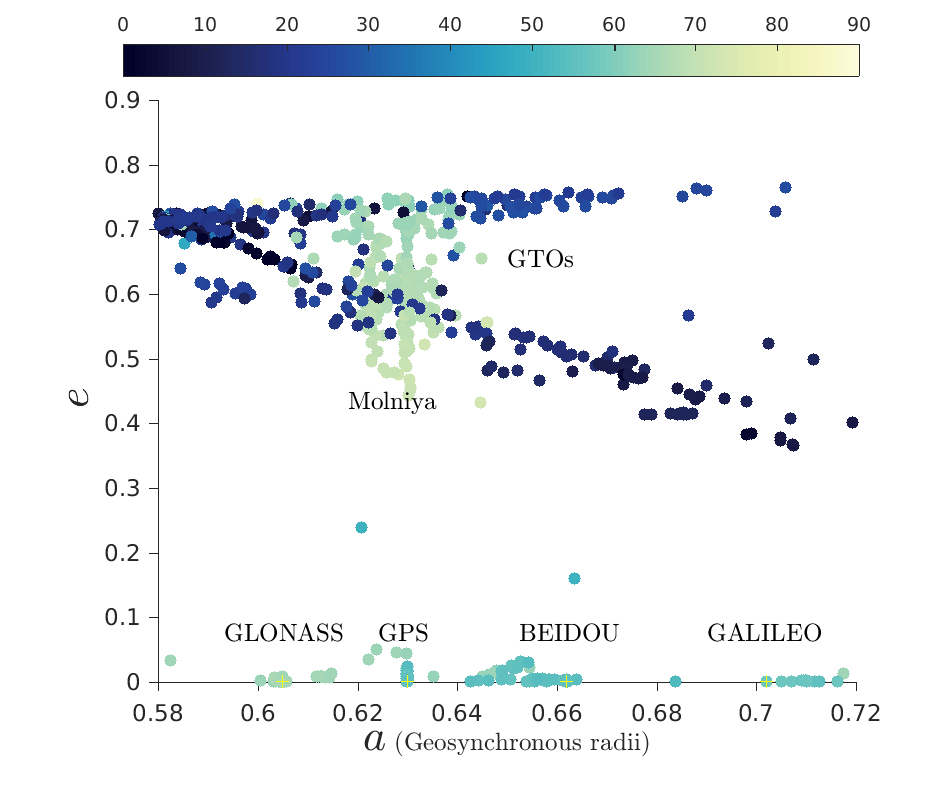}
    \end{subfigure} 
    \begin{subfigure}[b]{0.4\textwidth}
      \includegraphics[width=\textwidth]{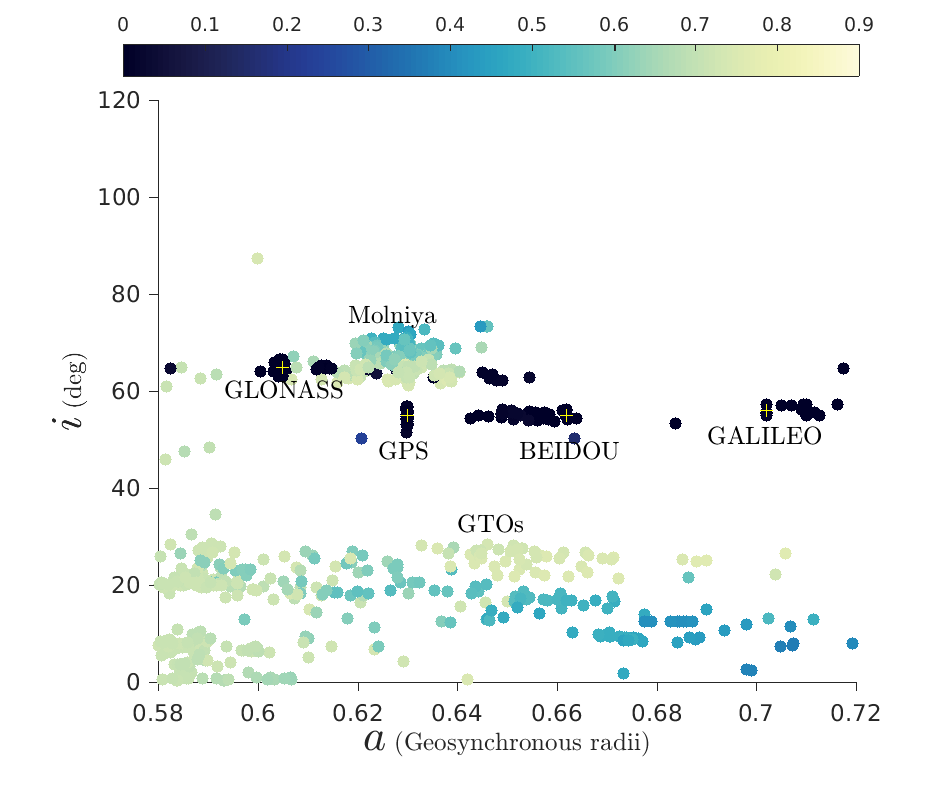}
    \end{subfigure}  
    \begin{subfigure}[b]{0.4\textwidth}
      \includegraphics[width=\textwidth]{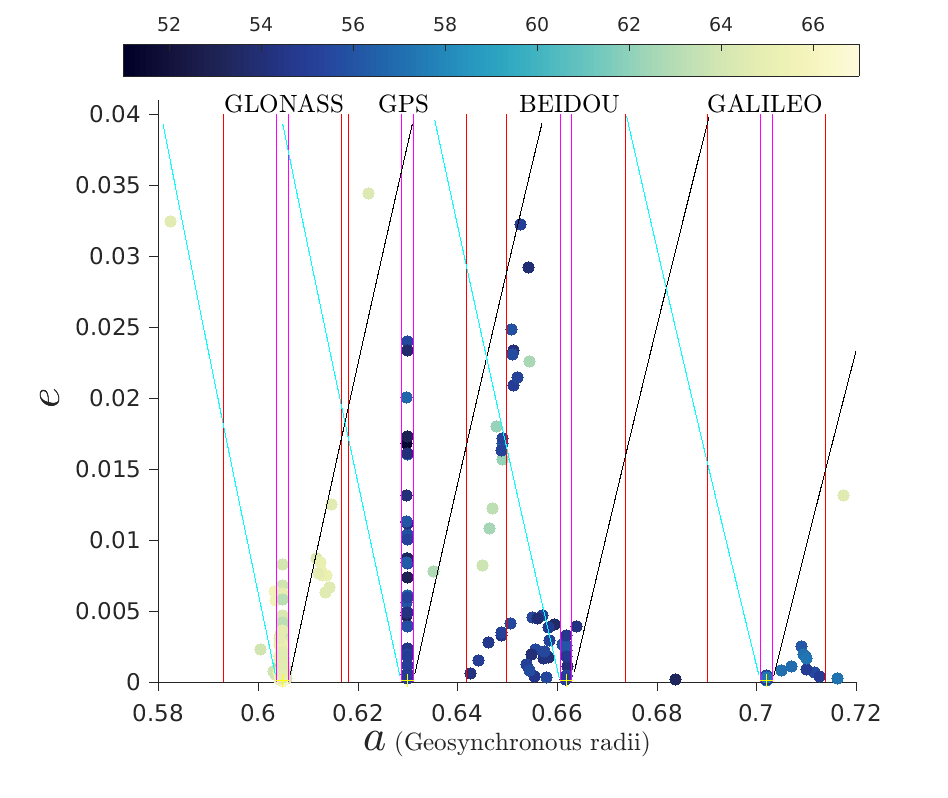}
    \end{subfigure}                             
    \begin{subfigure}[b]{0.4\textwidth}
      \includegraphics[width=\textwidth]{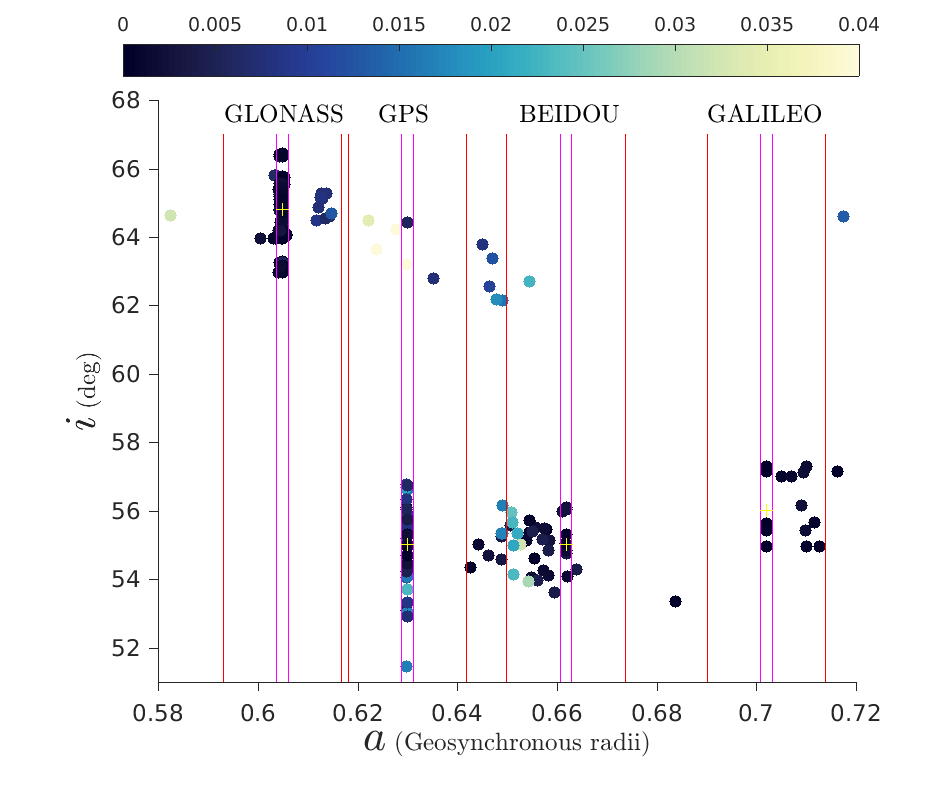}
    \end{subfigure}
    \caption{The cataloged resident space objects in the semi-major axis--eccentricity (left) and semi-major axis--inclination 
    (right) space, where the colorbar corresponds to the ``missing'' of the three elements in each two-dimensional projection (dark blue 
    corresponds to low values, while yellow indicates high values). The top figures correspond to $e\in\left[0,0.9\right]$, whereas 
    the bottom figures focus around the GNSS orbital space. Vertical lines correspond to limits in $a$, $a_{nom}\pm50$~km (magenta) and 
    $a_{nom}\pm500$~km (red), while slanted lines denote limits in apogee/perigee altitude, for each graveyard zone (see definition in the text). 
    (Resident Space Object Catalog, www.space-track.org; assessed 25 Oct.\ 2016)}
    \label{fig:cataloge_objects}
\end{figure}

Figure \ref{fig:cataloge_objects} shows the current population at GNSS altitudes ($a\in\left[0.58:0.72\right]~a_{GEO}$) and 
includes operational satellites and space debris with size larger than $10$~cm, in the $a-e$ (left) and $a-i$ (right) space; the 
colorbar corresponds to the ``missing'' element in each 2-D projection. The top diagrams show the population for 
$e=\left[0:0.9\right]$ and $i=\left[0:90^{\circ}\right]$, while the bottom diagrams focus around the GNSS groups. 
We refer the reader to \citet{dS18} for a recent more detailed study of the long-term dynamics of the GTO and Molniya populations.  \\

In the bottom diagrams of Figure \ref{fig:cataloge_objects}, the population with $a\in\left[0.58:0.72\right]~a_{GEO}$ and 
$e\in\left[0:0.04\right]$ is shown, which consists of $296$ bodies in total. Part of them are GNSS operational satellites and the 
rest are upper-stage launchers and space debris. All bodies have $i\in\left(51^{\circ},67^{\circ}\right)$ and a small value of 
effective area-to-mass ratio, $C_{R}A/m$, where $A$ is the cross-sectional area of the object, $m$ is its mass and $C_{R}$ is the reflectivity 
coefficient. According to the Resident Space Object Catalog\footnote{The Resident Space Object Catalog is provided by 
JSpOC (Joint Space Operations Center), www.space-track.org; assessed at 25/10/2016}, within $i_{nom}\pm2^{\circ}$ and $a_{nom}\pm500$~km, 
where the subscript `nom' hereafter stands for the nominal group value of an element\footnote{See Table \ref{tab:init_cond2} for $a_{nom}$ 
and $i_{nom}$ values.}, there exist 183, 35, 34 and 21 objects in the GLONASS, GPS, BEIDOU and GALILEO group, respectively. The operational 
satellites are within $a_{nom}\pm50$~km.  \\

\subsection{Set-up and dynamical model}
\label{subsec:22}
One of the purposes of this study is to find reentry and graveyard orbits that could be useful in the design of EoL strategies. To 
this end, we perform ``$200$yr-long'' simulations over a large grid of initial conditions, as defined bellow.
We use a dynamical model that accounts for the gravitational potential of Earth up to degree and order 2 (i.e., 
$J_{20}$, $J_{22}$), the Moon and Sun as perturbing bodies, and direct solar radiation pressure (SRP), using the ``cannonball 
model'' \citep{cM99}. We do not include shadowing effects that we assume to be negligible far away from the LEO region. Atmospheric drag plays 
a major role in the evolution of low-altitude satellites. On the other hand, it is negligible in the MEO/GNSS region, as bodies 
with low to moderate eccentricities cannot reach low-enough altitudes\footnote{Of course, atmospheric drag is relevant for GTO and 
Molniya evolution, as expounded on in \citet{dS18}.}, and hence we do not include it in our model.  \\

For our numerical integrations, we use our SWIFT-SAT integrator, which is based on the mixed-variable symplectic integrator of 
\citet{jWmH91}, as included in the SWIFT package of \citet{hLmD94}. SWIFT-SAT uses the full equations of motion and is suitable 
for dynamical studies of bodies with negligible mass, orbiting an oblate central body and perturbed by other massive bodies. In 
addition, SWIFT-SAT is able to incorporate weakly dissipative effects. We refer the reader to 
\citet{aR19} for a more in-depth discussion of SWIFT-SAT and particular validations that were performed to ensure effective 
performance.  \\

In this study, we focus on a region of semi-major axes close to the GNSS. First, we adopt a wide grid of initial conditions 
in $a$, $e$ and $i$ (hereafter denoted as \textit{MEO-general}), as shown in Table~\ref{tab:init_cond1}; it is actually more refined 
than the grid used in the ``LEO-to-GEO'' study presented in \citet{aR19}. As the initial orbit orientation angles ($\Omega$ and $\omega$) 
also affect secular evolution, we chose to study $16$ different configurations\footnote{Note that the satellite's initial node and 
perigee angles were taken relative to the equatorial lunar values at the corresponding epoch.}. Finally, the initial mean anomaly was always 
set to $M=0$.  \\

We are also interested in mapping the graveyard solutions around the nominal GNSS values of $a$, $e$ and $i$. We set 
the nominal eccentricity for all GNSS groups to $e_{nom}=10^{-4}$. We also set the nominal semi-major axis and inclination 
values for each GNSS group as shown in Table~\ref{tab:init_cond2}. We assume the ``protected'' region for each GNSS group to be 
within $\pm 50~$km from $a_{nom}$. Accordingly, we define the graveyard regions to be in the range $a\in\left[a_{nom}-500,a_{nom}-50\right]$~km 
(hereafter denoted as REGION I) and $a\in\left[a_{nom}+50,a_{nom}+500\right]$~km (hereafter denoted as REGION II). In addition to this, 
the perigee/apogee limiting altitudes of an acceptable graveyard solution should be such that they do not cross any neighboring protected 
region. According to this definition, graveyard orbits with $e> 0.018$ do not exist. In Table~\ref{tab:init_cond2}, the limits for $a$, 
perigee (denoted as $q$) and apogee (denoted as $Q$)\footnote{We use 
$a_{GLO},~a_{GPS},~a_{BEI},~a_{GAL}$ to denote $a_{nom}$ values of GLONASS, GPS, BEIDOU, and GALILEO respectively.} are shown and are 
also described graphically in the bottom panels of Fig.~\ref{fig:cataloge_objects}. Hence, an accepted graveyard solution should not violate 
any boundary during the whole 200 years of evolution. The grid of initial conditions used for our \textit{GNSS-graveyard} study follows the above 
definitions, with a mesh of $da=5\cdot10^{-4}a_{GEO}\simeq20$~km and $de=5\cdot10^{-4}$. We studied three values of initial inclination for 
each group, namely $i_{nom}$ and $i_{nom}\pm0.5$, and we repeat the calculations for the same set of $\Omega$, $\omega$, and $M$ values as in 
our \textit{MEO-general} study.  \\

Both for the \textit{MEO-general} study and in the \textit{GNSS-graveyard} study, all computations were performed for two preselected epochs 
(JD 2458475.2433, denoted hereafter as ``Epoch 2018'', and JD 2459021.78, hereafter ``Epoch 2020'', see \citet{aR19}) and for two different 
values of $C_{R}A/m$; a typical one for spent upper stages and space debris (0.015 m$^2$/kg) and an augmented one (1 m$^2$/kg), assumed to 
represent a satellite equipped with a large sail (or, a smaller debris); the augmented $A/m$ value was used only in our \textit{MEO-general} 
study. The time span for integration was $120$~yr (\textit{MEO-general}) and $200$~yr (\textit{GNSS-graveyard}) respectively, the time-step 
of integration was $dt=4\cdot10^{-3}$~sidereal days, and our Earth reentry limit was set to $R_{E}+400$~km. In total, a set of $\sim$ 6 
million initial conditions were propagated.  \\

\begin{table}[htp!]
	\captionsetup{justification=justified}
	\centering
\caption{Grids of initial conditions for the \textit{MEO-general} study using SWIFT-SAT, 
	for dynamical maps in $a - e$ phase space, as a function of $i$.}
\label{tab:init_cond1}
 \begin{tabular}{ll}
\hline\noalign{\smallskip}
 $a$ $(a_{GEO})$ & $0.600 - 0.710$\\
      $\Delta a$ & $0.0025$\\ 
             $e$ & $0 - 0.88$\\ 
      $\Delta e$ & $0.02$\\ 
$i$ $(^\circ)$ &  $0 - 90$\\
$\Delta i$ $(^\circ)$ & $2$\\
$\Delta \Omega$ ($^\circ$) & $\left\{ 0, 90, 180, 270 \right\}$\\
$\Delta \omega$ ($^\circ$) & $\left\{ 0, 90, 180, 270 \right\}$\\ 
$C_R (A/m)$ (m$^2$/kg) & $\left\{ 0.015, 1 \right\}$\\
\hline\noalign{\smallskip}
 \end{tabular}
\end{table}

\begin{table}[htp!]
	\captionsetup{justification=justified}
	\centering
\caption{Grids of initial conditions for the \textit{GNSS-graveyard} study using SWIFT-SAT, 
	for dynamical maps in $a - e$ phase space, as a function of $i$. 
	Values of $\Omega$, and $\omega$ were set as shown in Table \ref{tab:init_cond1}, and $C_R (A/m)=0.015$ m$^2$/kg.}
\label{tab:init_cond2}
 \begin{tabular}{lcc}
\hline\noalign{\smallskip}
  & \textbf{GLONASS} & \textbf{GPS}\\
 $a_{nom}$ $\left(\right.$km~$\left.\left|~a_{GEO}\right)\right.$ & $25509.64~|~0.605$ & $26561.18~|~0.630$\\
 $i_{nom}$ $(^\circ)$ &  $64.8$ & $55$ \vspace{0.2cm}\\
 \multirow{3}{*}{REGION I} & $a\in\left[a_{GLO}-500,a_{GLO}-50\right]$~km & $a\in\left[a_{GPS}-500,a_{GPS}-50\right]$~km\\
 & $q>0$  & $q\ge a_{GLO}+50$km\\
 & $Q\le a_{GLO}-50$km  & $Q\le a_{GPS}-50$km  \vspace{0.2cm}\\
 \multirow{3}{*}{REGION II} &  $a\in\left[a_{GLO}+50,a_{GLO}+500\right]$~km & $a\in\left[a_{GPS}+50,a_{GPS}+500\right]$~km\\  
 & $q\ge a_{GLO}+50$km & $q\ge a_{GPS}+50$km\\
 & $Q\le a_{GPS}-50$km & $Q\le a_{BEI}-50$km \vspace{0.3cm}\\
 \hline\noalign{\smallskip}
   &  \textbf{BEIDOU}  & \textbf{GALILEO}\\
 $a_{nom}$ $(km~|~a_{GEO})$ &  $27906.14~|~0.662$ & $29601.31~|~0.702$\\
 $i_{nom}$ $(^\circ)$ &  $55$ & $56$ \vspace{0.2cm}\\
  \multirow{3}{*}{REGION I}  & $a\in\left[a_{BEI}-500,a_{BEI}-50\right]$~km & $a\in\left[a_{GAL}-500,a_{GAL}-50\right]$~km\\
 & $q\ge a_{GPS}+50$km & $q\ge a_{BEI}+50$km\\
 & $Q\le a_{BEI}-50$km  & $Q\le a_{GAL}-50$km  \vspace{0.2cm}\\
 \multirow{3}{*}{REGION II} &  $a\in\left[a_{BEI}+50,a_{BEI}+500\right]$~km & $a\in\left[a_{GAL}+50,a_{GAL}+500\right]$~km\\  
 &  $q\ge a_{BEI}+50$km & $q\ge a_{GAL}+50$km\\
 &  $Q\le a_{GAL}-50$km & $Q\le a_{GAL}+2500$km \vspace{0.3cm}\\
\hline\noalign{\smallskip}
 \end{tabular}
\end{table}

%-------------------------------------------------------------------------------------------------------------------------------------- 
%          Dynamical Atlas
% -------------------------------------------------------------------------------------------------------------------------------------- 
\section{Dynamical Atlas}
\label{sec:3}

\subsection{MEO-general study}
\label{subsec:MEO}

The results of our numerical simulations performed for the MEO region are presented here in the form of dynamical maps. In each 2-D 
map, the initial grid in $\left(a,e\right)$ for a specific initial inclination, $i$, a given set of orientation angles, epoch and 
$C_{R}A/m$ value is presented. Color-coded is the dynamical lifetime of a trajectory (i.e.,\ the time until it reaches our reentry limit in 
$q$) or the eccentricity indicator $De = (e_{max}-e_0)/(e_c-e_0)$. This quantity has the property of varying between 0 and 1 (in the 
region $e<e_c$) and is a direct measure of the eccentricity variation offered by the dynamics, relative what is needed to achieve atmospheric 
reentry \citep{iGetal16}. Hence, $De\rightarrow 1$ means that an initial eccentricity, $e_{0}$, grows to a maximum value, $e_{max}$, that 
is greater or equal to the critical value for reentry, $e_c$.  \\

\begin{figure}[htp!]
  \centering
     \begin{subfigure}[b]{0.67\textwidth}
      \caption{$\bm{\Delta}\bm{\Omega} = {\bf 0}$, $\bm{\Delta}\bm{\omega} = {\bf 0}$}
      \includegraphics[width=.49\textwidth]{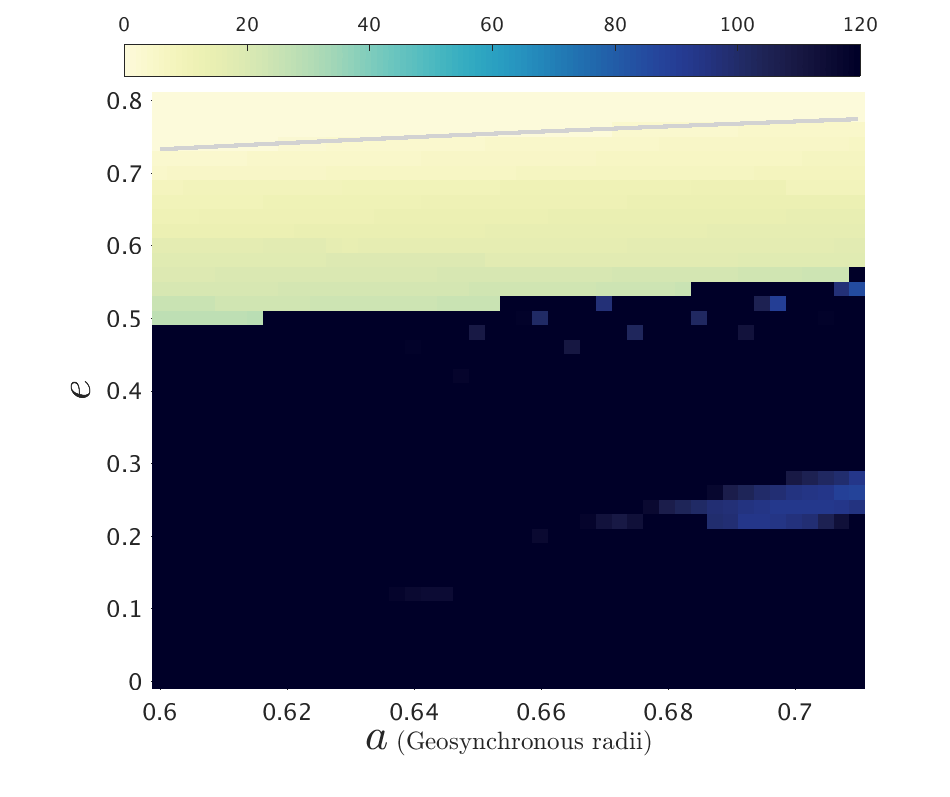} 
      \includegraphics[width=.49\textwidth]{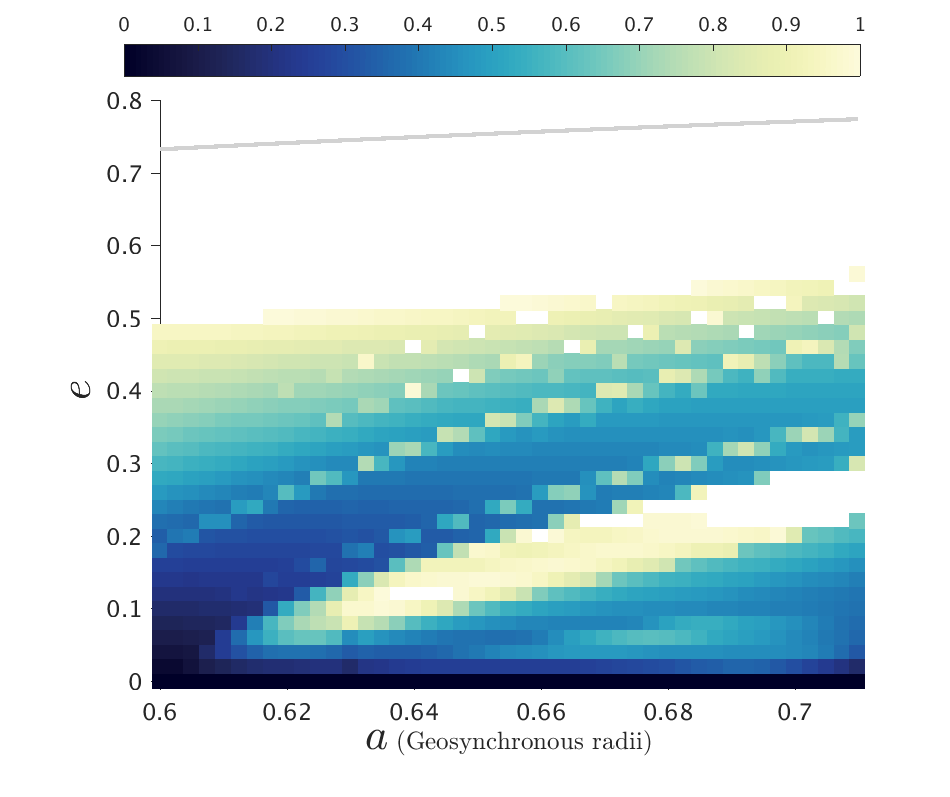}
    \end{subfigure}  
    \begin{subfigure}[b]{0.67\textwidth}
      \caption{$\bm{\Delta}\bm{\Omega} = {\bf 0}$, $\bm{\Delta}\bm{\omega} = {\bf 90^\circ}$}
      \includegraphics[width=.49\textwidth]{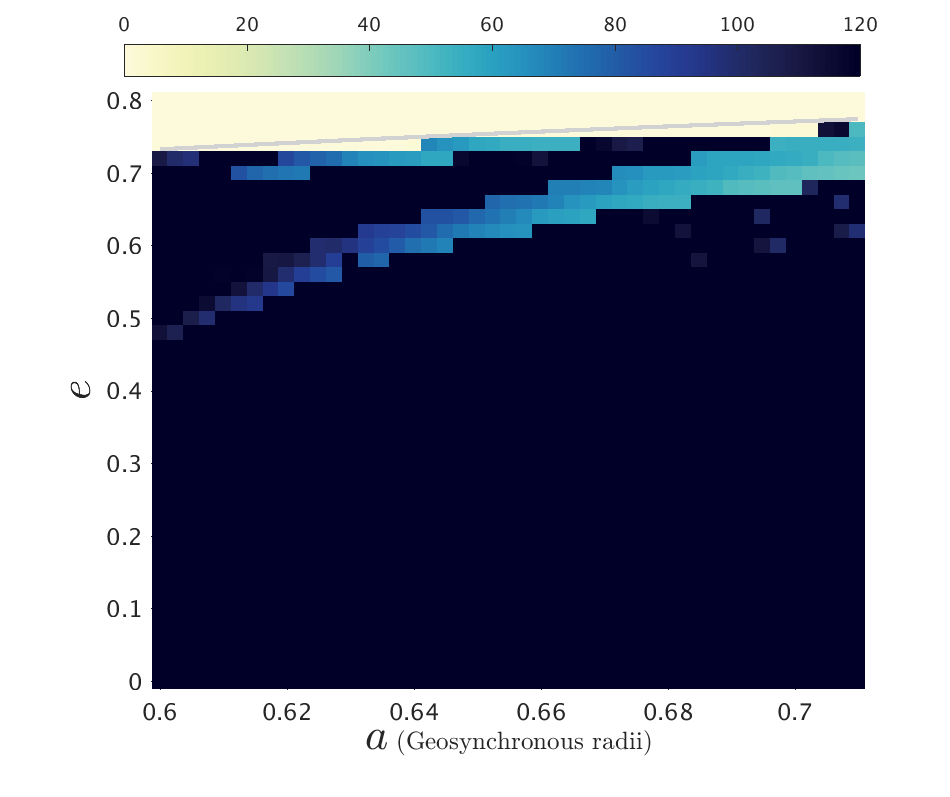}
      \includegraphics[width=.49\textwidth]{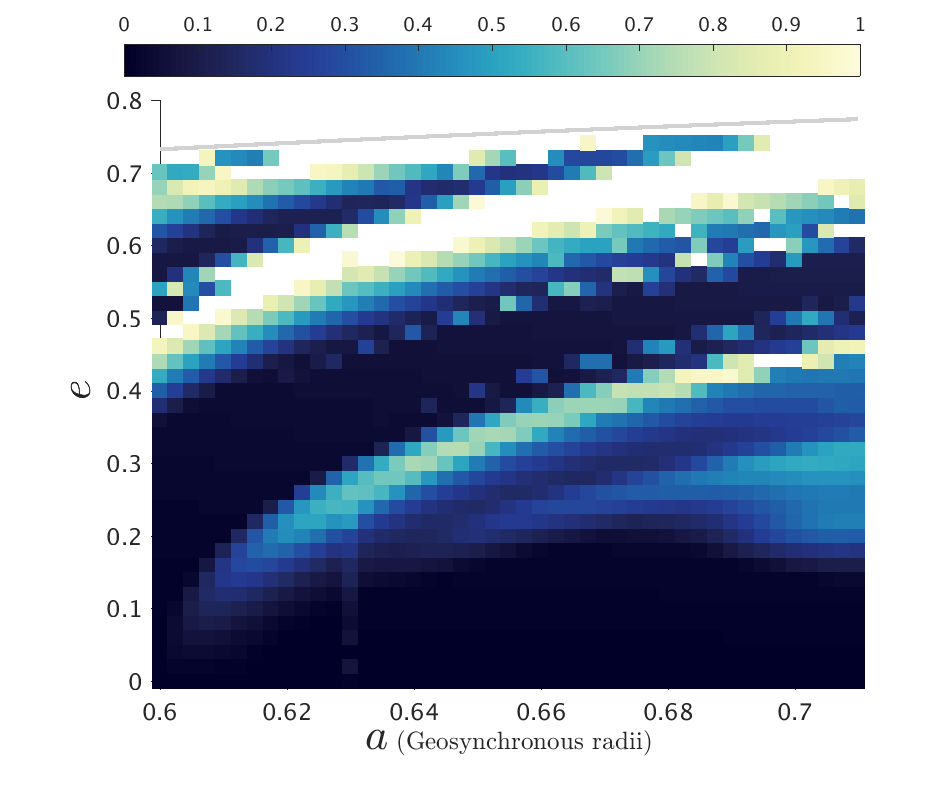}
    \end{subfigure} 
    \begin{subfigure}[b]{0.67\textwidth}
      \caption{$\bm{\Delta}\bm{\Omega} = {\bf 90^\circ}$, $\bm{\Delta}\bm{\omega} = {\bf 0}$}
      \includegraphics[width=.49\textwidth]{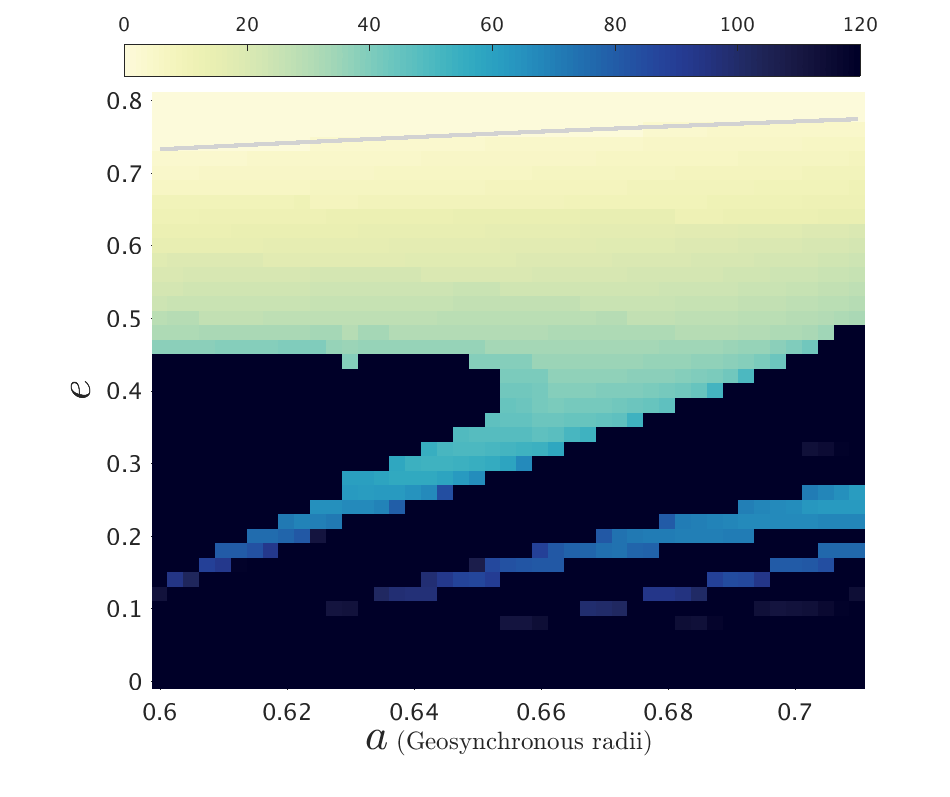}
      \includegraphics[width=.49\textwidth]{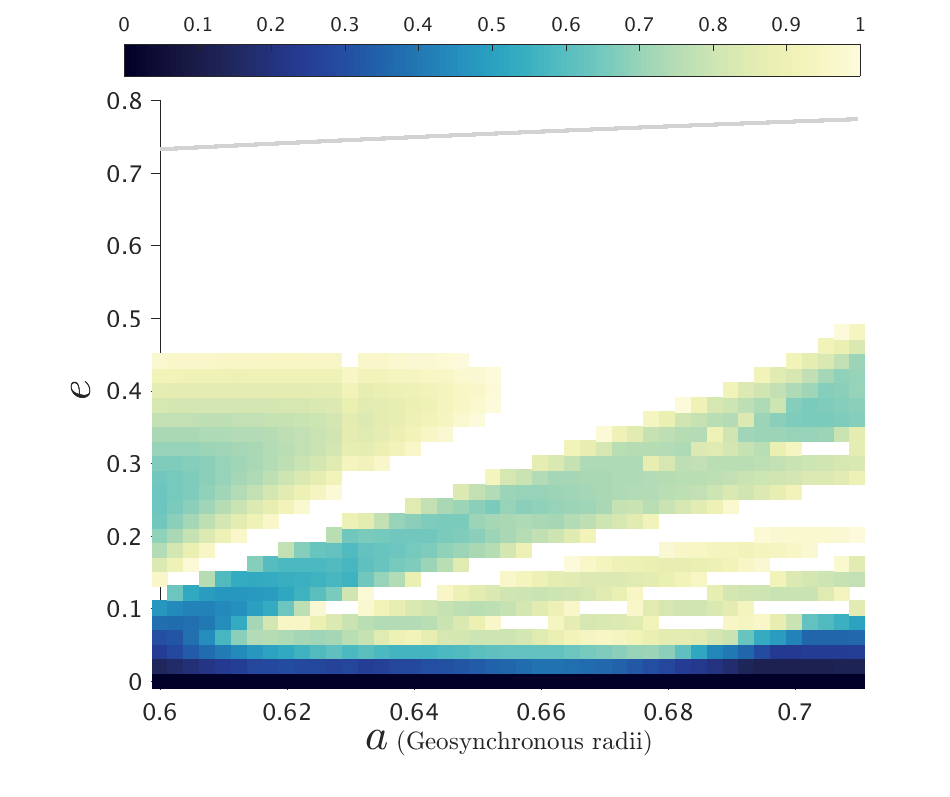}
    \end{subfigure}  
    \begin{subfigure}[b]{0.67\textwidth}
      \caption{$\bm{\Delta}\bm{\Omega} = {\bf 90^\circ}$, $\bm{\Delta}\bm{\omega} = {\bf 90^\circ}$}
      \includegraphics[width=.49\textwidth]{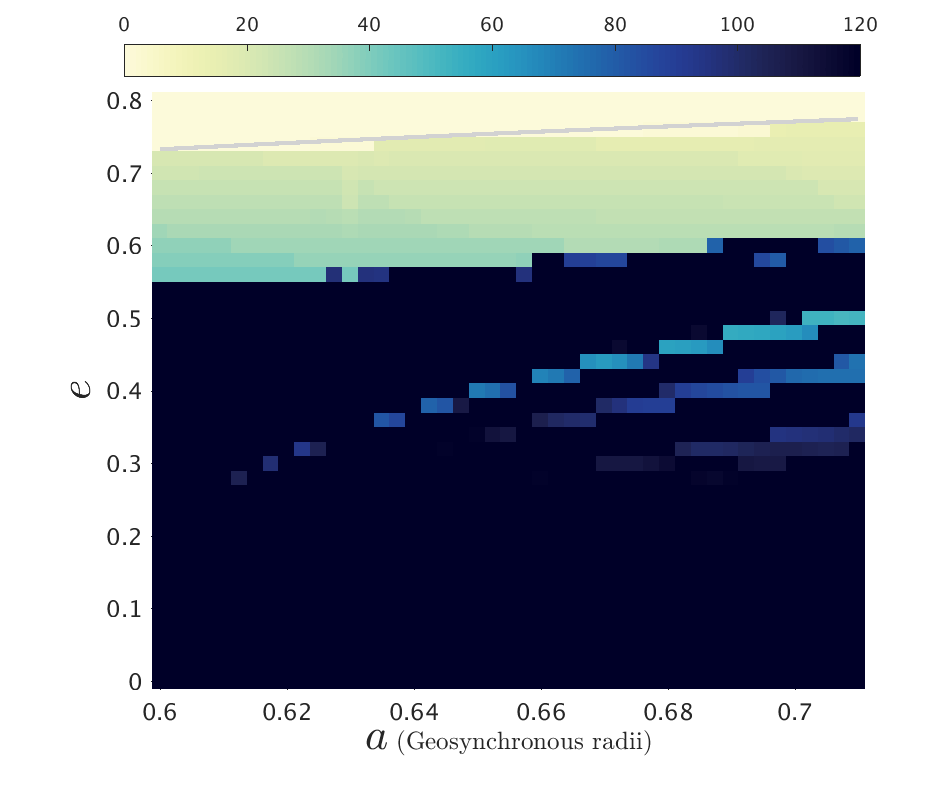} 
      \includegraphics[width=.49\textwidth]{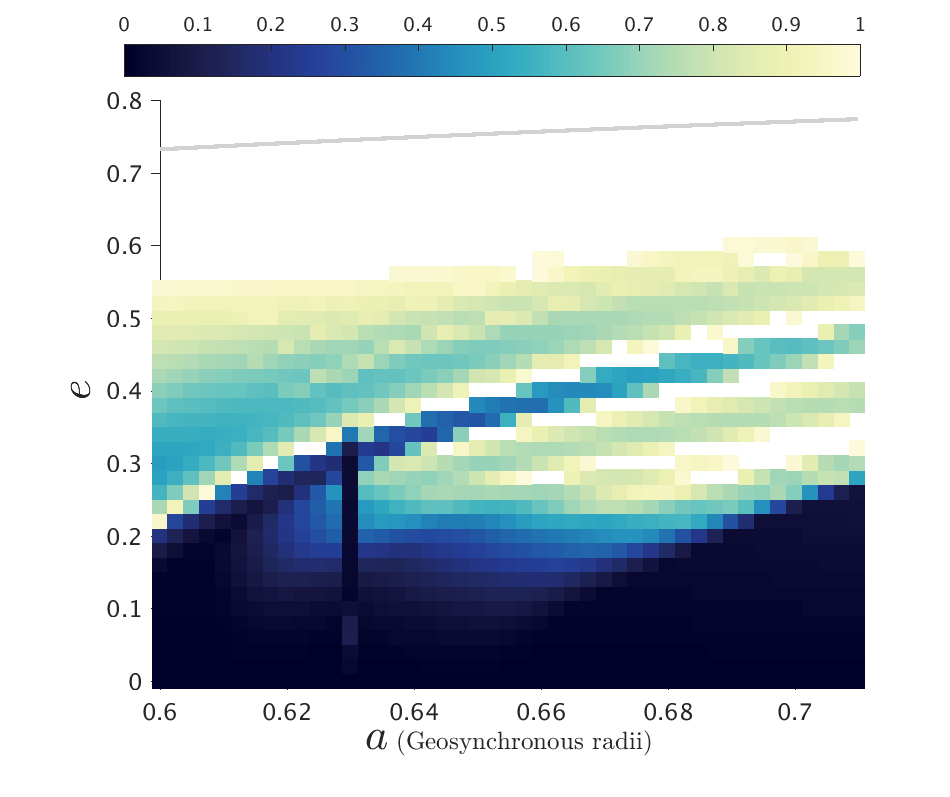}
    \end{subfigure}     
  \caption{Lifetime and $De$ maps of the \textit{MEO-general} phase space for $\bm{i_{o}} = {\bf 56^\circ}$,  
  for Epoch 2018, and for $C_{R}A/m=0.015$ m$^2$/kg.
  The colorbar for the lifetime maps is from 0 to 120 years and 
  that of the $De$ maps is from 0 to 1.}
  \label{fig:MEO_inc56_srp1a}
\end{figure}

\begin{figure}[htp!]
  \centering
    \begin{subfigure}[b]{0.67\textwidth}
      \caption{$\bm{\Delta}\bm{\Omega} = {\bf 180^\circ}$, $\bm{\Delta}\bm{\omega} = {\bf 0}$}
      \includegraphics[width=.49\textwidth]{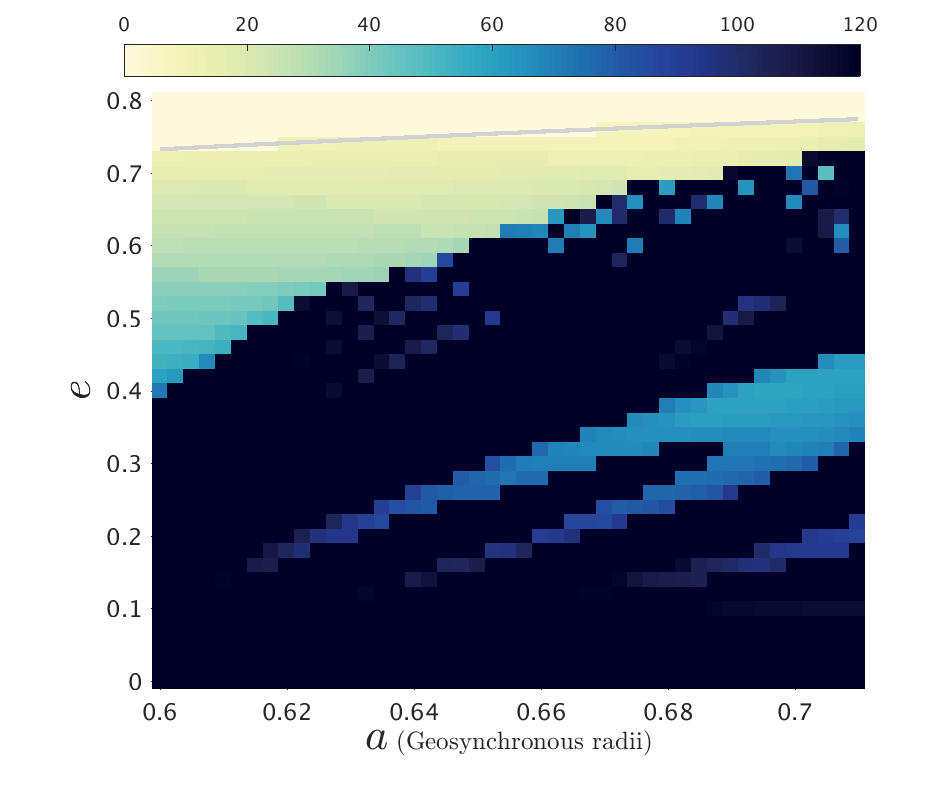}
      \includegraphics[width=.49\textwidth]{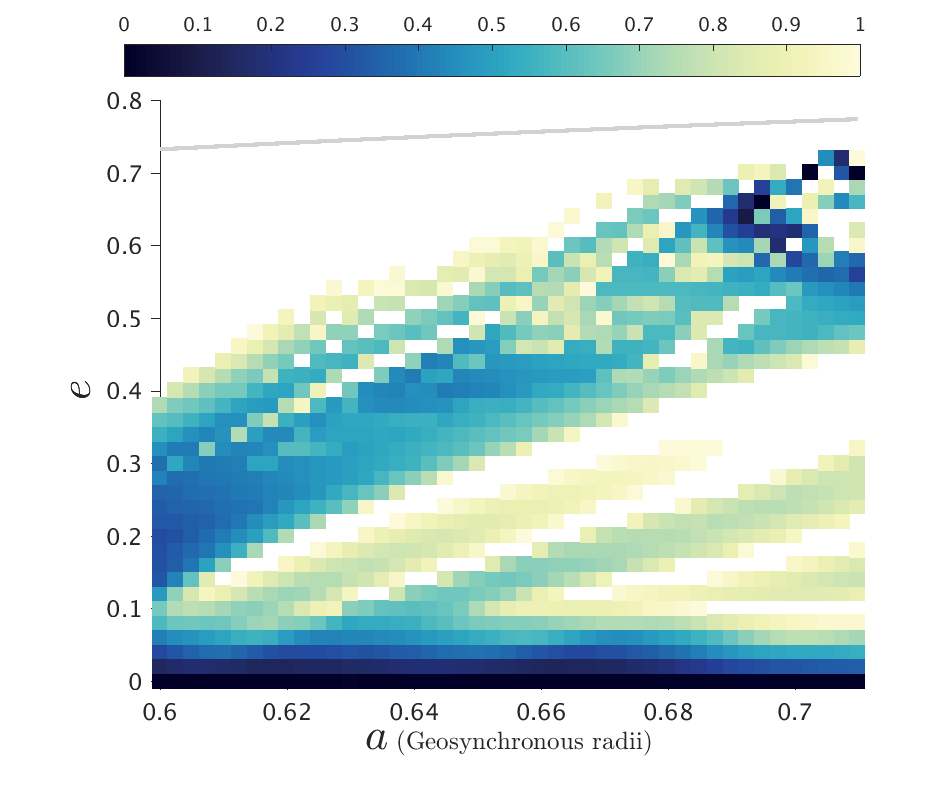}
    \end{subfigure}     
    \begin{subfigure}[b]{0.67\textwidth}
      \caption{$\bm{\Delta}\bm{\Omega} = {\bf 180^\circ}$, $\bm{\Delta}\bm{\omega} = {\bf 90^\circ}$}
      \includegraphics[width=.49\textwidth]{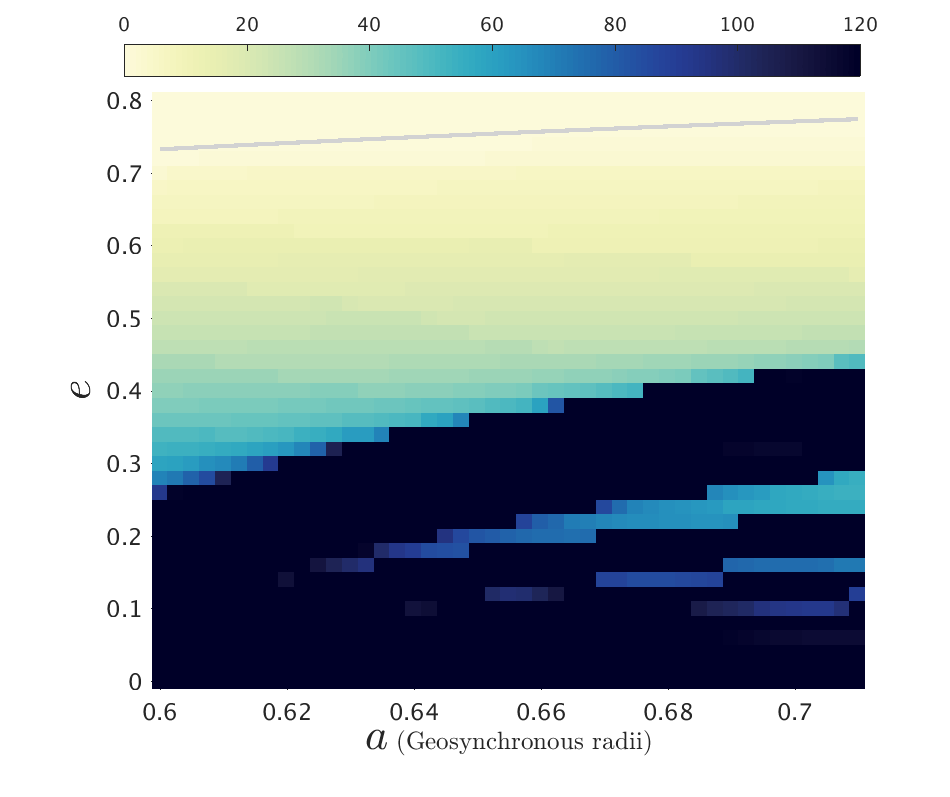}
      \includegraphics[width=.49\textwidth]{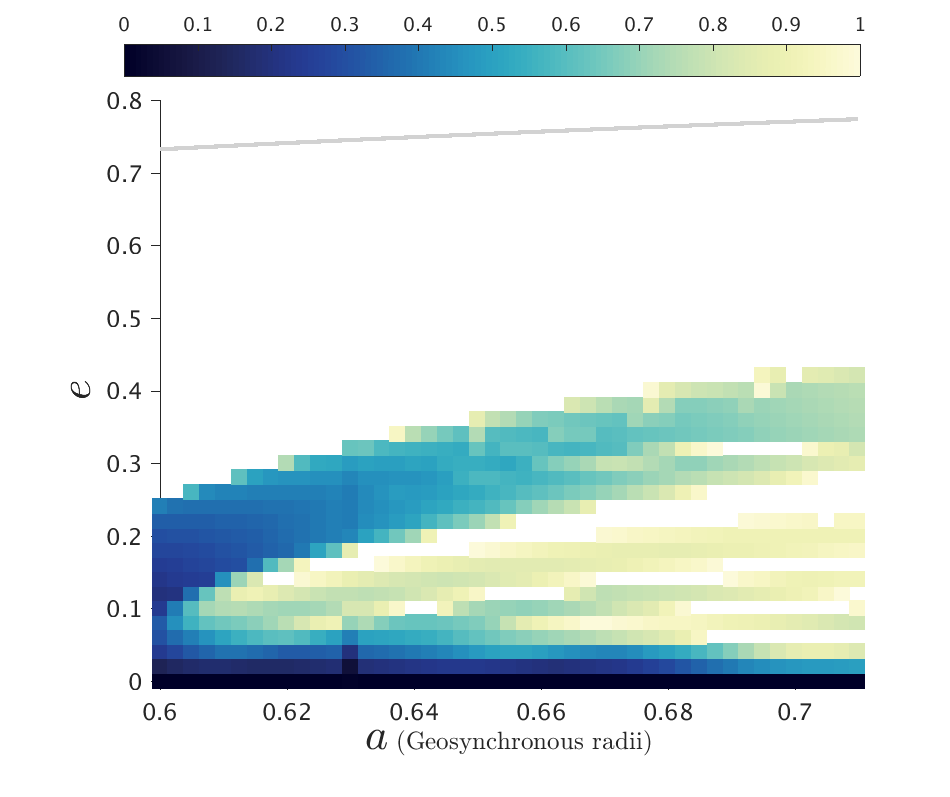}
    \end{subfigure}
    \begin{subfigure}[b]{0.67\textwidth}
      \caption{$\bm{\Delta}\bm{\Omega} = {\bf 270^\circ}$, $\bm{\Delta}\bm{\omega} = {\bf 0}$}
      \includegraphics[width=.49\textwidth]{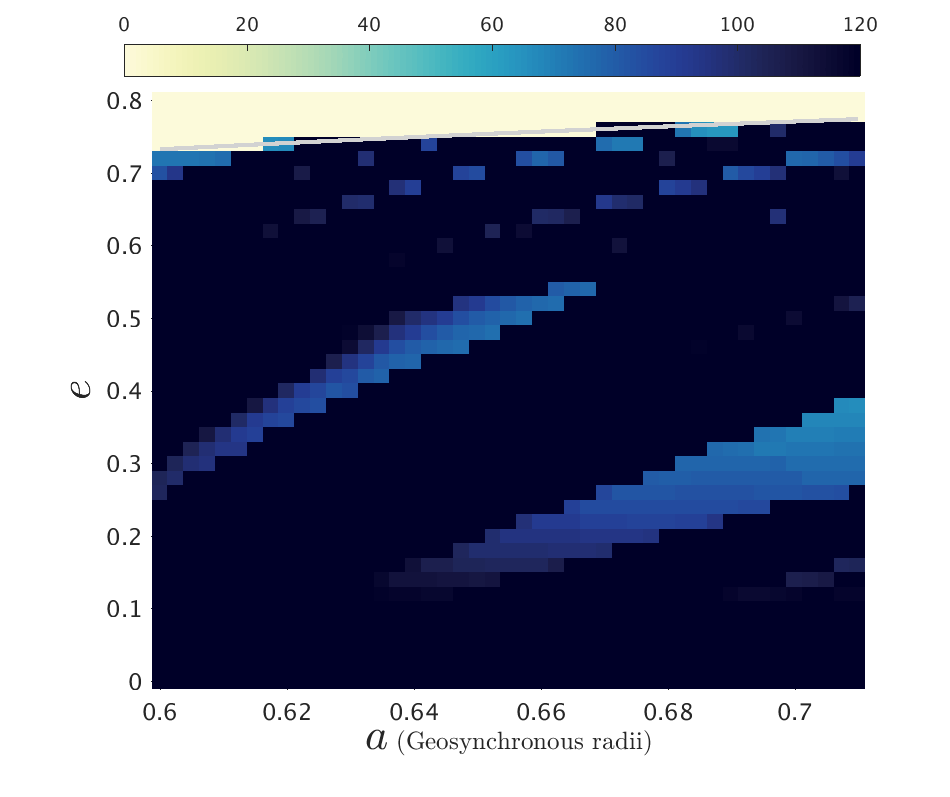}
      \includegraphics[width=.49\textwidth]{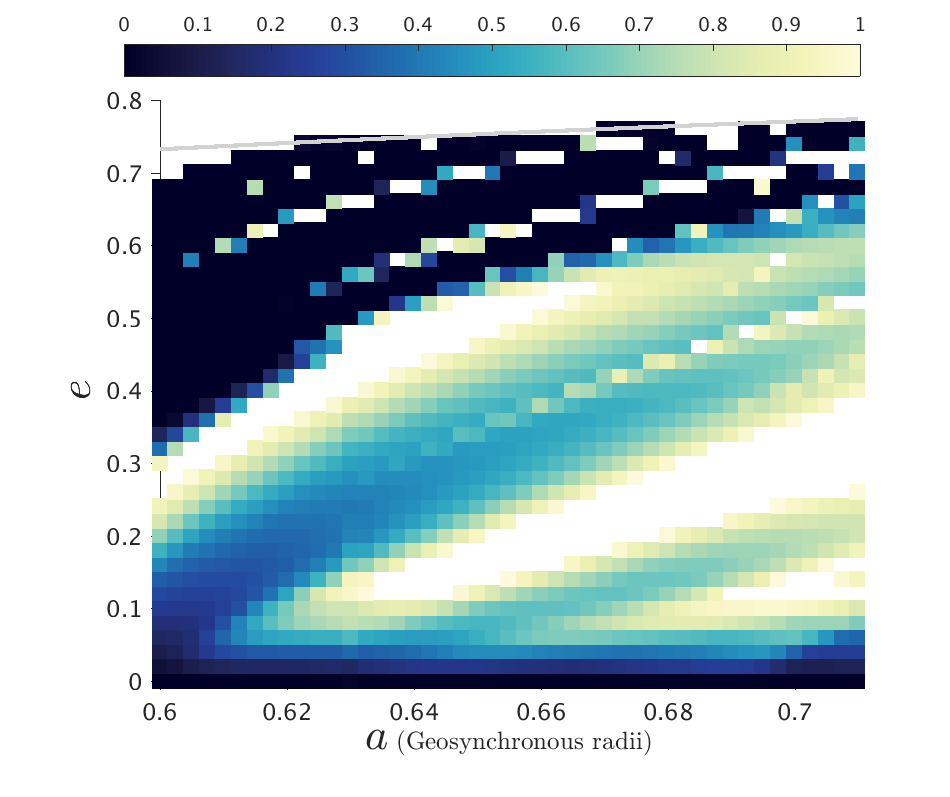}
    \end{subfigure} 
    \begin{subfigure}[b]{0.67\textwidth}
      \caption{$\bm{\Delta}\bm{\Omega} = {\bf 270^\circ}$, $\bm{\Delta}\bm{\omega} = {\bf 90^\circ}$}
      \includegraphics[width=.49\textwidth]{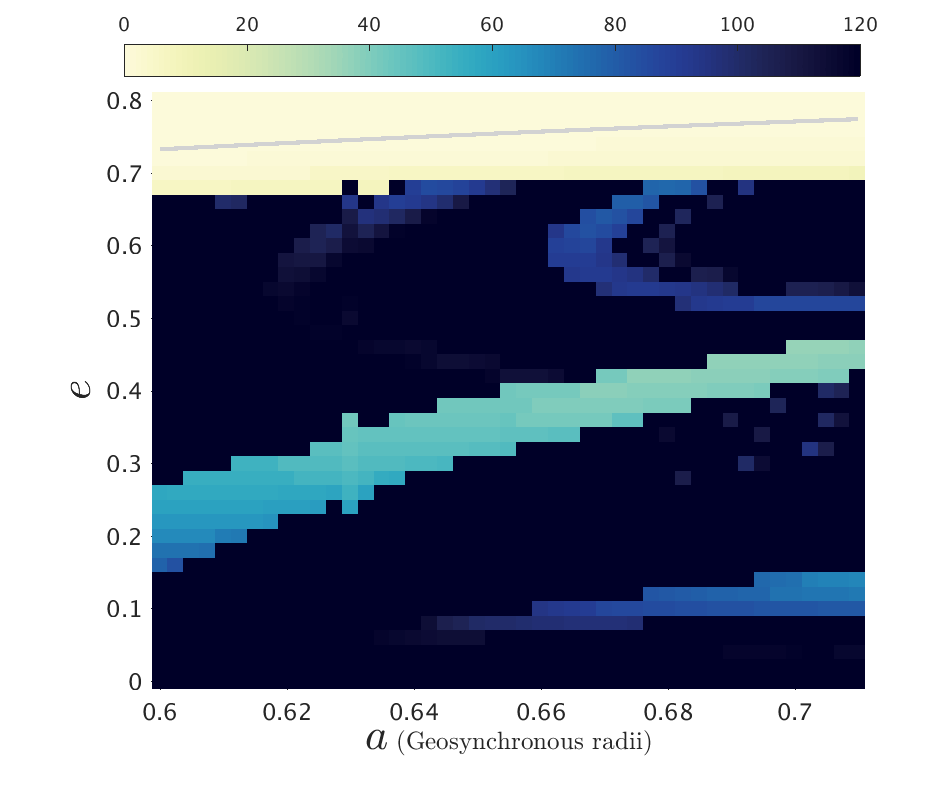}
      \includegraphics[width=.49\textwidth]{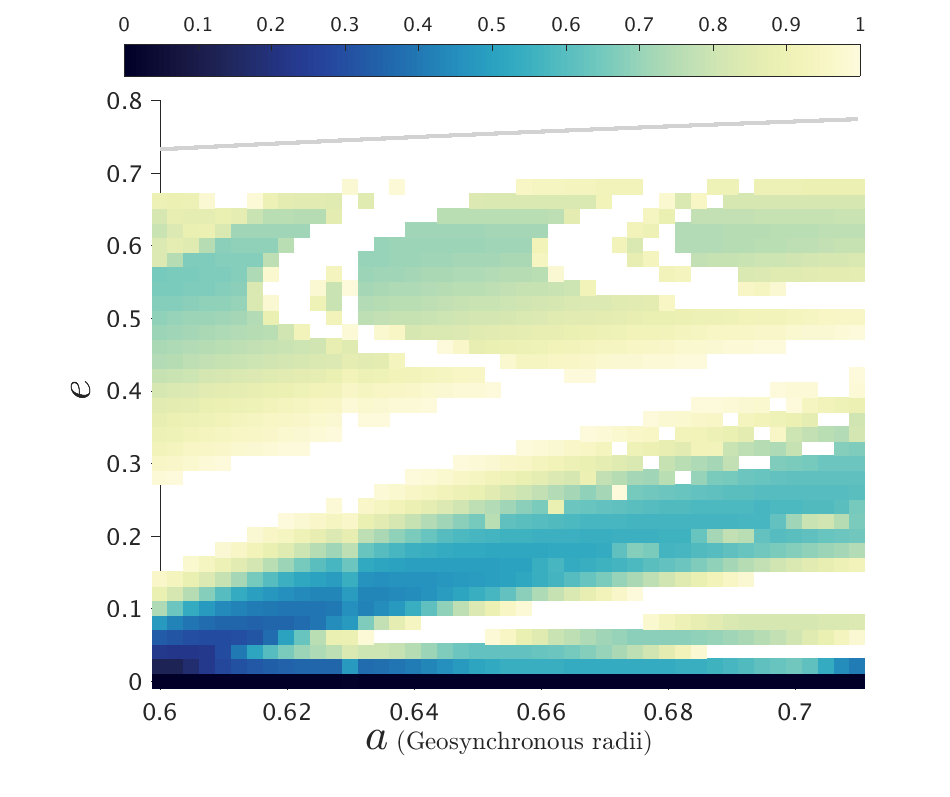}
    \end{subfigure}    
  \caption{Lifetime and $De$ maps of the \textit{MEO-general} phase space for $\bm{i_{o}} = {\bf 56^\circ}$,  
  for Epoch 2018, and for $C_{R}A/m=0.015$ m$^2$/kg.
  The colorbar for the lifetime maps is from 0 to 120 years and 
  that of the $De$ maps is from 0 to 1.}
  \label{fig:MEO_inc56_srp1b}
\end{figure}

\begin{figure}[htp!]
  \centering
    \begin{subfigure}[b]{0.67\textwidth}
      \caption{$\bm{\Delta}\bm{\Omega} = {\bf 0}$, $\bm{\Delta}\bm{\omega} = {\bf 0}$}
      \includegraphics[width=.49\textwidth]{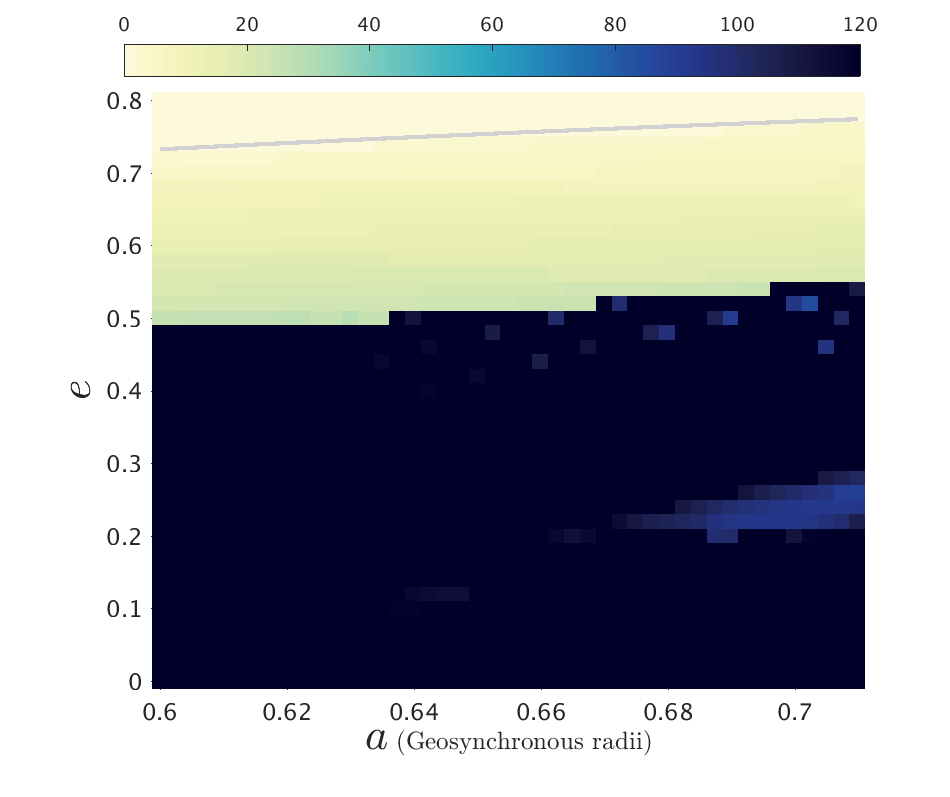} 
      \includegraphics[width=.49\textwidth]{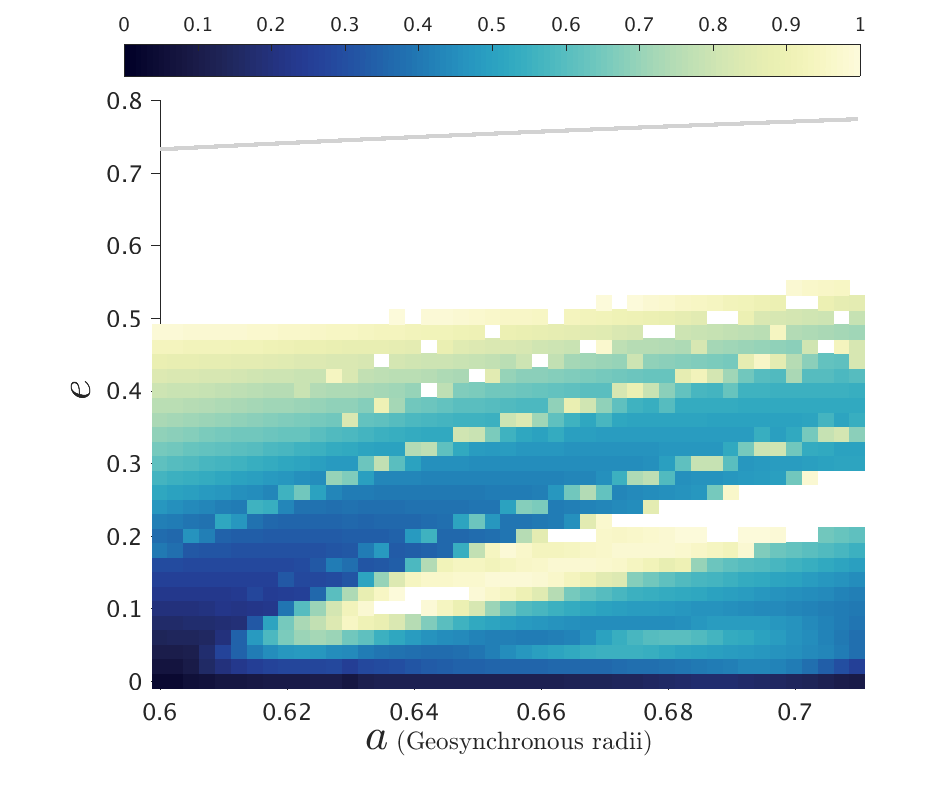}
    \end{subfigure}  
    \begin{subfigure}[b]{0.67\textwidth}
      \caption{$\bm{\Delta}\bm{\Omega} = {\bf 0}$, $\bm{\Delta}\bm{\omega} = {\bf 90^\circ}$}
      \includegraphics[width=.49\textwidth]{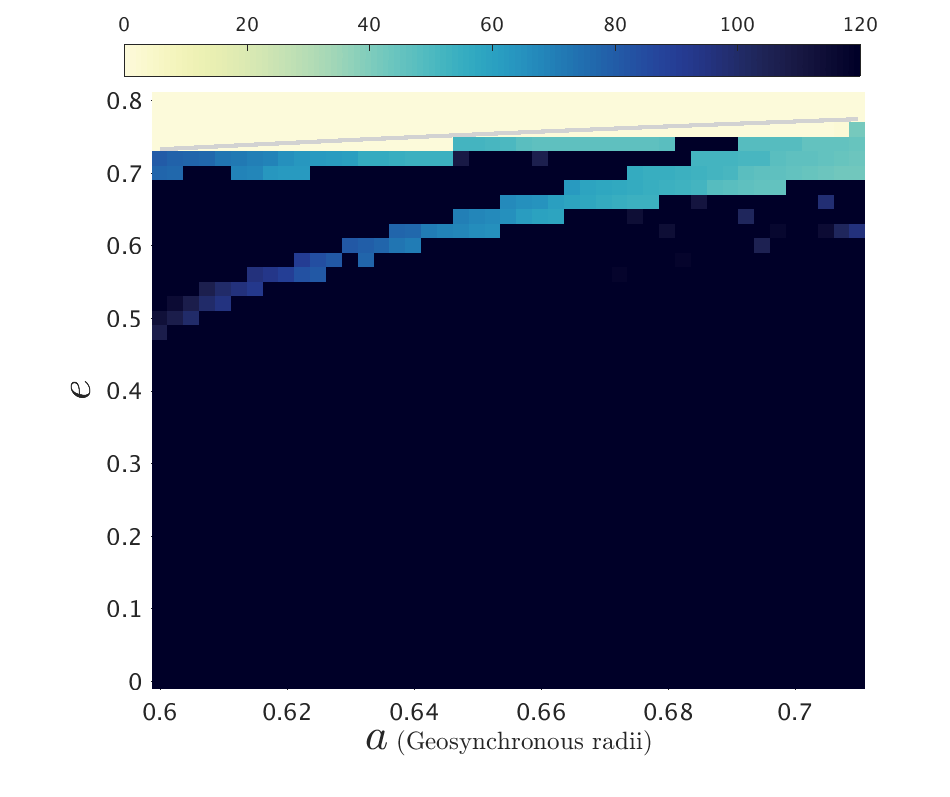}
      \includegraphics[width=.49\textwidth]{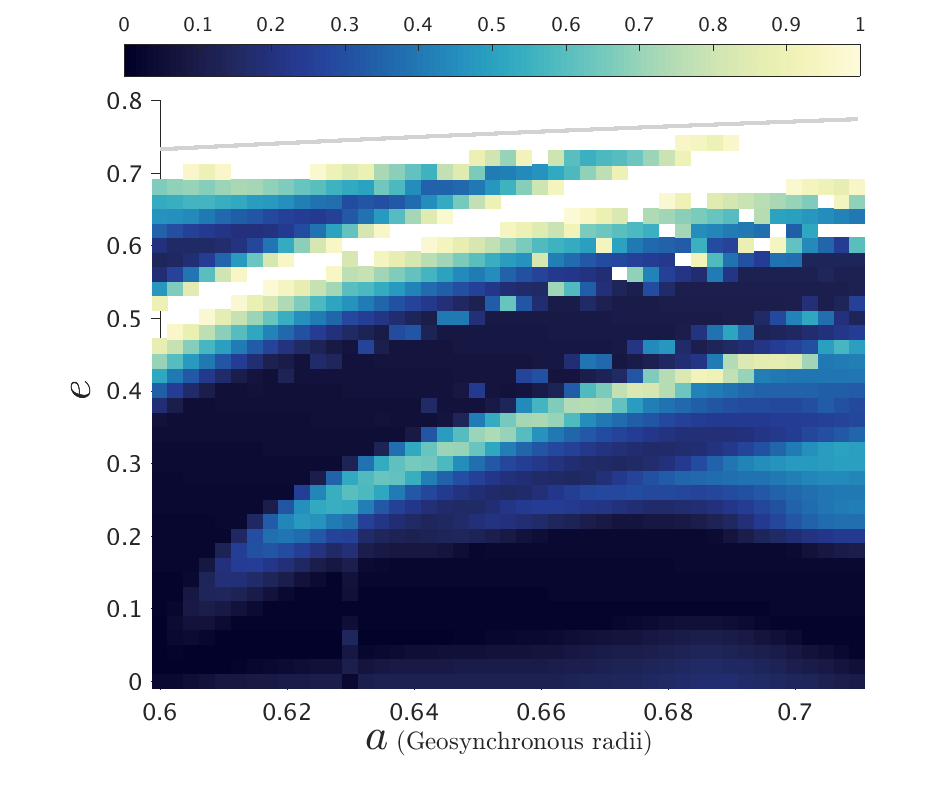}
    \end{subfigure} 
    \begin{subfigure}[b]{0.67\textwidth}
      \caption{$\bm{\Delta}\bm{\Omega} = {\bf 90^\circ}$, $\bm{\Delta}\bm{\omega} = {\bf 0}$}
      \includegraphics[width=.49\textwidth]{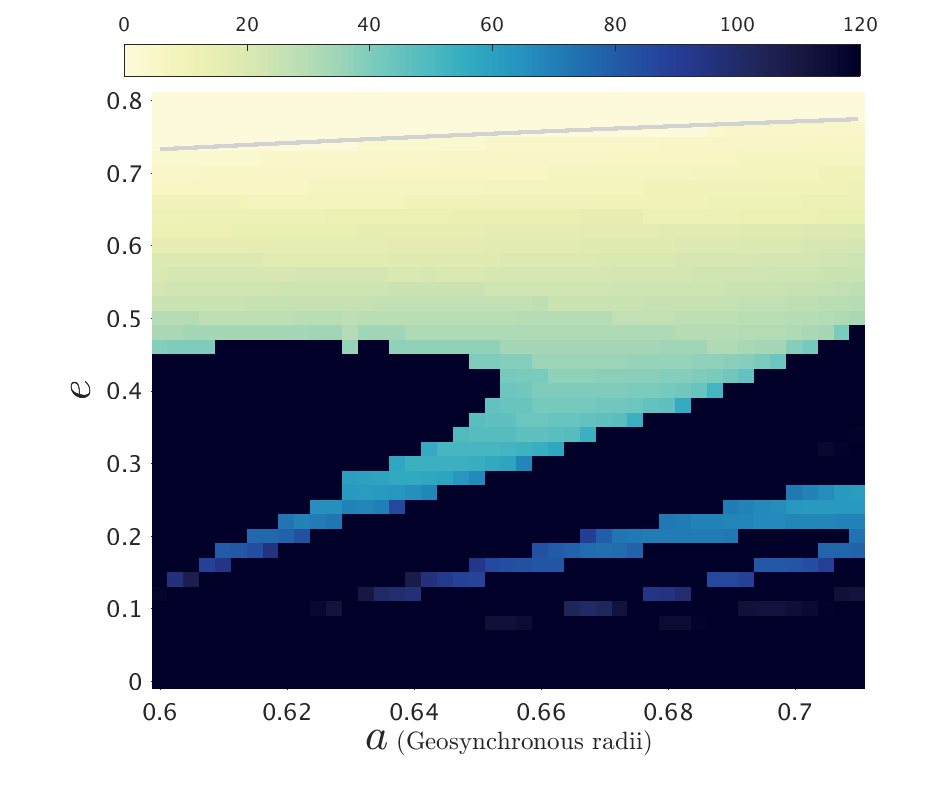}
      \includegraphics[width=.49\textwidth]{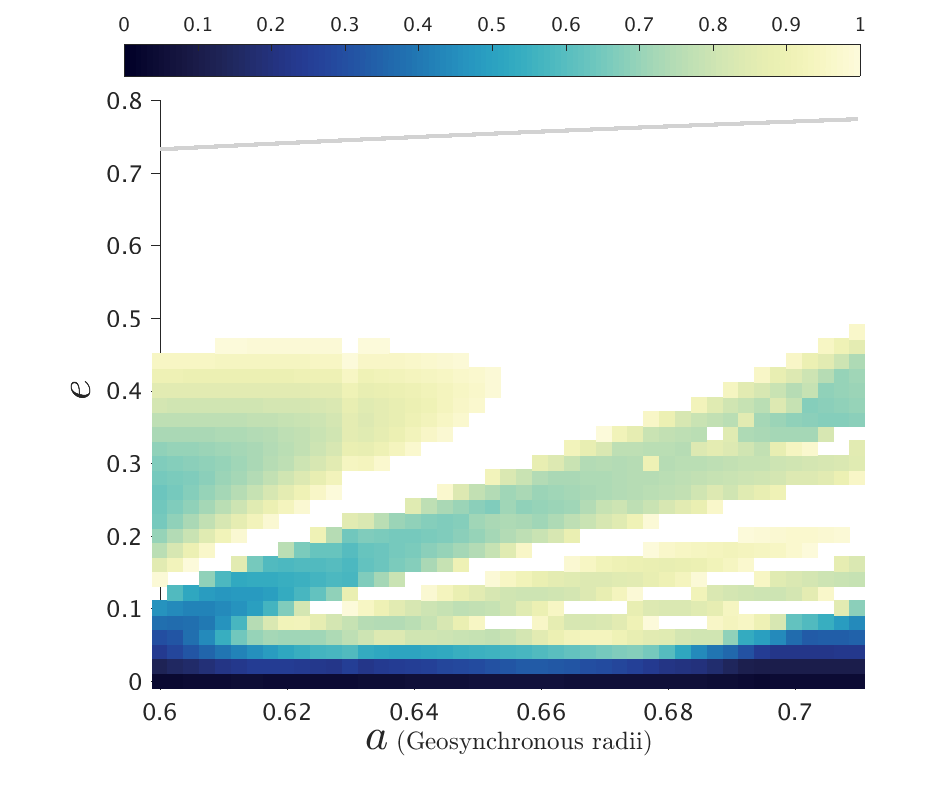}
    \end{subfigure}  
    \begin{subfigure}[b]{0.67\textwidth}
      \caption{$\bm{\Delta}\bm{\Omega} = {\bf 90^\circ}$, $\bm{\Delta}\bm{\omega} = {\bf 90^\circ}$}
      \includegraphics[width=.49\textwidth]{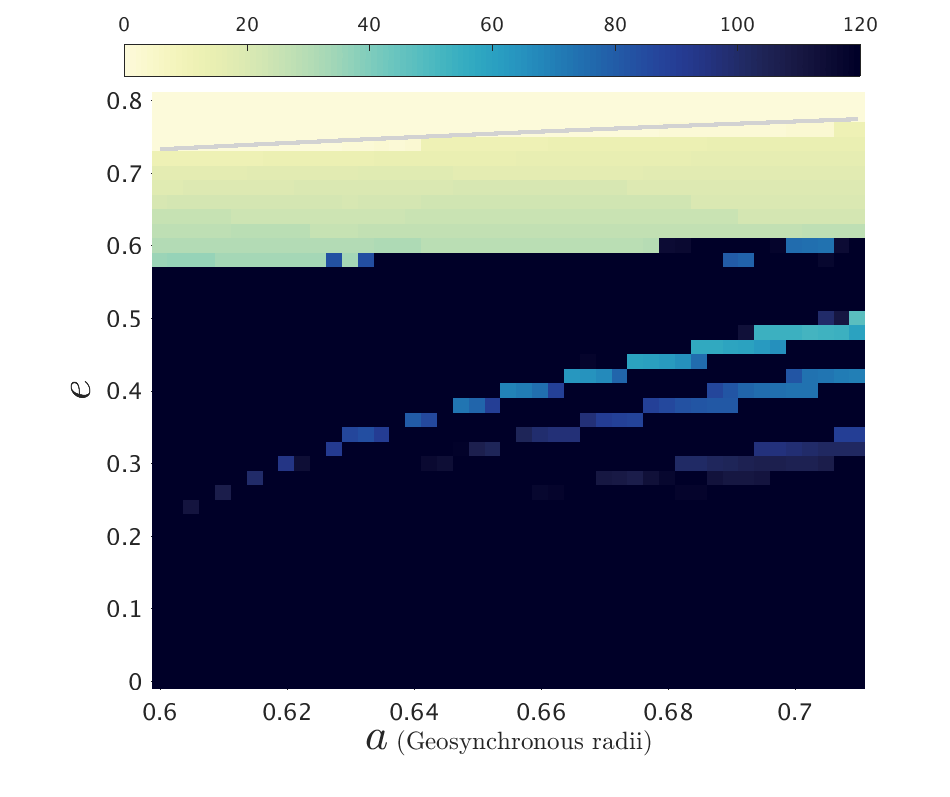} 
      \includegraphics[width=.49\textwidth]{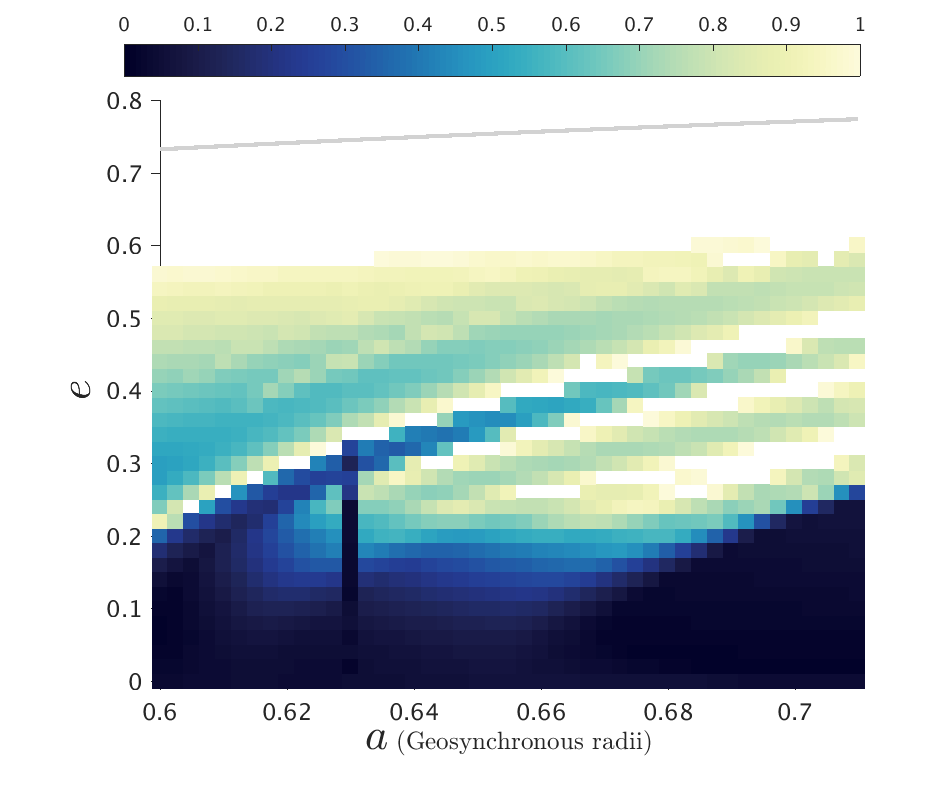}
    \end{subfigure}  
  \caption{Lifetime and $De$ maps of the \textit{MEO-general} phase space for $\bm{i_{o}} = {\bf 56^\circ}$,  
  for Epoch 2018, and for $C_{R}A/m=1$ m$^2$/kg.
  The colorbar for the lifetime maps is from 0 to 120 years and 
  that of the $De$ maps is from 0 to 1.}
  \label{fig:MEO_inc56_srp2a}
\end{figure}

\begin{figure}[htp!]
  \centering  
    \begin{subfigure}[b]{0.67\textwidth}
      \caption{$\bm{\Delta}\bm{\Omega} = {\bf 180^\circ}$, $\bm{\Delta}\bm{\omega} = {\bf 0}$}
      \includegraphics[width=.49\textwidth]{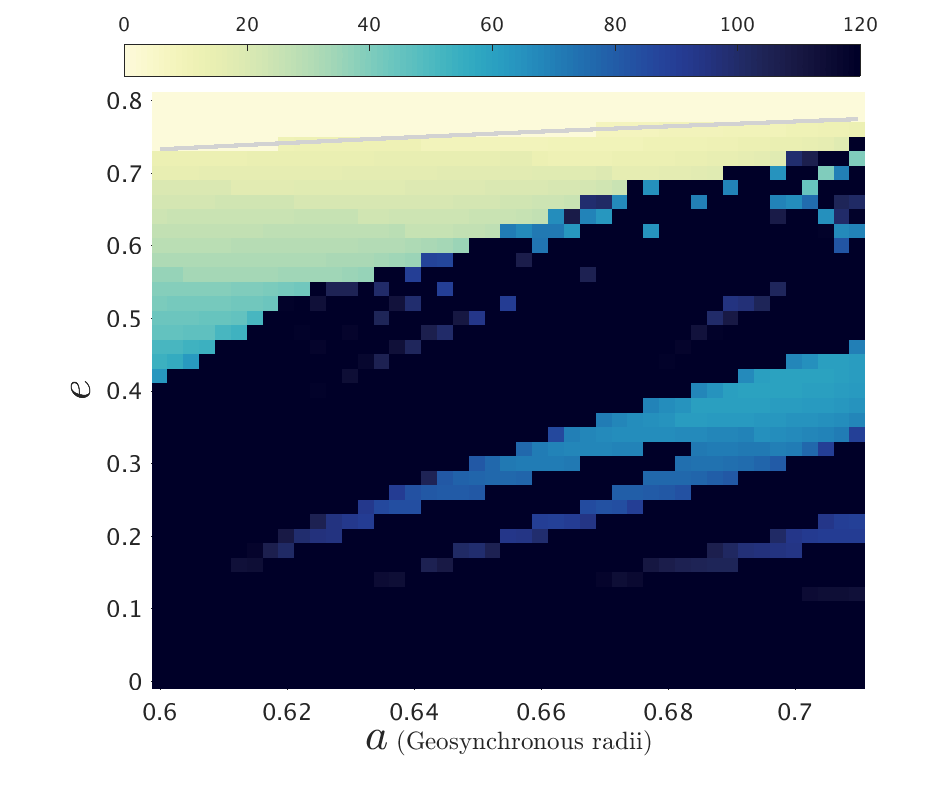}
      \includegraphics[width=.49\textwidth]{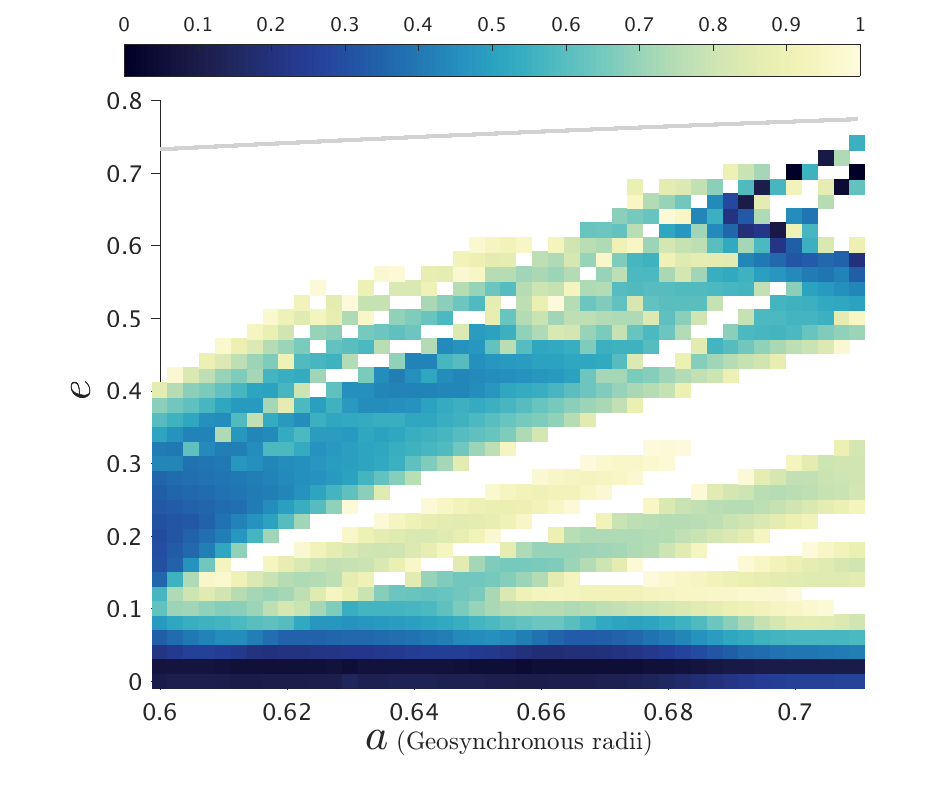}
    \end{subfigure}     
    \begin{subfigure}[b]{0.67\textwidth}
      \caption{$\bm{\Delta}\bm{\Omega} = {\bf 180^\circ}$, $\bm{\Delta}\bm{\omega} = {\bf 90^\circ}$}
      \includegraphics[width=.49\textwidth]{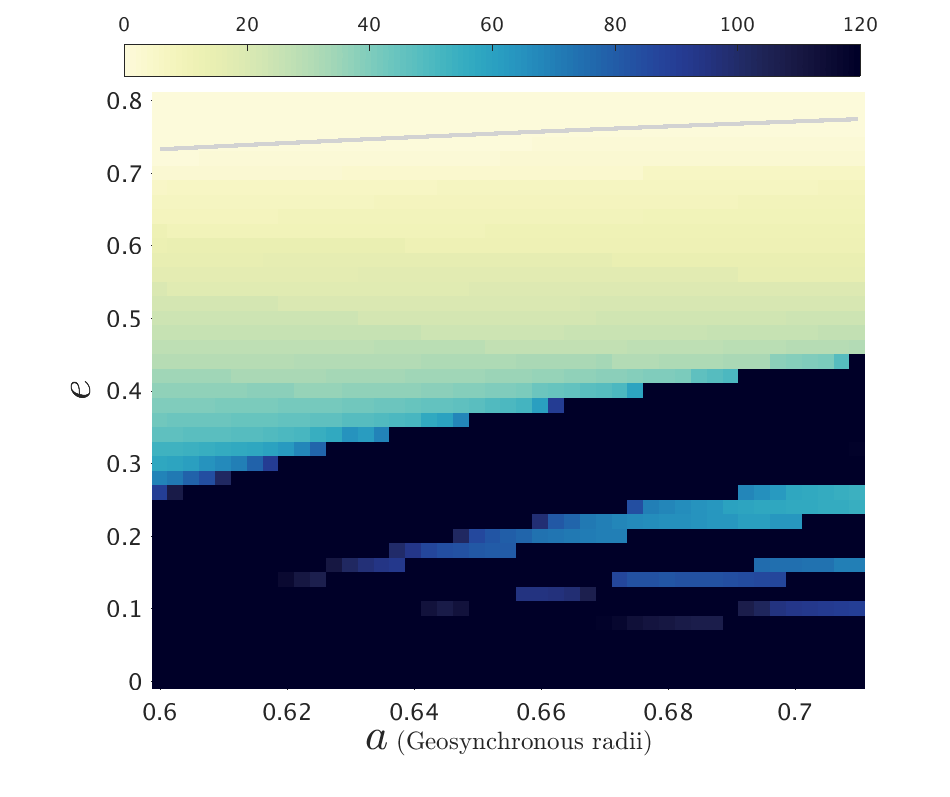}
      \includegraphics[width=.49\textwidth]{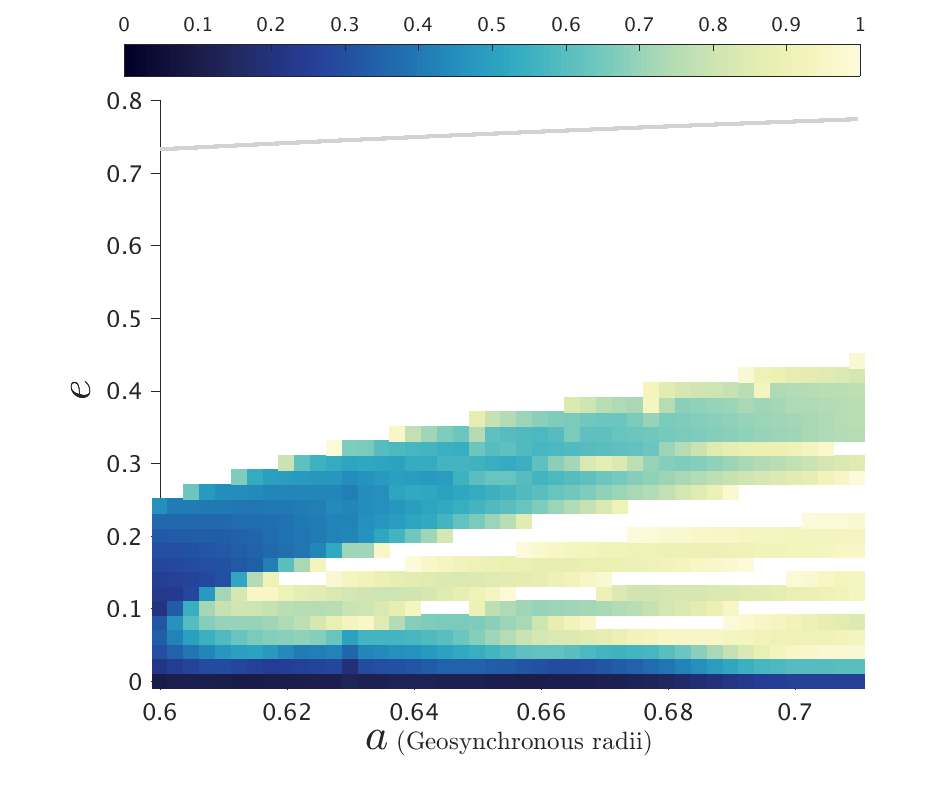}
    \end{subfigure}
    \begin{subfigure}[b]{0.67\textwidth}
      \caption{$\bm{\Delta}\bm{\Omega} = {\bf 270^\circ}$, $\bm{\Delta}\bm{\omega} = {\bf 0}$}
      \includegraphics[width=.49\textwidth]{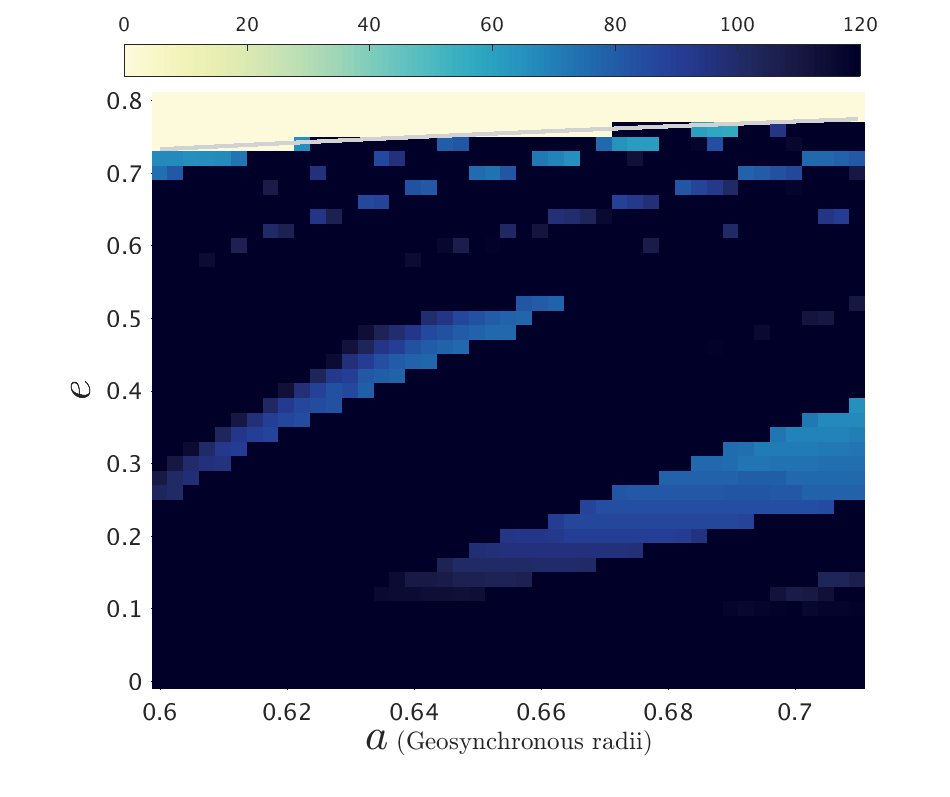}
      \includegraphics[width=.49\textwidth]{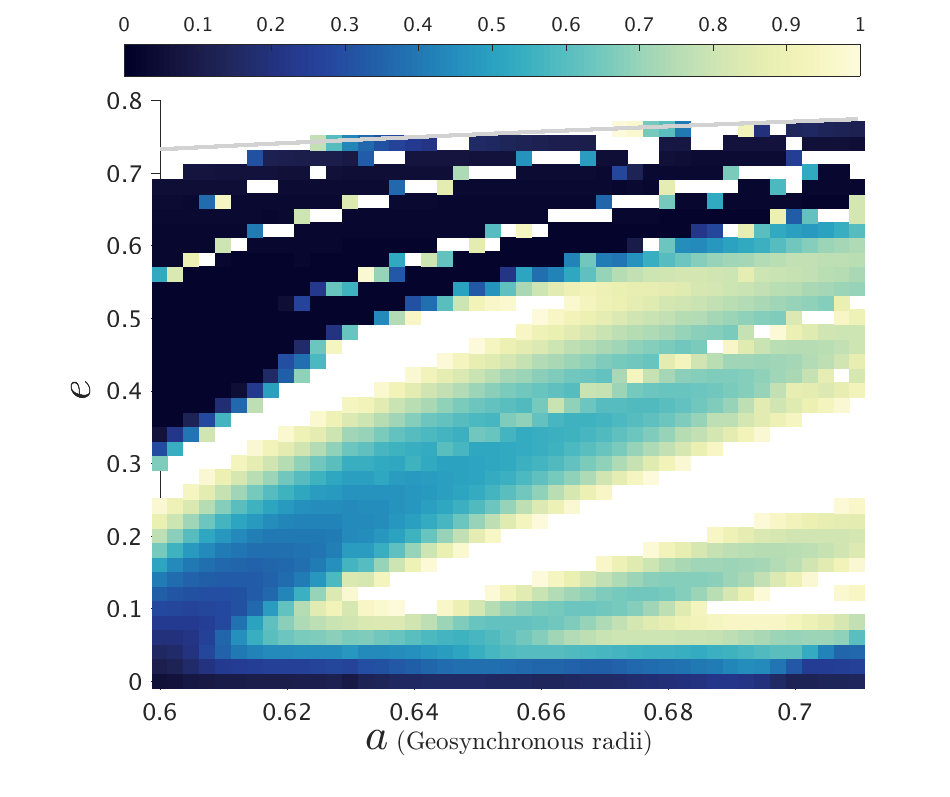}
    \end{subfigure} 
    \begin{subfigure}[b]{0.67\textwidth}
      \caption{$\bm{\Delta}\bm{\Omega} = {\bf 270^\circ}$, $\bm{\Delta}\bm{\omega} = {\bf 90^\circ}$}
      \includegraphics[width=.49\textwidth]{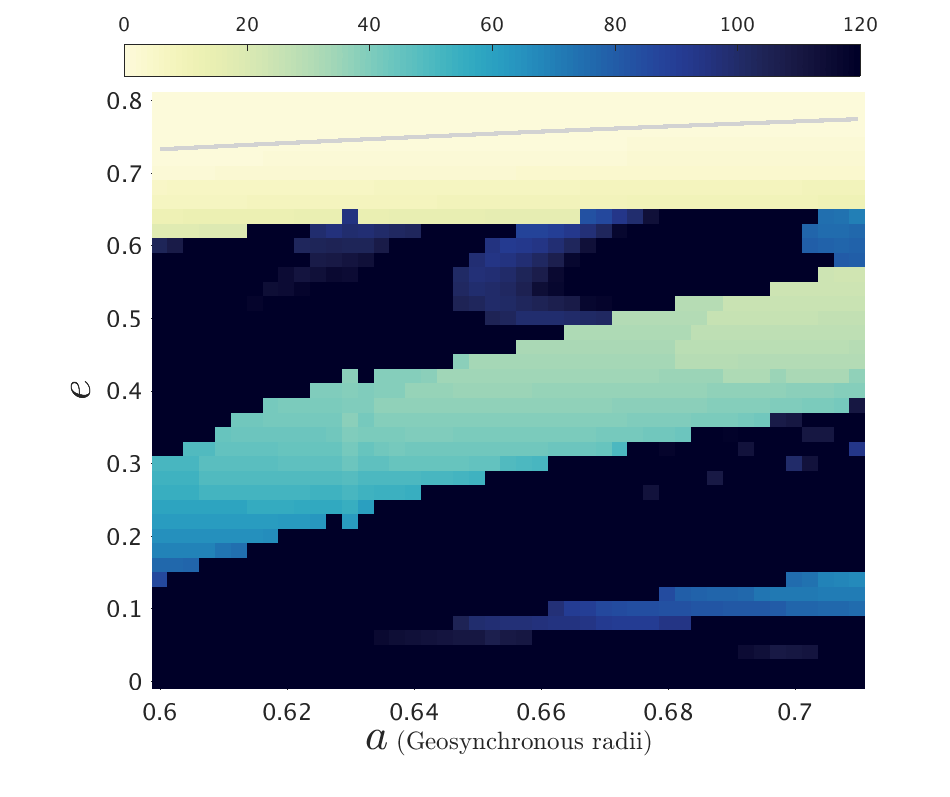}
      \includegraphics[width=.49\textwidth]{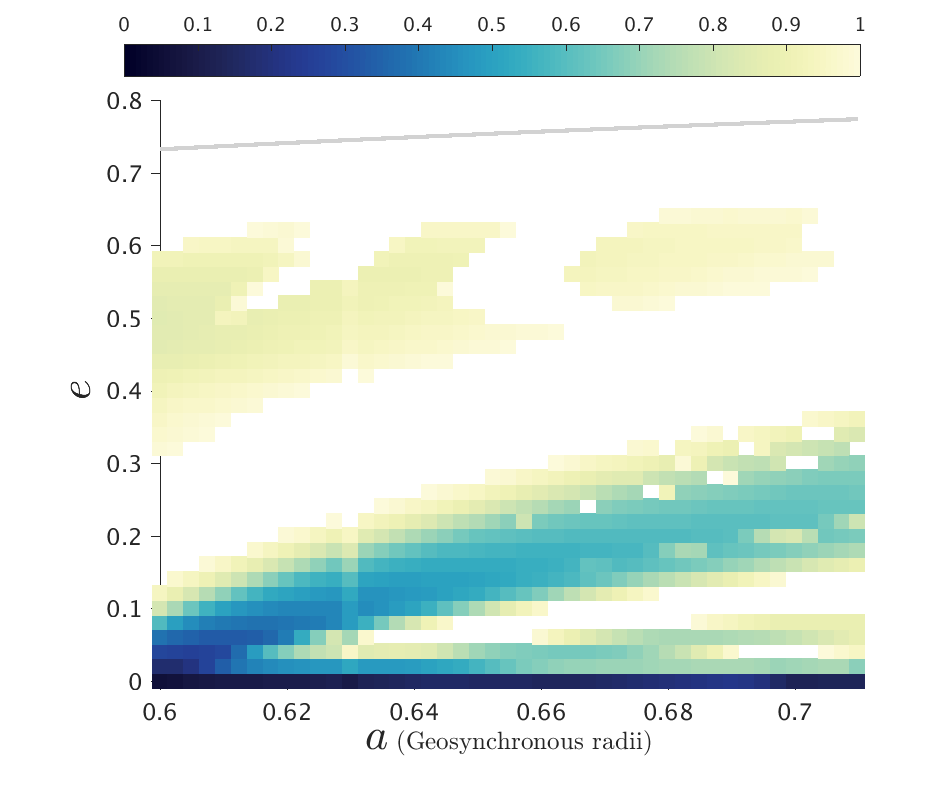}
    \end{subfigure}    
  \caption{Lifetime and $De$ maps of the \textit{MEO-general} phase space for $\bm{i_{o}} = {\bf 56^\circ}$,  
  for Epoch 2018, and for $C_{R}A/m=1$ m$^2$/kg.
  The colorbar for the lifetime maps is from 0 to 120 years and 
  that of the $De$ maps is from 0 to 1.}
  \label{fig:MEO_inc56_srp2b}
\end{figure}

\begin{figure}[htp!]
  \centering
    \begin{subfigure}[b]{0.67\textwidth}
      \caption{$\bm{\Delta}\bm{\Omega} = {\bf 0}$, $\bm{\Delta}\bm{\omega} = {\bf 0}$}
      \includegraphics[width=.49\textwidth]{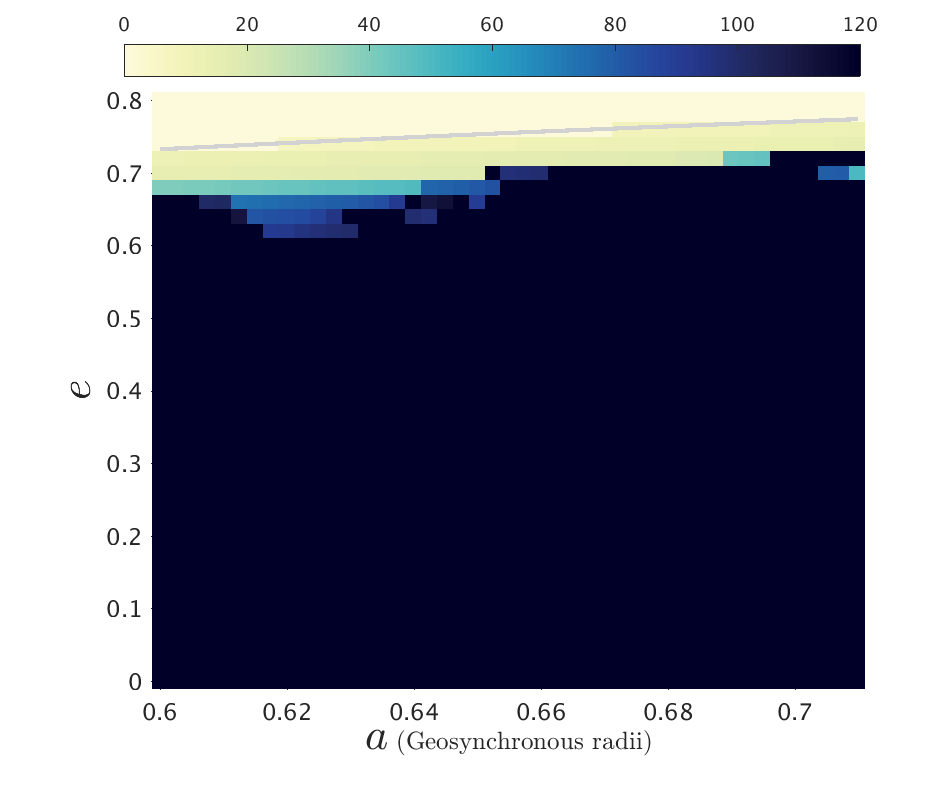} 
      \includegraphics[width=.49\textwidth]{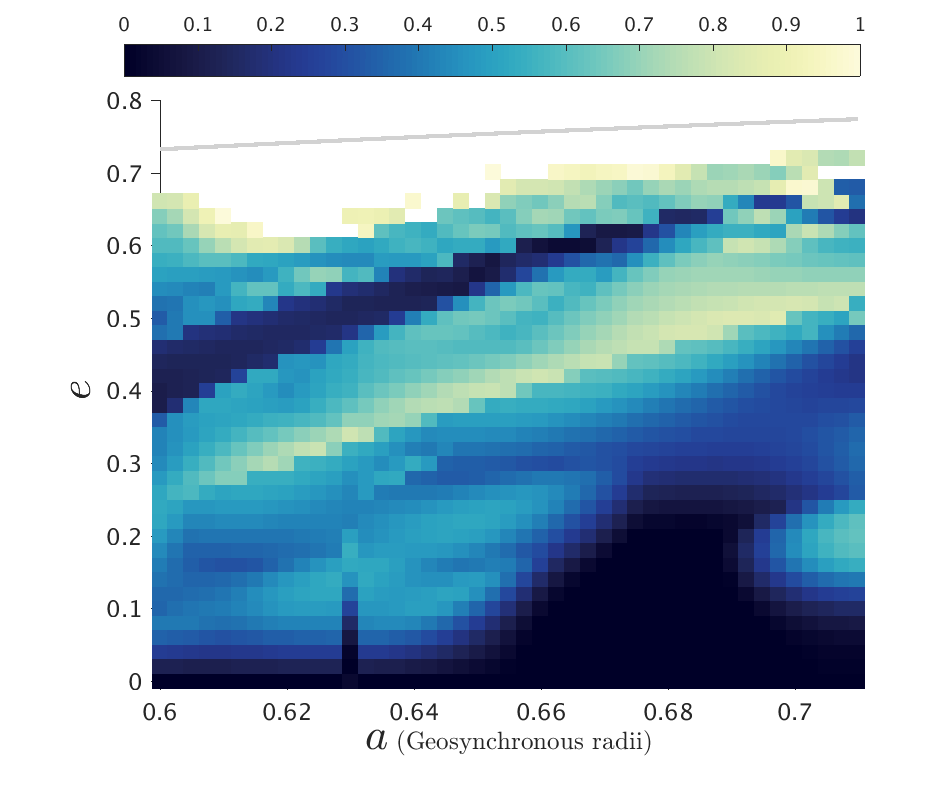}
    \end{subfigure}  
    \begin{subfigure}[b]{0.67\textwidth}
      \caption{$\bm{\Delta}\bm{\Omega} = {\bf 0}$, $\bm{\Delta}\bm{\omega} = {\bf 90^\circ}$}
      \includegraphics[width=.49\textwidth]{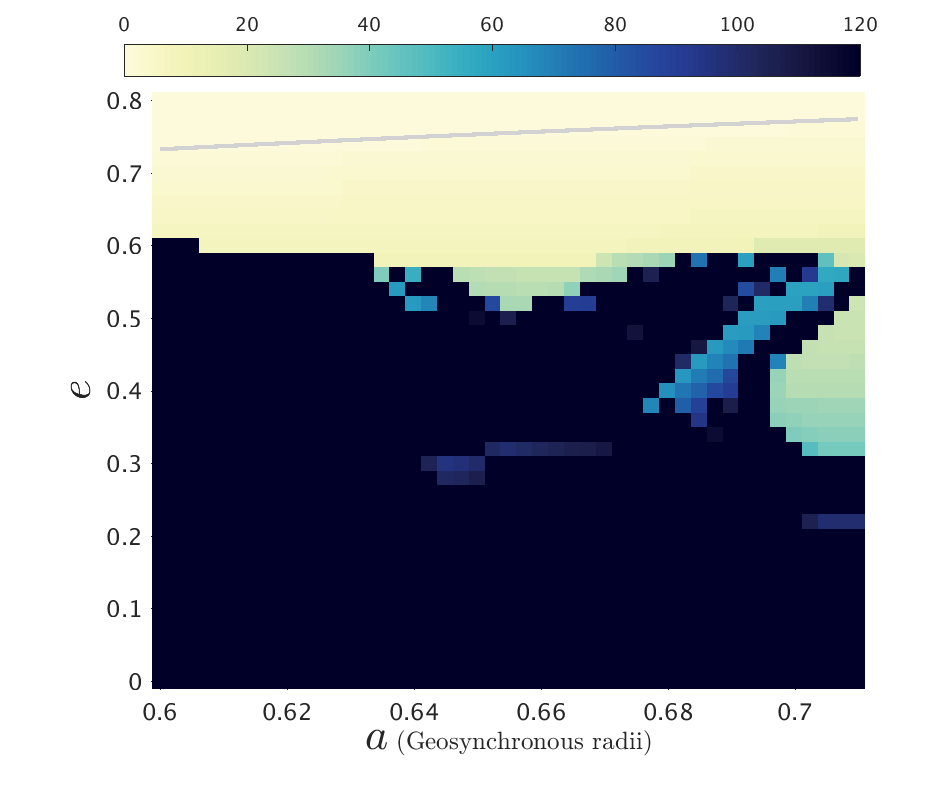}
      \includegraphics[width=.49\textwidth]{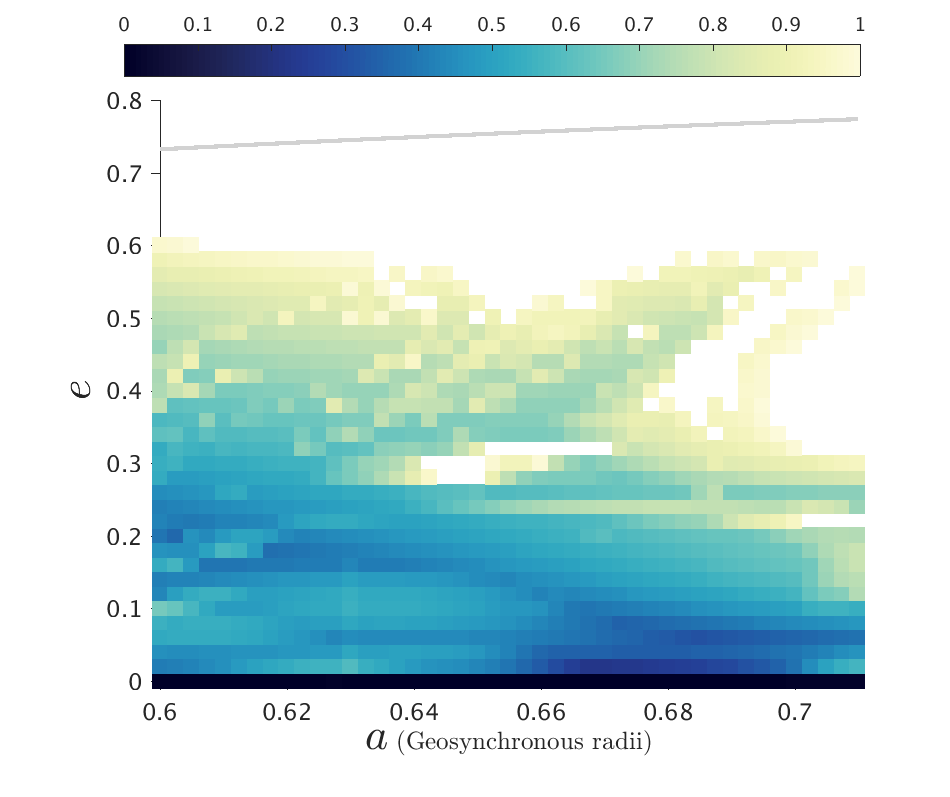}
    \end{subfigure} 
    \begin{subfigure}[b]{0.67\textwidth}
      \caption{$\bm{\Delta}\bm{\Omega} = {\bf 90^\circ}$, $\bm{\Delta}\bm{\omega} = {\bf 0}$}
      \includegraphics[width=.49\textwidth]{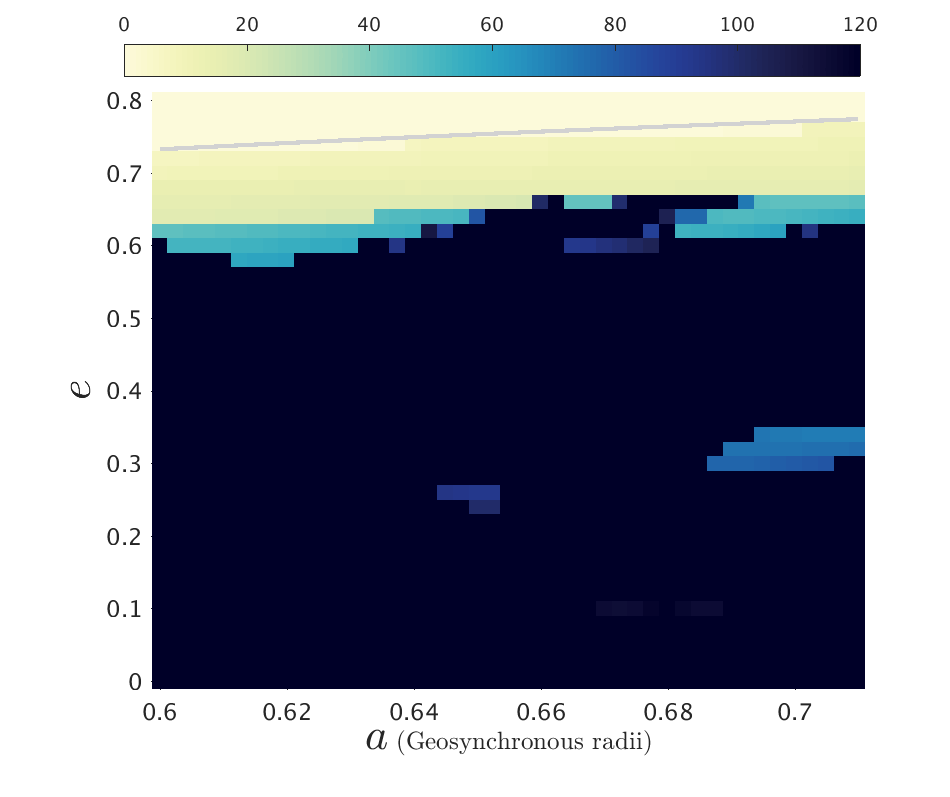}
      \includegraphics[width=.49\textwidth]{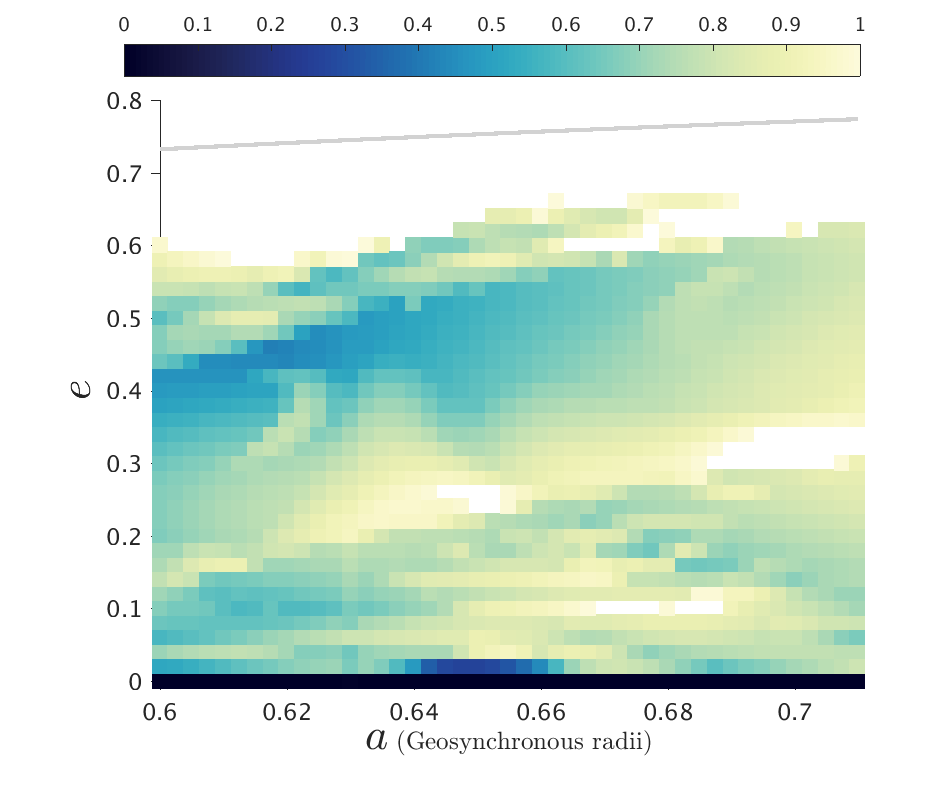}
    \end{subfigure}  
    \begin{subfigure}[b]{0.67\textwidth}
      \caption{$\bm{\Delta}\bm{\Omega} = {\bf 90^\circ}$, $\bm{\Delta}\bm{\omega} = {\bf 90^\circ}$}
      \includegraphics[width=.49\textwidth]{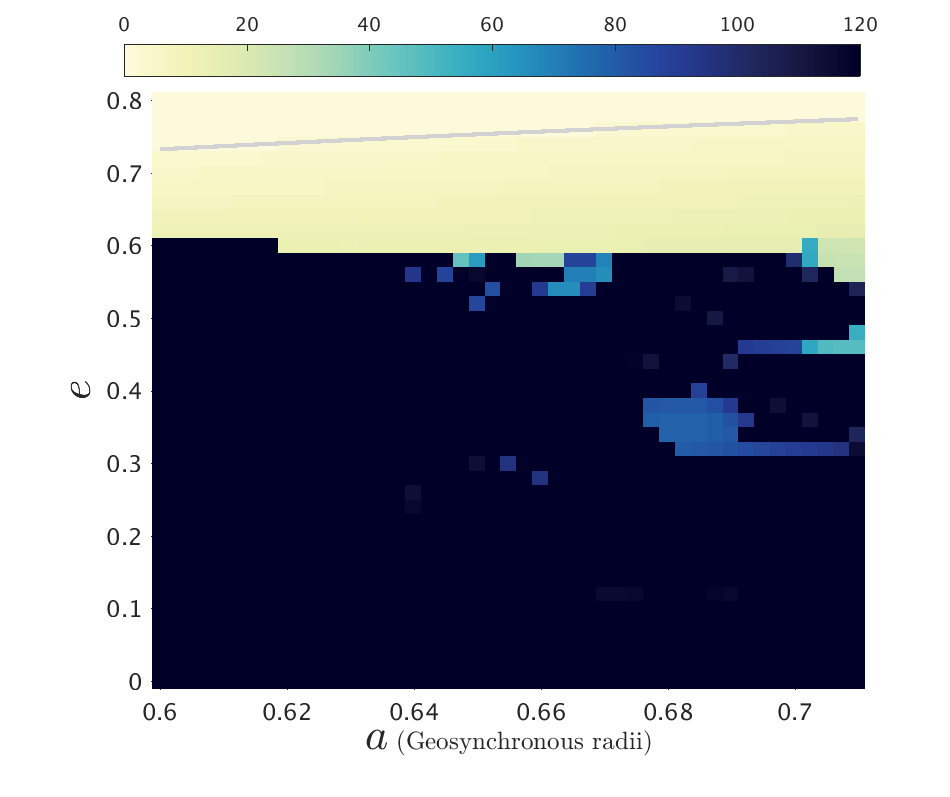} 
      \includegraphics[width=.49\textwidth]{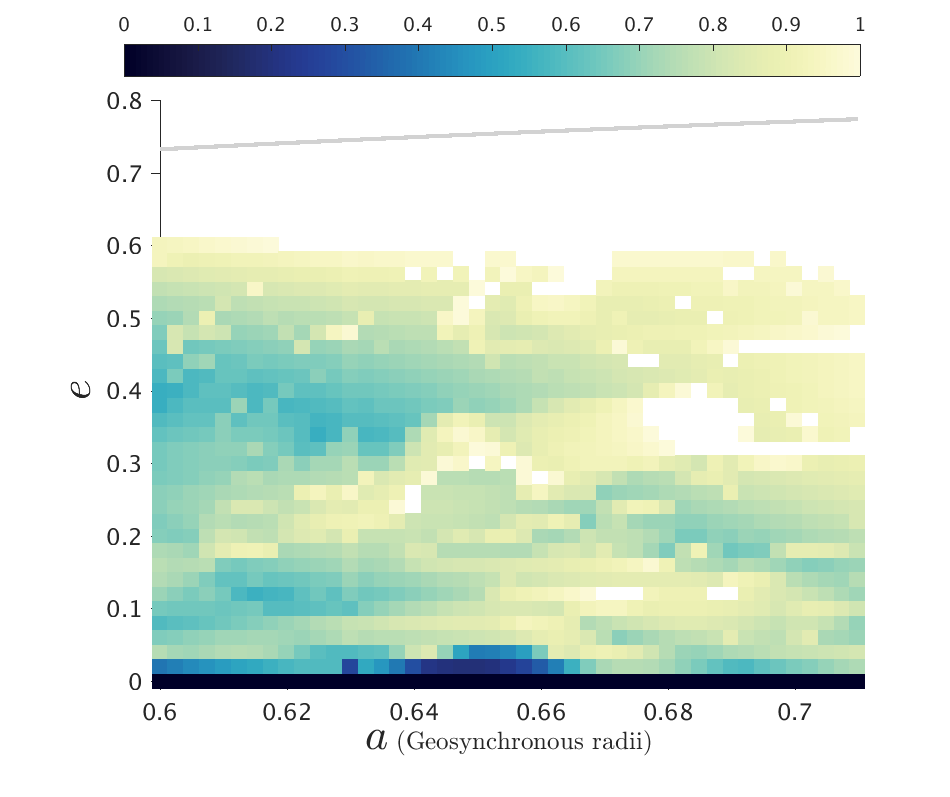}
    \end{subfigure}      
  \caption{Lifetime and $De$ maps of the \textit{MEO-general} phase space for $\bm{i_{o}} = {\bf 64^\circ}$,  
  for Epoch 2018, and for $C_{R}A/m=0.015$ m$^2$/kg.
  The colorbar for the lifetime maps is from 0 to 120 years and 
  that of the $De$ maps is from 0 to 1.}
  \label{fig:MEO_inc64_ep18a}
\end{figure}

\begin{figure}[htp!]
  \centering
    \begin{subfigure}[b]{0.67\textwidth}
      \caption{$\bm{\Delta}\bm{\Omega} = {\bf 180^\circ}$, $\bm{\Delta}\bm{\omega} = {\bf 0}$}
      \includegraphics[width=.49\textwidth]{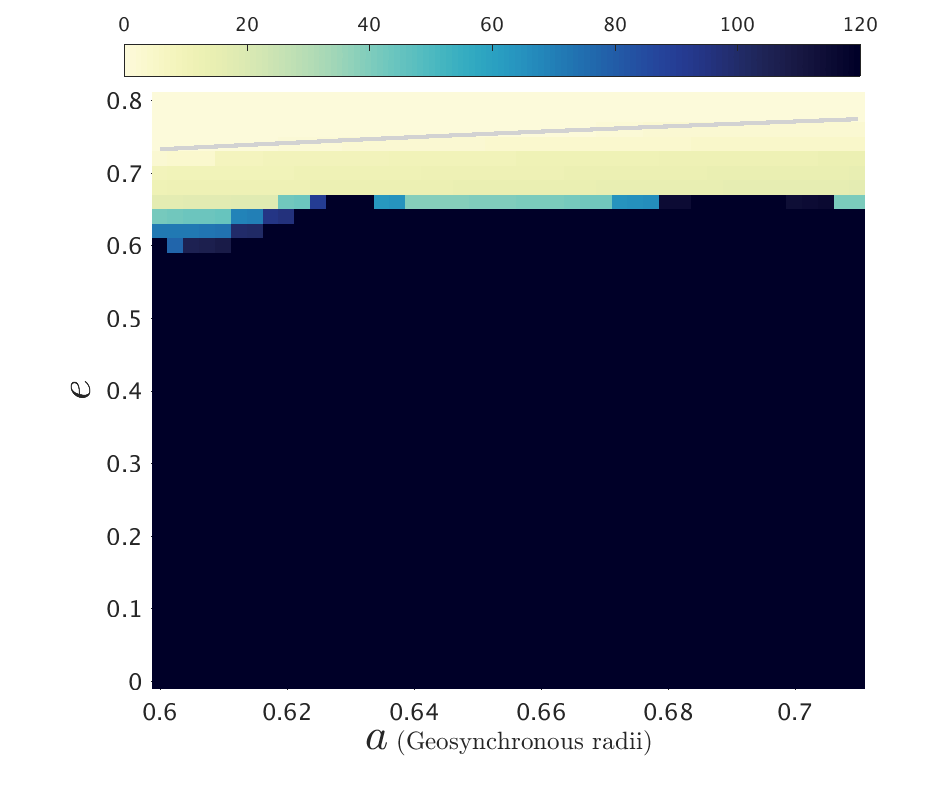}
      \includegraphics[width=.49\textwidth]{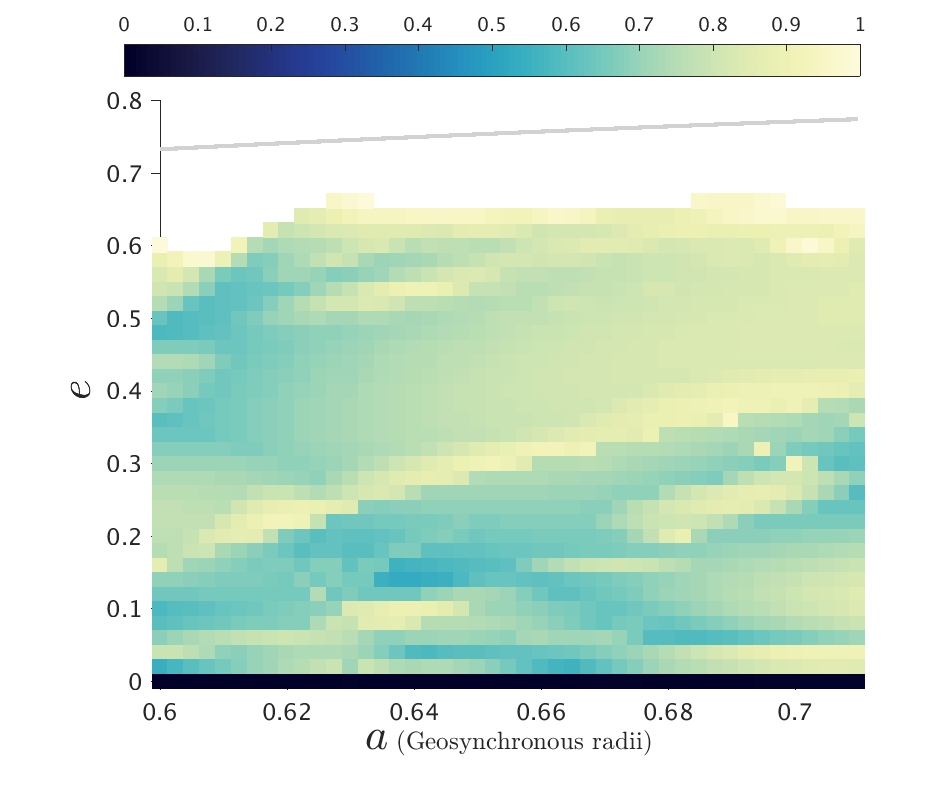}
    \end{subfigure}     
    \begin{subfigure}[b]{0.67\textwidth}
      \caption{$\bm{\Delta}\bm{\Omega} = {\bf 180^\circ}$, $\bm{\Delta}\bm{\omega} = {\bf 90^\circ}$}
      \includegraphics[width=.49\textwidth]{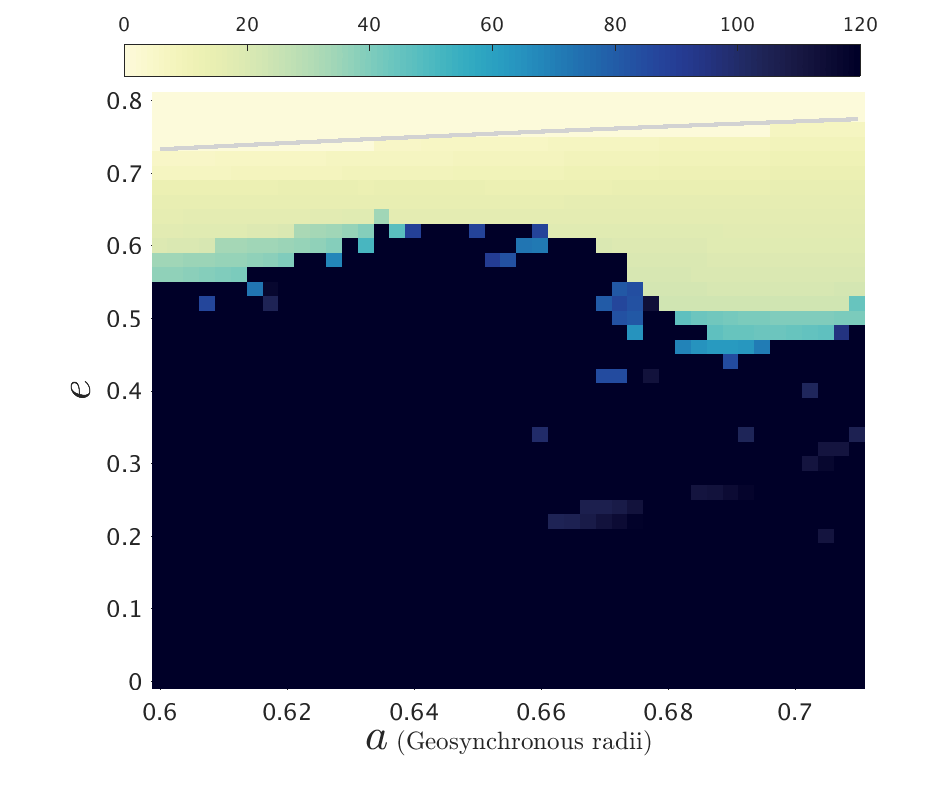}
      \includegraphics[width=.49\textwidth]{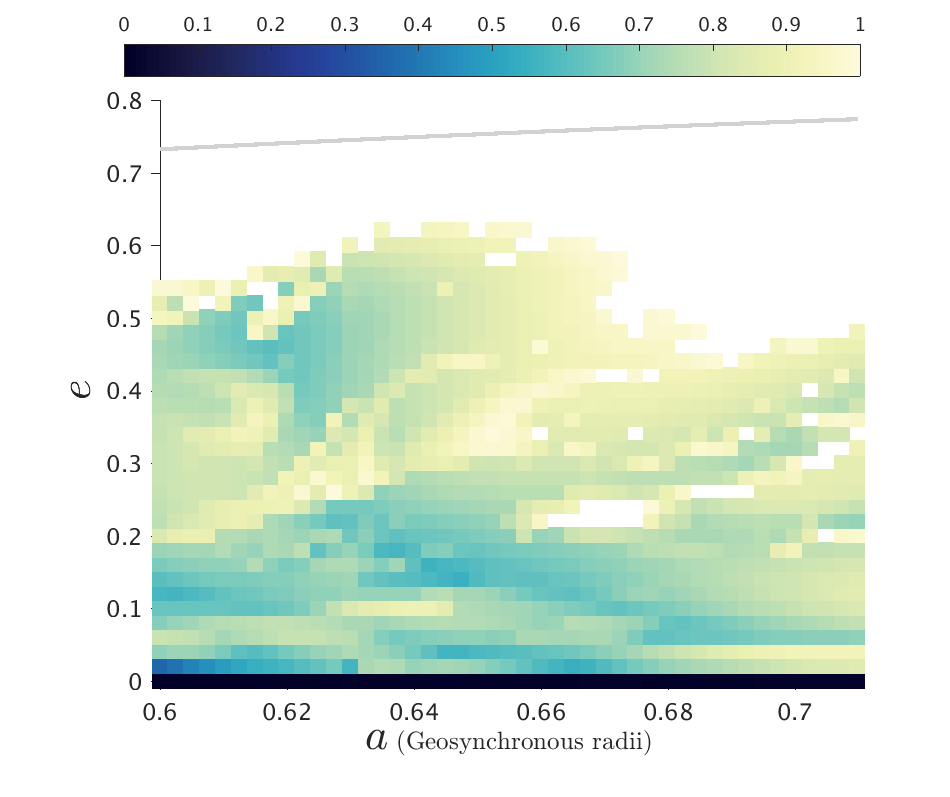}
    \end{subfigure}
    \begin{subfigure}[b]{0.67\textwidth}
      \caption{$\bm{\Delta}\bm{\Omega} = {\bf 270^\circ}$, $\bm{\Delta}\bm{\omega} = {\bf 0}$}
      \includegraphics[width=.49\textwidth]{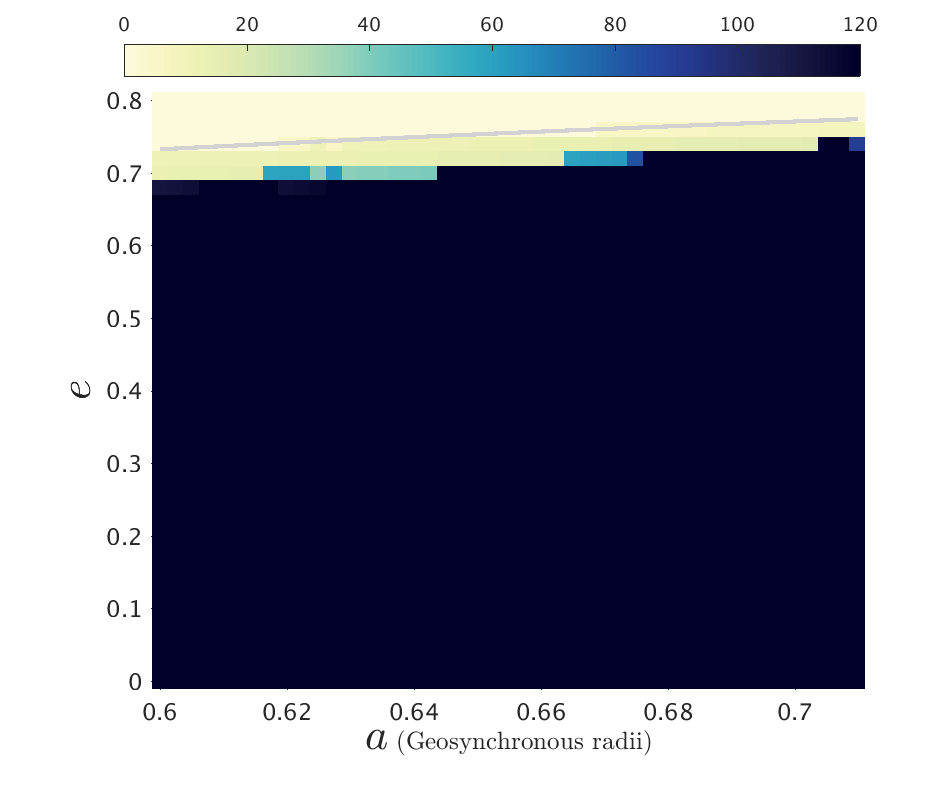}
      \includegraphics[width=.49\textwidth]{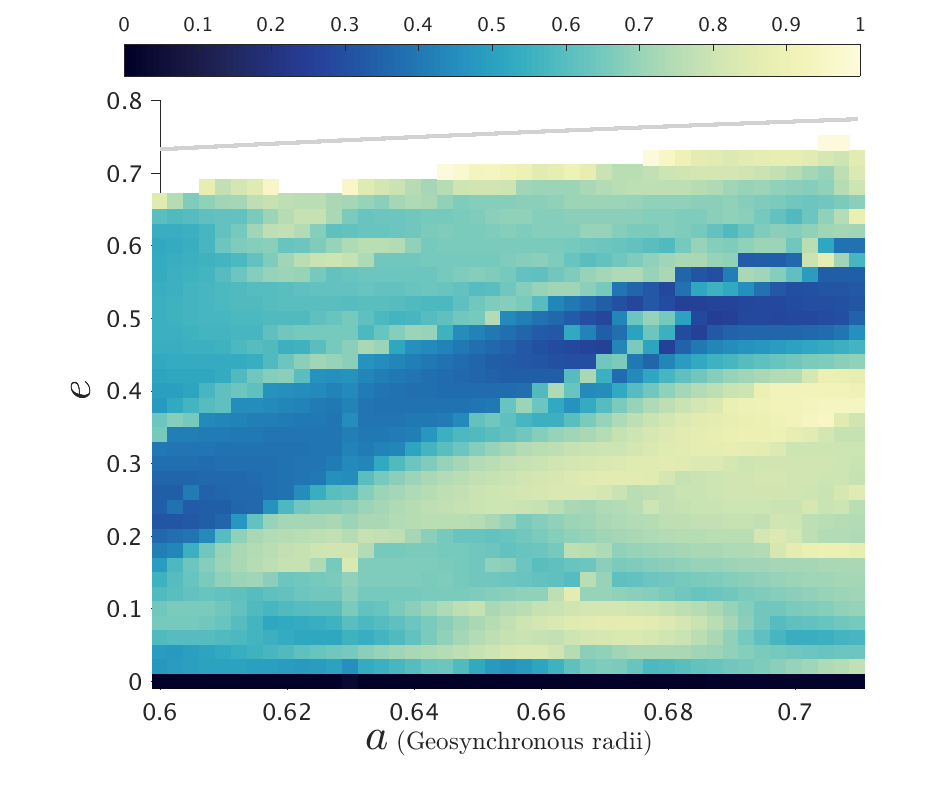}
    \end{subfigure} 
    \begin{subfigure}[b]{0.67\textwidth}
      \caption{$\bm{\Delta}\bm{\Omega} = {\bf 270^\circ}$, $\bm{\Delta}\bm{\omega} = {\bf 90^\circ}$}
      \includegraphics[width=.49\textwidth]{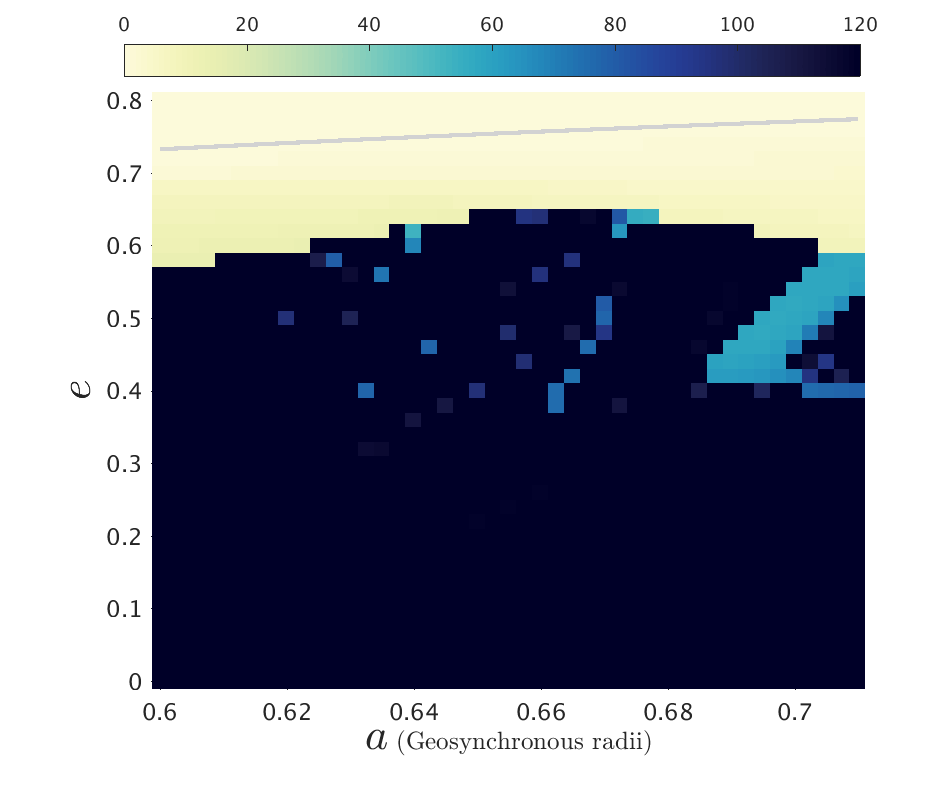}
      \includegraphics[width=.49\textwidth]{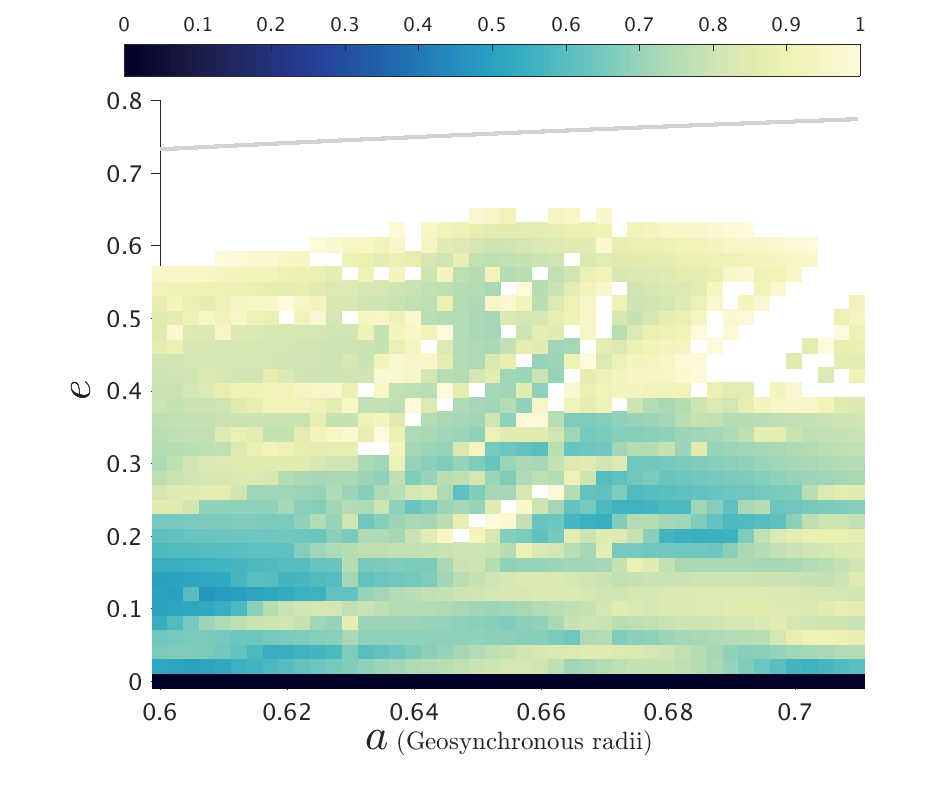}
    \end{subfigure}    
  \caption{Lifetime and $De$ maps of the \textit{MEO-general} phase space for $\bm{i_{o}} = {\bf 64^\circ}$,  
  for Epoch 2018, and for $C_{R}A/m=0.015$ m$^2$/kg.
  The colorbar for the lifetime maps is from 0 to 120 years and 
  that of the $De$ maps is from 0 to 1.}
  \label{fig:MEO_inc64_ep18b}
\end{figure}

\begin{figure}[htp!]
  \centering
      \begin{subfigure}[b]{0.67\textwidth}
      \caption{$\bm{\Delta}\bm{\Omega} = {\bf 0}$, $\bm{\Delta}\bm{\omega} = {\bf 0}$}
      \includegraphics[width=.49\textwidth]{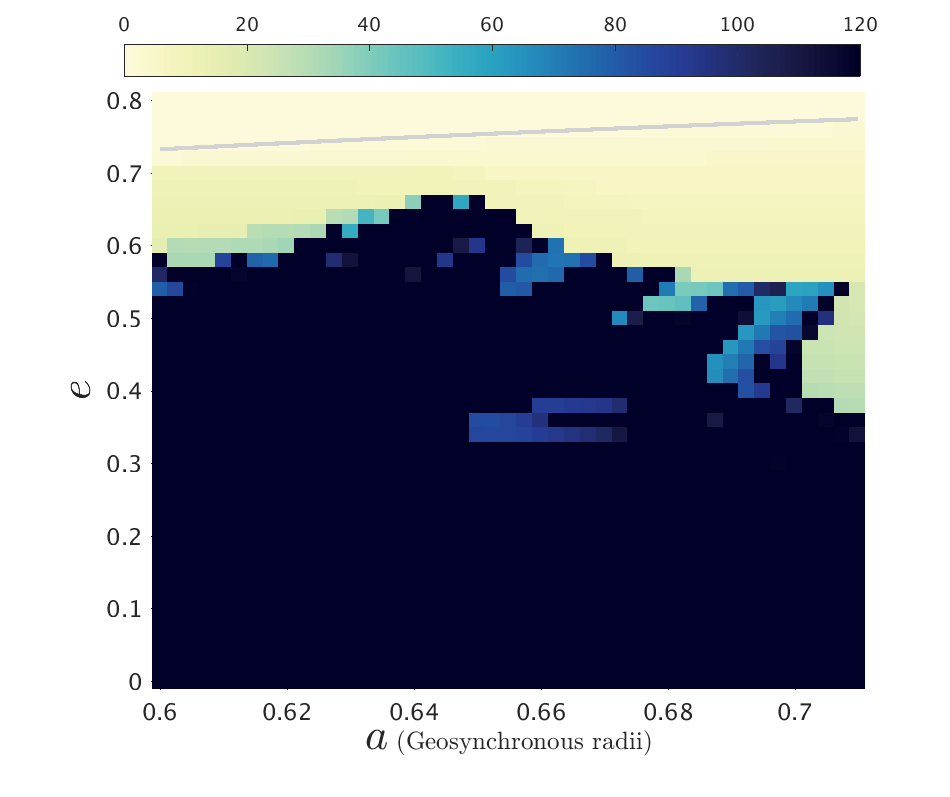} 
      \includegraphics[width=.49\textwidth]{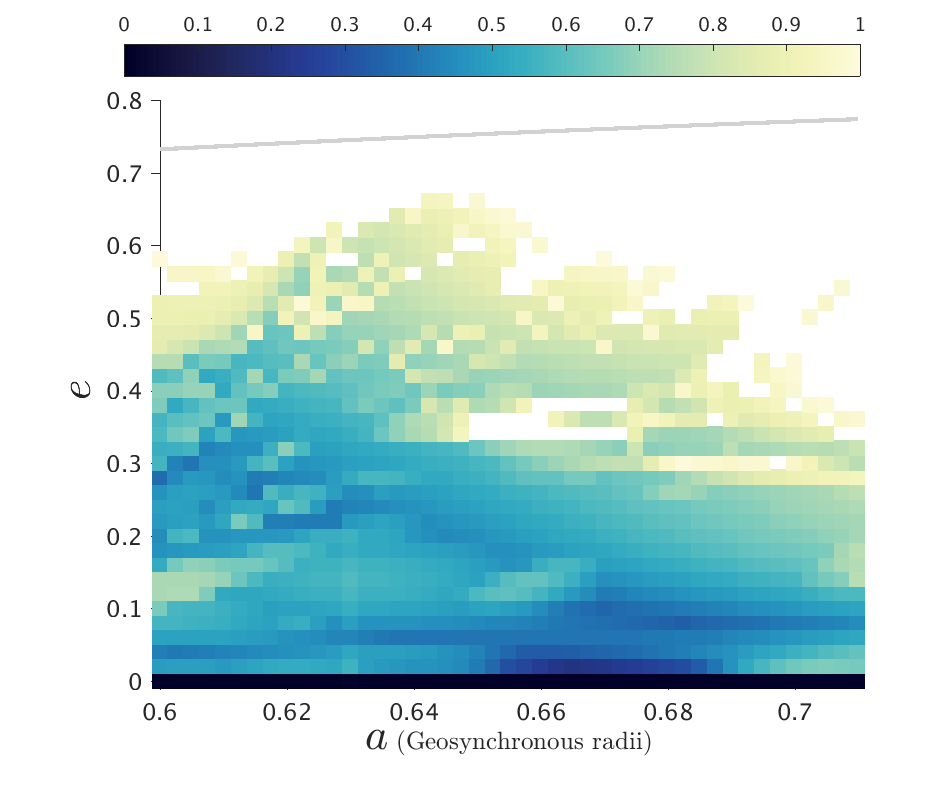}
    \end{subfigure}  
    \begin{subfigure}[b]{0.67\textwidth}
      \caption{$\bm{\Delta}\bm{\Omega} = {\bf 0}$, $\bm{\Delta}\bm{\omega} = {\bf 90^\circ}$}
      \includegraphics[width=.49\textwidth]{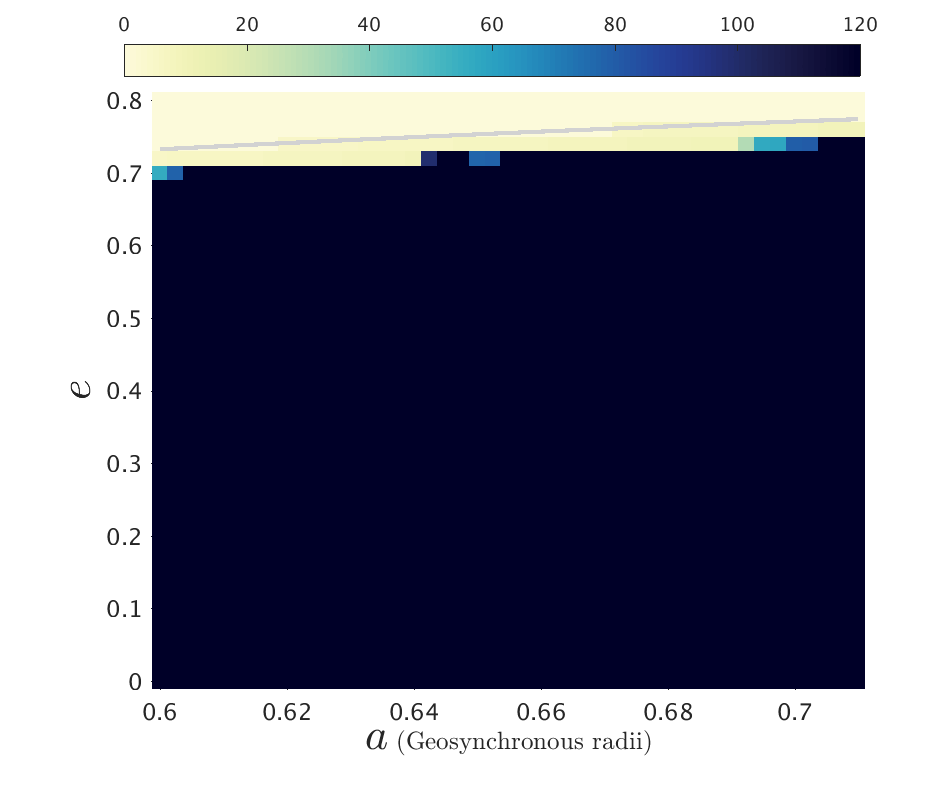}
      \includegraphics[width=.49\textwidth]{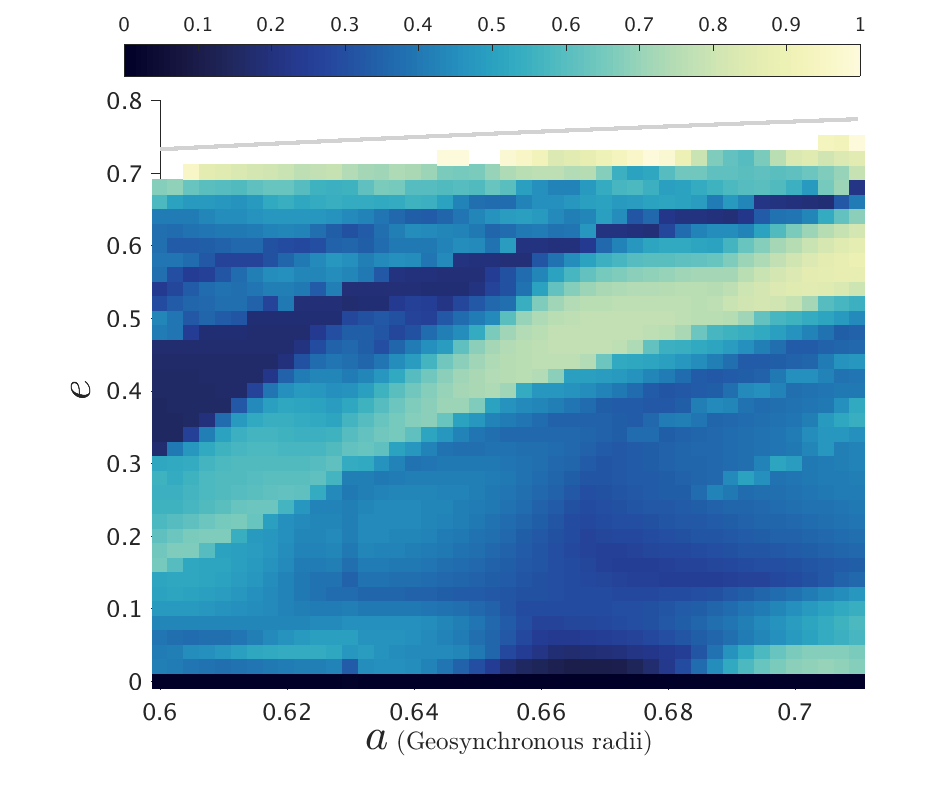}
    \end{subfigure} 
    \begin{subfigure}[b]{0.67\textwidth}
      \caption{$\bm{\Delta}\bm{\Omega} = {\bf 90^\circ}$, $\bm{\Delta}\bm{\omega} = {\bf 0}$}
      \includegraphics[width=.49\textwidth]{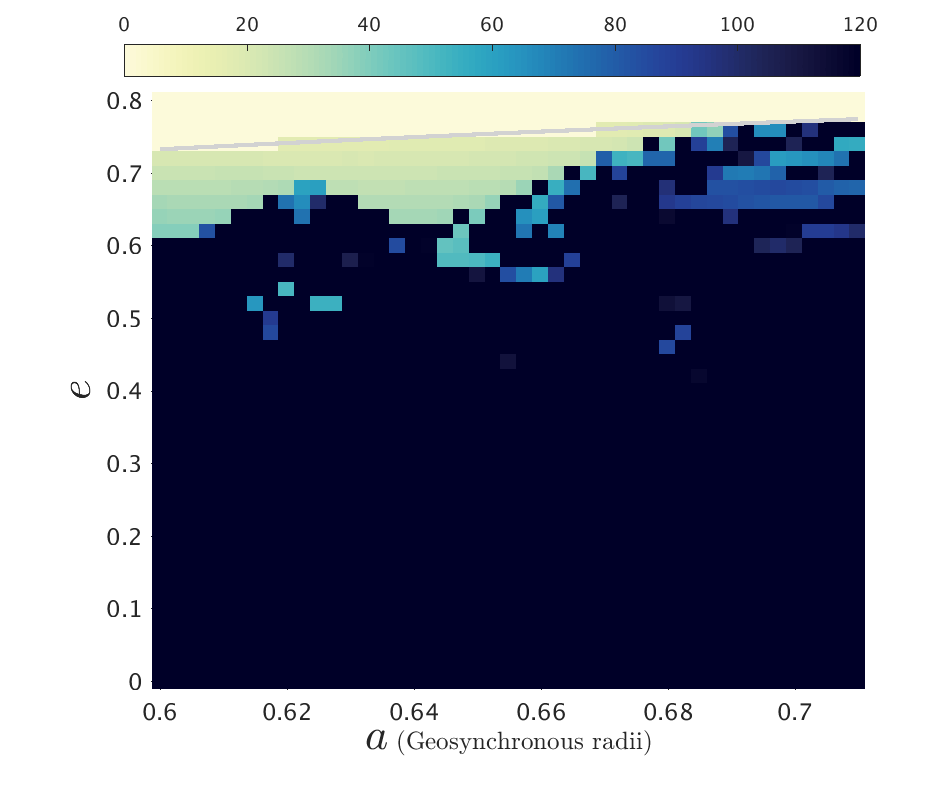}
      \includegraphics[width=.49\textwidth]{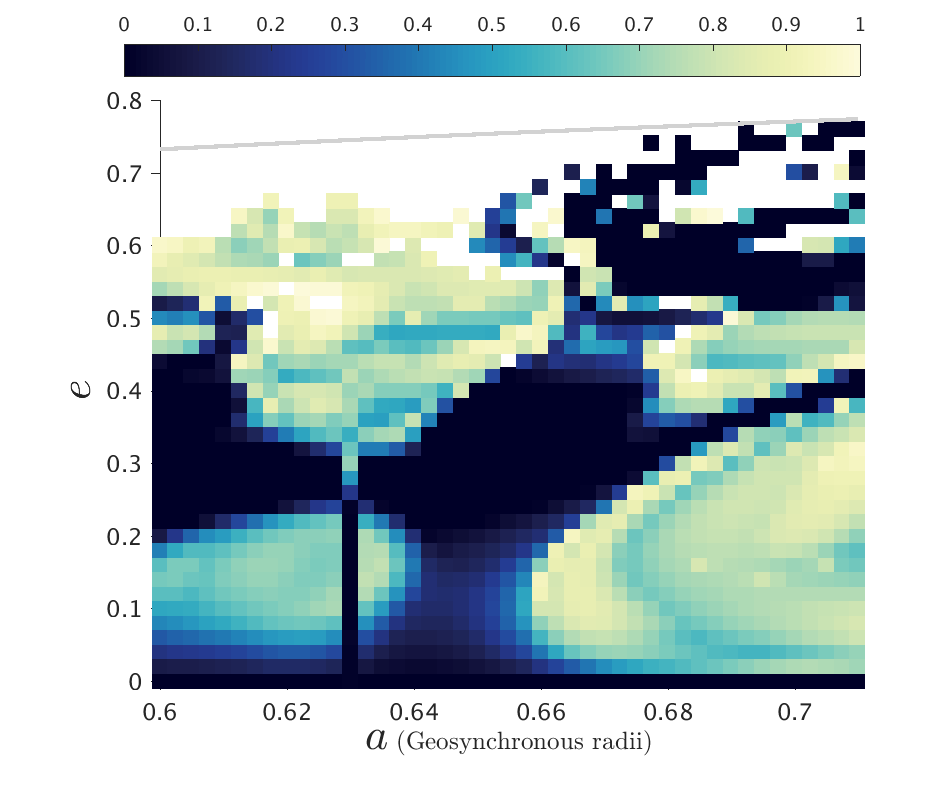}
    \end{subfigure}  
    \begin{subfigure}[b]{0.67\textwidth}
      \caption{$\bm{\Delta}\bm{\Omega} = {\bf 90^\circ}$, $\bm{\Delta}\bm{\omega} = {\bf 90^\circ}$}
      \includegraphics[width=.49\textwidth]{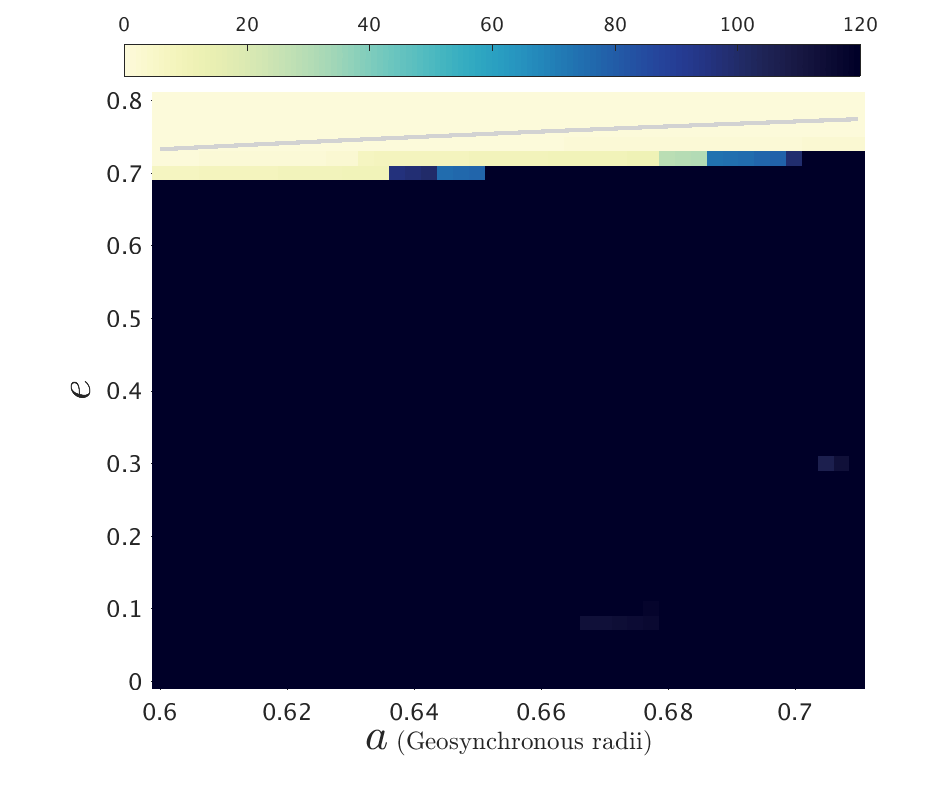} 
      \includegraphics[width=.49\textwidth]{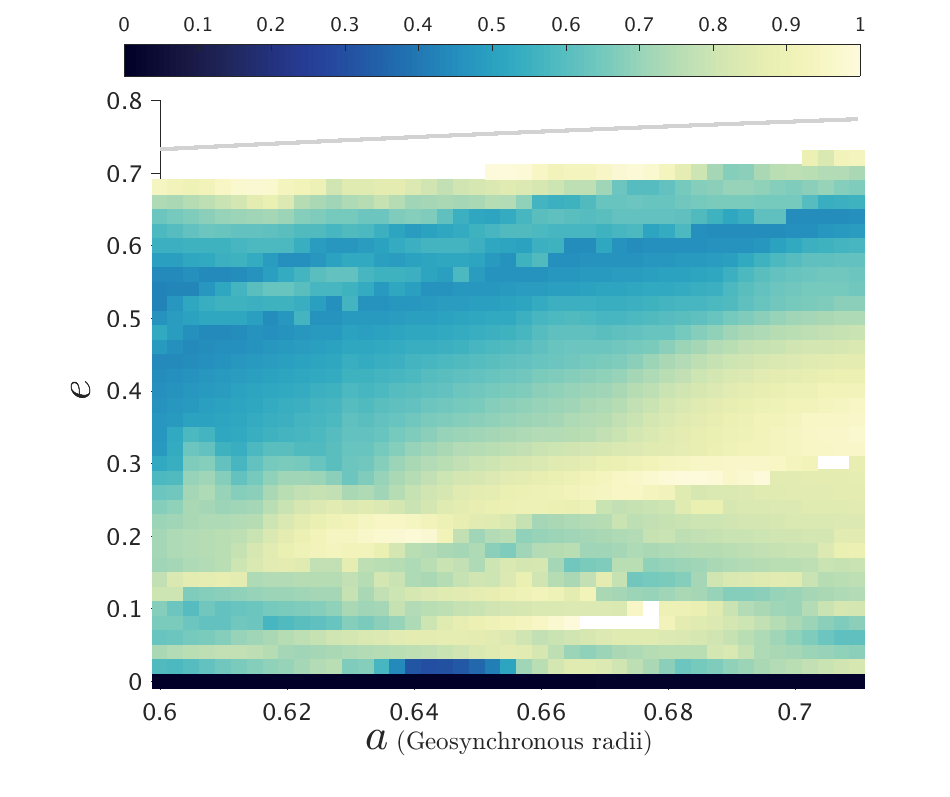}
    \end{subfigure}      
  \caption{Lifetime and $De$ maps of the \textit{MEO-general} phase space for $\bm{i_{o}} = {\bf 64^\circ}$,  
  for Epoch 2020, and for $C_{R}A/m=0.015$ m$^2$/kg.
  The colorbar for the lifetime maps is from 0 to 120 years and 
  that of the $De$ maps is from 0 to 1.}
  \label{fig:MEO_inc64_ep20a}
\end{figure}

\begin{figure}[htp!]
  \centering 
    \begin{subfigure}[b]{0.67\textwidth}
      \caption{$\bm{\Delta}\bm{\Omega} = {\bf 180^\circ}$, $\bm{\Delta}\bm{\omega} = {\bf 0}$}
      \includegraphics[width=.49\textwidth]{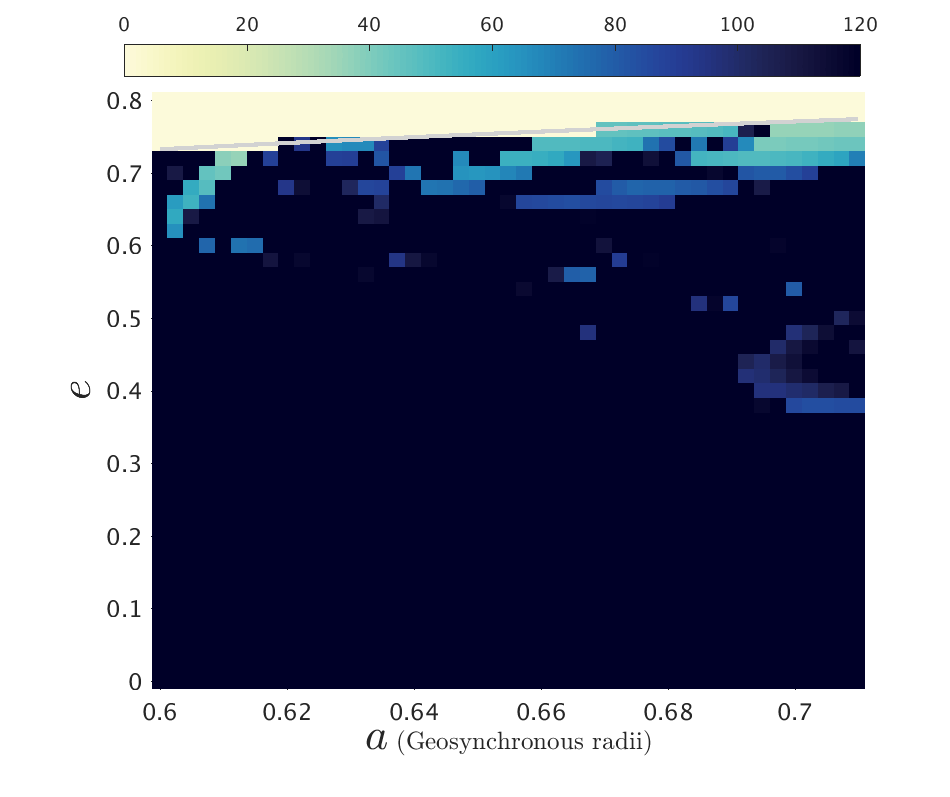}
      \includegraphics[width=.49\textwidth]{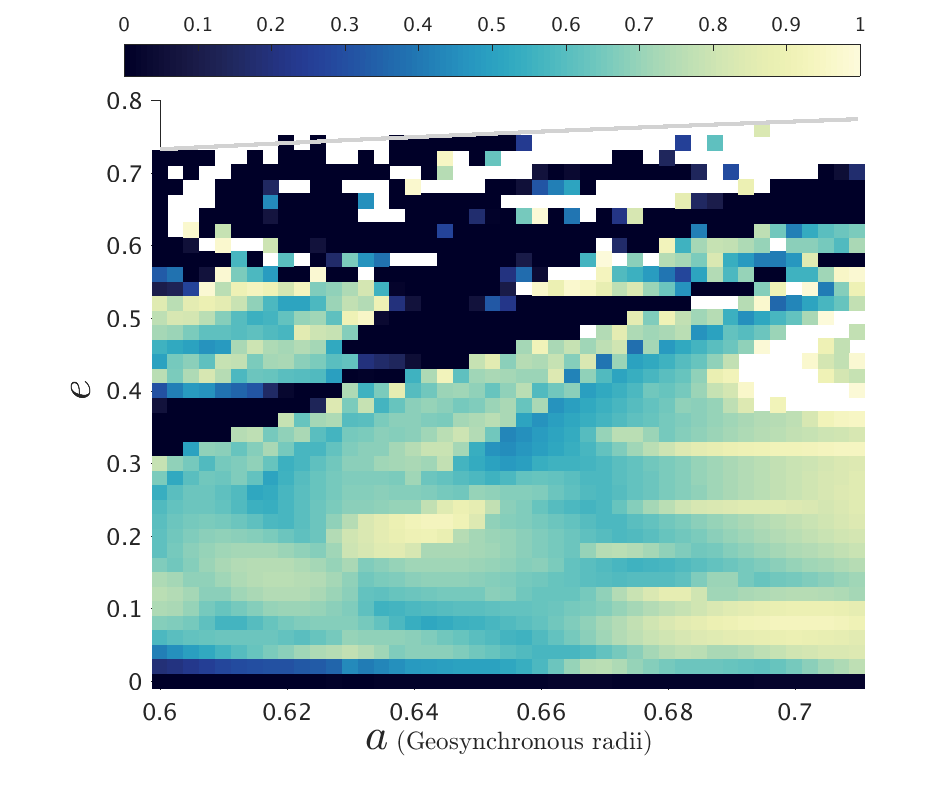}
    \end{subfigure}     
    \begin{subfigure}[b]{0.67\textwidth}
      \caption{$\bm{\Delta}\bm{\Omega} = {\bf 180^\circ}$, $\bm{\Delta}\bm{\omega} = {\bf 90^\circ}$}
      \includegraphics[width=.49\textwidth]{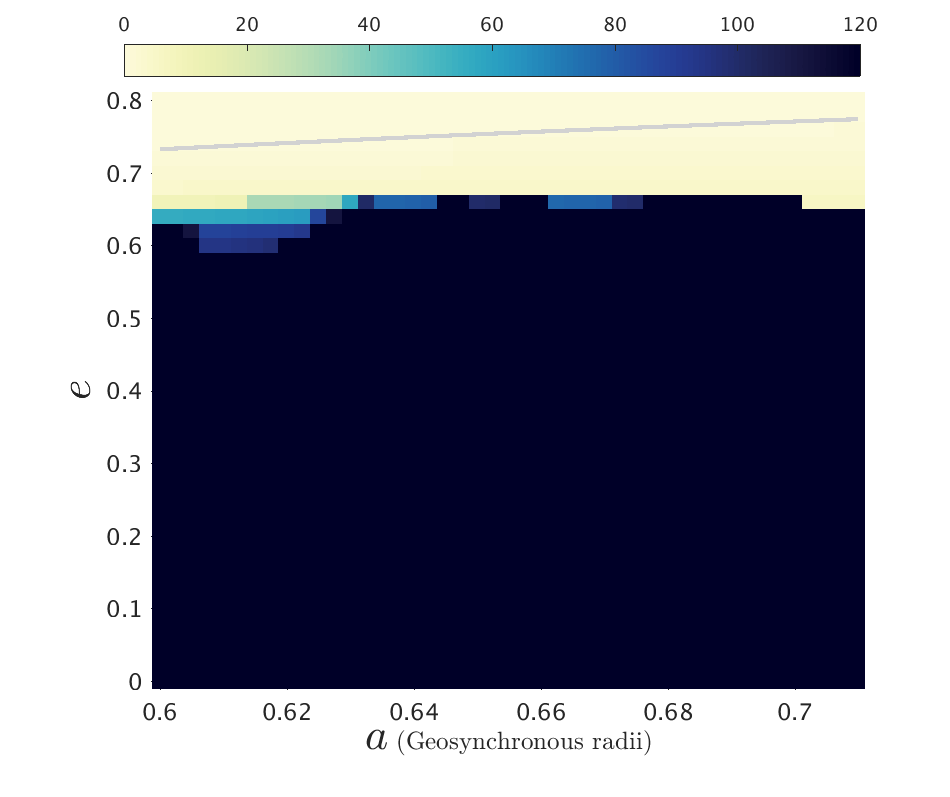}
      \includegraphics[width=.49\textwidth]{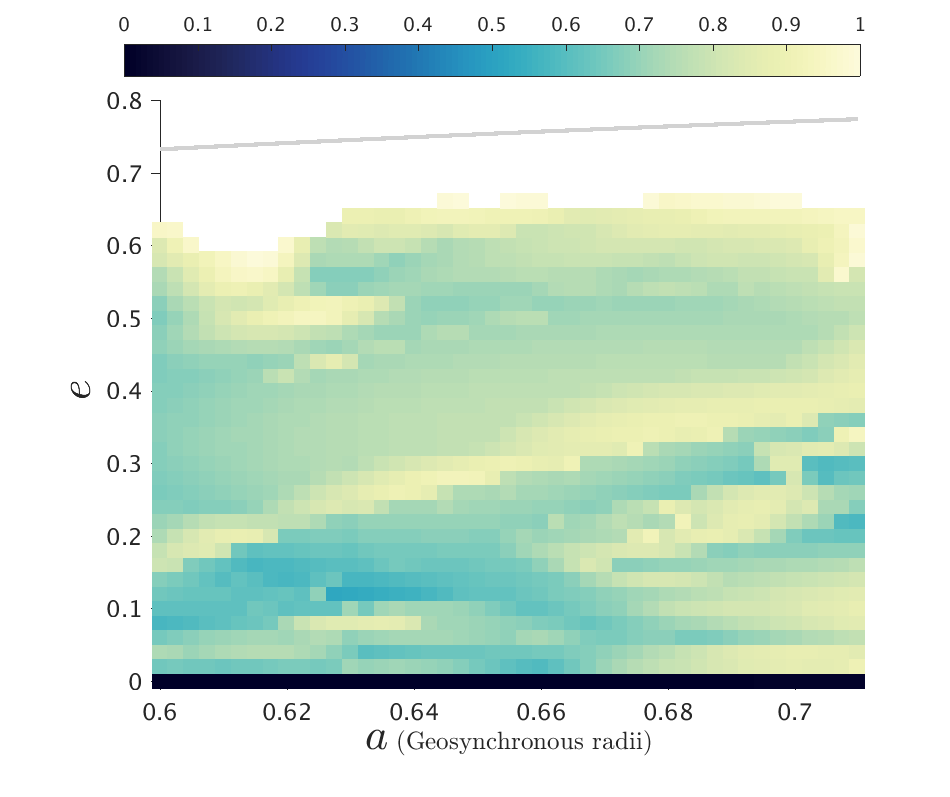}
    \end{subfigure}
    \begin{subfigure}[b]{0.67\textwidth}
      \caption{$\bm{\Delta}\bm{\Omega} = {\bf 270^\circ}$, $\bm{\Delta}\bm{\omega} = {\bf 0}$}
      \includegraphics[width=.49\textwidth]{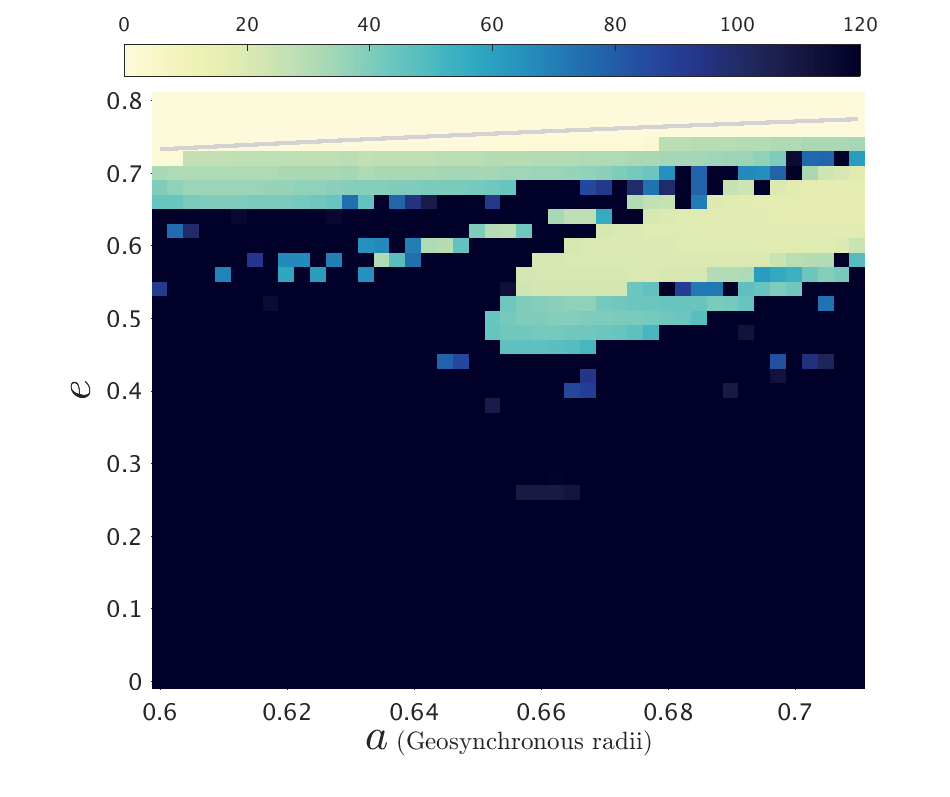}
      \includegraphics[width=.49\textwidth]{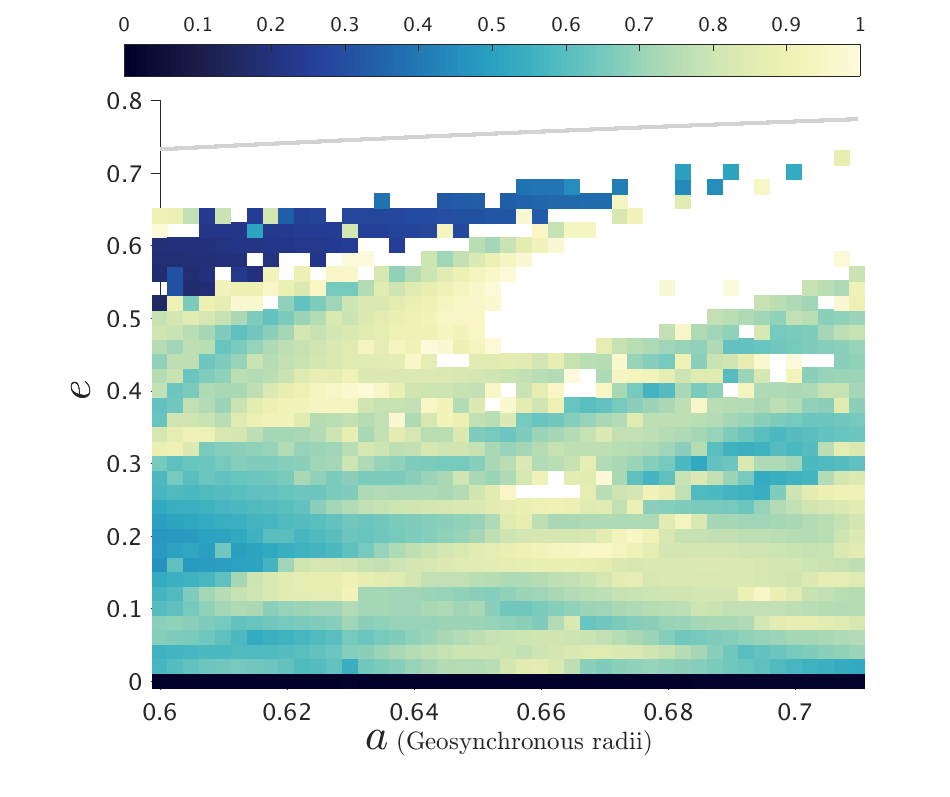}
    \end{subfigure} 
    \begin{subfigure}[b]{0.67\textwidth}
      \caption{$\bm{\Delta}\bm{\Omega} = {\bf 270^\circ}$, $\bm{\Delta}\bm{\omega} = {\bf 90^\circ}$}
      \includegraphics[width=.49\textwidth]{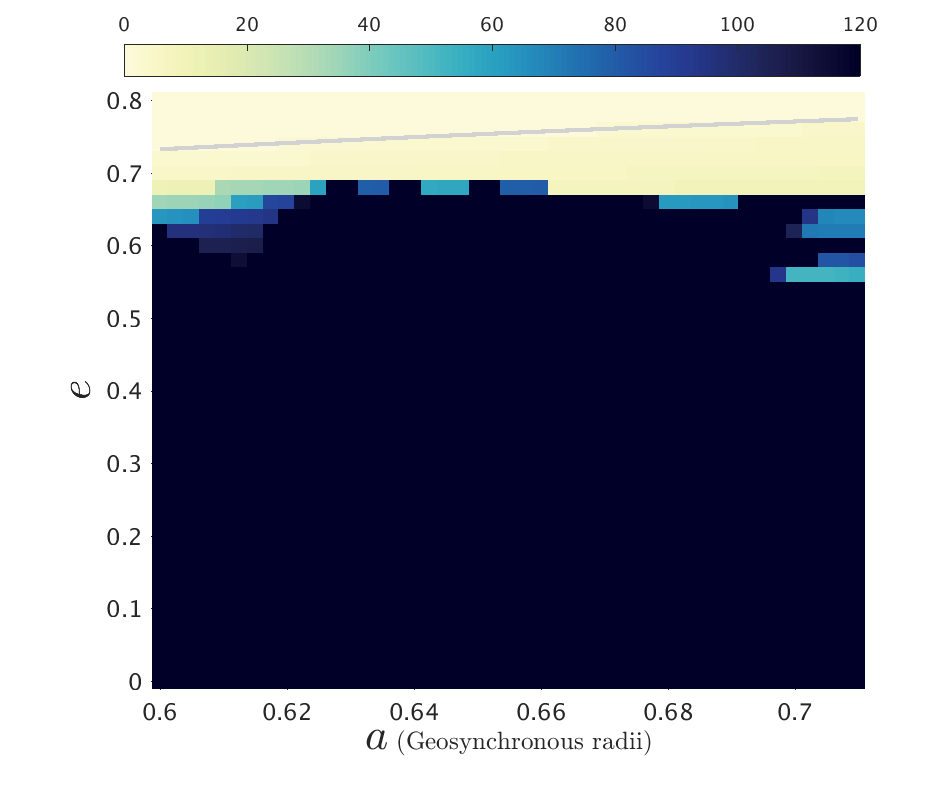}
      \includegraphics[width=.49\textwidth]{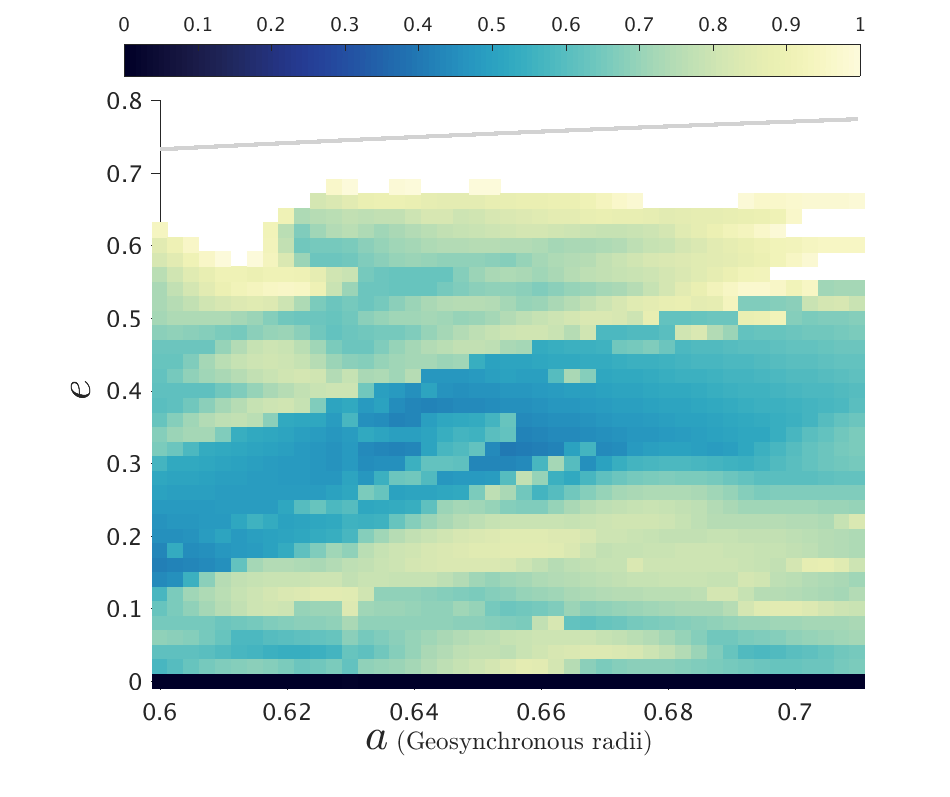}
    \end{subfigure}    
  \caption{Lifetime and $De$ maps of the \textit{MEO-general} phase space for $\bm{i_{o}} = {\bf 64^\circ}$,  
  for Epoch 2020, and for $C_{R}A/m=0.015$ m$^2$/kg.
  The colorbar for the lifetime maps is from 0 to 120 years and 
  that of the $De$ maps is from 0 to 1.}
  \label{fig:MEO_inc64_ep20b}
\end{figure}

\begin{figure}[htp!]
  \centering
    \begin{subfigure}[b]{0.49\textwidth}
      \includegraphics[width=\textwidth]{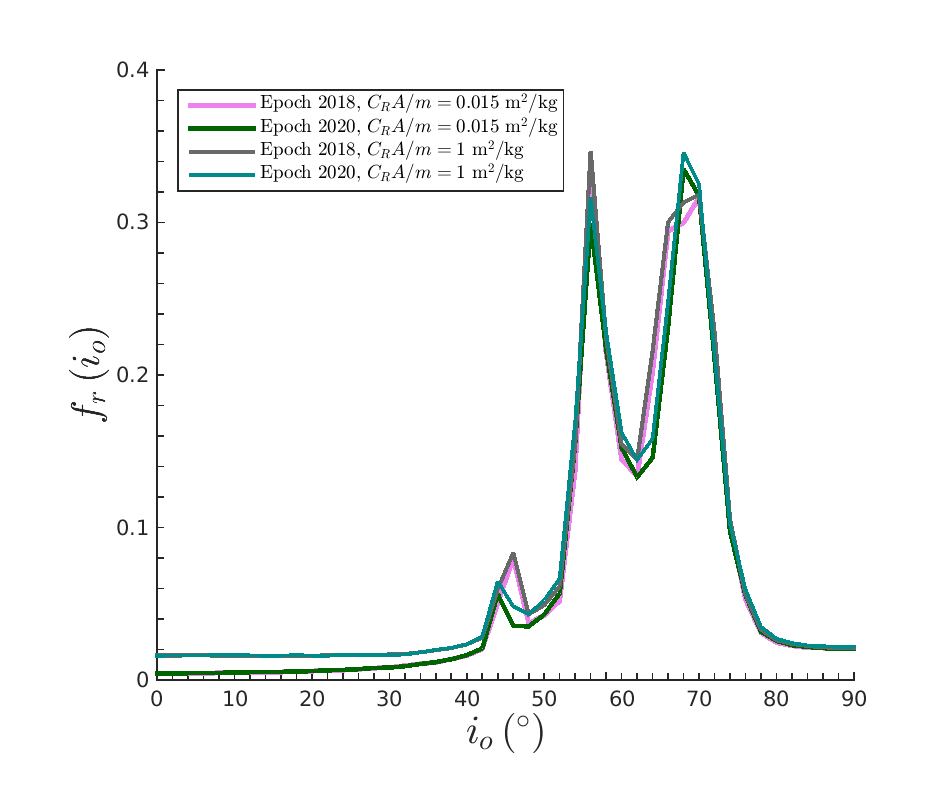}
    \end{subfigure} 
    \begin{subfigure}[b]{0.49\textwidth}
      \includegraphics[width=\textwidth]{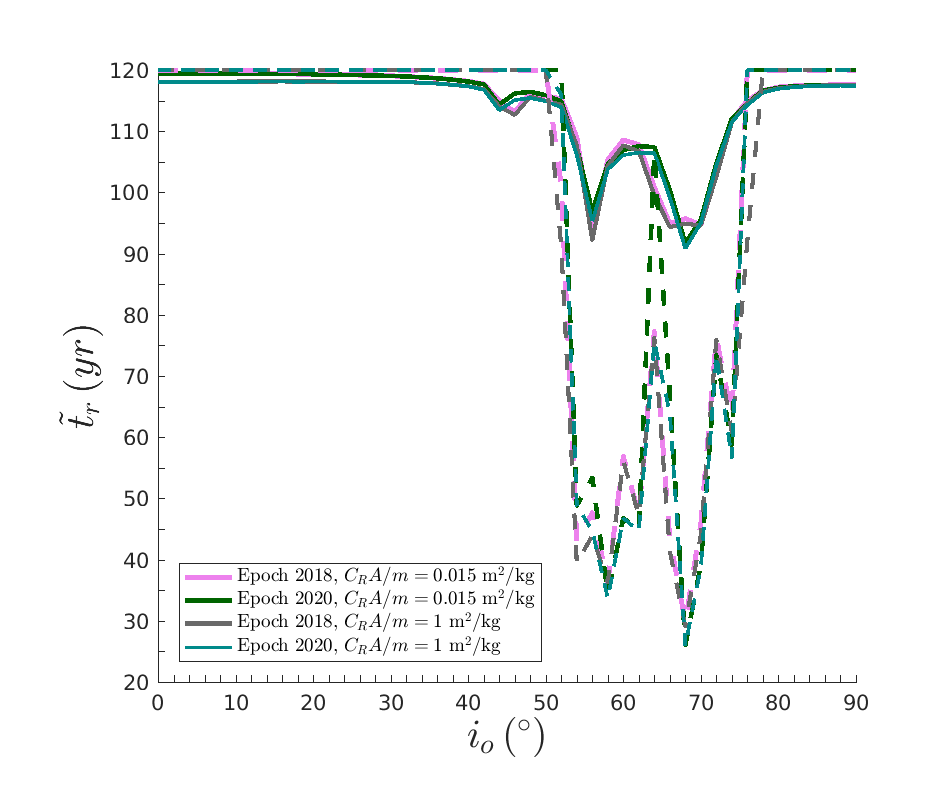}
    \end{subfigure}  
    \caption{Frequency, $f_{r}$, (left) and mean dynamical lifetime, $\tilde{t}_{r}$, (right) of the reentry population with 
    $q>R_{E}+400$km, as function of initial inclination (see text for details on their calculation).  
    Solid and dashed lines in $i_{o}-\tilde{t}_{r}$ diagram denote the mean dynamical lifetime of the entire reentry population and 
    the minimum dynamical lifetime of the reentry population with initial $e\le0.3$, respectively.
    Different colors denote different initial epoch and $C_{R}A/m$; as it is indicated by the legend.}
    \label{fig:MEO_fr}
\end{figure}

Figures \ref{fig:MEO_inc56_srp1a}-\ref{fig:MEO_inc64_ep20b} show a subset of our results for $i_{o}=56^{\circ}$ and $64^{\circ}$. 
In the \ref{app:1}, a subset of our results for various inclinations is also shown. The results of our complete atlas, for all 
inclinations, both epochs and both $C_{R}A/m$ values, can be found at the ReDSHIFT website \footnote{http://redshift-h2020.eu/results/}. 
Figure \ref{fig:MEO_fr} shows (on the left) the frequency $f_r$ of reentry solutions with $q>R_{E}+400$km, over the whole initial 
grid, and (on the right), the mean dynamical lifetime $\tilde{t}_{r}$ of the reentry population (solid lines) and 
the minimum lifetime of the reentry population with $e_0<0.3$ (dashed lines), as functions of the initial inclination, $i_{o}$. \\

For low to moderate inclinations (up to $\sim40^{\circ}$) as well as for high inclinations ($>\sim80^{\circ}$), the structure of 
the maps is quite smooth and very few reentry solutions can be found, even after $120$~yr. The typical values are $f_{r}<0.04$ and 
$\tilde{t}_{r}>115$~yr. Note that, at such inclinations, there is practically no strong secular resonances, and hence no strong 
instabilities \citep{aR19}. For inclinations between $\sim40^{\circ}$ and $\sim80^{\circ}$ the structure is more 
complicated. Figure \ref{fig:MEO_fr} reveals three distinct inclination regions where $f_{r}$ shows relative maxima, at 
$i_{o}=$46$^{\circ}$, 56$^{\circ}$, and 68$^{\circ}$. Note that these curves show `angle-averaged' results (i.e., all values of $\Omega$ and $\omega$ are combined for each $i_{o}$). Numerous studies in the past decade or so 
\citep{cC00,aJrG05,aR08,jD15,aR15,eA16,aCcG16,iGetal16,aR17} have highlighted the importance of lunisolar secular
resonances near GNSS inclinations. Resonances lead to eccentricity growth (decrease of perigee distance), whereas resonance 
overlapping introduces chaos in their orbits, on top of any regular, secular excitation. Those phenomena can lead a nearly 
circular GNSS orbit to reentry on centennial timescales. 
Our results suggest that $f_r\left(i_{o}\right)$ and 
$\tilde{t}_{r}\left(i_{o}\right)$ are roughly independent on $C_{R}A/m$. 
This is strongly indicative of the absence of dynamical influence of solar (semi-secular) resonances, i.e.\ a minimal effect of 
SRP. Moreover, the curves also seem to be roughly the same, independent of the initial epoch chosen. Again, these are 'angle-
averaged' results, and this indicates that the initial choice of $\Omega$ and $\omega$ is (on average) not important on the long 
run. However, note that the effect of lunisolar resonances on $e$ and $i$ do depend on the initial values orientation of these two 
angles, but also on a `third angle' (i.e.\ epoch), namely the lunar (ecliptic) node $\left(\Omega_{M}\right)$. The two initial 
epochs chosen in this study are relatively close to each other, and hence $\Omega_{M}$ does not differ by much 
$\left(\sim30^{\circ}\right)$; in equatorial coordinates, $\Omega_{M,eq}$ is almost the same in both epochs. Hence, even if -- as 
shown in our dynamical maps -- the results can differ substantially for the two epochs, we cannot 
really conclude that Figure \ref{fig:MEO_fr} is 'epoch-invariant'. In fact, integrating for more (and more distant) epochs could reveal a near-
periodic variation of $f_{r}$ and/or $\tilde{t}_{r}$, but we expect that the central values of the peaks will be roughly 
preserved, as they represent the location (in $i$) of known lunisolar resonances. 
Another feature, noticeable in the $De$ maps, is a thin, vertical band at the location of the 2:1 commensurability with 
the Earth's rotation rate ($a\approx 0.63~a_{GEO}$), where the variation of $De$ may appear as `discontinuous' with respect to the 
neighboring values of $a$. This resonance leads to coupled oscillations in $a$ and $e$, but is quite narrow in $a$ so that, at our 
grid resolution, it has the width of only once cell. Depending on the choice of orientation angles, the eccentricities of resonant 
particles will remain low, if they are close to the stable equilibrium point of the resoannce, characterised by a value of the resonant 
angle $\sigma=\lambda-2\lambda_{E}+\varpi\approx0$, or will increase significantly, if $\sigma\approx\pi$ (the unstable fixed point of 
the resonance). However, the overall dynamical structure of the neighborhood is not severely affected.  \\

According to our results, for inclinations near the nominal values for the the GPS, BEIDOU and GALILEO groups and for low 
$C_{R}A/m$, $\sim 35\%$ of our initial conditions can re-enter, with a mean dynamical lifetime of $\sim90$~yr. There is, though, a 
strong dependence on the orientation of the secular angles. Nevertheless, reentry orbits with lifetimes $\sim40$~yr can occur even 
for eccentricities smaller than $\sim0.15$. For inclinations near the GLONASS nominal value, only $\sim15\%$ of the initial 
conditions reenter, with a mean dynamical lifetime of $\sim90$~yr. Despite the fact that the reentry particles with initial $e<0.3$ and 
dynamical lifetimes of $\sim30$~yr exist, disposal dynamical hatches appear generally at higher eccentricities, which could make 
it much harder to actually use as disposal strategy. Note that $De$ for non-escaping orbits that are found close to reentry 
hatches is $De>0.8$, which indicates a significant increase of eccentricity and, hence, possible reentry at times somehow longer 
than $120$~years. When an augmented $C_{R}A/m$ is used, the reentry regions widen and the lifetimes decrease by a few years on 
average at all inclinations, but the overall structure of the maps is preserved.  \\

\subsection{GNSS-graveyard study}
\label{subsec:GNSS}

This part of our study was performed assuming $C_{R}A/m=0.015$ m$^2$/kg only. The grid of initial conditions that remain within the 
defined graveyard regions for a timespan of $200$~yr is presented in each dynamical map. Again, each $\left(a,e\right)$ map corresponds to a given 
set of $\left(\Omega,\omega\right)$, epoch and $i$ value. Color-coded is the maximum eccentricity that the orbit reaches during the 200 years 
of evolution.  \\

\begin{figure}[htp!]
  \centering
    \begin{subfigure}[b]{0.67\textwidth}
      \caption{$\bm{\Delta}\bm{\Omega} = {\bf 0}$, $\bm{\Delta}\bm{\omega} = {\bf 0}$}
      \includegraphics[width=.49\textwidth]{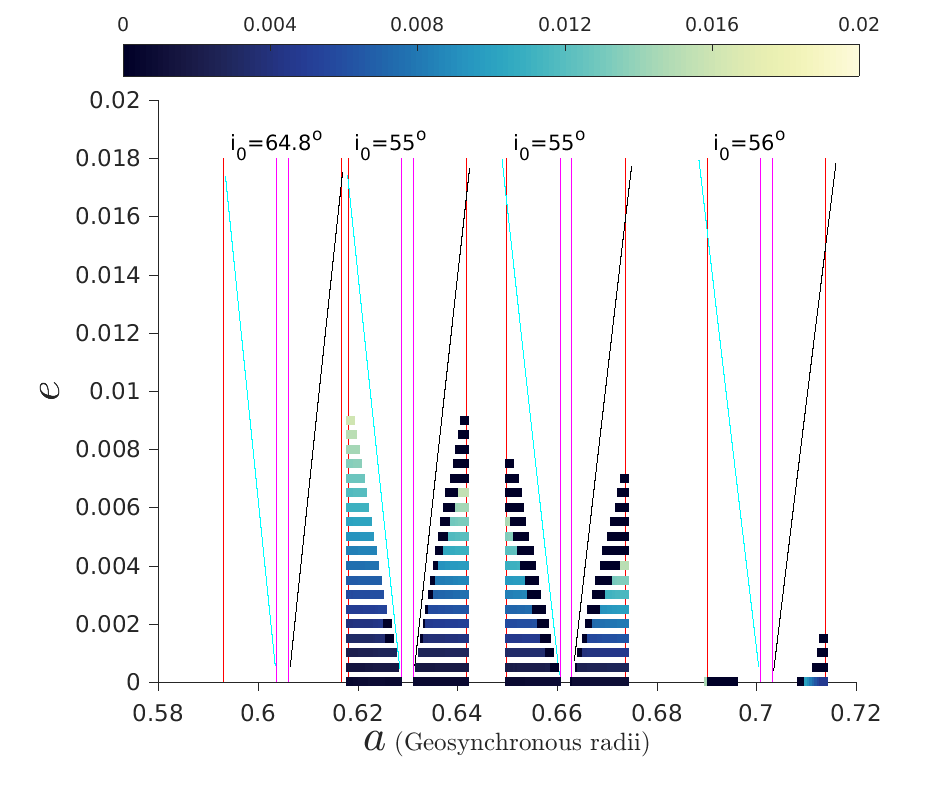} 
      \includegraphics[width=.49\textwidth]{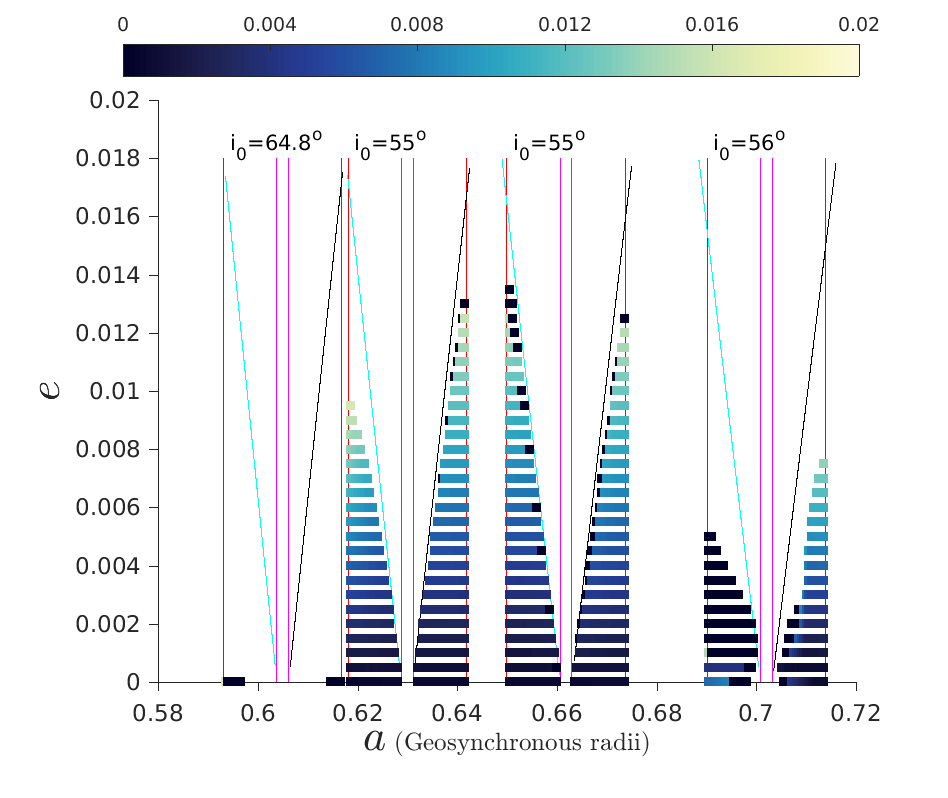}
    \end{subfigure}  
    \begin{subfigure}[b]{0.67\textwidth}
      \caption{$\bm{\Delta}\bm{\Omega} = {\bf 0}$, $\bm{\Delta}\bm{\omega} = {\bf 90^\circ}$}
      \includegraphics[width=.49\textwidth]{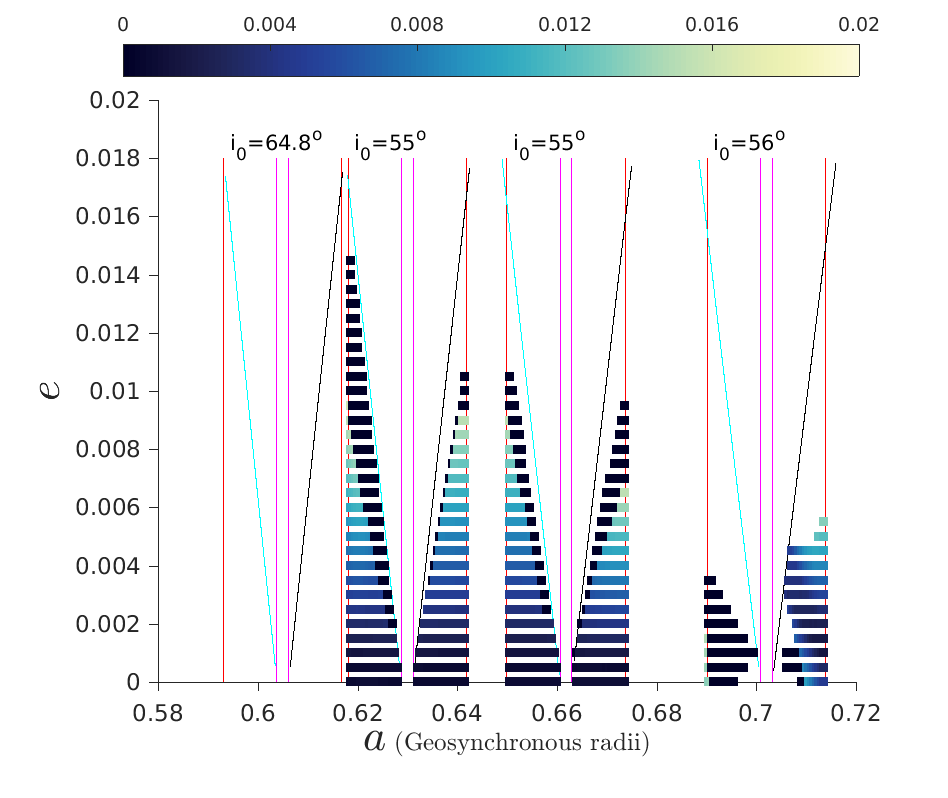} 
      \includegraphics[width=.49\textwidth]{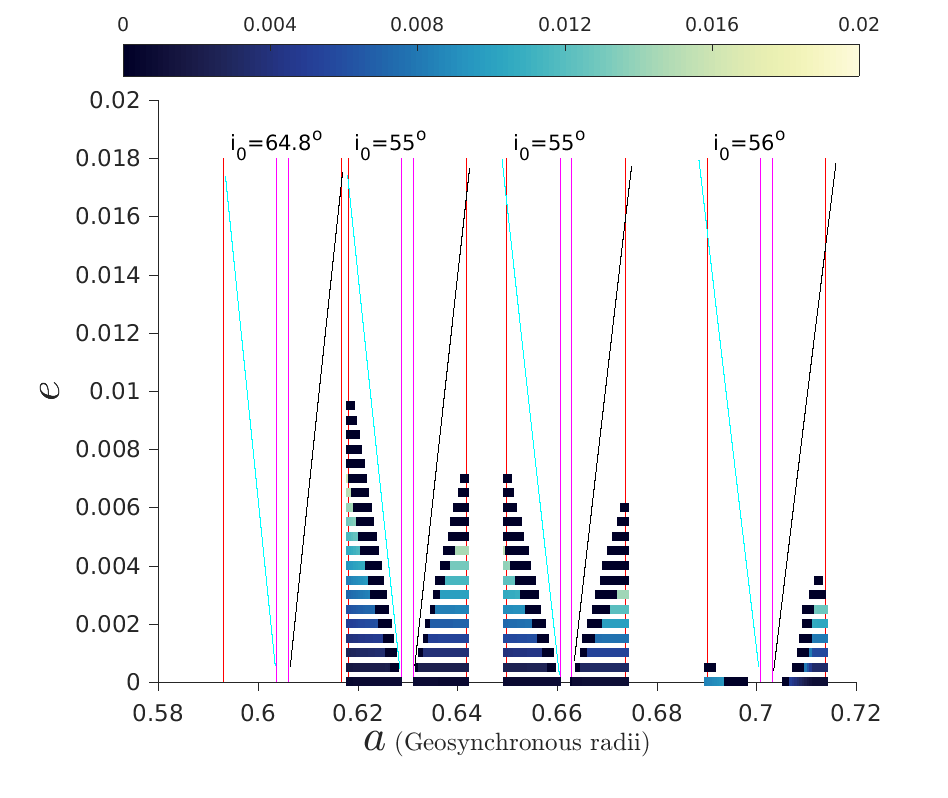}
    \end{subfigure} 
    \begin{subfigure}[b]{0.67\textwidth}
      \caption{$\bm{\Delta}\bm{\Omega} = {\bf 90^\circ}$, $\bm{\Delta}\bm{\omega} = {\bf 0}$}
      \includegraphics[width=.49\textwidth]{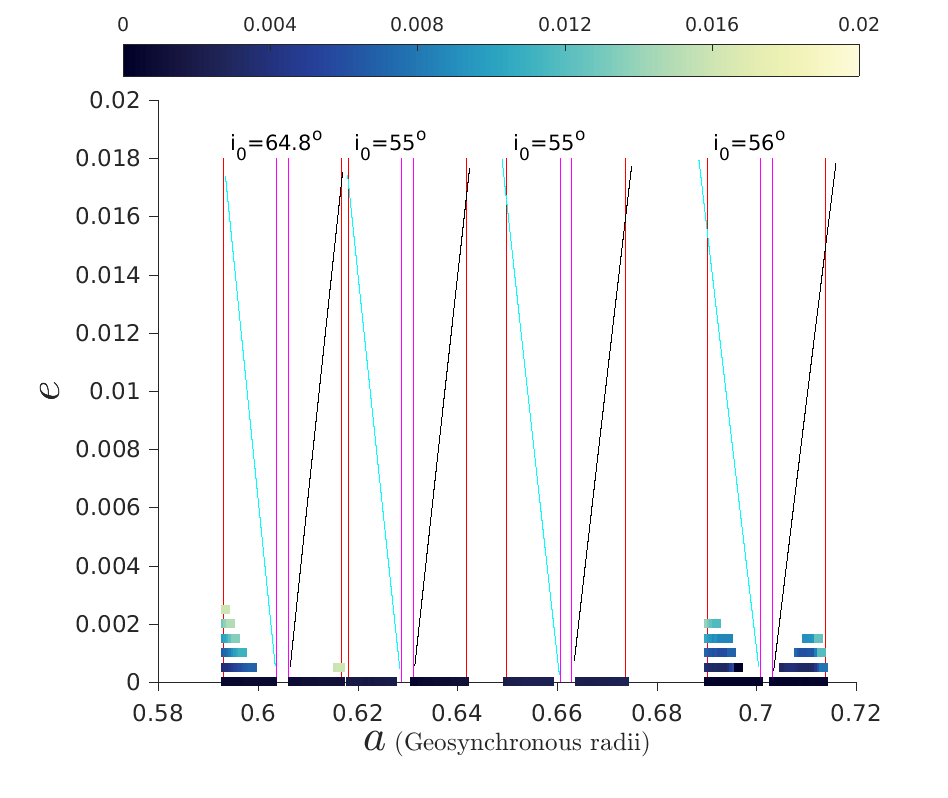} 
      \includegraphics[width=.49\textwidth]{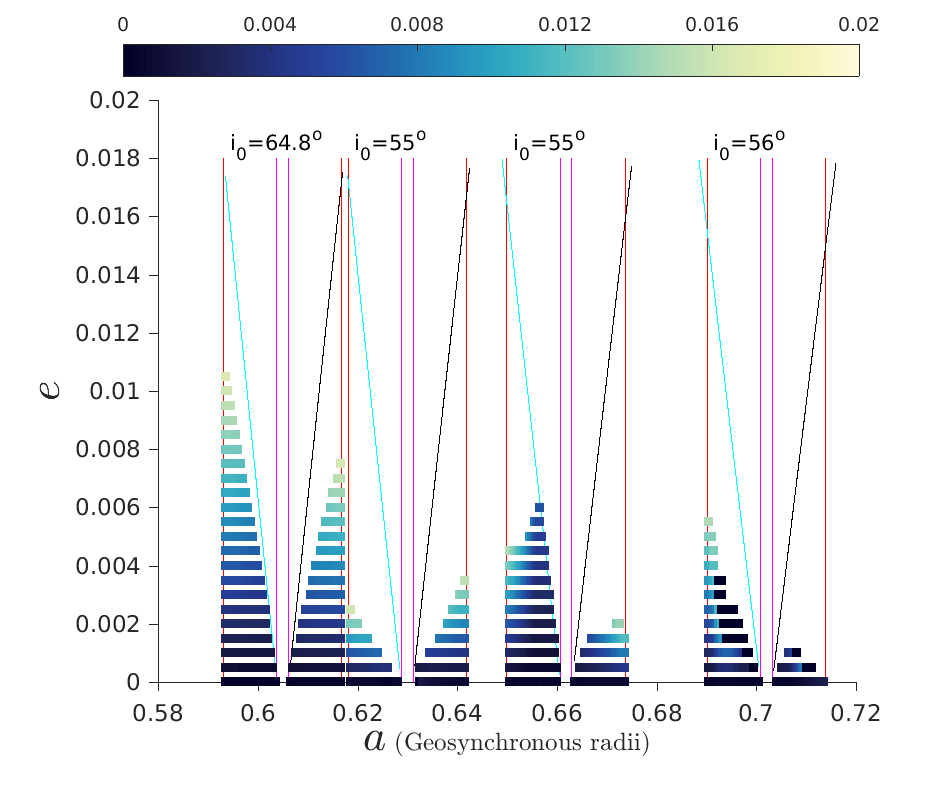}
    \end{subfigure}  
    \begin{subfigure}[b]{0.67\textwidth}
      \caption{$\bm{\Delta}\bm{\Omega} = {\bf 90^\circ}$, $\bm{\Delta}\bm{\omega} = {\bf 90^\circ}$}
      \includegraphics[width=.49\textwidth]{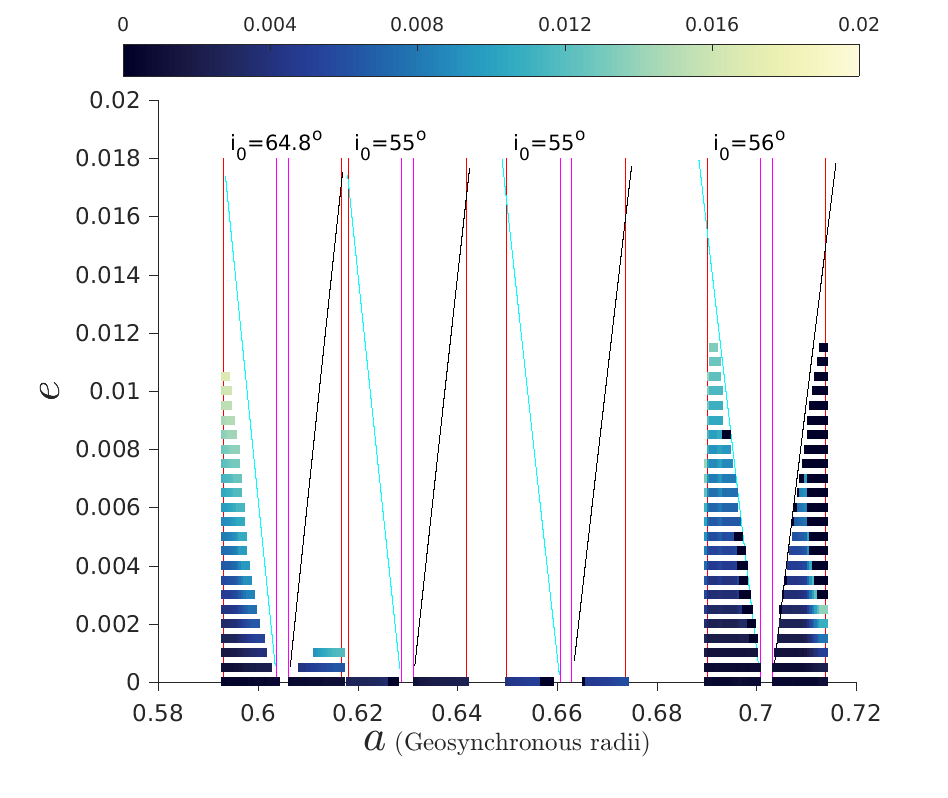} 
      \includegraphics[width=.49\textwidth]{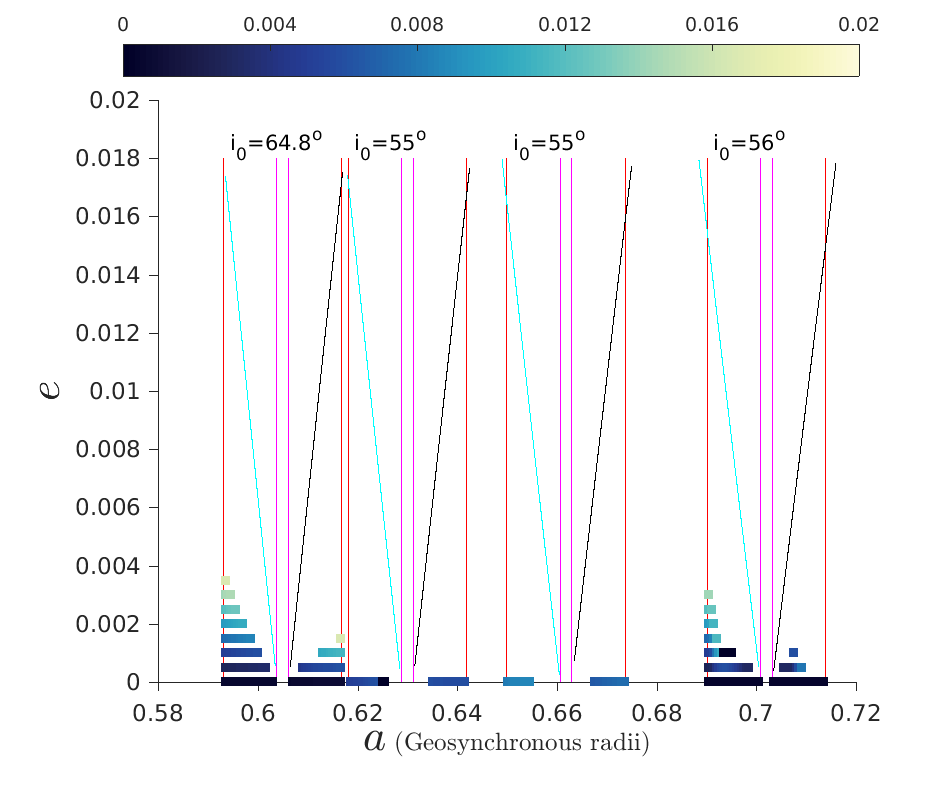}
    \end{subfigure}     
  \caption{Maximum eccentricity maps of the \textit{GNSS-graveyard} phase space for $\bm{i_{o}} = {\bf i_{nom}}$,  
  for Epoch 2018 (left) and Epoch 2020 (right), and for $C_{R}A/m=0.015$ m$^2$/kg. 
  $i_{nom}=64.8^{\circ}$ for GLONASS, $55^{\circ}$ for GPS and BEIDOU, and $56^{\circ}$ for GALILEO. 
  The colorbar for maximum eccentricity maps is from 0 to 0.02.}
  \label{fig:GRAV_srp1a}
\end{figure}

\begin{figure}[htp!]
  \centering
    \begin{subfigure}[b]{0.67\textwidth}
      \caption{$\bm{\Delta}\bm{\Omega} = {\bf 180^\circ}$, $\bm{\Delta}\bm{\omega} = {\bf 0}$}
      \includegraphics[width=.49\textwidth]{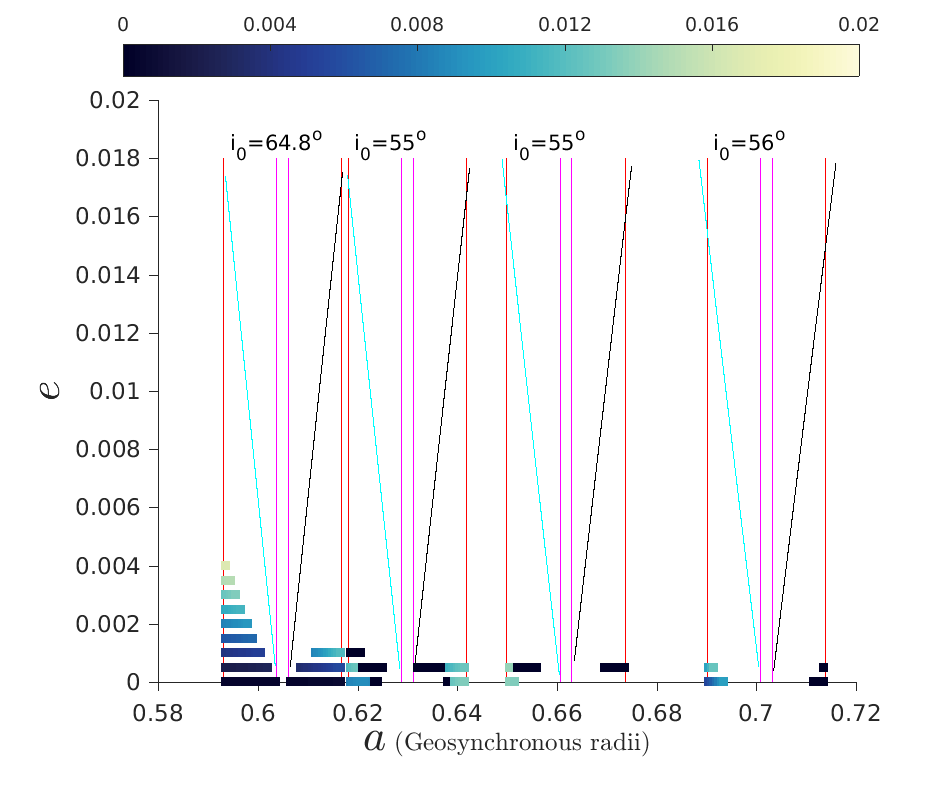}
      \includegraphics[width=.49\textwidth]{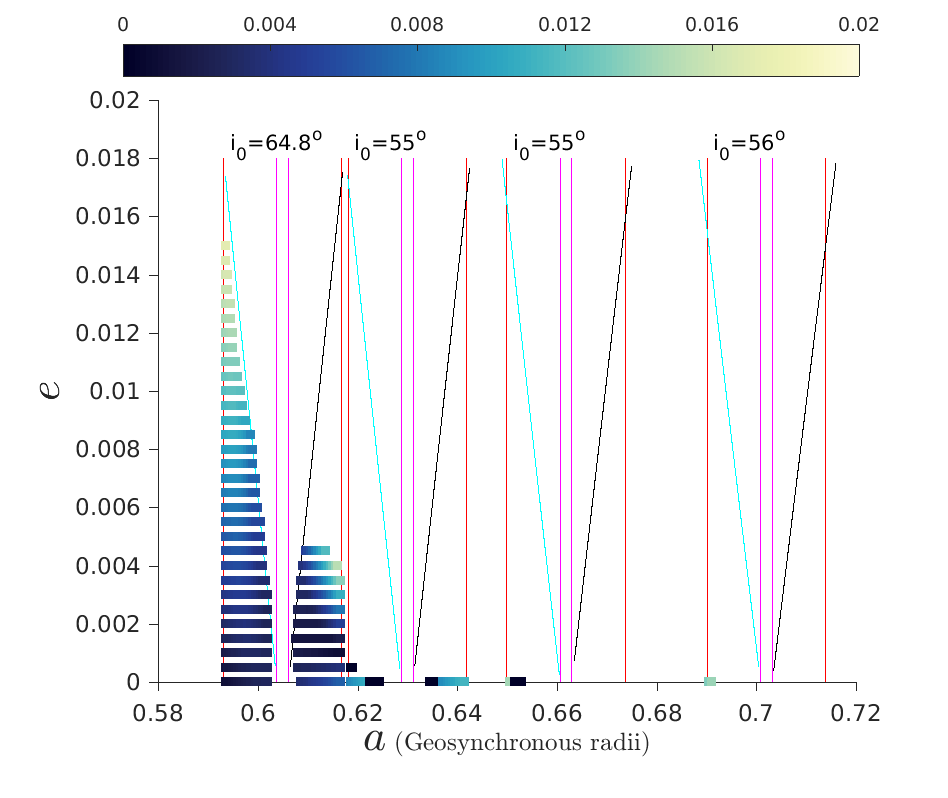}
    \end{subfigure}     
    \begin{subfigure}[b]{0.67\textwidth}
      \caption{$\bm{\Delta}\bm{\Omega} = {\bf 180^\circ}$, $\bm{\Delta}\bm{\omega} = {\bf 90^\circ}$}
      \includegraphics[width=.49\textwidth]{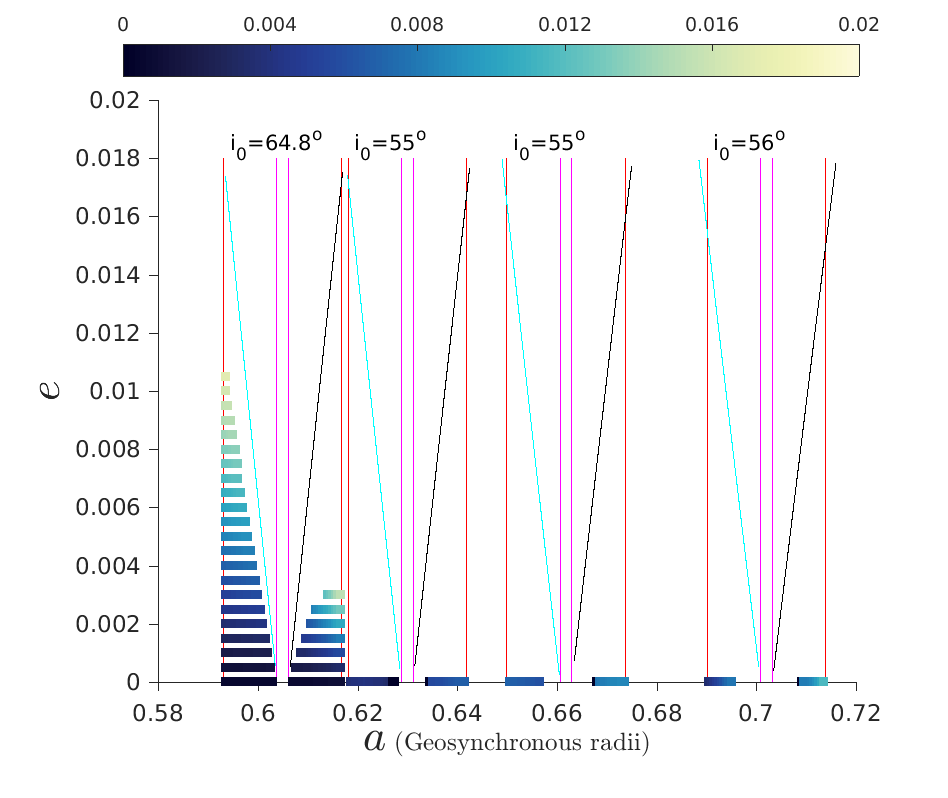}
      \includegraphics[width=.49\textwidth]{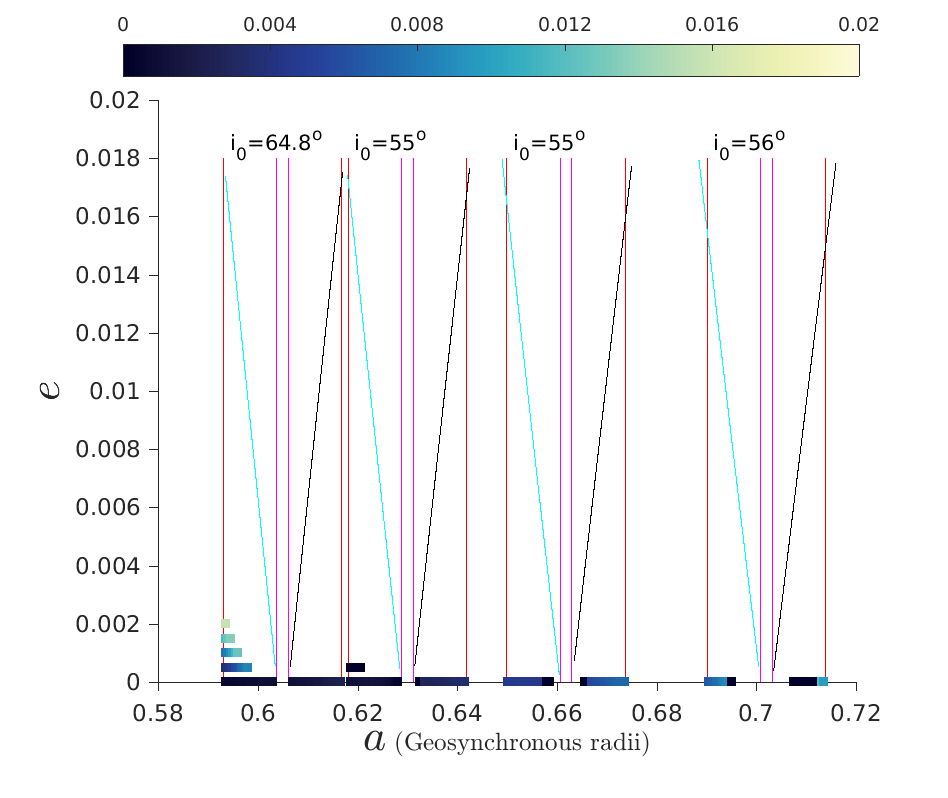}
    \end{subfigure}
    \begin{subfigure}[b]{0.67\textwidth}
      \caption{$\bm{\Delta}\bm{\Omega} = {\bf 270^\circ}$, $\bm{\Delta}\bm{\omega} = {\bf 0}$}
      \includegraphics[width=.49\textwidth]{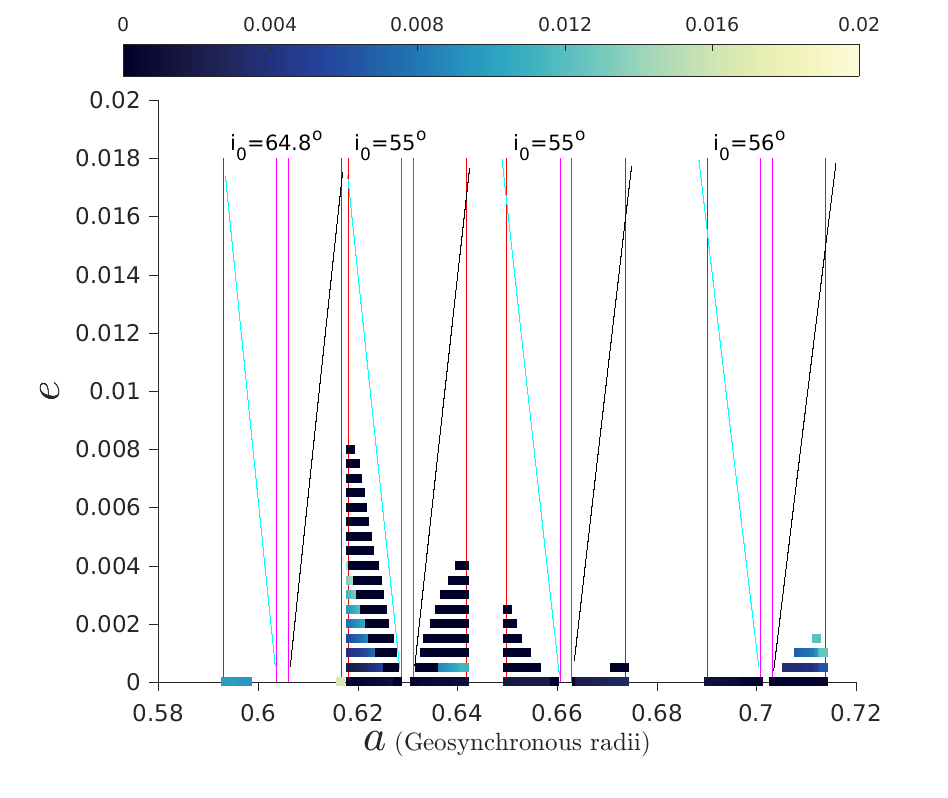}
      \includegraphics[width=.49\textwidth]{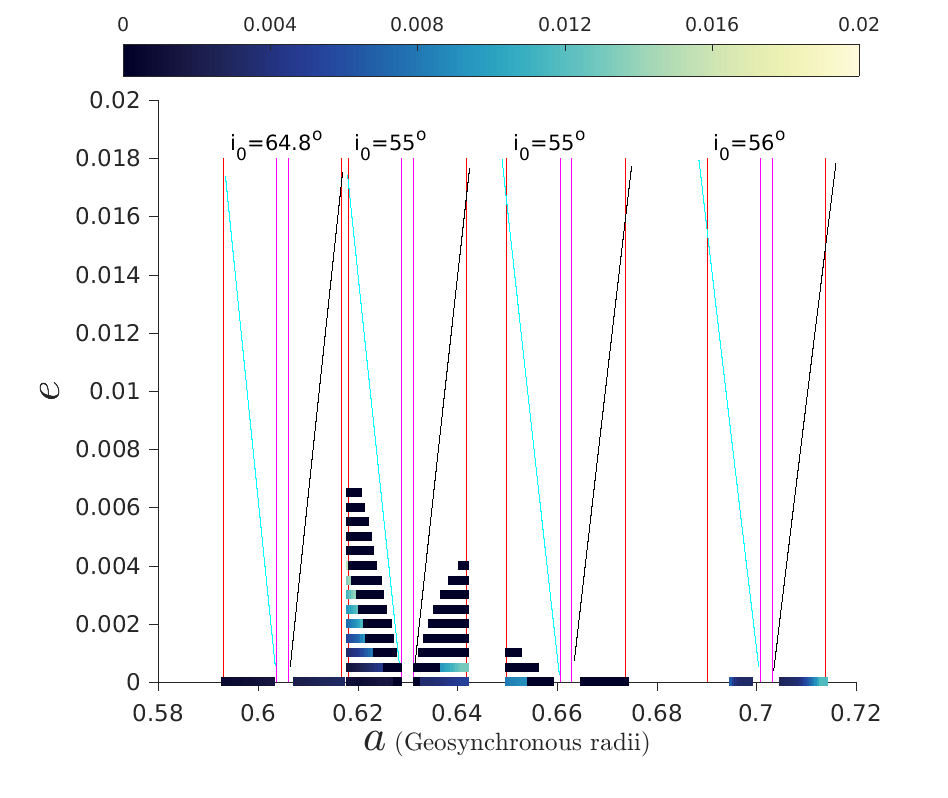}
    \end{subfigure} 
    \begin{subfigure}[b]{0.67\textwidth}
      \caption{$\bm{\Delta}\bm{\Omega} = {\bf 270^\circ}$, $\bm{\Delta}\bm{\omega} = {\bf 90^\circ}$}
      \includegraphics[width=.49\textwidth]{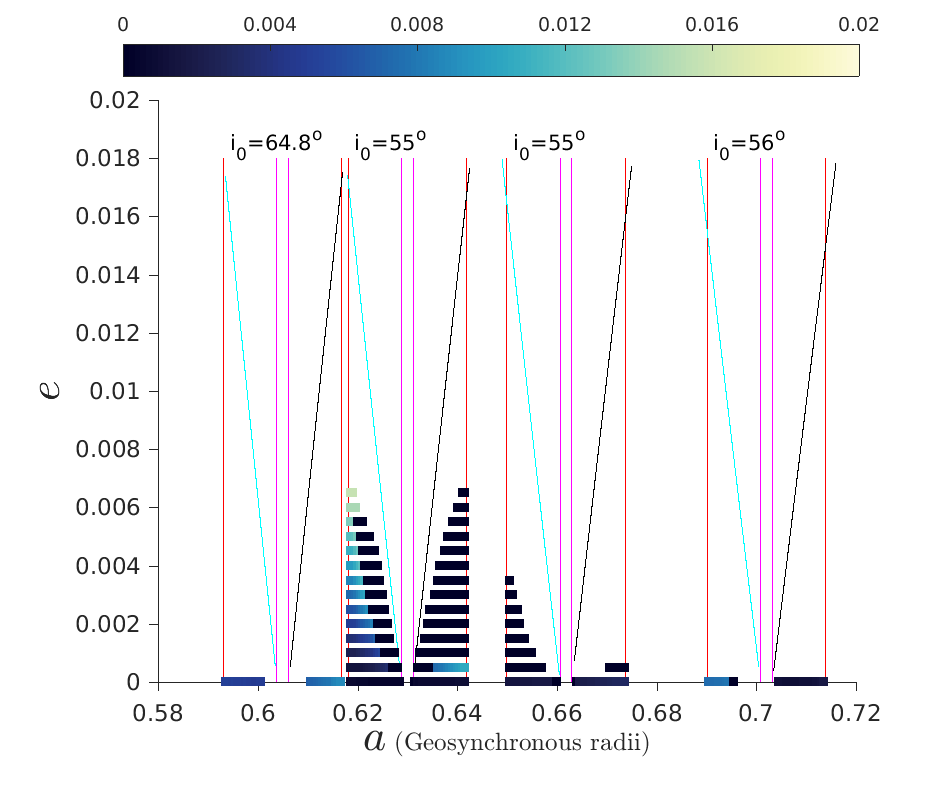}
      \includegraphics[width=.49\textwidth]{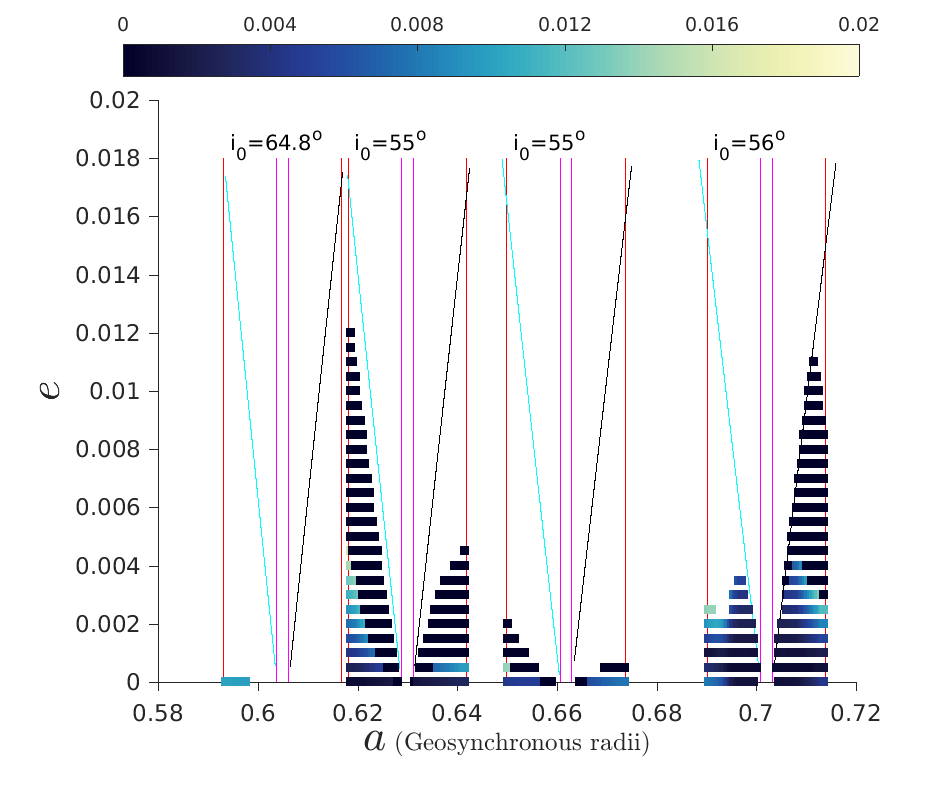}
    \end{subfigure}    
  \caption{Maximum eccentricity maps of the \textit{GNSS-graveyard} phase space for $\bm{i_{o}} = {\bf i_{nom}}$,  
  for Epoch 2018 (left) and Epoch 2020 (right), and for $C_{R}A/m=0.015$ m$^2$/kg. 
  $i_{nom}=64.8^{\circ}$ for GLONASS, $55^{\circ}$ for GPS and BEIDOU, and $56^{\circ}$ for GALILEO. 
  The colorbar for maximum eccentricity maps is from 0 to 0.02.}
  \label{fig:GRAV_srp1b}
\end{figure}

Figures \ref{fig:GRAV_srp1a}-\ref{fig:GRAV_srp1b} show a subset of our results; more results can be found in \ref{app:2}. 
The number of test-orbits that 
remain in the defined graveyard regions for $200$~yr depends on the initial secular angle configuration and epoch. It is 
obvious from the dynamical maps that eccentricity of surviving graveyards does not increase more than $\sim 0.02$; otherwise the 
orbit would violate our definition of graveyard. However, the structure of the maps is not entirely smooth. Table 
\ref{tab:stats_grav} shows the mean values of the fraction of the initial population that constitute acceptable graveyards over 
all combinations of $\left(\Omega,\omega\right)$ studied. Overall, $\sim20-40\%$ of the initial test population constitute 
acceptable graveyards. Most of the particles with initial $e<0.001$ have survived for $200$~yr time-span, which is in consistent  
with the results shown in \cite{jR15}.  \\

\begin{table}[h!]
	\captionsetup{justification=justified}
	\centering
\caption{Percentage of the initial population that remains within the defined graveyard region for $200$~yr. Mean values, $m_{i}$, 
for $i_{nom}$, and 
$i_{nom}\pm0.5$ over the chosen set of $\left(\Omega,\omega\right)$ configurations.}
\label{tab:stats_grav}
 \begin{tabular}{lccc}
 \hline\noalign{\smallskip}
         &  $m_{i_{nom}-0.5^{\circ}}$ & $m_{i_{nom}}$ & $m_{i_{nom}+0.5^{\circ}}$\\
 GLONASS &  $ 3.83$ & $20.80$ & $40.79$\\
 GPS     &  $30.78$ & $34.21$ & $36.27$\\ 
 BEIDOU  &  $20.74$ & $25.49$ & $27.67$\\ 
 GALILEO &  $20.78$ & $31.76$ & $35.02$\\
 \hline\noalign{\smallskip}
 \end{tabular}
\end{table}

%-------------------------------------------------------------------------------------------------------------------------------------- 
%          DISPOSAL MANEUVERS
%-------------------------------------------------------------------------------------------------------------------------------------- 
\section{Use of the Dynamical Atlas for satellite disposal strategies}
\label{sec:maneuv}

The dynamical study presented in Section \ref{sec:3} provides useful information for the design of satellite disposal strategies. 
In the extended MEO/GNSS region studied here, some interesting reentry hatches appear even at low eccentricities, and orbits 
initially placed there could be used for the reentry of operational satellites when they become inactive. Additionally, we found 
numerous stable graveyard solutions for $200$~yr, which also can be used for clearing the operational GNSS regions. However, it is 
practically impossible for near circular orbits to reach those regions without some active assistance (i.e.,\ only by natural 
dynamics), even after very long timescales. Moreover, the possibility of a non-operational satellite to cause problems in the 
remaining operational ones decreases, if it is  removed fast from its operational region after the end of its mission. Hence, an 
appropriate disposal strategy is required utilizing $\Delta V$ maneuvers.  \\

Near-optimal transfer orbits can be found, by requiring low $\Delta V$ budget and/or small lifetime (or, {\it waiting time}) of the final, 
reentry orbit. Since the 1960s, single- and multiple-impulse methods were studied \citep{fG69,jM79}. In general, with a given limiting 
$\Delta V$, it is possible to reach all orbits situated in a certain volume of the three-dimensional space of the variations 
$\Delta a$, $\Delta e$, $\Delta i$, called the \textit{reachable domain}. Optimal transfers correspond to the boundary of this  
domain. Quite involved methods for determining this boundary have been presented recently \citep{dX10,mH14}.  \\

In this study we do not seek optimal transfers in the strict sense defined above. Instead, we are interested in finding 
the best reentry and/or graveyard solution among our pre-computed evolutions, given a {\it starting orbit}. As our grid of initial 
conditions is considerably dense in $(a,e,i)$ but sparse in $\Omega,\omega$, we limit our search to co-planar transfers, which 
means we are looking in our maps, using the same values of $i$ and $\Omega$ as for the starting orbit, but allow for changes in 
$a$, $e$, and $\omega$. For every solution found, we compute the required $\Delta V$ of a single-burn or a two-burn transfer.
\footnote{We follow the procedure described in Chapters 3.2 and 3.3 of \citet{mS97} for single-burn and two-burn (Hohmann-type) 
maneuvers, respectively.} Note that a single-burn transfer can only be performed if the starting and final orbits intersect. \\

\begin{figure}[htp!]
  \centering
    \begin{subfigure}[b]{0.45\textwidth}
      \caption{$a_{GLO}$}
      \includegraphics[width=.99\textwidth]{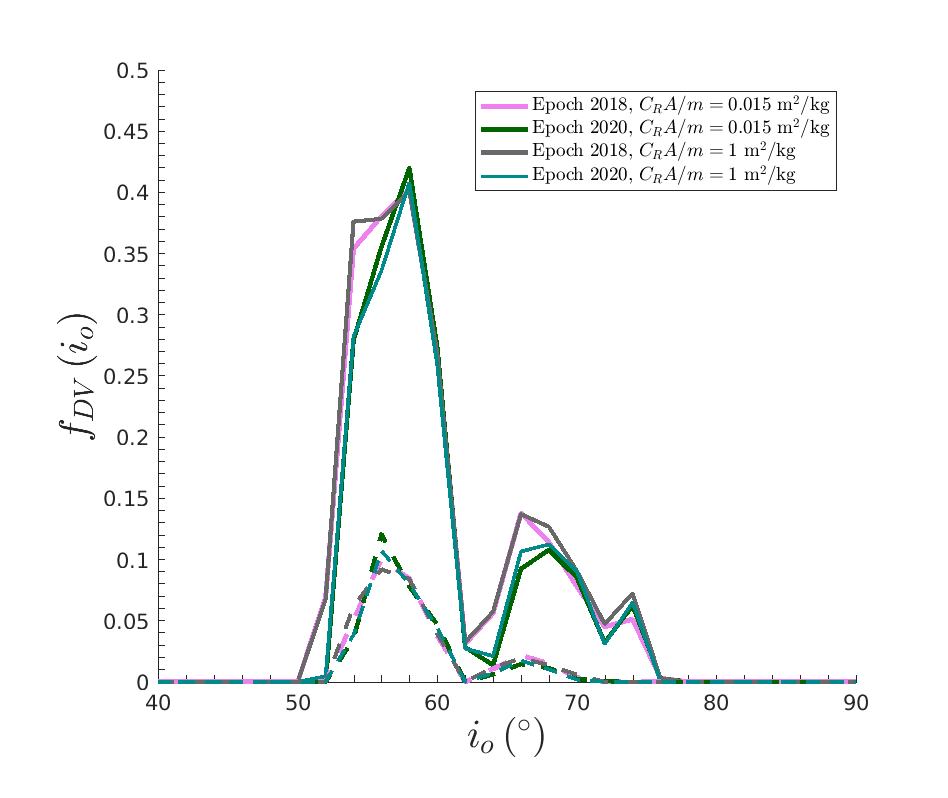}
    \end{subfigure}
     \begin{subfigure}[b]{0.45\textwidth}
      \caption{$a_{GPS}$}
      \includegraphics[width=.99\textwidth]{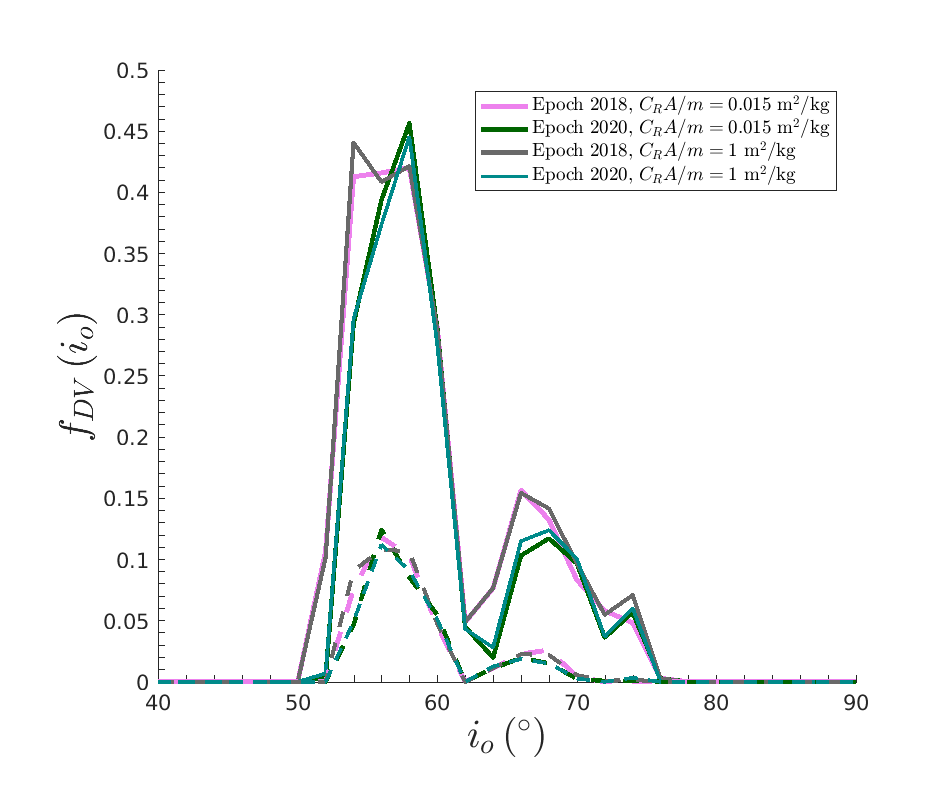}
    \end{subfigure}  
    \begin{subfigure}[b]{0.45\textwidth}
      \caption{$a_{BEI}$}
      \includegraphics[width=.99\textwidth]{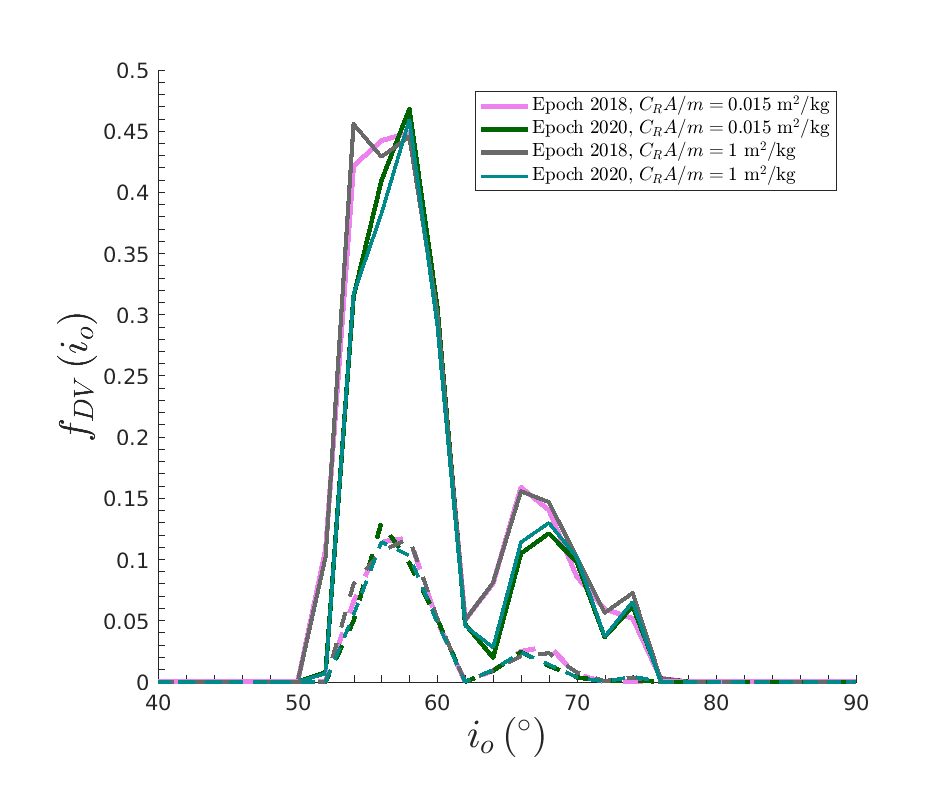}
    \end{subfigure}
    \begin{subfigure}[b]{0.45\textwidth}
      \caption{$a_{GAL}$}
      \includegraphics[width=.99\textwidth]{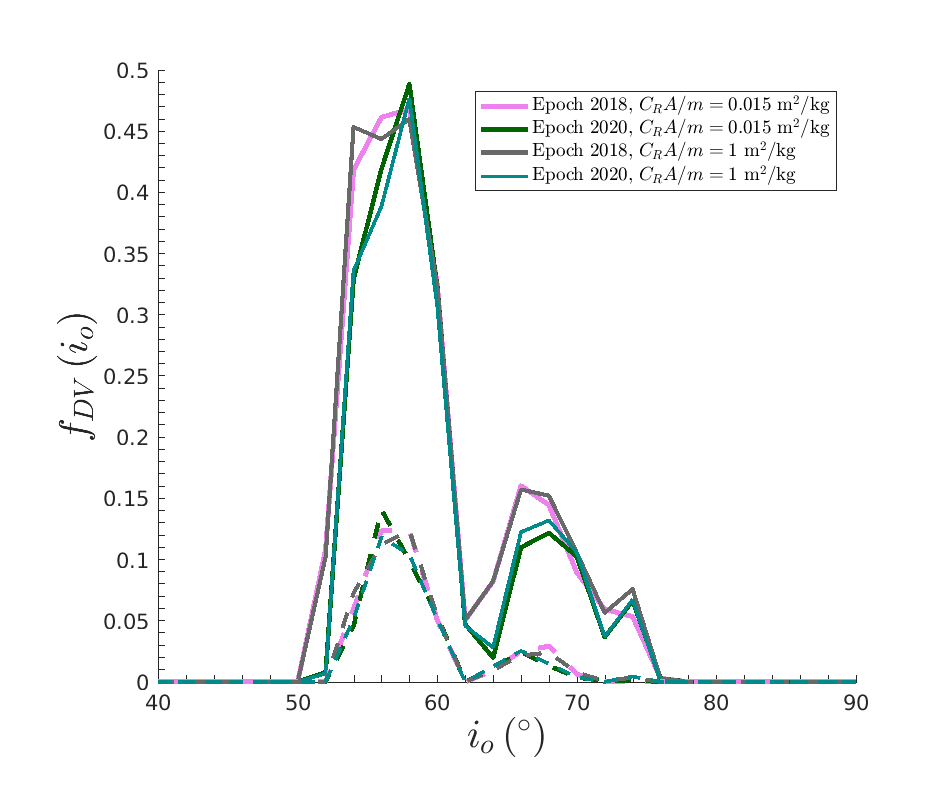}
    \end{subfigure}  
    \caption{Frequency of the reentry particles with a limited $\Delta V$, $f_{DV}$,  
    as function of initial inclination (see text for details on their calculation). The color scheme is the same as in Figure 
    \ref{fig:MEO_fr}. The dashed lines refer to an upper limit of $\Delta V=300$~m/s, whereas the solid ones refer to an upper limit of 
    $\Delta V=600$~m/s. For each figure we assumed a starting orbit with fixed $\left(a,e,i\right)$ and various values of 
    $\left(\Omega,\omega\right)$; $\left(a_{GLO},e_{nom},i_{o}\right)$ (top left), $\left(a_{GPS},e_{nom},i_{o}\right)$ (top right), 
    $\left(a_{BEI},e_{nom},i_{o}\right)$ (bottom left), $\left(a_{GAL},e_{nom},i_{o}\right)$ (bottom right).}
  \label{fig:reentry_fDV}
\end{figure}

\begin{figure}[htp!]
  \centering
    \begin{subfigure}[b]{0.45\textwidth}
      \caption{$a_{GLO}$}
      \includegraphics[width=.99\textwidth]{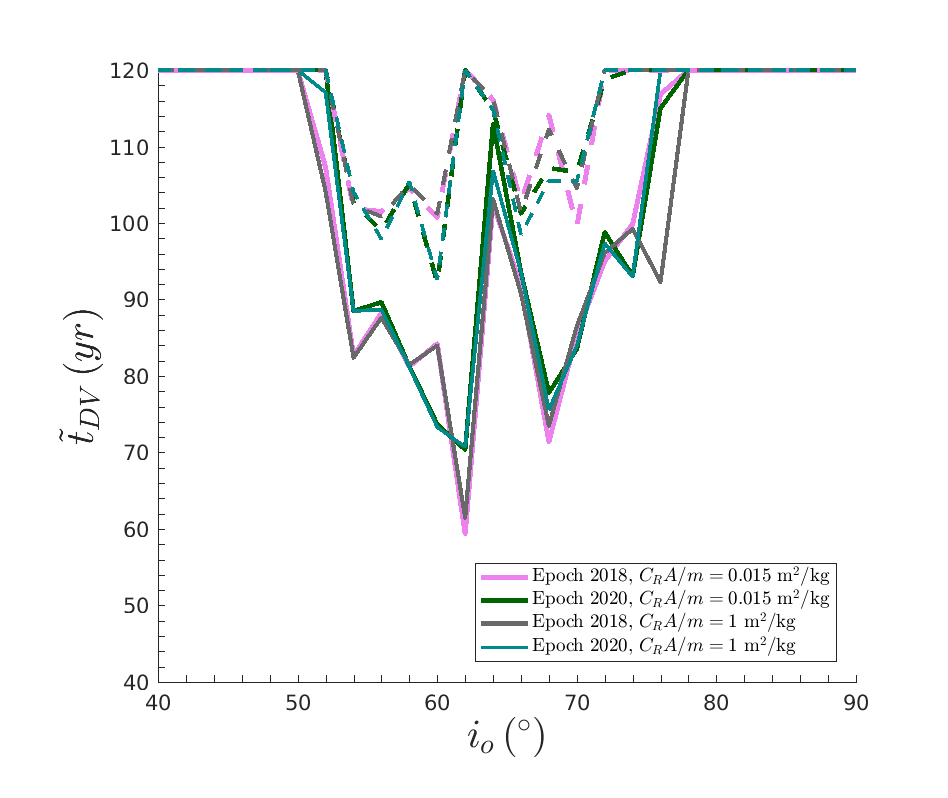}
    \end{subfigure}
     \begin{subfigure}[b]{0.45\textwidth}
      \caption{$a_{GPS}$}
      \includegraphics[width=.99\textwidth]{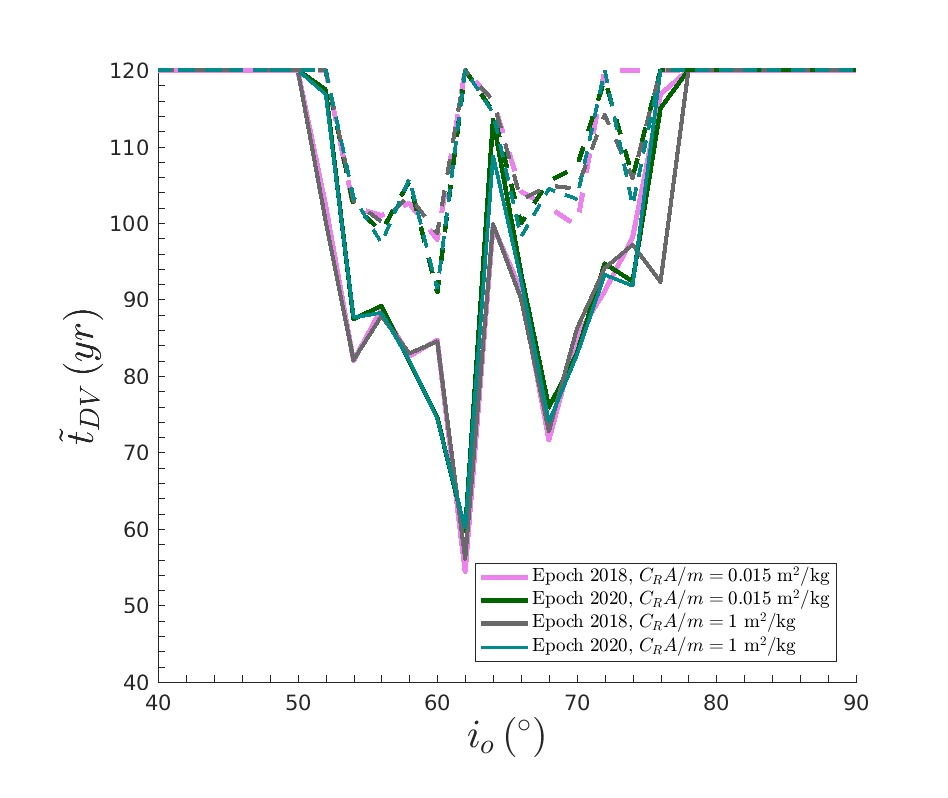}
    \end{subfigure}  
    \begin{subfigure}[b]{0.45\textwidth}
      \caption{$a_{BEI}$}
      \includegraphics[width=.99\textwidth]{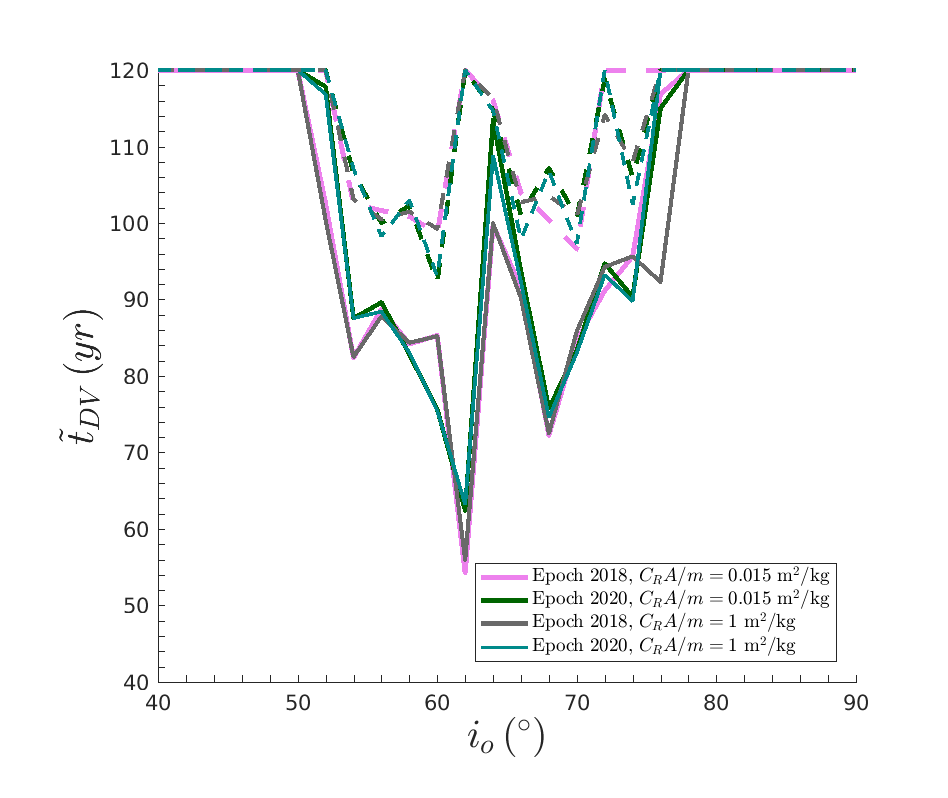}
    \end{subfigure}
    \begin{subfigure}[b]{0.45\textwidth}
      \caption{$a_{GAL}$}
      \includegraphics[width=.99\textwidth]{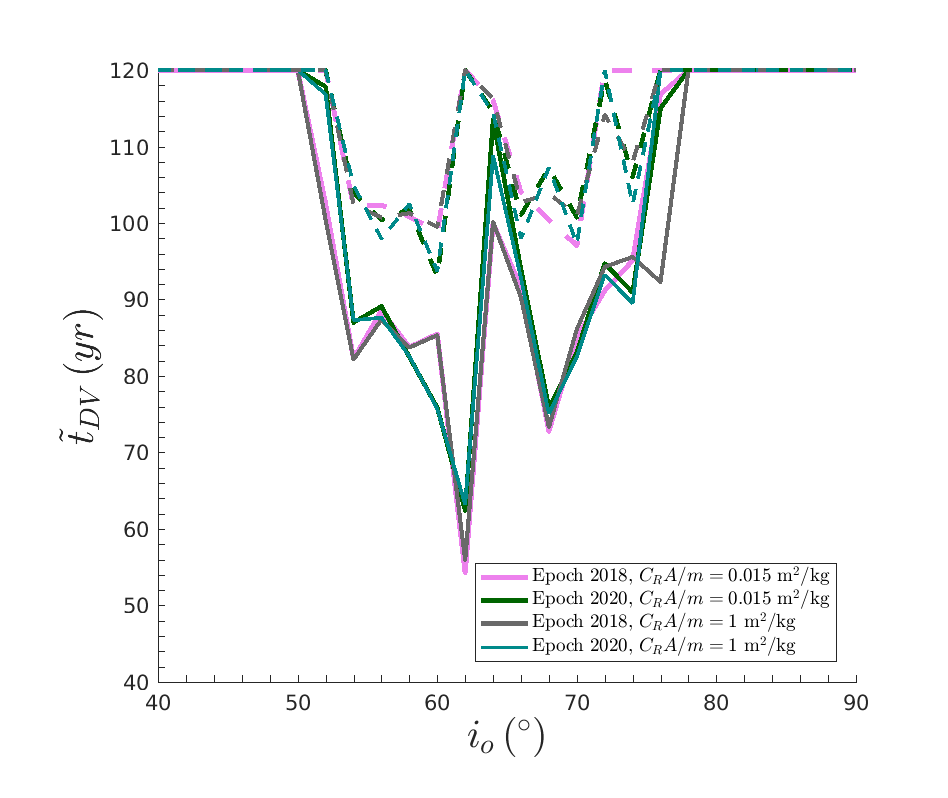}
    \end{subfigure}  
    \caption{Mean dynamical lifetime of the reentry particles with a limited $\Delta V$, $\tilde{t}_{DV}$, 
    as function of initial inclination (see text for details on their calculation). The 
    color and line scheme is the same as in Figure \ref{fig:reentry_fDV}.}
  \label{fig:reentry_tDV}
\end{figure}

We computed the required $\Delta V$ for these types of maneuvers, starting from orbits with given $\left(a,e,i\right)$ and for all 
combinations of $\left(\Omega,\omega\right)$. We chose nearly circular initial orbits ($e=10^{-4}$) with 
$a=a_{GLO},a_{GPS},a_{BEI}$ and $a_{GAL}$ and any inclination $i=i_{o}$ used in our \textit{MEO-general} grid. For each starting 
orbit we target all respective reentry solutions, as found in the \textit{MEO-general} study. Figure \ref{fig:reentry_fDV} shows 
the frequency of the reentry population that can be reached with maximum $\Delta V= 300$~m/s (dashed lines) or 
$600$~m/s (solid lines), as functions of $i_{o}$. The higher value of maximum $\Delta V$ used here is taken in 
compliance to \citet{jR15} and \citet{rAfS18}. Figure \ref{fig:reentry_tDV} shows the mean dynamical lifetime of the reentry solutions 
reached with maximum $\Delta V=300$~m/s or $\Delta V=600$~m/s, as functions of $i_{o}$. The color scheme is the same as in Figure 
\ref{fig:MEO_fr}. For low to moderate inclinations (up to $\sim45^{\circ}$) and for really high inclinations ($>\sim80^{\circ}$), 
no reentry solution can be reached from an initially quasi-circular orbit, even with $\Delta V\le600$~m/s. For inclinations around 
$56^{\circ}$, $\sim45-50\%$ of the reentry solutions, with a mean dynamical lifetime varies $\sim80-90$~yr, can be reached 
with $\Delta V\le600$~m/s. When an upper limit of $\Delta V=300$~m/s is assumed, only $\sim15\%$ of the reentry populations could 
be used for disposing an initially quasi-circular satellite, whereas the mean dynamical lifetime increases to $\sim100$~yr. For 
inclinations near $\sim64^{\circ}$, $\sim15\%$ and $\sim3\%$ of the reentry solutions can be reached with $\Delta V\le600$~m/s and 
$\Delta V\le300$~m/s, respectively, the corresponding mean dynamical lifetime being $\sim75$~yr and $\sim100$~yr. Note that, as in 
the general MEO case, the results do not vary a lot with the choice of initial epoch and $C_{R}A/m$.  \\

\begin{figure}[htp!]
  \centering
    \begin{subfigure}[b]{0.45\textwidth}
      \caption{GLONASS}
      \includegraphics[width=.99\textwidth]{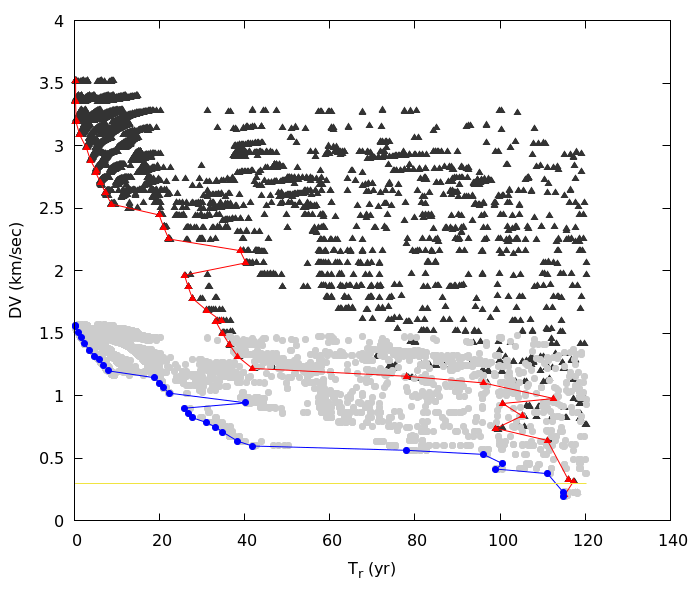}
    \end{subfigure}
     \begin{subfigure}[b]{0.45\textwidth}
      \caption{GPS}
      \includegraphics[width=.99\textwidth]{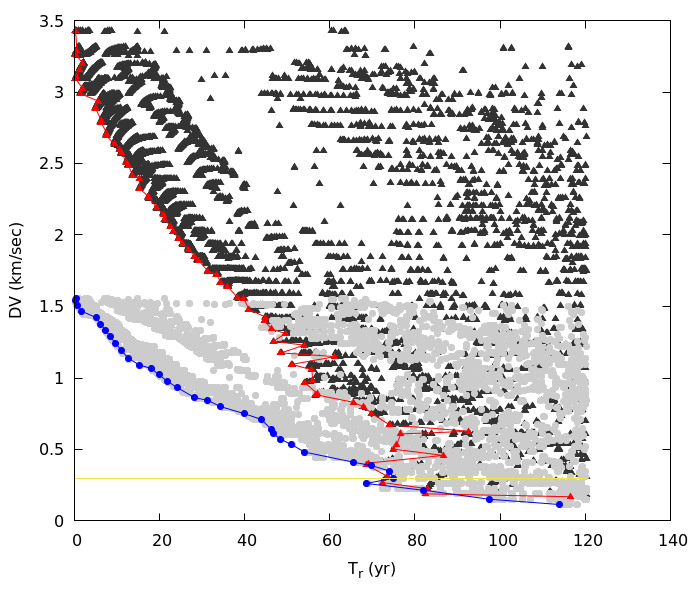}
    \end{subfigure}  
    \begin{subfigure}[b]{0.45\textwidth}
      \caption{BEIDOU}
      \includegraphics[width=.99\textwidth]{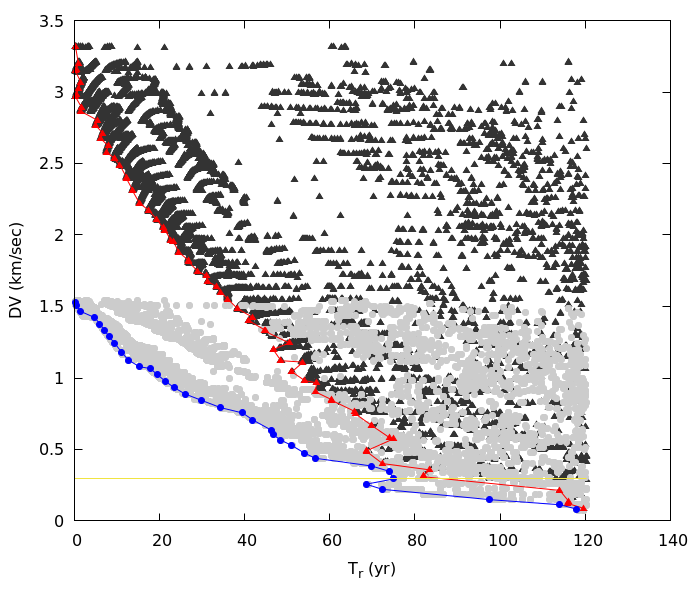}
    \end{subfigure}
    \begin{subfigure}[b]{0.45\textwidth}
      \caption{GALILEO}
      \includegraphics[width=.99\textwidth]{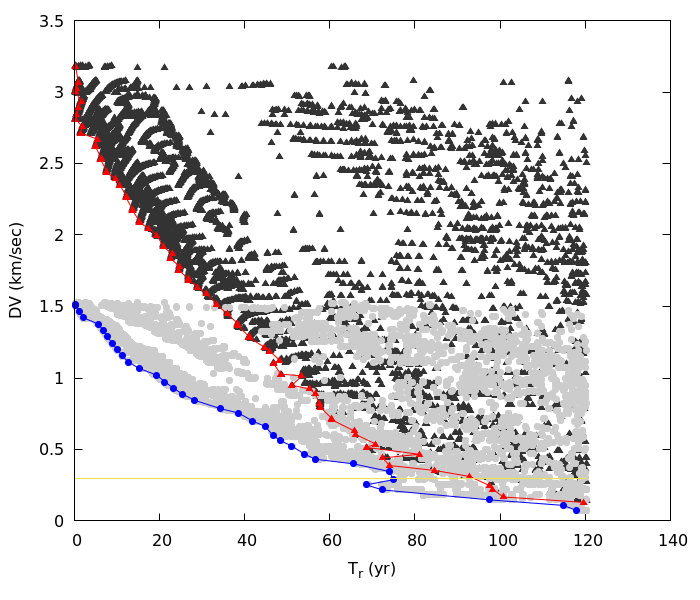}
    \end{subfigure}    
  \caption{$\Delta V$-lifetime maps for reentry solutions found around typical 
  GLONASS (top left), GPS (top right), BEIDOU (bottom left) and GALILEO (bottom right) orbits. 
  Light gray points correspond to solutions using a two-burn method (bi-elliptic transfers).
  Dark gray points correspond to solutions using a single-burn method. 
  Blue and red lines are the Pareto fronts of the two-burn and single-burn method, respectively.
  Yellow line corresponds to $\Delta V=300$~m/sec.}
  \label{fig:reentry_DV_life_graphs}
\end{figure}

\begin{figure}[htp!]
  \centering
    \begin{subfigure}[b]{0.45\textwidth}
      \caption{GLONASS}
      \includegraphics[width=.99\textwidth]{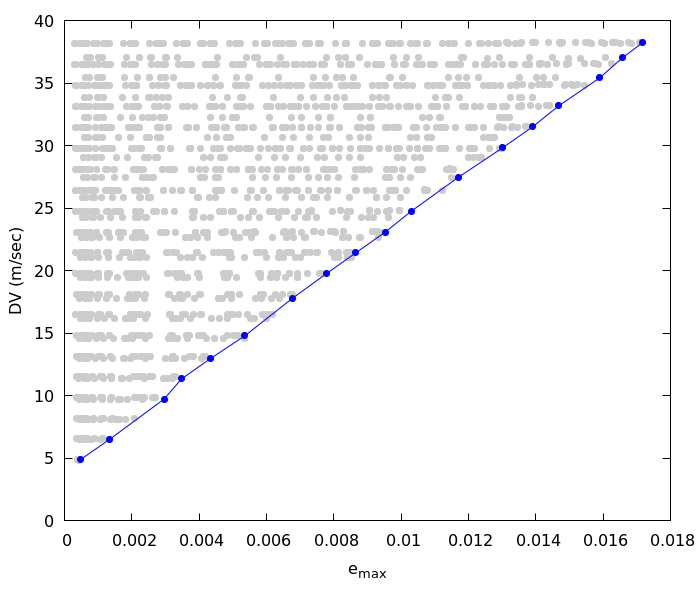}
    \end{subfigure}
     \begin{subfigure}[b]{0.45\textwidth}
      \caption{GPS}
      \includegraphics[width=.99\textwidth]{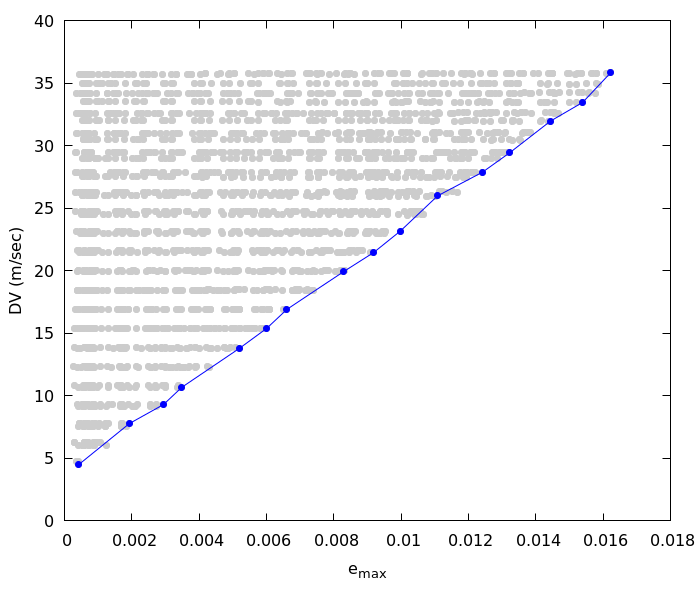}
    \end{subfigure}  
    \begin{subfigure}[b]{0.45\textwidth}
      \caption{BEIDOU}
      \includegraphics[width=.99\textwidth]{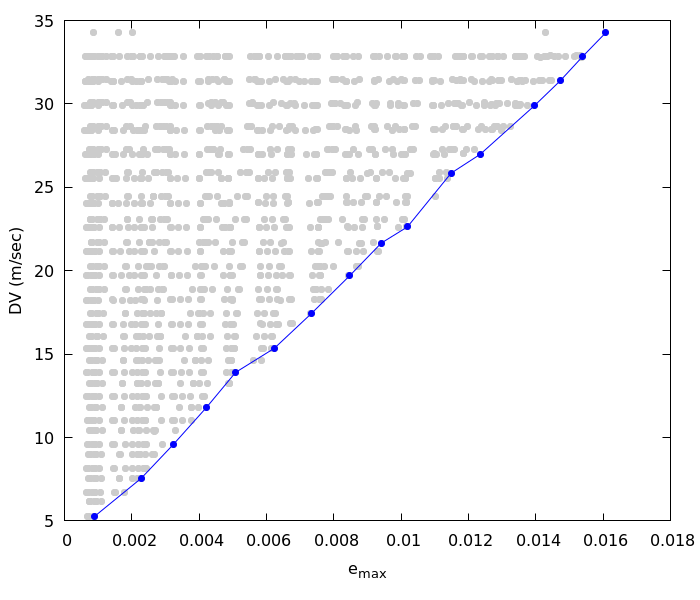}
    \end{subfigure}
    \begin{subfigure}[b]{0.45\textwidth}
      \caption{GALILEO}
      \includegraphics[width=.99\textwidth]{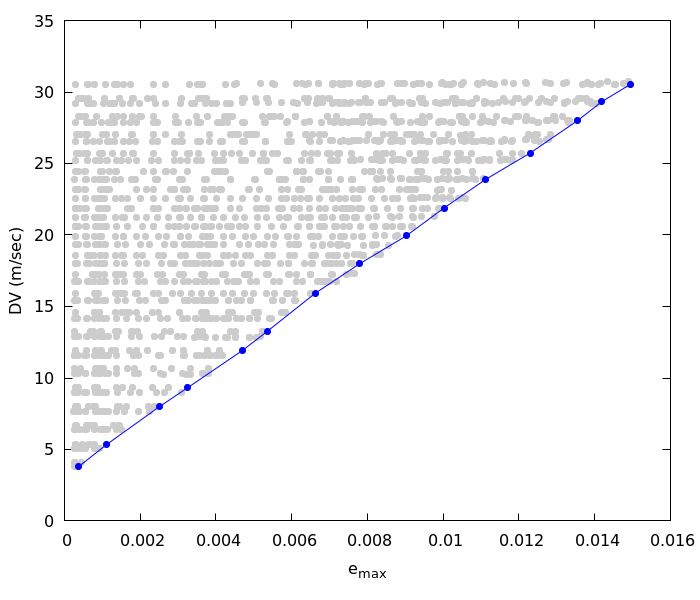}
    \end{subfigure}    
  \caption{$\Delta V-e_{max}$ maps for graveyard solutions found around typical 
  GLONASS (top left), GPS (top right), BEIDOU (bottom left) and GALILEO (bottom right) orbits  
  using a two-burn method (bi-elliptic transfer for nearly zero $e$, i.e.,\ Hohmann-like). Blue lines are the Pareto fronts of the two-burn method.}
  \label{fig:grav_DV_emax_graphs}
\end{figure}

Focusing at the GNSS constellations, we performed the same study starting from typical GNSS orbits with $e_0=10^{-4}$, and varying 
orientations. We targeted final orbits among our reentry solutions database, and graveyard solutions found in our \textit{GNSS-
graveyard} study. Given our MEO grid's resolution, when looking for reentry solutions we set the inclination of the 
starting orbit to $i_{o}=56^{\circ}$ (for GPS, BEIDOU and GALILEO) or $i_{o}=64^{\circ}$ (GLONASS).When looking for graveyard 
solutions, we set $i_{o}=i_{nom}$ for each group. Figure ~\ref{fig:reentry_DV_life_graphs} shows the $\Delta V$ required vs the 
lifetime of the reentry orbit and Figure~\ref{fig:grav_DV_emax_graphs} shows the $\Delta V$ required vs $e_{max}$ of the 
graveyard orbit\footnote{Figure \ref{fig:reentry_DV_life_graphs} shows results for Epoch 2018 and 
$A/m=0.015$~m$^2/$kg. In \citet{aR19} similar diagrams for Epoch 2020 were presented}. 
For the reentry solutions, one can see two `V-shaped' clouds of points in every graph; these correspond to single- and 
two-burn (bi-elliptic) transfers. 
The lower envelopes (red and blue curves) of these V-shaped clouds approximately constist of the respective `optimal' 
solutions with respect to lifetime, i.e.,\ approximately constitute what are known as \textit{atmospheric reentry Pareto fronts} 
\citep{rAfS18}. 
Reentry solutions with small lifetimes (of decadal order) cannot be reached due to very high cost ($\Delta V 
\geq 300$~m/sec). Note also that the cost of a single-burn transfer is twice of 
that for a two-burn transfer for lifetimes smaller than $\sim80$~yr (for GPS, BEIDOU and GALILEO) or $\sim110$~yr (for GLONASS). 
There exist reentry solutions with $\Delta V<300$~m/sec and lifetime $\sim80$~yr, but they are not equally numerous at every 
$\left(\Omega,\omega\right)$ configuration. \\

For the graveyard solutions (Fig. \ref{fig:grav_DV_emax_graphs}), one can see `triangle-shaped' clouds of points; 
those all correspond to two-burn transfers, as graveyard orbits cannot cross the operational GNSS regions by definition. The lower 
envelop (blue curve) consists of the respective optimal solutions, where $\Delta V$ is roughly proportional to $e_{max}$. In 
general, the $\Delta V$ needed to transfer to a graveyard orbit is quite small, $\sim5-40$~m/sec, as opposed to a reentry orbit.  
Note that similar results were found by \cite{dMrA16} for the GALILEO case.
Moreover, the lower $e_{max}$, the more likely for a graveyard orbit to remain `stable' for times longer than $200$~yr. Of course, 
it is not clear how many disposed satellites one could safely store in these narrow graveyards bands, and such computations likely 
require a more accurate dynamical model.  \\

\begin{figure}[htp!]
  \centering
    \begin{subfigure}[b]{0.32\textwidth}
      \includegraphics[width=.99\textwidth]{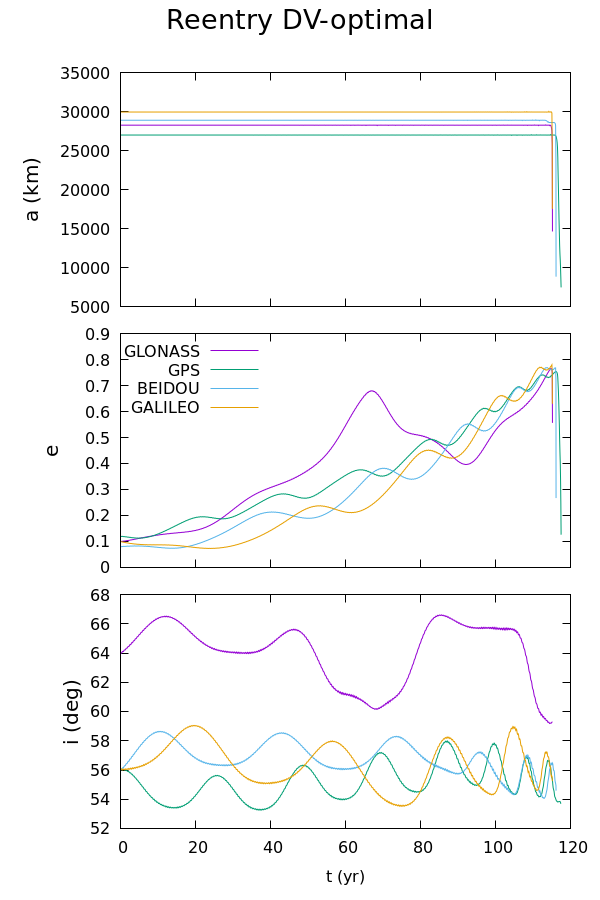}
    \end{subfigure}  
    \begin{subfigure}[b]{0.32\textwidth}
      \includegraphics[width=.99\textwidth]{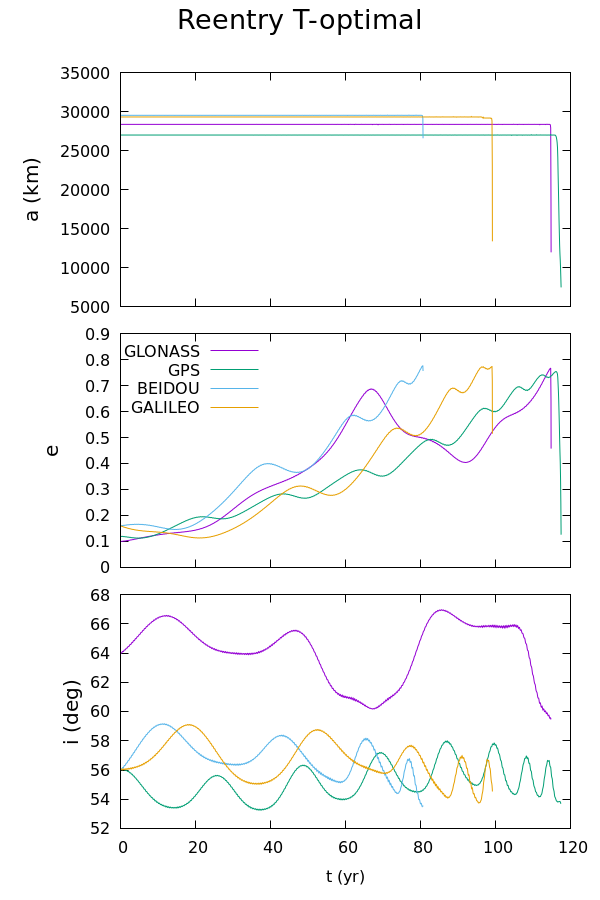}
    \end{subfigure}
     \begin{subfigure}[b]{0.32\textwidth}
      \includegraphics[width=.99\textwidth]{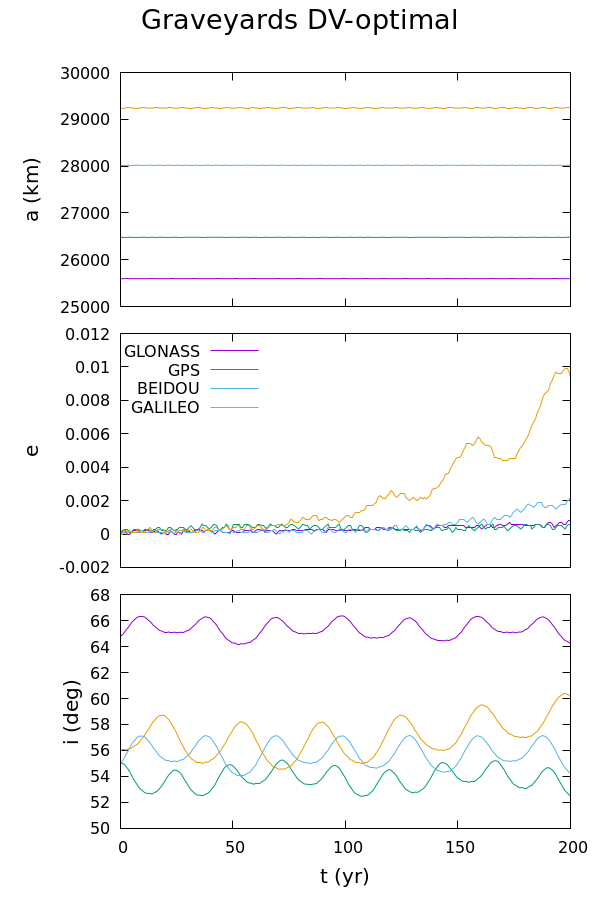}
    \end{subfigure}
  \caption{Evolution of $a$, $e$, and $i$ of $\Delta V$-optimal (left) and $T$-optimal (middle) reentry solutions 
  and of $\Delta V$-optimal (right) graveyard solutions, 
  for four typical starting orbits, one for each GNSS group.}
  \label{fig:example_reentry_grav_evol}
\end{figure}

One of the goals of the ReDSHIFT is the development of a software toolkit that would enable an informed design of ``debris compliant'' 
missions, including the design of an appropriate passive removal strategy, given an operational orbit and $\Delta V$ budget. We refer 
the reader to \citet{ReDSHIFT18} for a in-depth discussion about the software toolkit. The results presented in this paper have been collected 
in the form of a database of pre-computed solutions, to be used as input in the software toolkit. Of course, the database is expandable and 
we hope to keep expanding it in the future. An example of the typical output that this toolkit should give for a MEO mission is shown 
in Figures \ref{fig:reentry_DV_life_graphs},\ref{fig:grav_DV_emax_graphs},\ref{fig:example_reentry_grav_evol}.  \\

In Table \ref{tab:example_reentry_o1}, the initial conditions for a set of starting orbits are given. Corresponding to these starting 
orbits, the `optimal' solutions are shown in Table \ref{tab:example_reentry_o2} where, along with ($a,e$) values, the $\Delta V$ spent 
on the transfer and the waiting time $T$ of reentry are shown. Two reentry solutions are given for each starting orbit, labeled as {\it  
$\Delta V$-optimal} and {\it $T$-optimal}, corresponding to a minimum $\Delta V$ or a minimum $T$. Similarly, in Table 
\ref{tab:example_graveyards_o2} the $\Delta V$-optimal graveyard solution for each starting orbit is given. 
In Figure \ref{fig:example_reentry_grav_evol} the time evolution of $a$, $e$ and $i$ of the $\Delta V$-optimal reentry (left), 
$T$-optimal reentry (middle) and $\Delta V$-optimal graveyard (right) solutions for each GNSS representative, is shown. Note that, for 
the purposes of the software toolkit, we have run again all our reentry solutions, adding atmospheric drag, 
with a simple density model described in \citet{dS18}. We verified that our chosen limit of $q=R_{E}+400$km in the drag-free case was 
adequate for identifying reentry solutions, while the differences found in reentry time between the former and the latter propagation 
are minute. One can clearly see the footprint of atmospheric drag at the final instances of the orbits shown in Fig 
\ref{fig:example_reentry_grav_evol}, where both $a$ and $e$ drop abruptly.  \\

The initial eccentricity of the `optimal' reentry solutions varies between $0.08-0.16$ and the lifetime for most of 
them is near $110$~yr. The $\Delta V$ budget required for these transfers varies in the range $150-300$~m/sec. Again, the 
results depend strongly on the choice of $\left(\Omega,\omega\right)$ for the starting orbit. Hence, it is possible to find `optimal' 
solutions with $\sim70$~yr lifetime and $\Delta V\sim300$~m/sec, as also shown in Figure \ref{fig:reentry_DV_life_graphs}. 
On the other hand, all $\Delta V$-optimal graveyard solutions start as circular orbits, and their eccentricities reach up to 
$\sim0.01$ within 200 years. Note that some of these evolutions suggest that eccentricities may in fact increase further and hence 
violate the boundaries of the operational zones, at times much longer than 200 years, however. 
The $\Delta V$-budget for transfer to these graveyards is $\leq 23$~m/sec. 
Note that the `optimal' solutions found here are among the set of evaluated solutions presented in Section \ref{sec:3}, 
but the real optimal transfer may correspond to a more favorable choice of $\omega$.
  \\

\begin{table}[htp!]
	\captionsetup{justification=justified}
	\centering
\caption{Initial choice of $a$, $e$, $i$, $\Omega$ and $\omega$ of assumed operational orbits.}
\label{tab:example_reentry_o1}
 \begin{tabular}{lccccc}
 \hline\noalign{\smallskip}
         & $a$ & $e$  & $i$ (deg) & $\Omega$ (deg) & $\omega$ (deg) \vspace{0.1cm}\\
 GLONASS & $a_{GLO}$ & $0.0001$ & $64.00/64.80$ & $102.83$ & $106.50$  \vspace{0.1cm}\\
                          
 GPS     & $a_{GPS}$ & $0.0001$ & $56.00/55.00$ & $~12.83$ & $106.50$ \vspace{0.1cm}\\
                         
 BEIDOU  & $a_{BEI}$ & $0.0001$ & $56.00/55.00$ & $102.83$ & $106.50$ \vspace{0.1cm}\\
 
 GALILEO & $a_{GAL}$ & $0.0001$ & $56.00$ & $192.83$ & $106.50$ \vspace{0.1cm}\\
 \hline\noalign{\smallskip}
 \end{tabular}
\end{table}

\begin{table}[htp!]
	\captionsetup{justification=justified}
	\centering
\caption{$\Delta V$-optimal and $T$-optimal reentry orbits for the assumed operational orbit}
\label{tab:example_reentry_o2}
 \begin{tabular}{lccccc}
 \hline\noalign{\smallskip}
                          & optimal     & $a\left(km\right)$ & $e$ & $\Delta V$ (m/sec) &  $T \left(yr\right)$\\
 \multirow{2}{*}{GLONASS} &  $\Delta V$ & $28249.99$ & $0.1000$ & $192.8$ & $115.22$\\
                          &  $~T$       & $28355.40$ & $0.1000$ & $193.7$ & $114.85$\vspace{0.1cm}\\
                          
 \multirow{2}{*}{GPS}     &  $\Delta V$ & $26985.07$ & $0.1200$ & $228.6$ & $117.52$\\
                          &  $~T$       & $26985.07$ & $0.1200$ & $228.6$ & $117.52$\vspace{0.1cm}\\
                          
 \multirow{2}{*}{BEIDOU}  &  $\Delta V$ & $28882.46$ & $0.0800$ & $148.6$ & $116.16$\\
                          &  $~T$       & $29514.92$ & $0.1600$ & $294.1$ & $~80.71$\vspace{0.1cm}\\
                          
 \multirow{2}{*}{GALILEO} &  $\Delta V$ & $29936.56$ & $0.1000$ & $180.9$ & $115.19$\\
                          &  $~T$       & $29304.10$ & $0.1600$ & $290.3$ & $~99.21$\vspace{0.1cm}\\
 \hline\noalign{\smallskip}
 \end{tabular}
\end{table}

\begin{table}[htp!]
	\captionsetup{justification=justified}
	\centering
\caption{$\Delta V$-optimal graveyard orbits for the assumed operational orbit}
\label{tab:example_graveyards_o2}
 \begin{tabular}{lccccc}
 \hline\noalign{\smallskip}
                          & optimal     & $a\left(km\right)$ & $e$ & $\Delta V$ (m/sec) & $e_{max}$\\
 \multirow{1}{*}{GLONASS} &  $\Delta V$ & $25593.65$ & $0.0000$ & $6.5$ & $0.00081$\vspace{0.1cm}\\
                    
 \multirow{1}{*}{GPS}     &  $\Delta V$ & $26479.10$ & $0.0000$ & $6.0$ & $0.00070$\vspace{0.1cm}\\
                            
 \multirow{1}{*}{BEIDOU}  &  $\Delta V$ & $28018.09$ & $0.0000$ & $7.6$ & $0.00228$\vspace{0.1cm}\\
                           
 \multirow{1}{*}{GALILEO} &  $\Delta V$ & $29240.85$ & $0.0000$ & $22.5$& $0.00997$\vspace{0.1cm}\\
 \hline\noalign{\smallskip}
 \end{tabular}
\end{table}

% -------------------------------------------------------------------------------------------------------------------------------------- 
%          CONCLUSION
% -------------------------------------------------------------------------------------------------------------------------------------- 
\newpage
\section{Conclusions}
\label{sec:concl}

We presented our results from a study on the long-term dynamics in the extended MEO/GNSS orbital region, using a dynamical 
model that consisted of the second degree and order geopotential, lunisolar perturbations, and SRP (cannonball model). In total, about 6 million 
orbits were propagated for a time equivalent to $120-200$~yr, for two different initial epochs, and two values of the $C_{R}A/m$ ratio. The 
aim of this study was to construct a detailed dynamical atlas and use it to locate suitable reentry and graveyard solutions, which could 
be useful for designing EoL strategies.  \\

As shown by our results, the most interesting and complex dynamical behavior appears for moderate-to-high inclinations 
($\sim40^{\circ}-70^{\circ}$), where the GNSS groups are actually placed. In particular, the number of reentry solutions seems to maximize 
around three particular inclination `zones' (around $i=$46, 56, and 68 degrees); this result seems to be roughly 
independent of $A/m$ as well as of the initial epoch, chosen here. However, as noted already, more initial epochs (and more distant ones), leading to diverse values of the lunar ascending node should be studied, before concluding that this structure is epoch-invariant. 
For the same inclination bands, the mean dynamical lifetime of reentry orbits minimizes. It is already known from previous studies that secular 
lunisolar resonances are actually dominating the dynamics at those inclinations. The variations of a satellite's eccentricity and inclination 
may lead to perigee decrease and eventually, atmospheric reentry. In the region around $56^{\circ}$, reentry dynamical hatches appear even 
for low-to-moderate eccentricities (e.g., reentry orbits with initial $e\sim 0.10$ and lifetimes of $\sim 60$~yr), with a strong dependence 
on secular angle configuration. On the other hand, around $64^{\circ}$ the reentry dynamical hatches appear generally for higher 
eccentricities. An enhanced $C_{R}A/m$ value extends the reentry regions and decreases the time required for reentry by a decade or so, 
but does not significantly alter the overall structure of the $\left(a,e\right)$ map. The reachability of all reentry solutions from a 
nearly-circular initial orbit, using single- and two-burn maneuvers, was studied (coaxial and coplanar elliptical orbits). For GNSS altitudes 
we find that the $\Delta V$ budget needed is roughly inversely proportional to the orbital lifetime (waiting time). Typically, reentry 
solutions with $\Delta V < 300$~m/s have lifetimes longer than $\sim70$yr. Note that this study did not focus on the computation of globaly optimal 
reentry solutions; this would require adopting a different strategy of optimizing $\Delta V$ (e.g.\ as in \cite{rAfS18}), or studying a much finner grid in ($\omega, \Omega$). Instead, we decided to focus on exporing the whole domain of inclinations, which had not 
been extensively studied so far, at the same time as studying the feasibility of using the dynamical maps for finding near-optimal reentry solutions. Clearly, our results need to be extended, especially in the current operational GNSS region. \\

A dedicated study for the graveyard regions around the four GNSS groups revealed that the percentage of bodies initially placed in 
graveyard regions and surviving for time spans of $200$~yr is $\sim20-40$\%. These regions are limited to a maximum eccentricity of $\sim 0.018$, 
but such mildly eccentric graveyard solutions exist. The results vary of course with secular angles orientation. However, as our definition 
of the graveyard regions is quite generous, the surviving solutions seem abundant and easily targeted by two-burn maneuvers with 
$\Delta V\in5-40$~m/sec. Note that $\Delta V$ is roughly proportional to the maximum eccentricity attained during the $200$~yr orbital 
evolution.  \\

One of the goals of the ReDSHIFT project is to provide a toolkit for designing ``debris-friendly'' passive EoL strategies for future 
satellites missions. Hence, the study presented here provides a database of solutions to be used in that toolkit. We intend to further 
expand this database, by propagating a more extensive grid of initial conditions, for various epochs and $\omega$-orientations. This will also allow a deeper understanding of the long-term extended MEO dynamics. We expect to be able to present our results in the near future.  \\

% -------------------------------------------------------------------------------------------------------------------------------------- 
%          ACKNOWLEDGEMENTS
% -------------------------------------------------------------------------------------------------------------------------------------- 
\section*{Acknowledgements}

This research is partially funded by the European Commission Horizon 2020 Framework Programme for Research and Innovation (2014-2020), 
under Grant Agreement 687500 (project ReDSHIFT; \url{http://redshift-h2020.eu/}). 
The work of D.K.~Skoulidou is also supported by General Secretariat for Research and Technology (GSRT) and Hellenic Foundation for Research 
and Innovation (HFRI).
We would like to acknowledge A. Rossi, C. Colombo, and the ReDSHIFT team for many discussions and for their internal review of this work. 
Special thanks goes to I. Gkolias, for discussions on maneuvers computations. 
Numerical results presented in this work have been produced using the Aristotle University of Thessaloniki (AUTh) Computer Infrastructure 
and Resources and the authors would like to acknowledge continuous support provided by the Scientific Computing Office.

% -------------------------------------------------------------------------------------------------------------------------------------- 
%          APPENDIX
% -------------------------------------------------------------------------------------------------------------------------------------- 
\begin{appendix}

\section{Dynamical maps of MEO-general grid}
\label{app:1}

\begin{figure}[htp!]
  \centering
    \begin{subfigure}[b]{0.45\textwidth}
      \caption{$\bm{i}_{o}={\bf 0}$}
      \includegraphics[width=.49\textwidth]{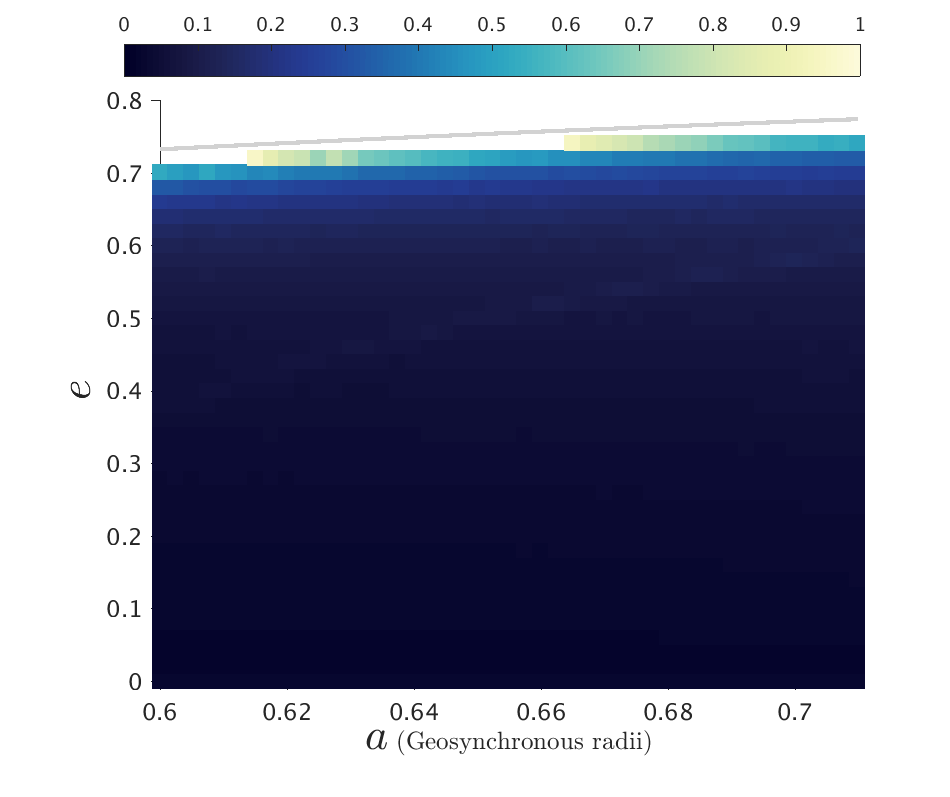} 
      \includegraphics[width=.49\textwidth]{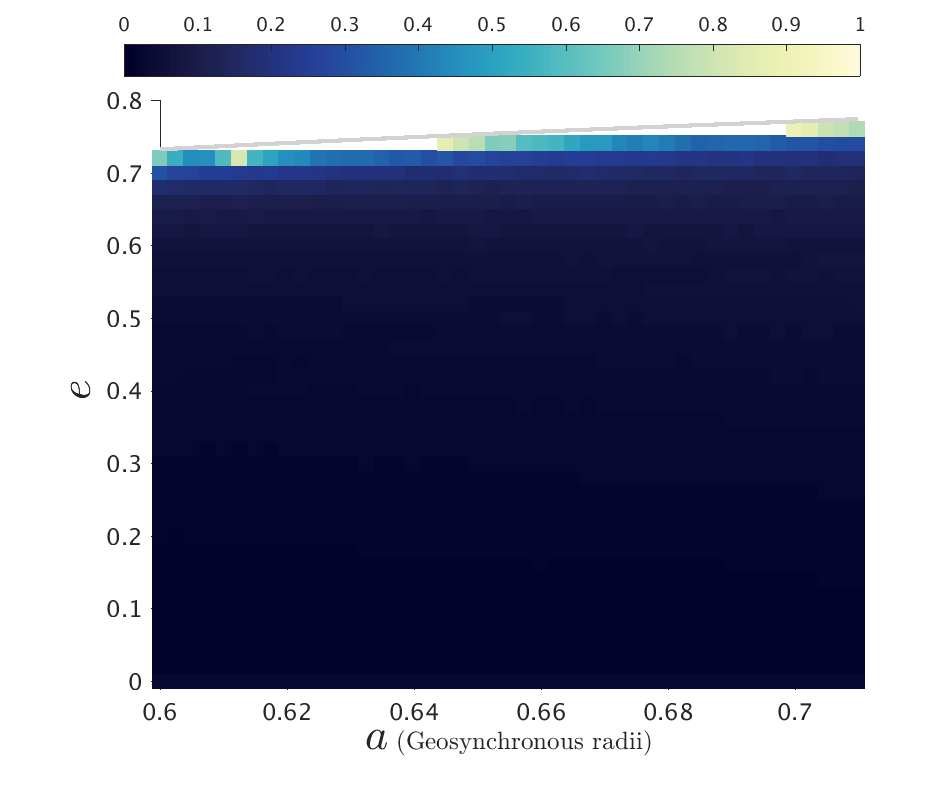}
    \end{subfigure} 
    \begin{subfigure}[b]{0.45\textwidth}
      \caption{$\bm{i}_{o}={\bf 28^{\circ}}$}
      \includegraphics[width=.49\textwidth]{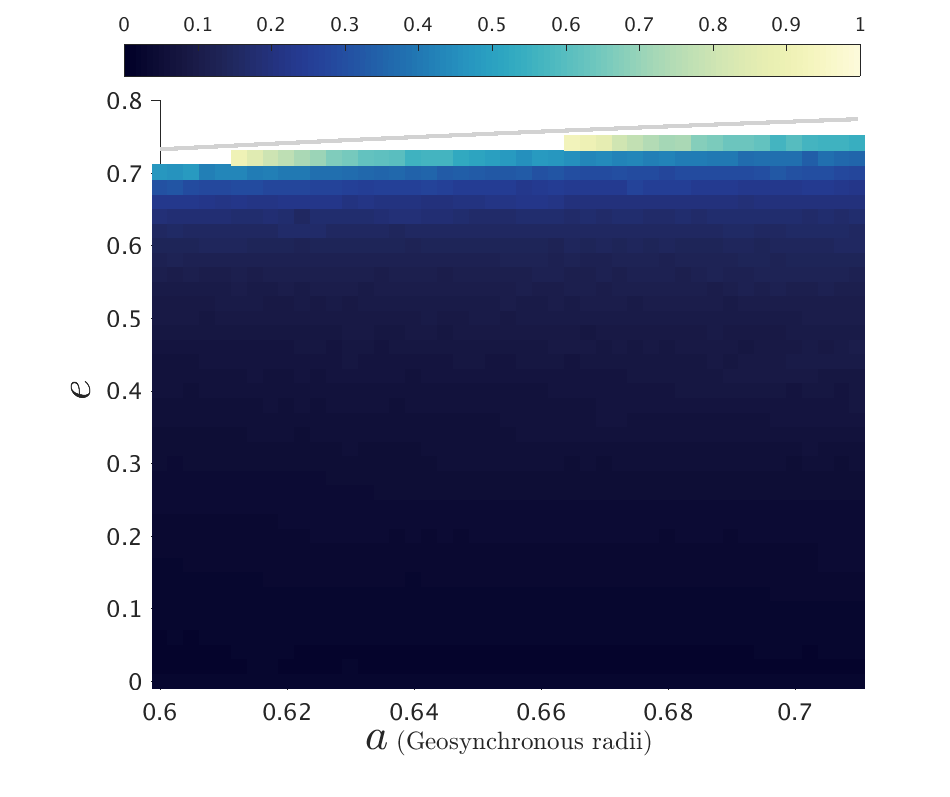} 
      \includegraphics[width=.49\textwidth]{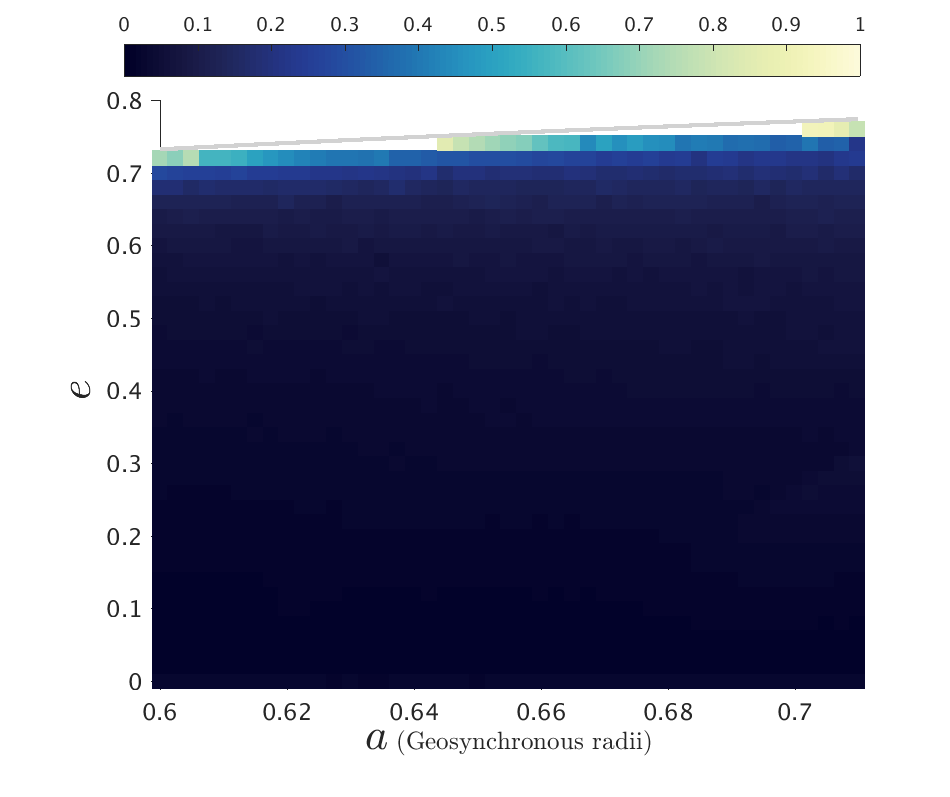}
    \end{subfigure}  
    \begin{subfigure}[b]{0.45\textwidth}
      \caption{$\bm{i}_{o}={\bf 44^{\circ}}$}
      \includegraphics[width=.49\textwidth]{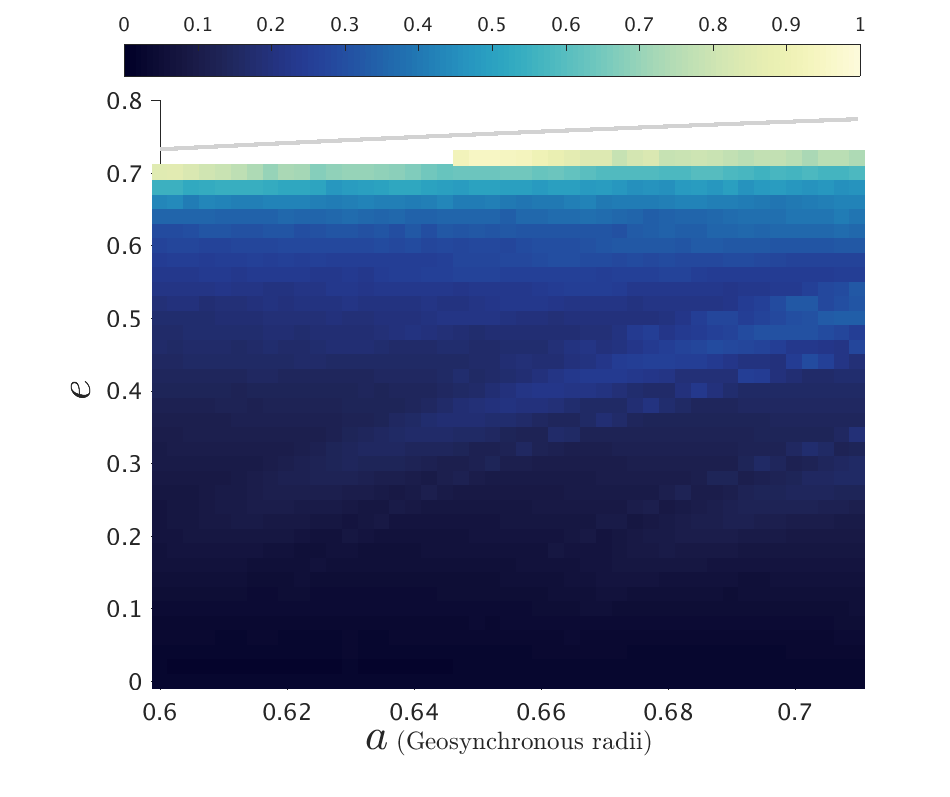} 
      \includegraphics[width=.49\textwidth]{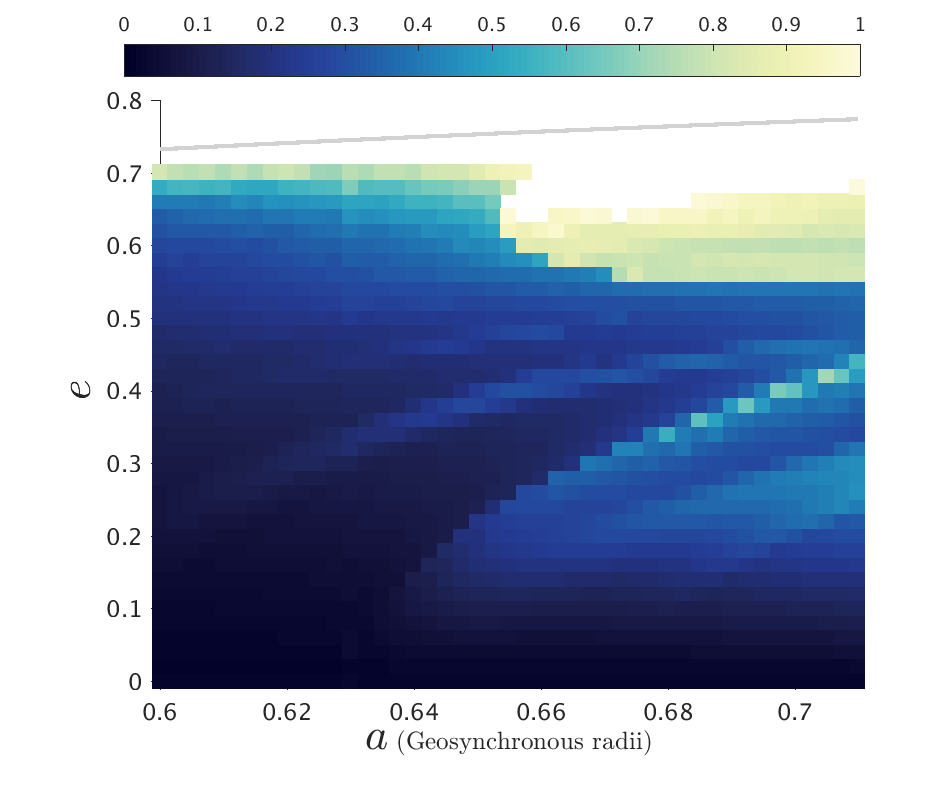}
    \end{subfigure}  
    \begin{subfigure}[b]{0.45\textwidth}
      \caption{$\bm{i}_{o}={\bf 46^{\circ}}$}
      \includegraphics[width=.49\textwidth]{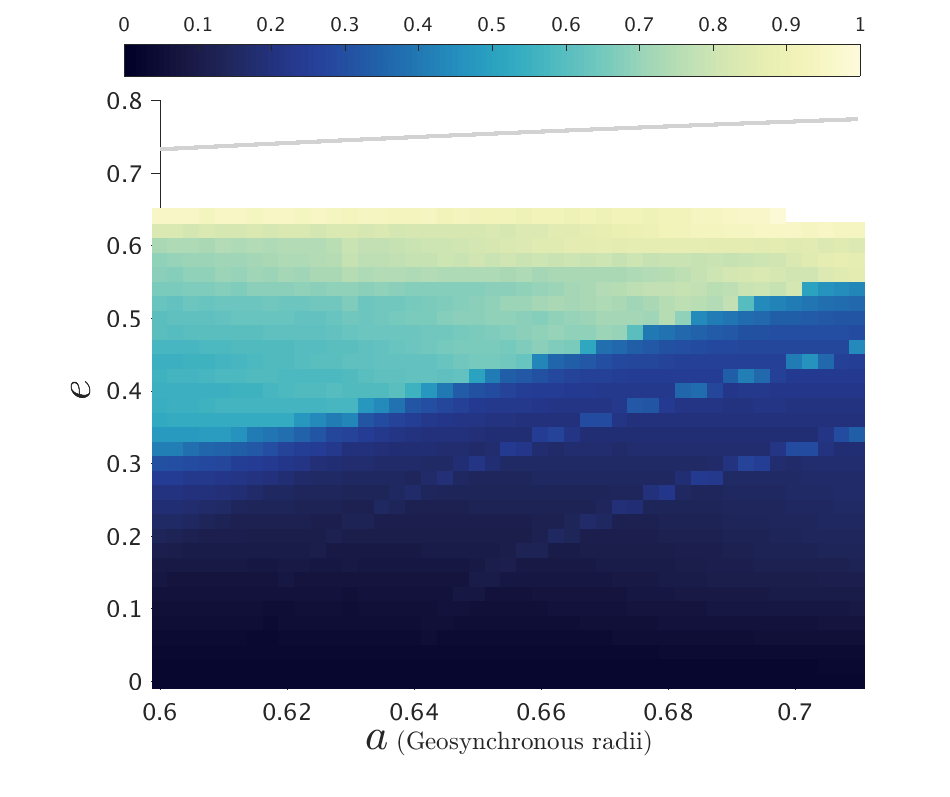} 
      \includegraphics[width=.49\textwidth]{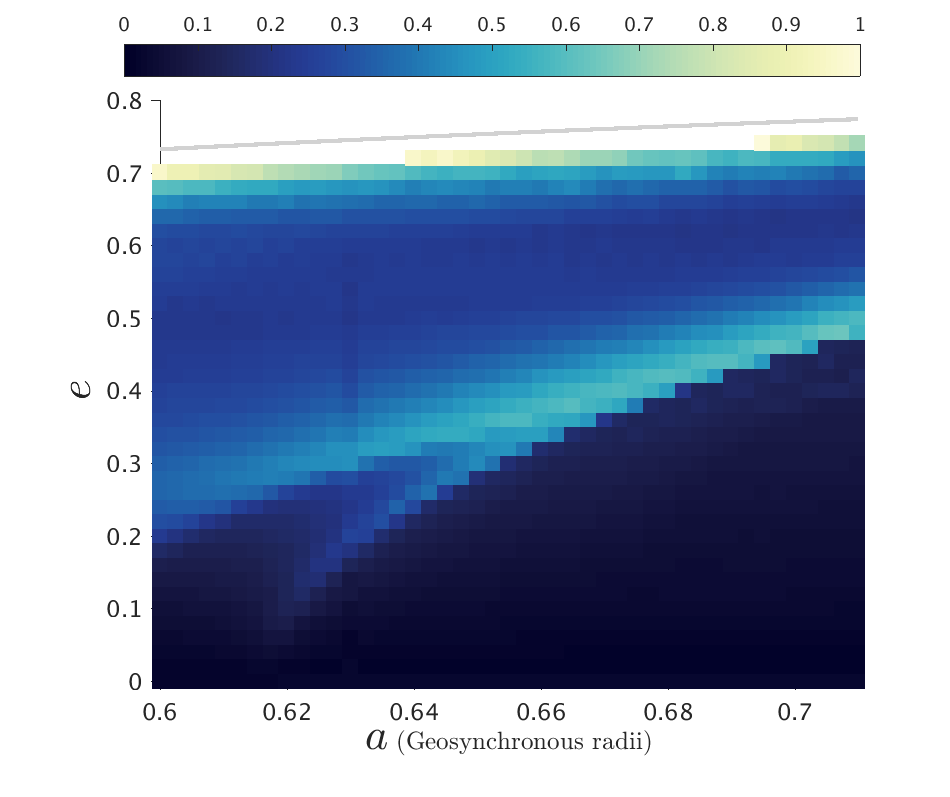}
    \end{subfigure}    
    \begin{subfigure}[b]{0.45\textwidth}
      \caption{$\bm{i}_{o}={\bf 54^{\circ}}$}
      \includegraphics[width=.49\textwidth]{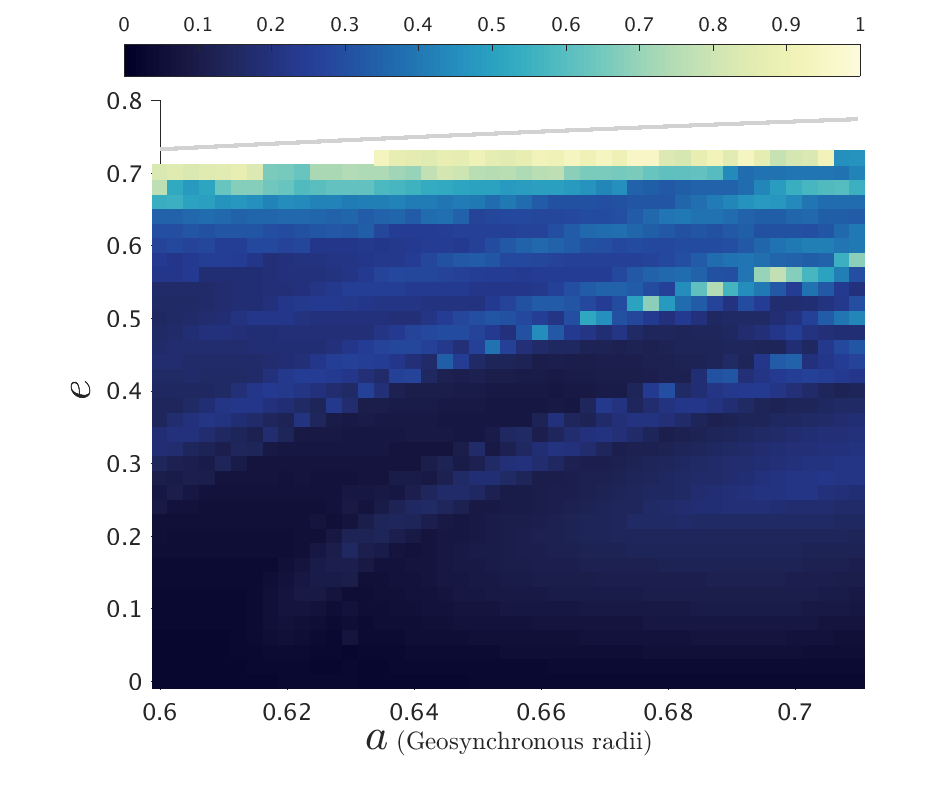} 
      \includegraphics[width=.49\textwidth]{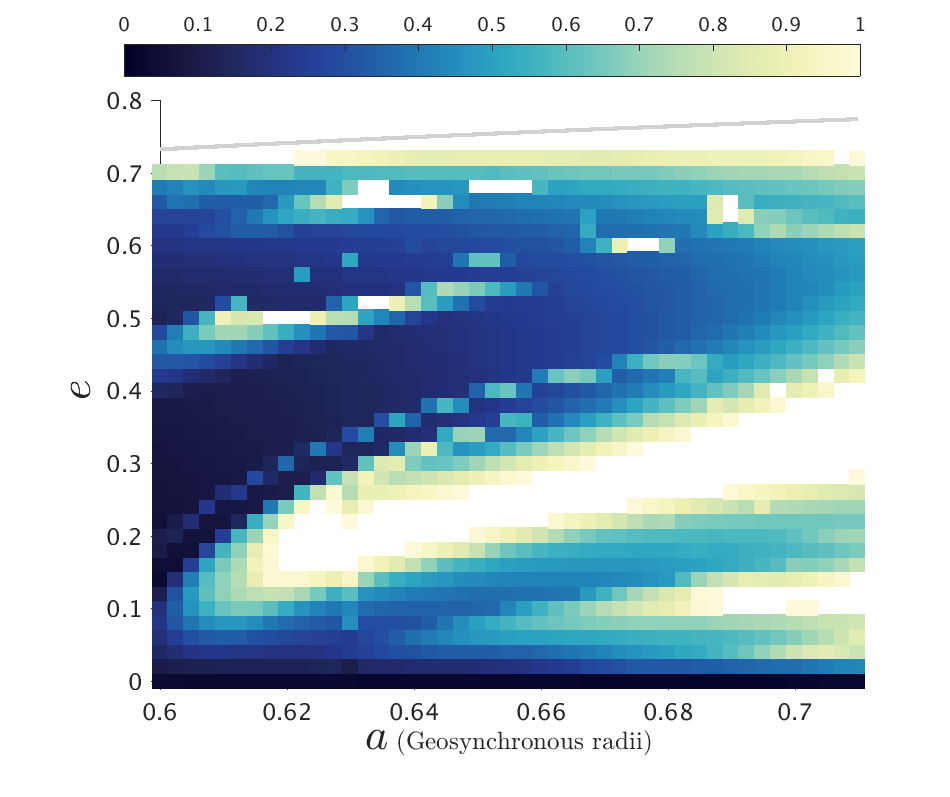}
    \end{subfigure}   
    \begin{subfigure}[b]{0.45\textwidth}
      \caption{$\bm{i}_{o}={\bf 58^{\circ}}$}
      \includegraphics[width=.49\textwidth]{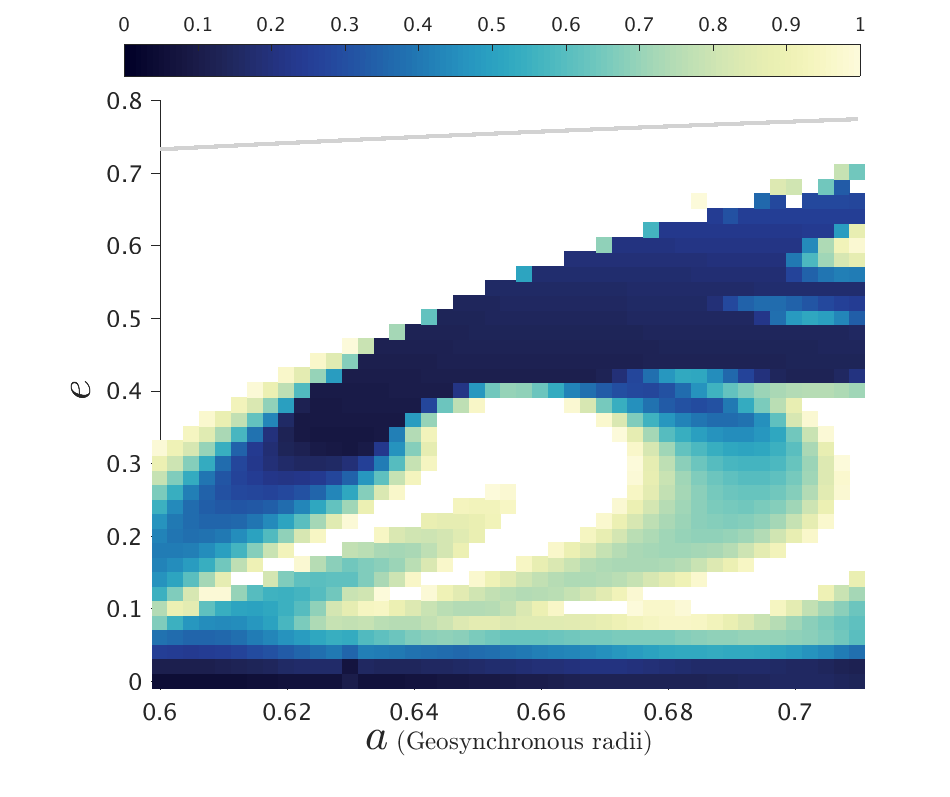} 
      \includegraphics[width=.49\textwidth]{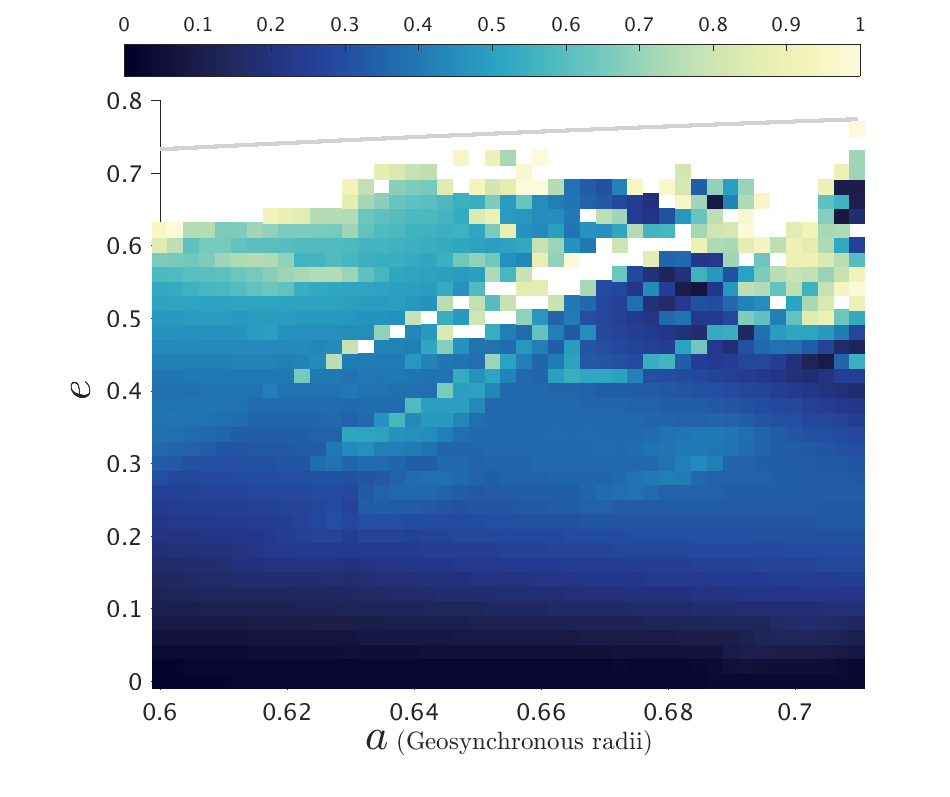}
    \end{subfigure}  
    \begin{subfigure}[b]{0.45\textwidth}
      \caption{$\bm{i}_{o}={\bf 66^{\circ}}$}
      \includegraphics[width=.49\textwidth]{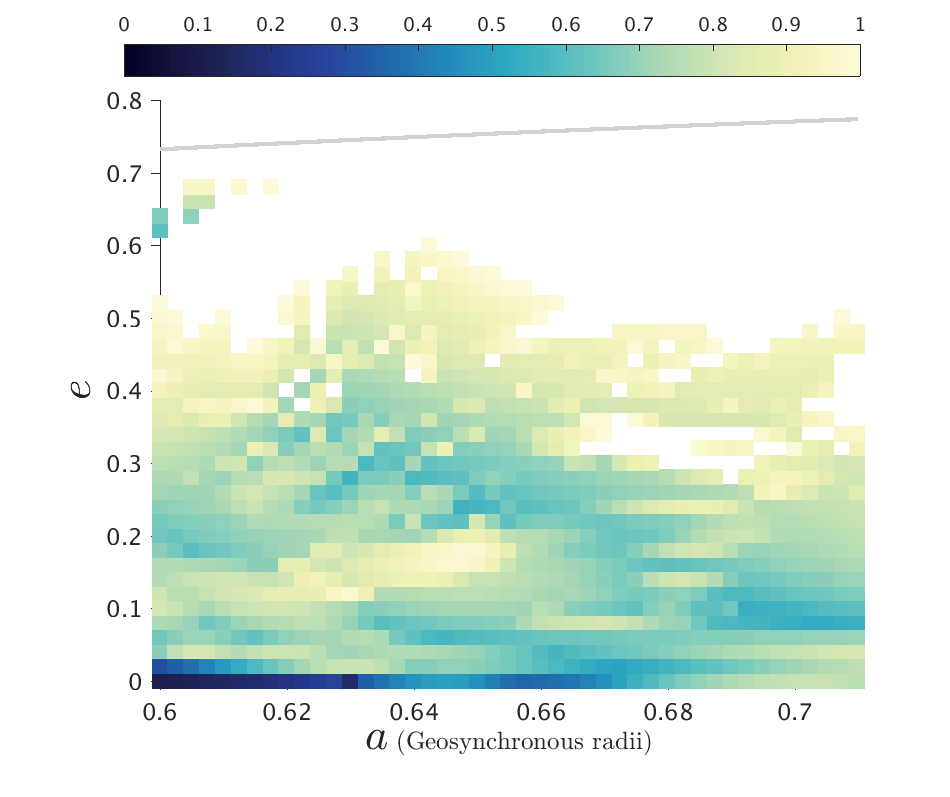} 
      \includegraphics[width=.49\textwidth]{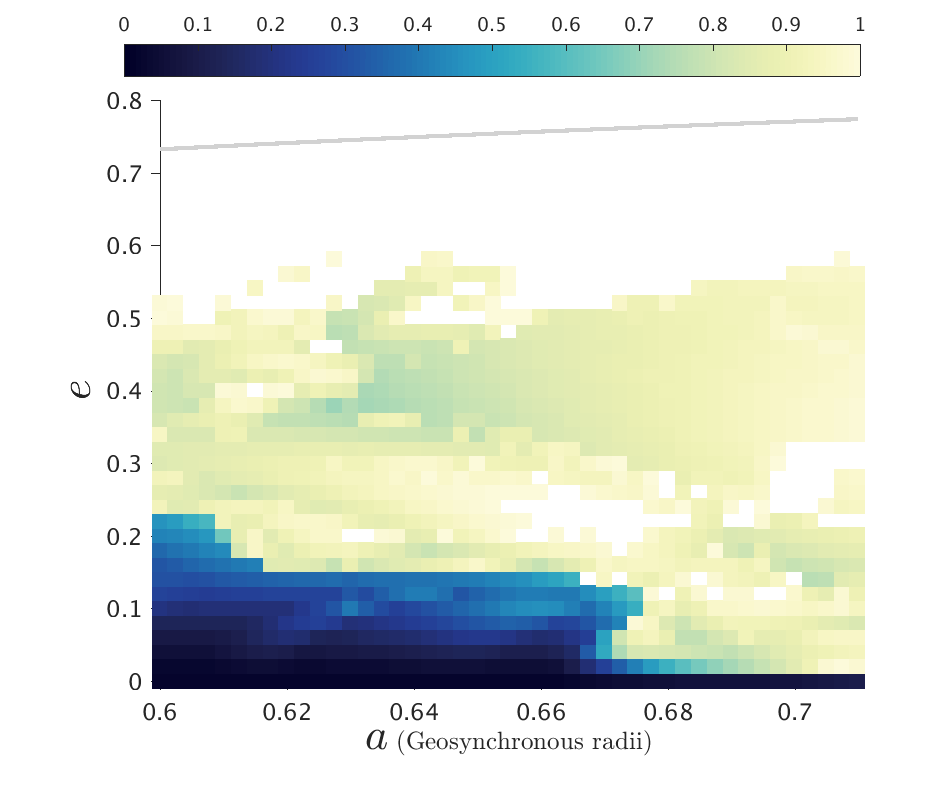}
    \end{subfigure}  
    \begin{subfigure}[b]{0.45\textwidth}
      \caption{$\bm{i}_{o}={\bf 68^{\circ}}$}
      \includegraphics[width=.49\textwidth]{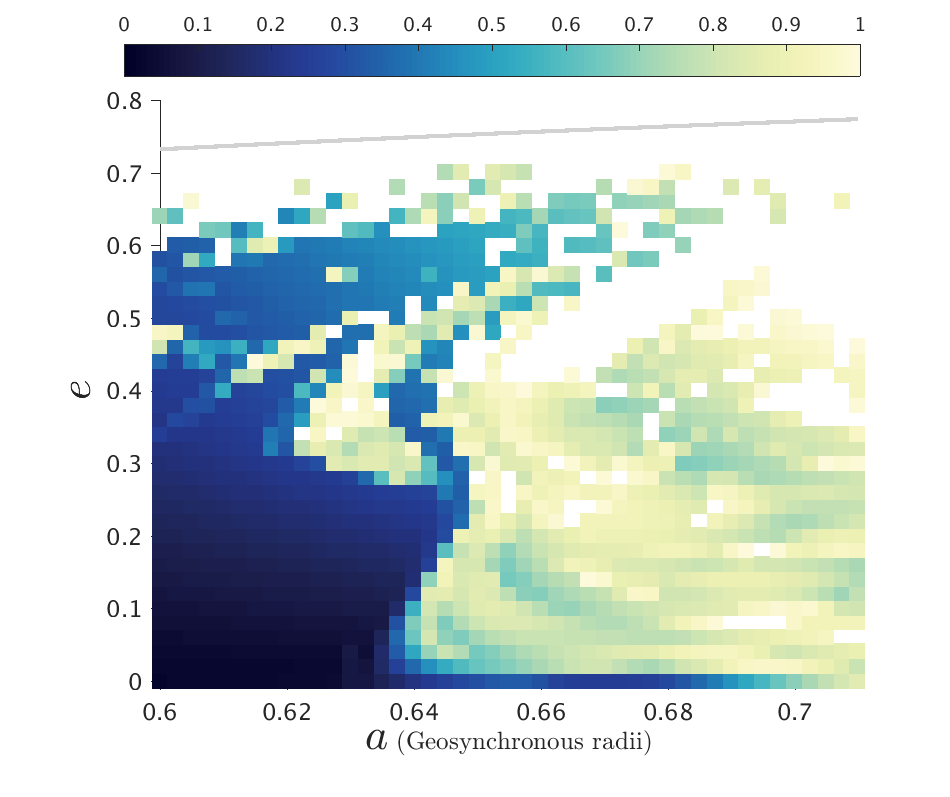} 
      \includegraphics[width=.49\textwidth]{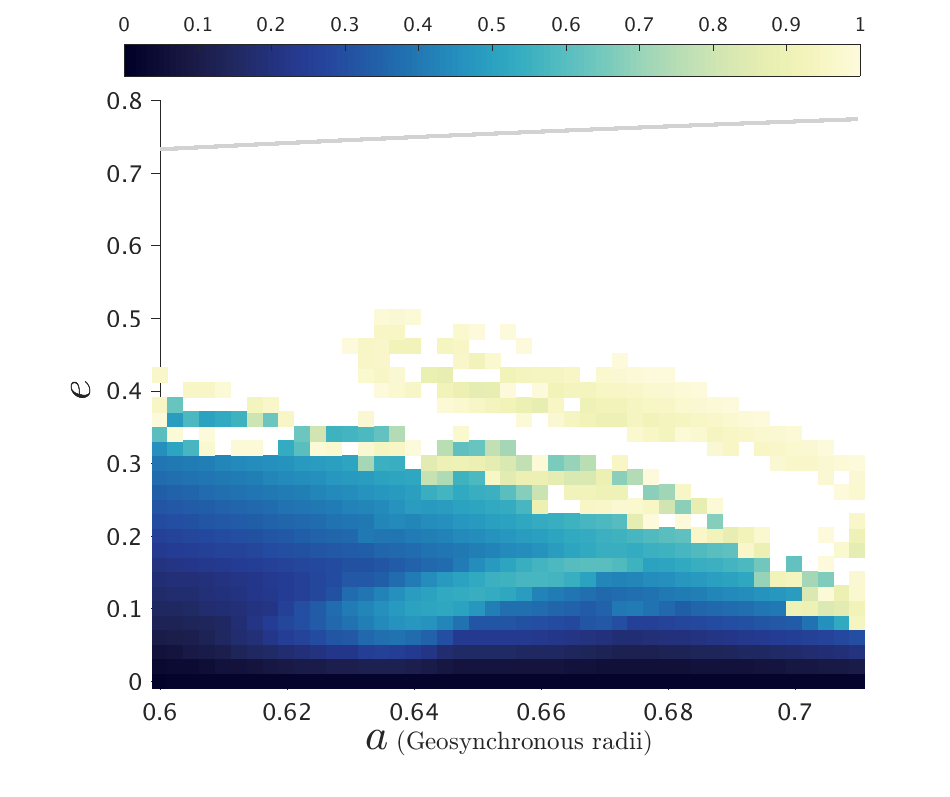}
    \end{subfigure}  
    \begin{subfigure}[b]{0.45\textwidth}
      \caption{$\bm{i}_{o}={\bf 70^{\circ}}$}
      \includegraphics[width=.49\textwidth]{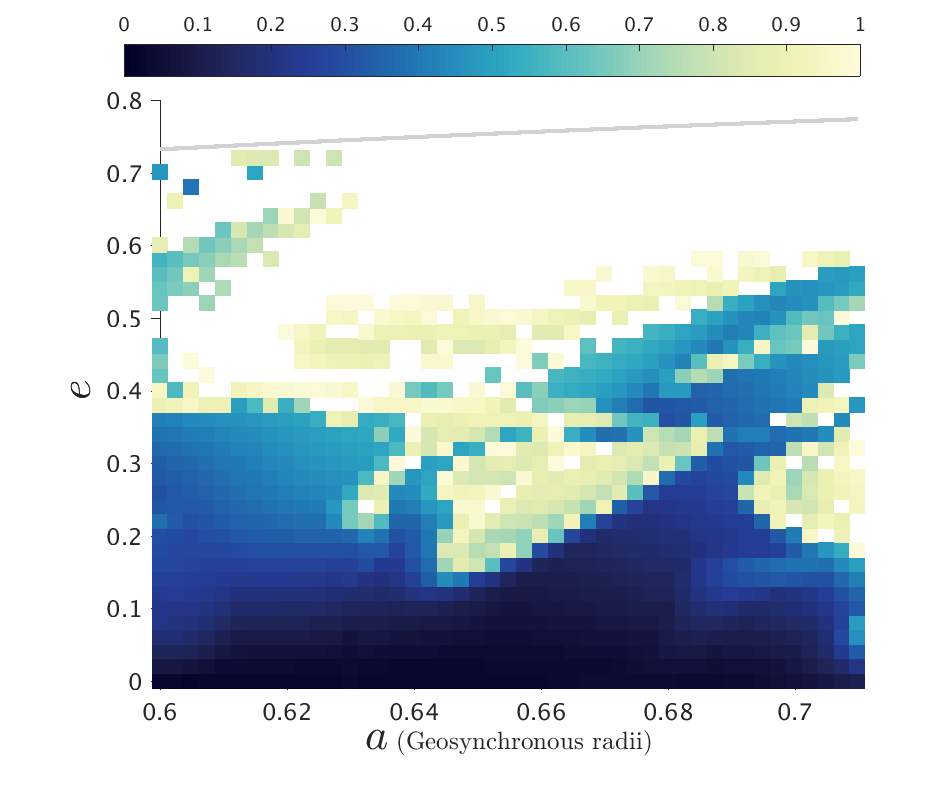} 
      \includegraphics[width=.49\textwidth]{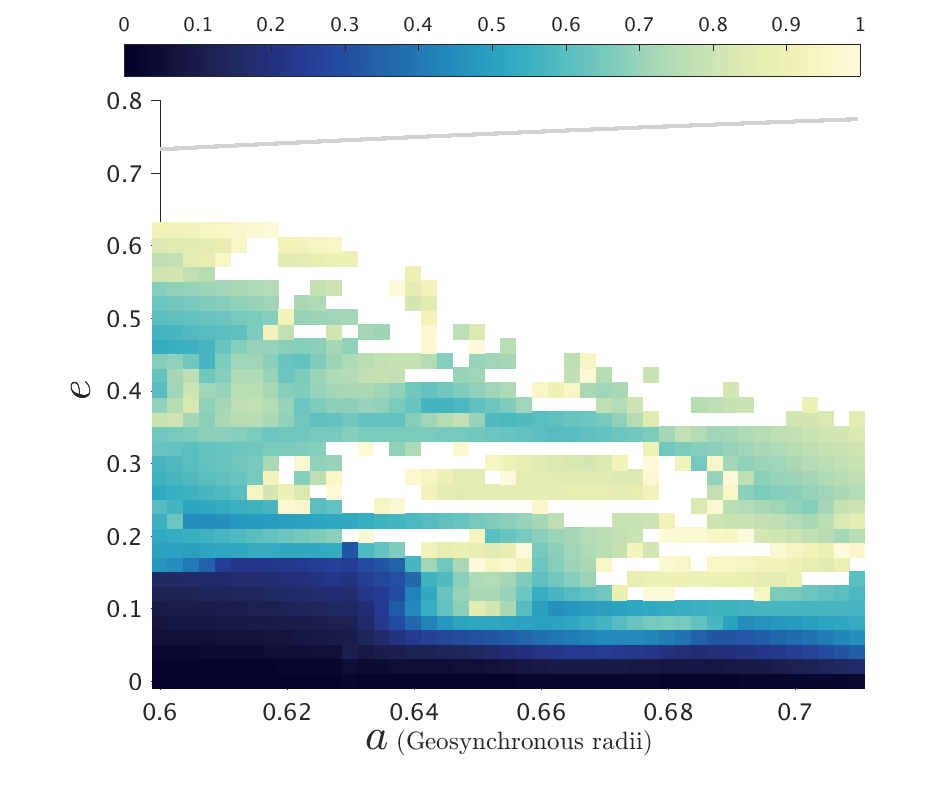}
    \end{subfigure}
    \begin{subfigure}[b]{0.45\textwidth}
      \caption{$\bm{i}_{o}={\bf 90^{\circ}}$}
      \includegraphics[width=.49\textwidth]{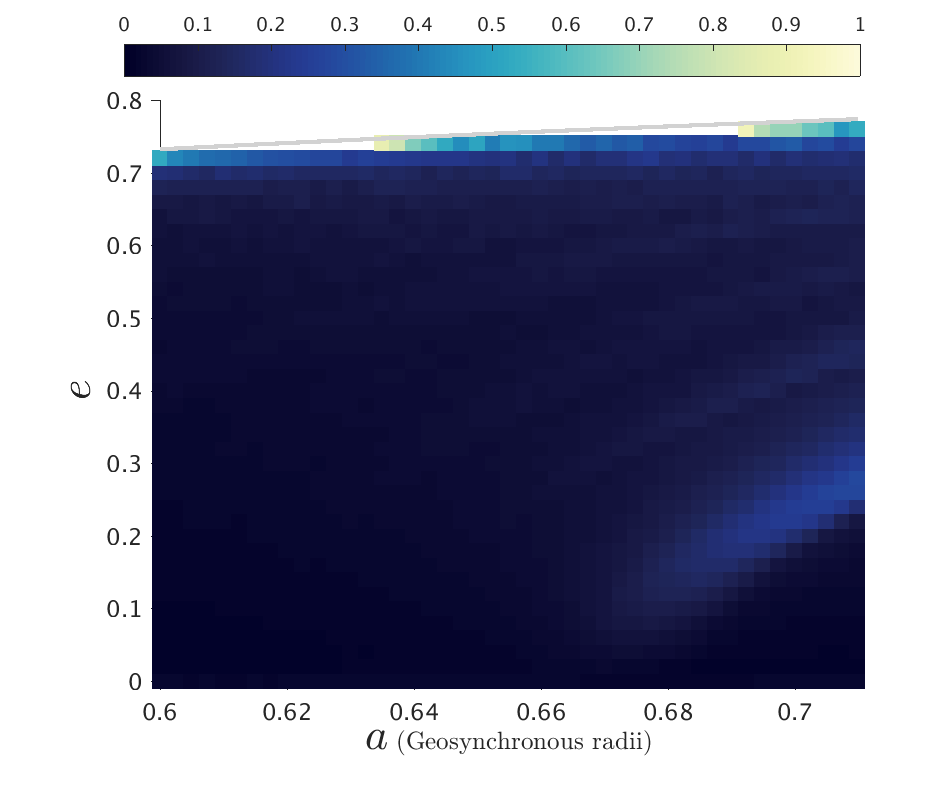} 
      \includegraphics[width=.49\textwidth]{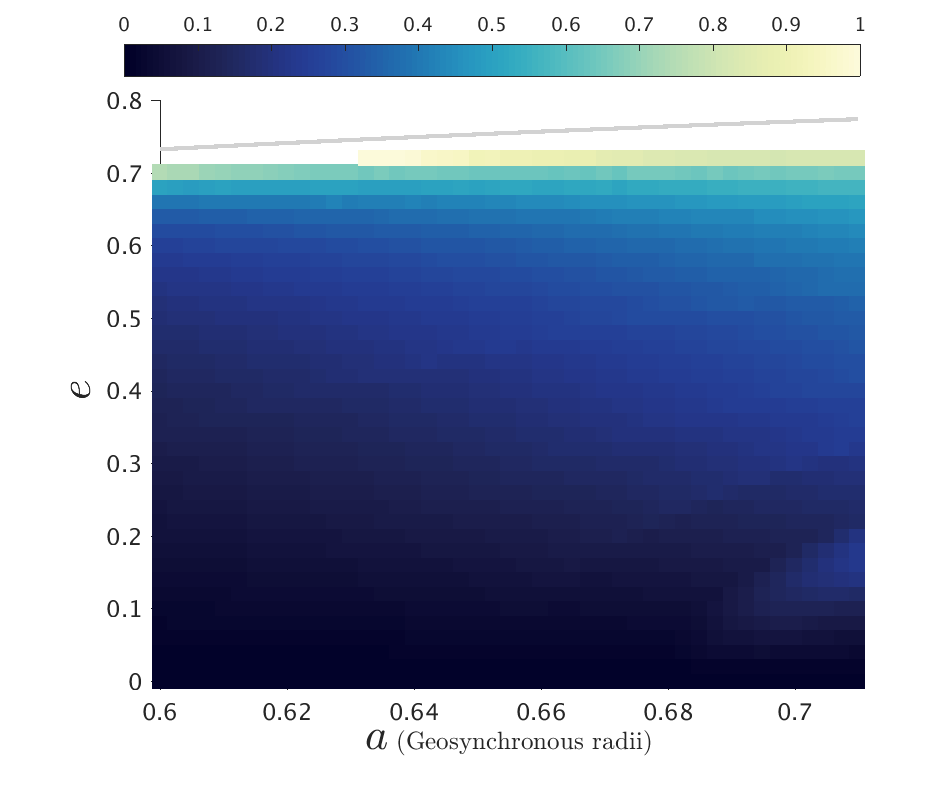}
    \end{subfigure}    
  \caption{$De$ maps of the \textit{MEO-general} phase space for various $\bm{i}_{o}$, 
  $\bm{\Delta}\bm{\Omega} = {\bf 0}$, $\bm{\Delta}\bm{\omega} = {\bf 270^\circ}$ (1st and 3rd columns) and 
  $\bm{\Delta}\bm{\Omega} = {\bf 90^\circ}$, $\bm{\Delta}\bm{\omega} = {\bf 0}$ (2nd and 4th columns),  
  for Epoch 2018, and for $C_{R}A/m=1$ m$^2$/kg.
  The colorbar for the $De$ maps is from 0 to 1, where the reentry particles were excluded (white).}
  \label{fig:MEO_gen_i}
\end{figure}

\begin{figure}[htp!]
  \centering
    \begin{subfigure}[b]{0.45\textwidth}
      \caption{$\bm{i}_{o}={\bf 0}$}
      \includegraphics[width=.49\textwidth]{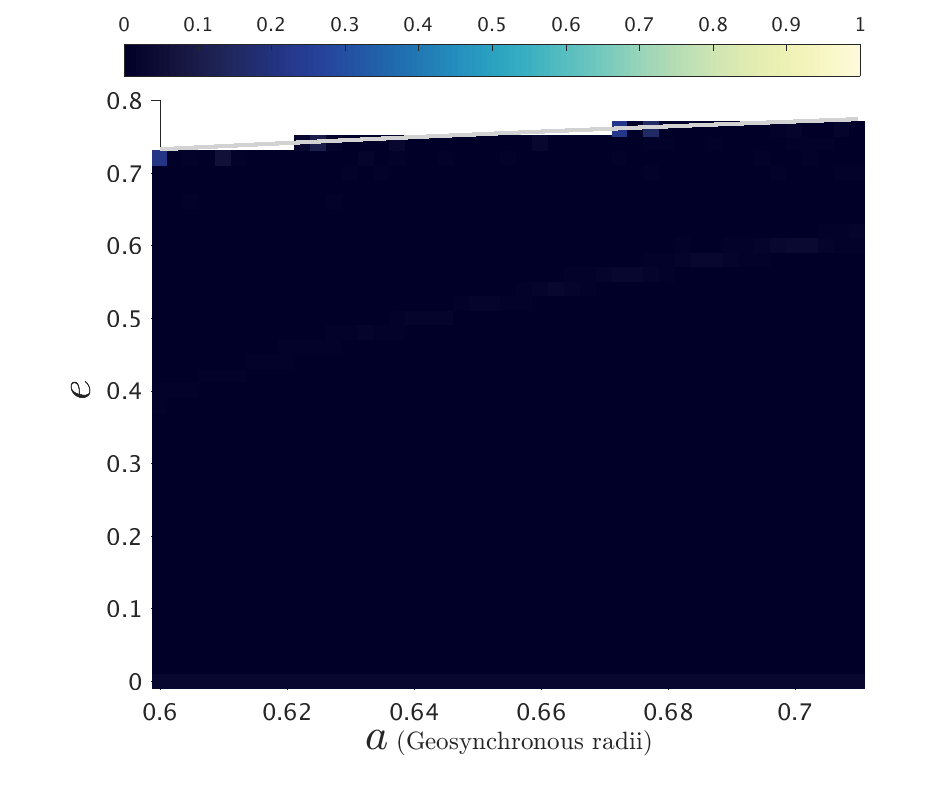} 
      \includegraphics[width=.49\textwidth]{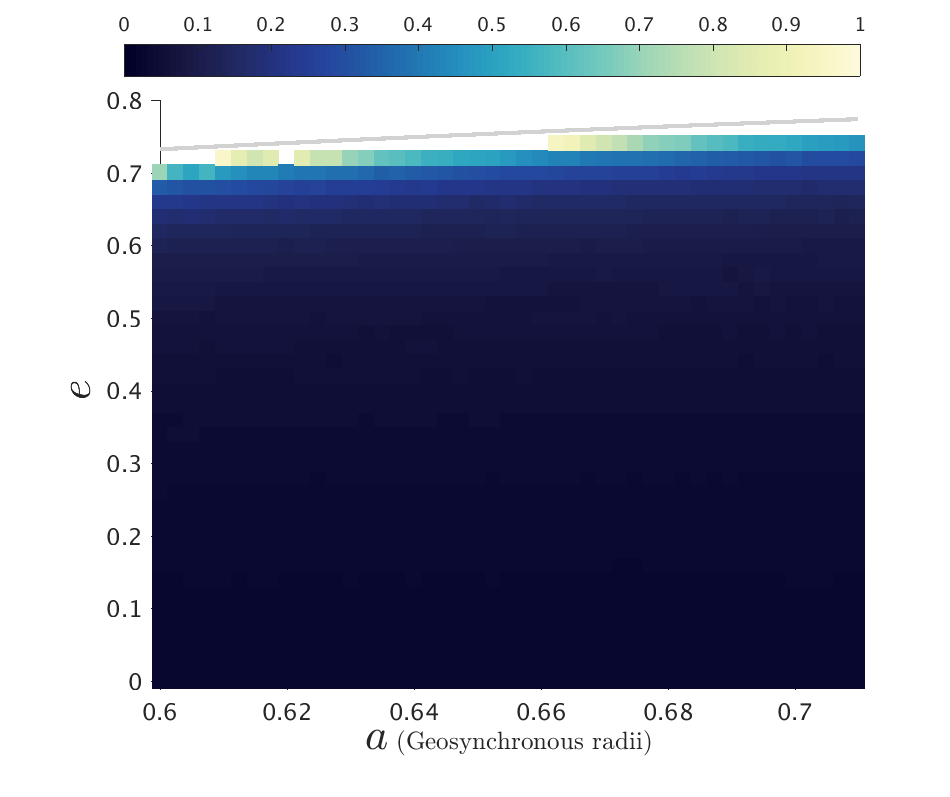}
    \end{subfigure} 
    \begin{subfigure}[b]{0.45\textwidth}
      \caption{$\bm{i}_{o}={\bf 28^{\circ}}$}
      \includegraphics[width=.49\textwidth]{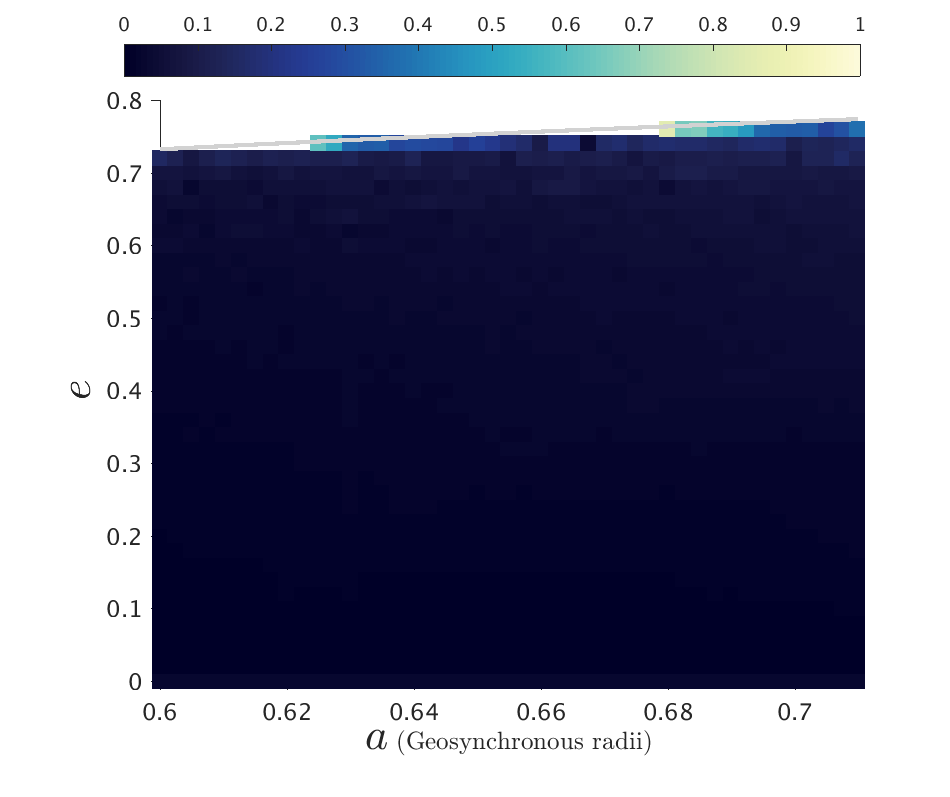} 
      \includegraphics[width=.49\textwidth]{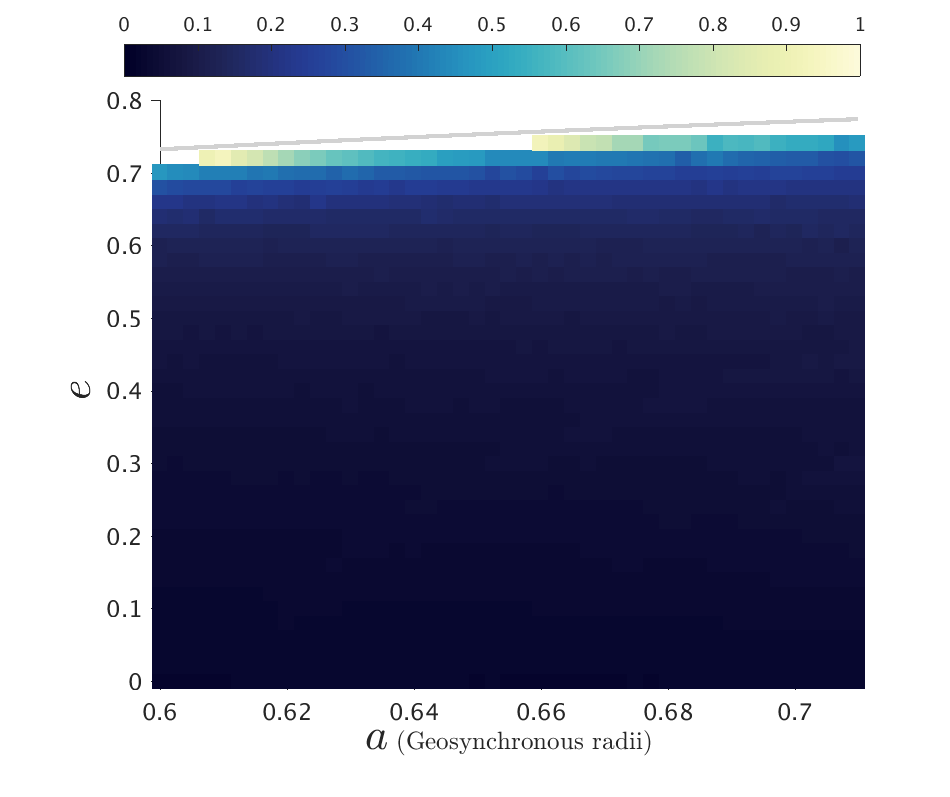}
    \end{subfigure}  
    \begin{subfigure}[b]{0.45\textwidth}
      \caption{$\bm{i}_{o}={\bf 44^{\circ}}$}
      \includegraphics[width=.49\textwidth]{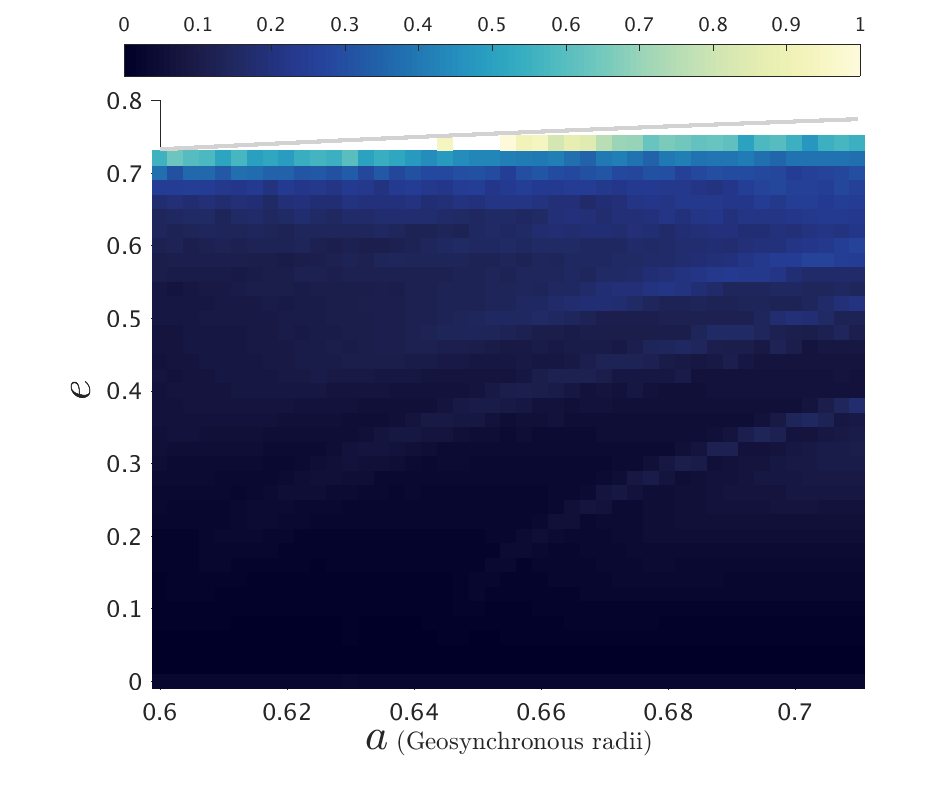} 
      \includegraphics[width=.49\textwidth]{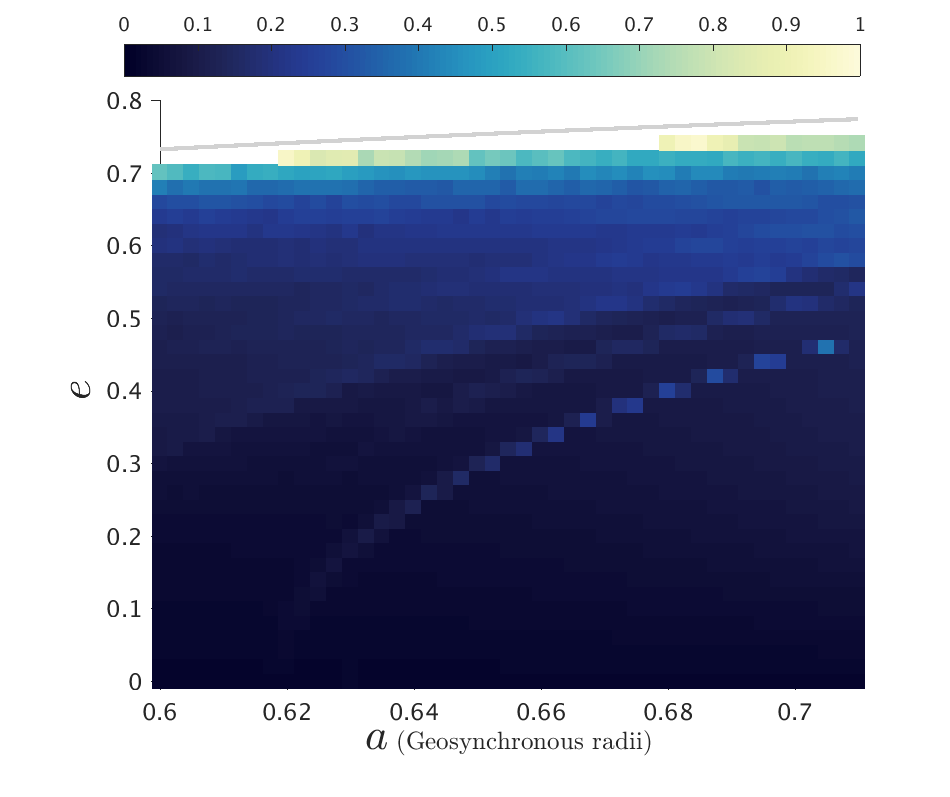}
    \end{subfigure}  
    \begin{subfigure}[b]{0.45\textwidth}
      \caption{$\bm{i}_{o}={\bf 46^{\circ}}$}
      \includegraphics[width=.49\textwidth]{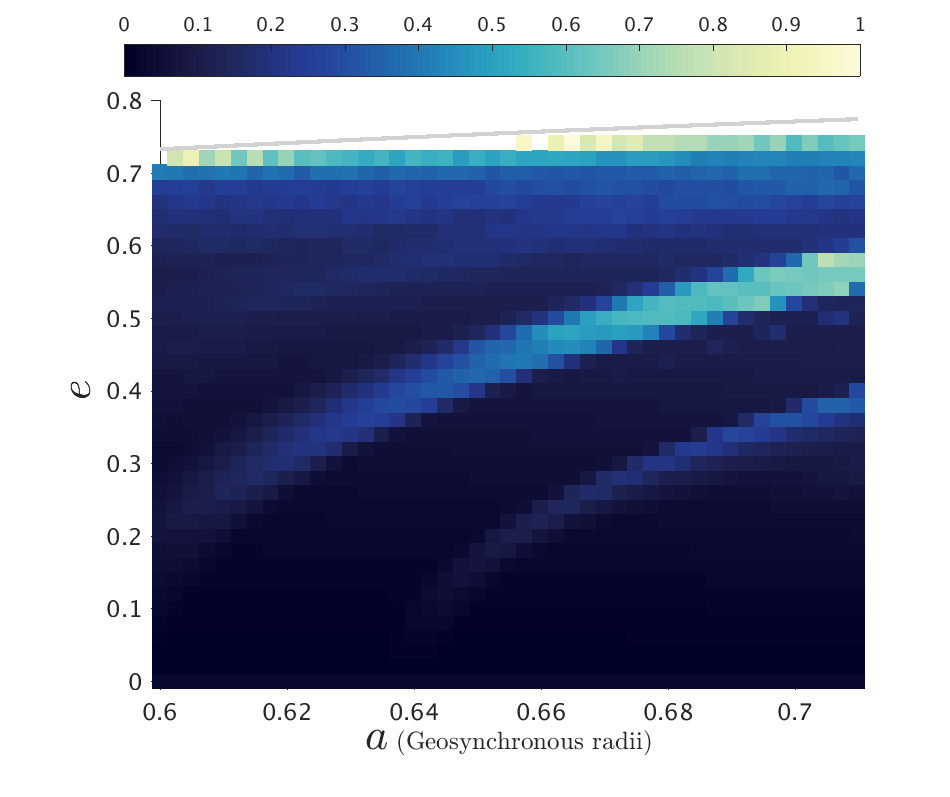} 
      \includegraphics[width=.49\textwidth]{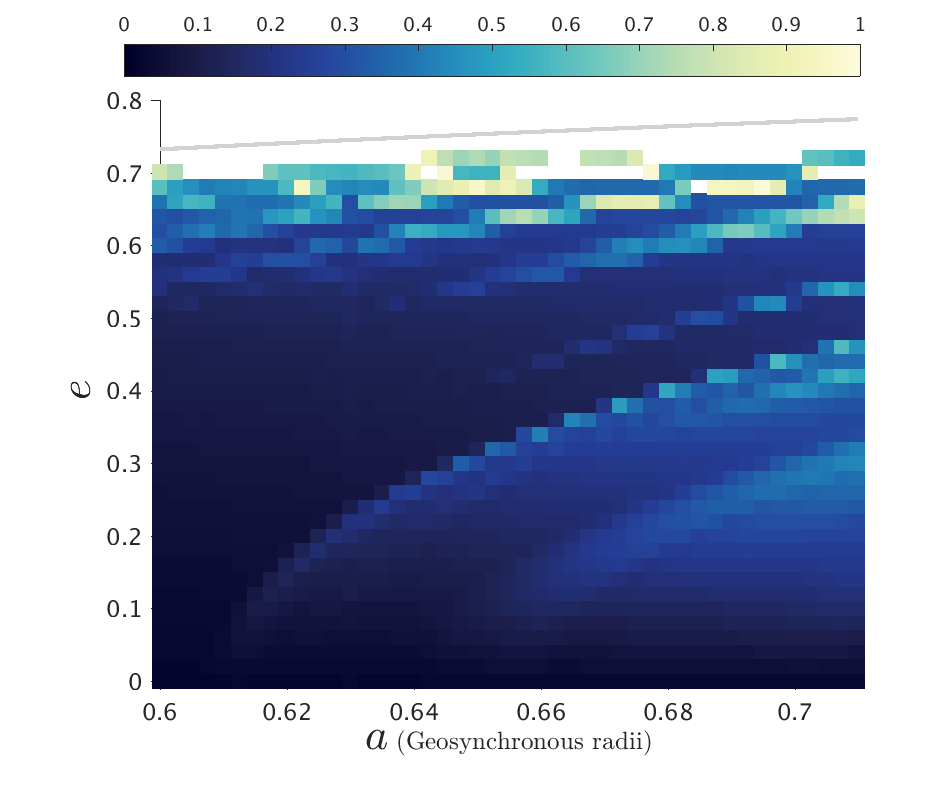}
    \end{subfigure}    
    \begin{subfigure}[b]{0.45\textwidth}
      \caption{$\bm{i}_{o}={\bf 54^{\circ}}$}
      \includegraphics[width=.49\textwidth]{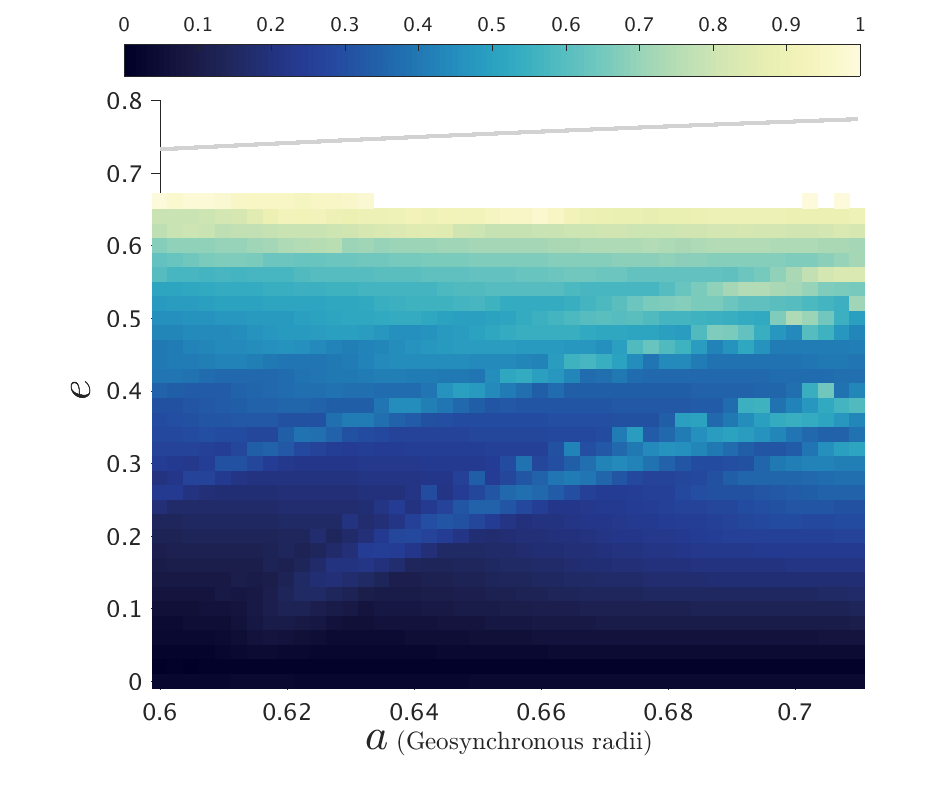} 
      \includegraphics[width=.49\textwidth]{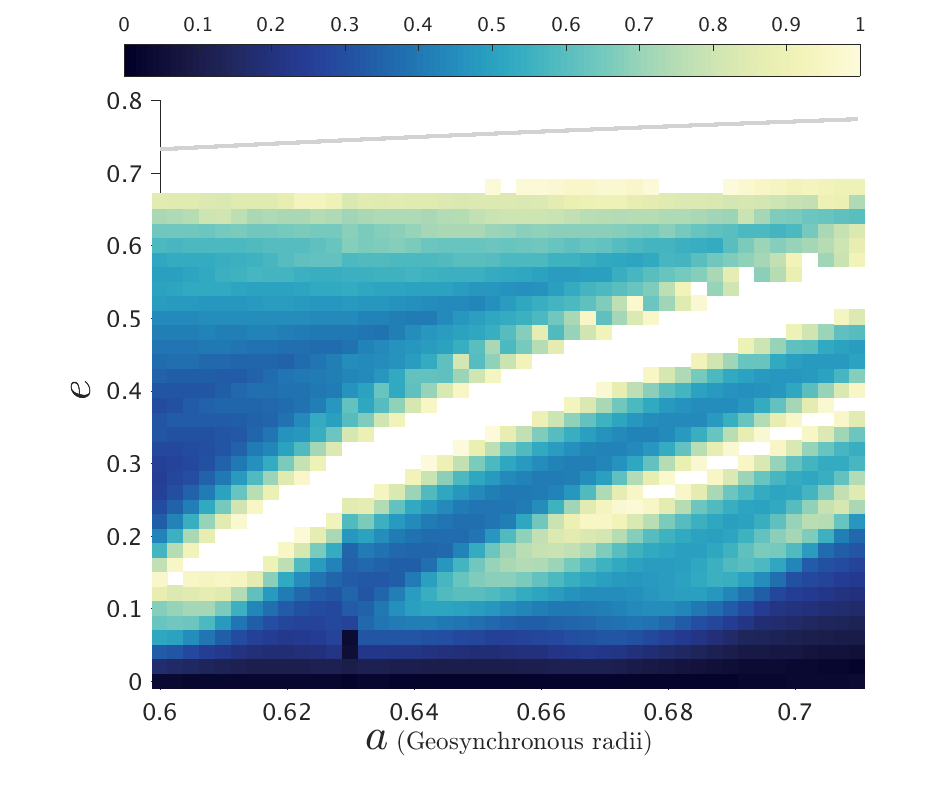}
    \end{subfigure}   
    \begin{subfigure}[b]{0.45\textwidth}
      \caption{$\bm{i}_{o}={\bf 58^{\circ}}$}
      \includegraphics[width=.49\textwidth]{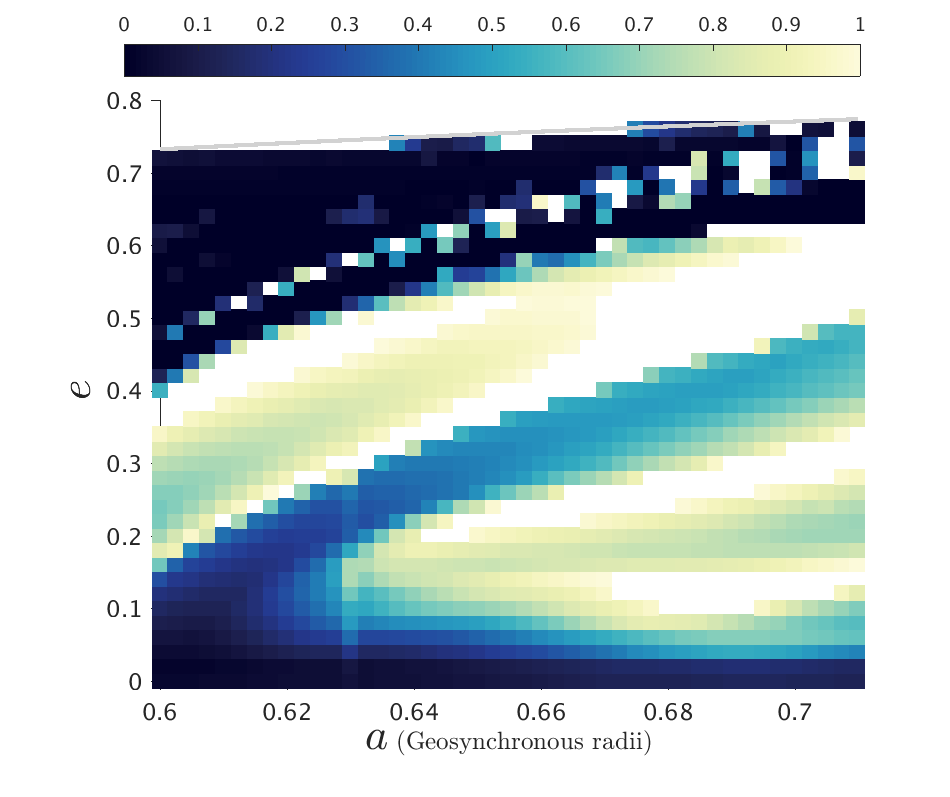} 
      \includegraphics[width=.49\textwidth]{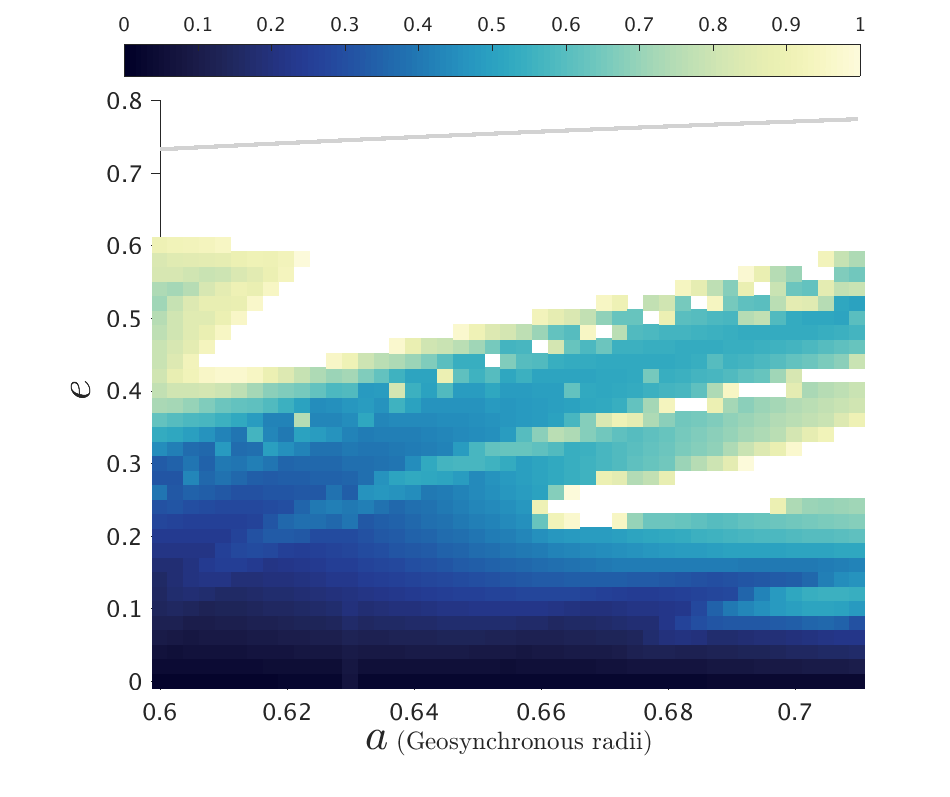}
    \end{subfigure}  
    \begin{subfigure}[b]{0.45\textwidth}
      \caption{$\bm{i}_{o}={\bf 66^{\circ}}$}
      \includegraphics[width=.49\textwidth]{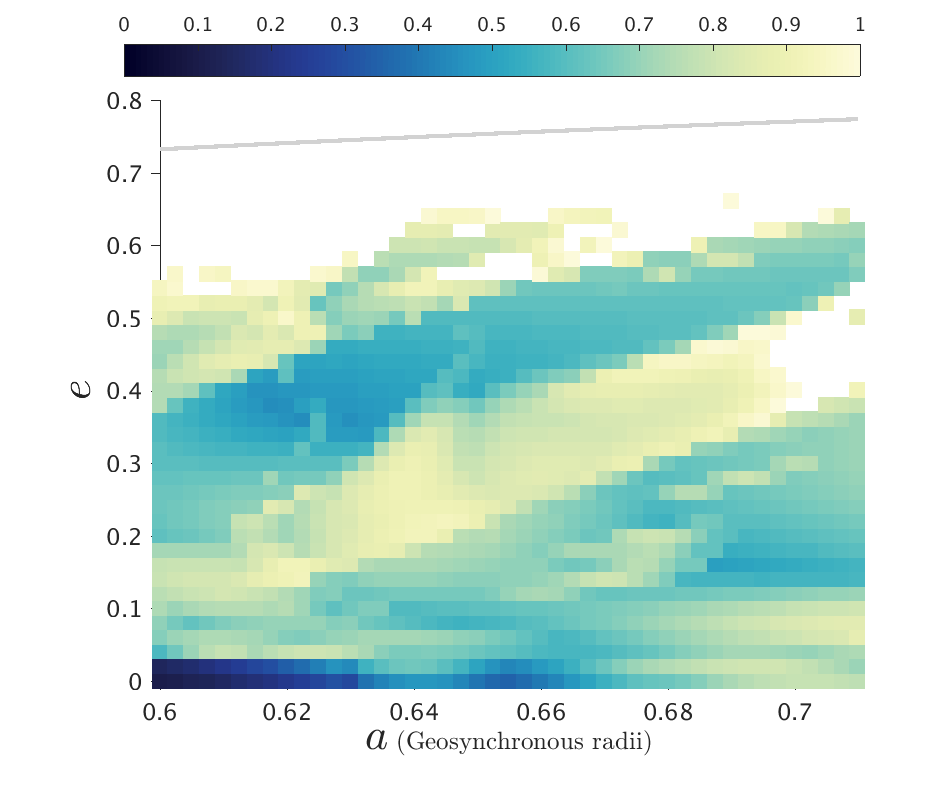} 
      \includegraphics[width=.49\textwidth]{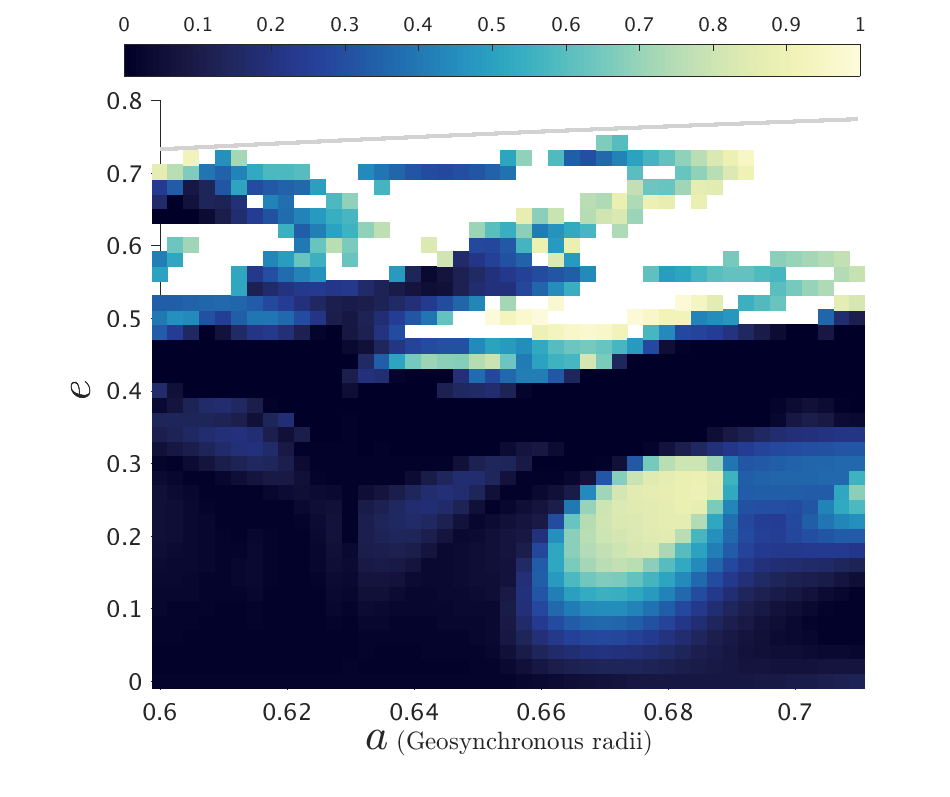}
    \end{subfigure}  
    \begin{subfigure}[b]{0.45\textwidth}
      \caption{$\bm{i}_{o}={\bf 68^{\circ}}$}
      \includegraphics[width=.49\textwidth]{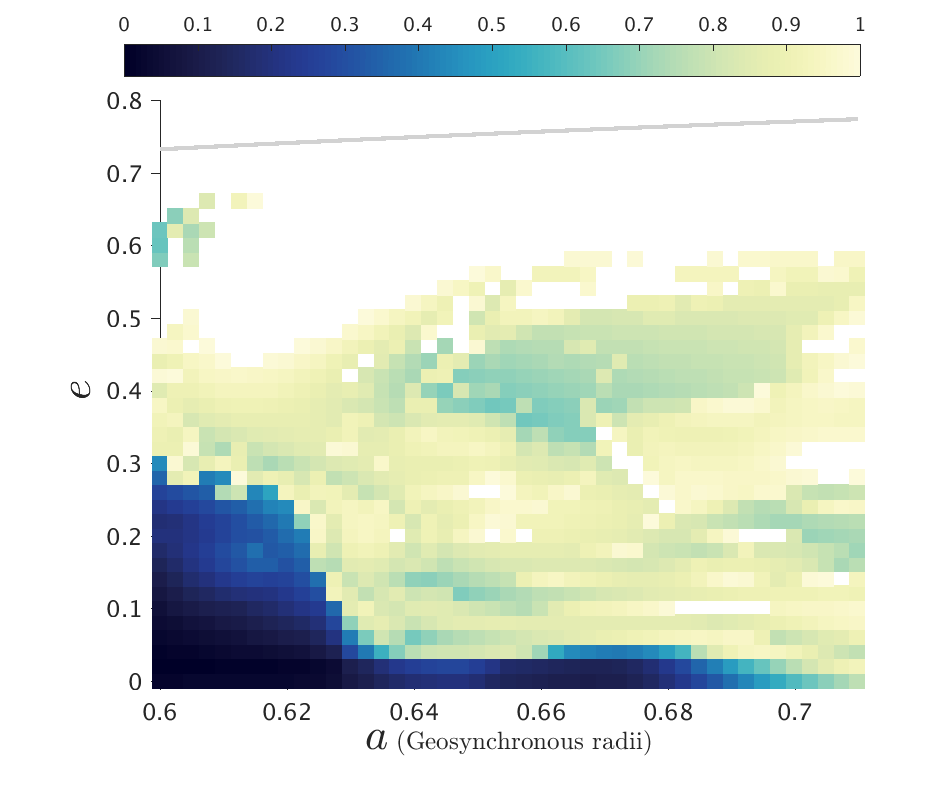} 
      \includegraphics[width=.49\textwidth]{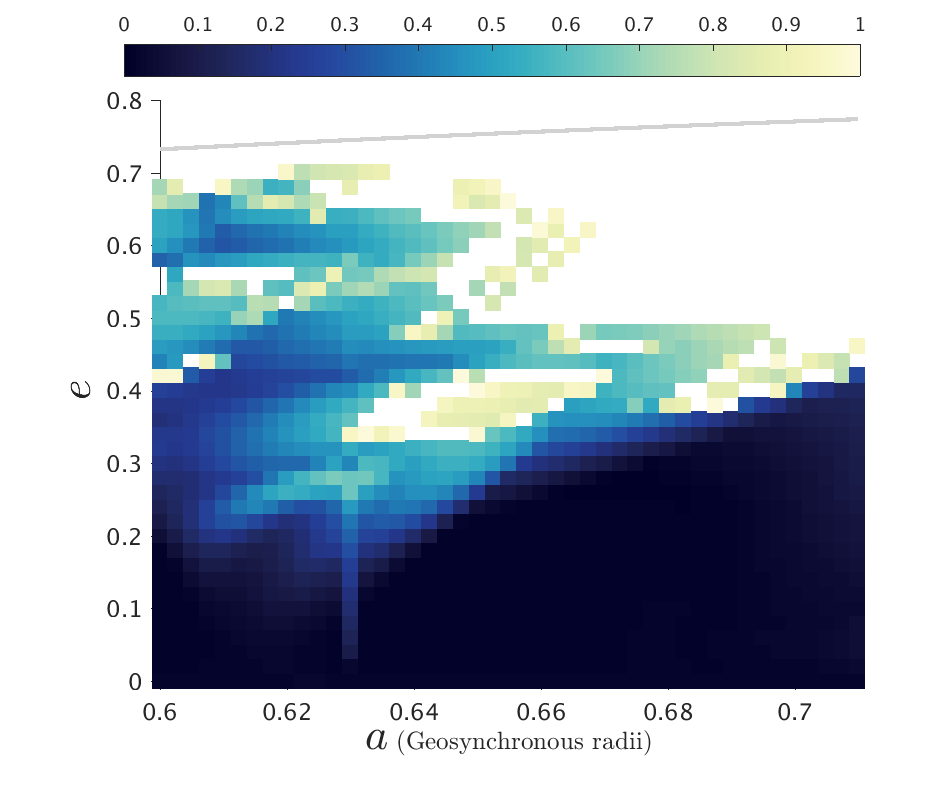}
    \end{subfigure}  
    \begin{subfigure}[b]{0.45\textwidth}
      \caption{$\bm{i}_{o}={\bf 70^{\circ}}$}
      \includegraphics[width=.49\textwidth]{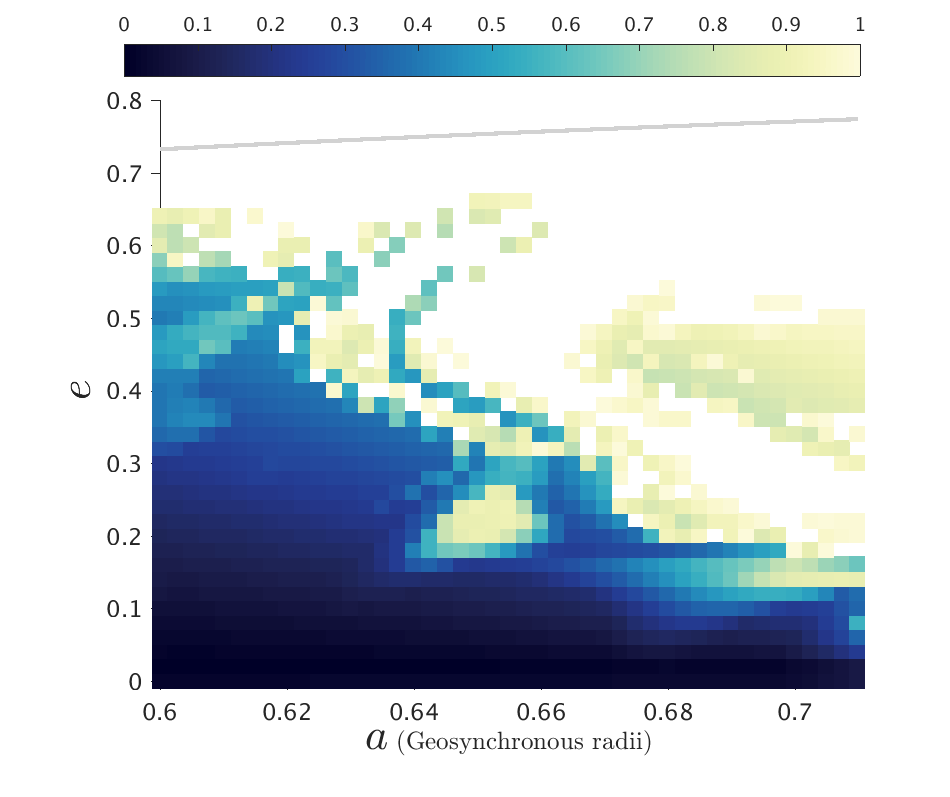} 
      \includegraphics[width=.49\textwidth]{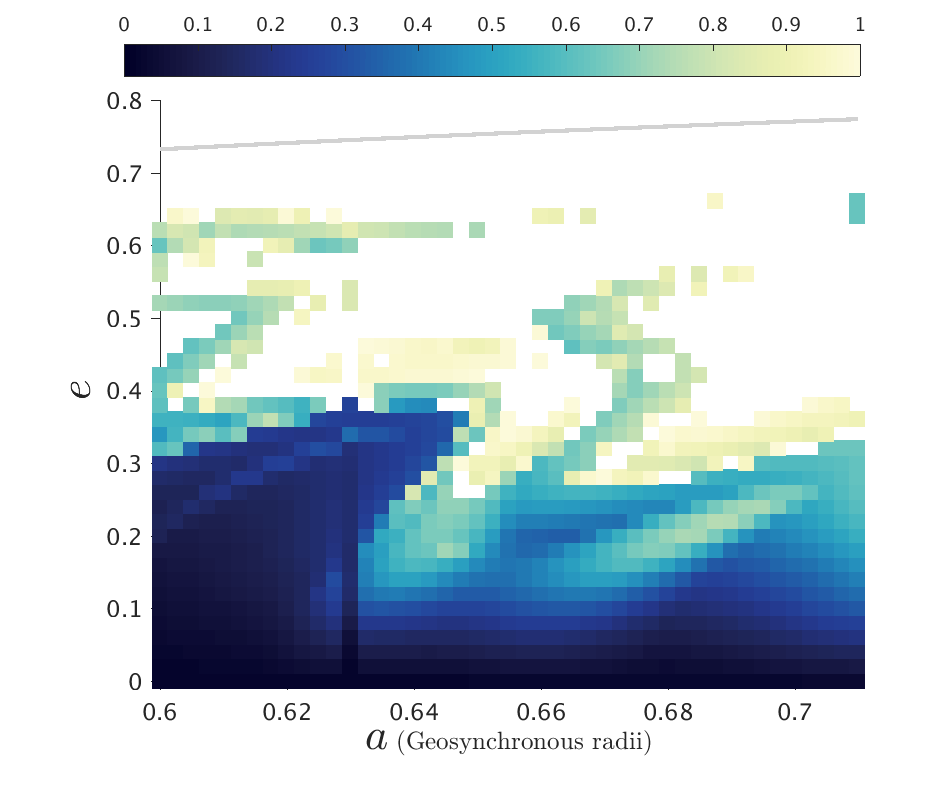}
    \end{subfigure}
    \begin{subfigure}[b]{0.45\textwidth}
      \caption{$\bm{i}_{o}={\bf 90^{\circ}}$}
      \includegraphics[width=.49\textwidth]{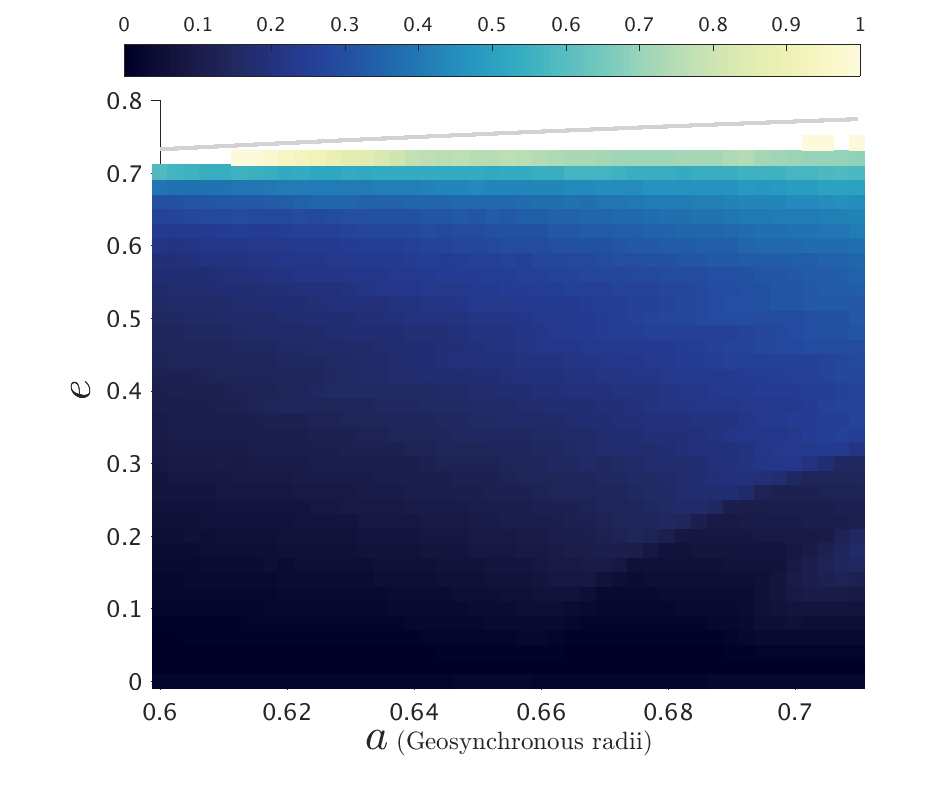} 
      \includegraphics[width=.49\textwidth]{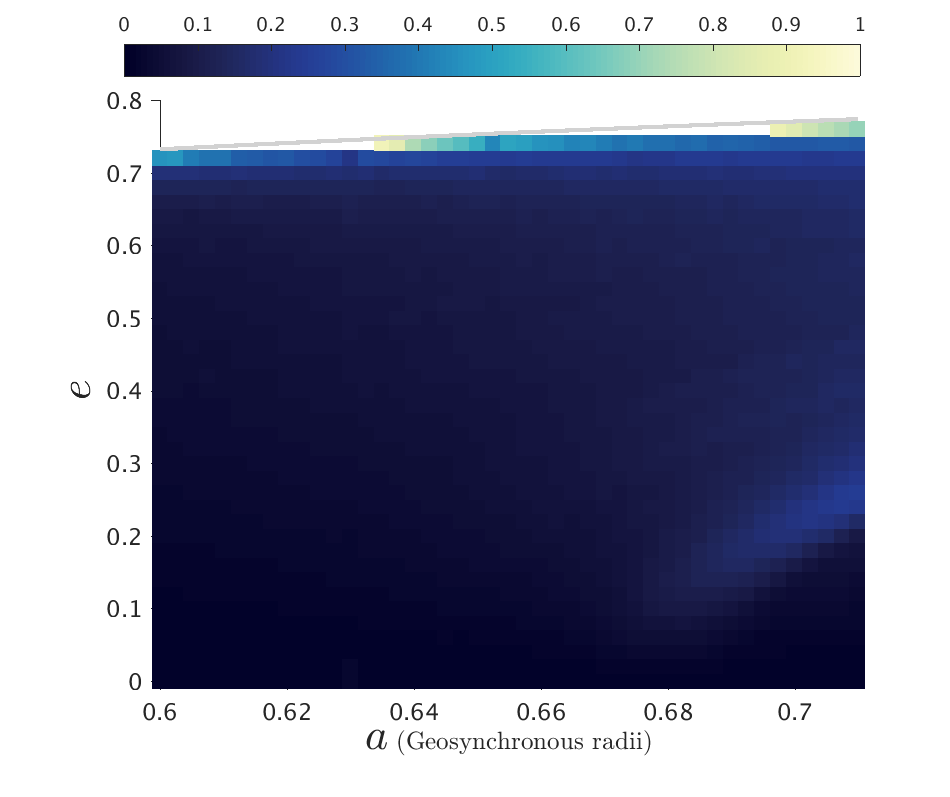}
    \end{subfigure}    
  \caption{$De$ maps of the \textit{MEO-general} phase space for various $\bm{i}_{o}$, 
  $\bm{\Delta}\bm{\Omega} = {\bf 0^\circ}$, $\bm{\Delta}\bm{\omega} = {\bf 270^\circ}$ (1st and 3rd columns) and 
  $\bm{\Delta}\bm{\Omega} = {\bf 90^\circ}$, $\bm{\Delta}\bm{\omega} = {\bf 0}$ (2nd and 4th columns),  
  for Epoch 2020, and for $C_{R}A/m=1$ m$^2$/kg.
  The colorbar for the $De$ maps is from 0 to 1, where the reentry particles were excluded (white).}
  \label{fig:MEO_gen_i2}
\end{figure}

\begin{figure}[htp!]
  \centering
    \begin{subfigure}[b]{0.45\textwidth}
      \caption{$\bm{i}_{o}={\bf 0}$}
      \includegraphics[width=.49\textwidth]{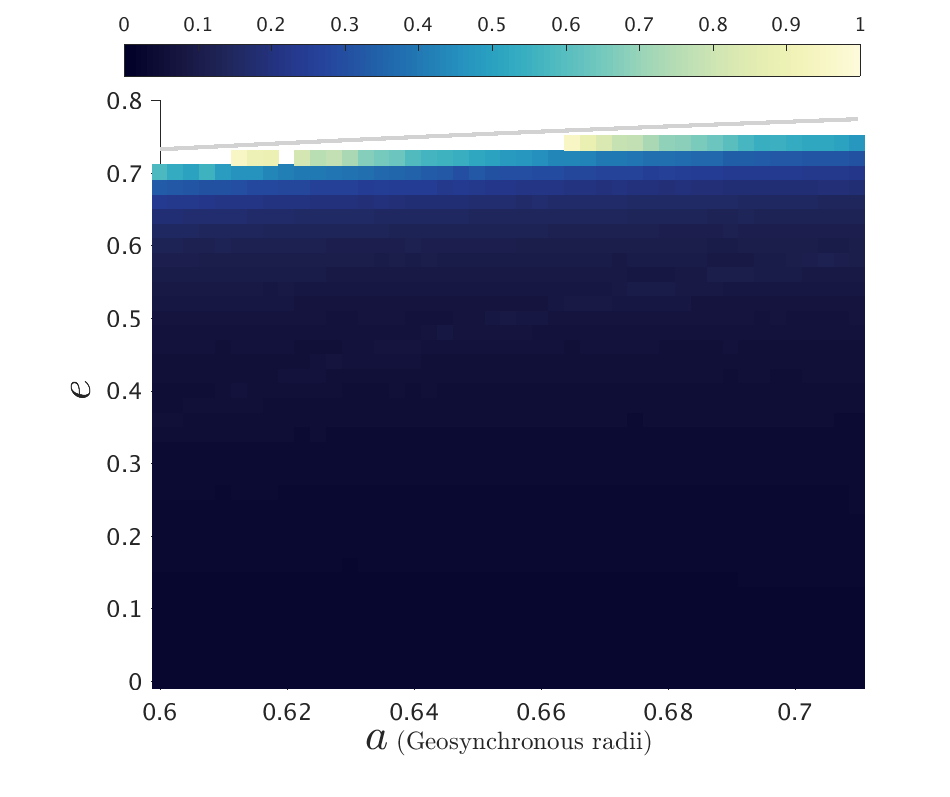} 
      \includegraphics[width=.49\textwidth]{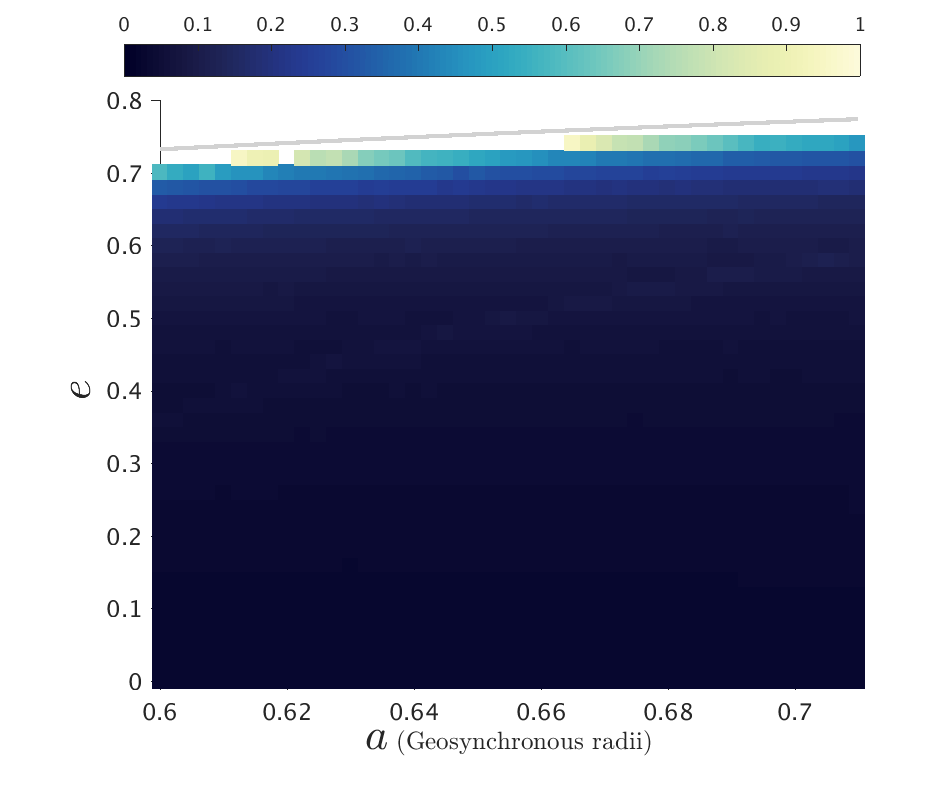}
    \end{subfigure} 
    \begin{subfigure}[b]{0.45\textwidth}
      \caption{$\bm{i}_{o}={\bf 28^{\circ}}$}
      \includegraphics[width=.49\textwidth]{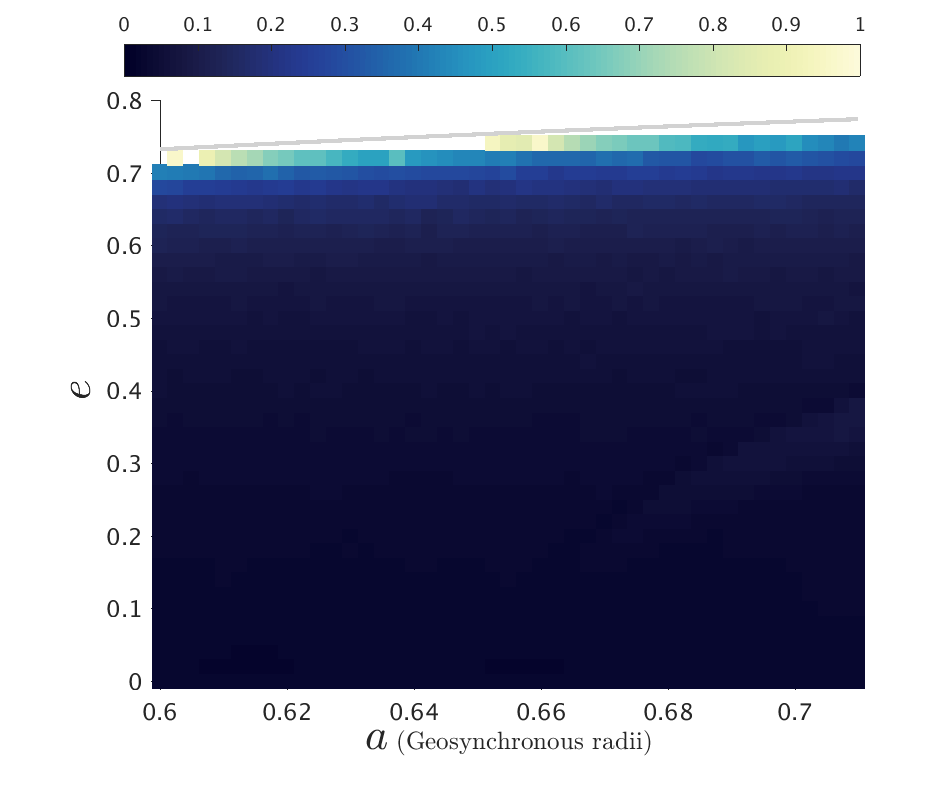} 
      \includegraphics[width=.49\textwidth]{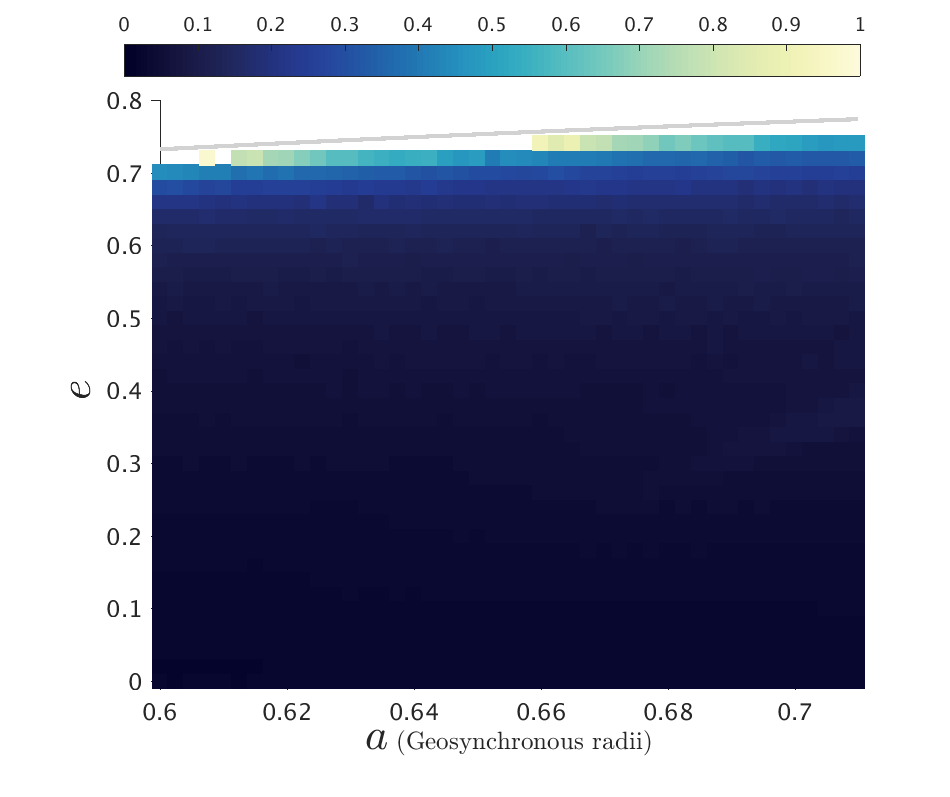}
    \end{subfigure}  
    \begin{subfigure}[b]{0.45\textwidth}
      \caption{$\bm{i}_{o}={\bf 44^{\circ}}$}
      \includegraphics[width=.49\textwidth]{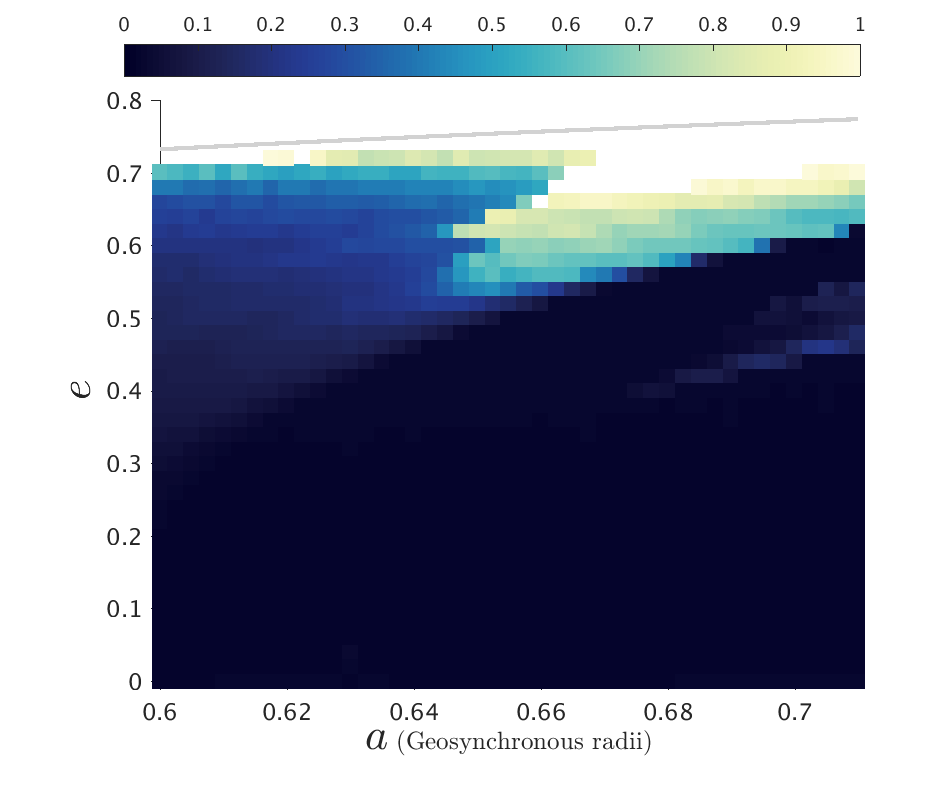} 
      \includegraphics[width=.49\textwidth]{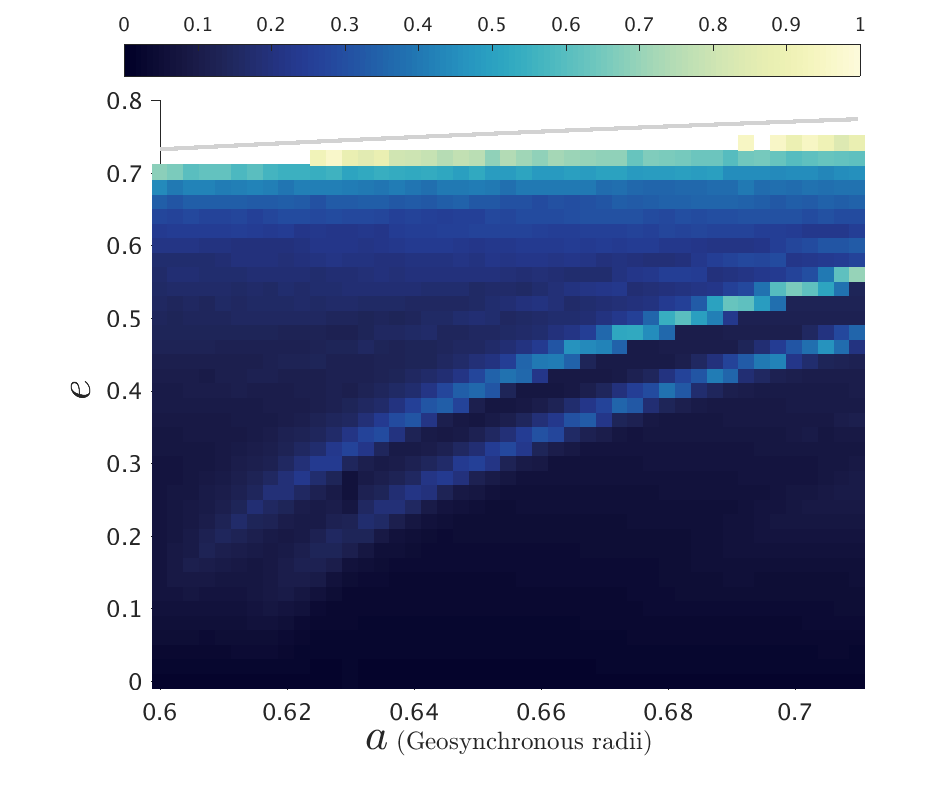}
    \end{subfigure}  
    \begin{subfigure}[b]{0.45\textwidth}
      \caption{$\bm{i}_{o}={\bf 46^{\circ}}$}
      \includegraphics[width=.49\textwidth]{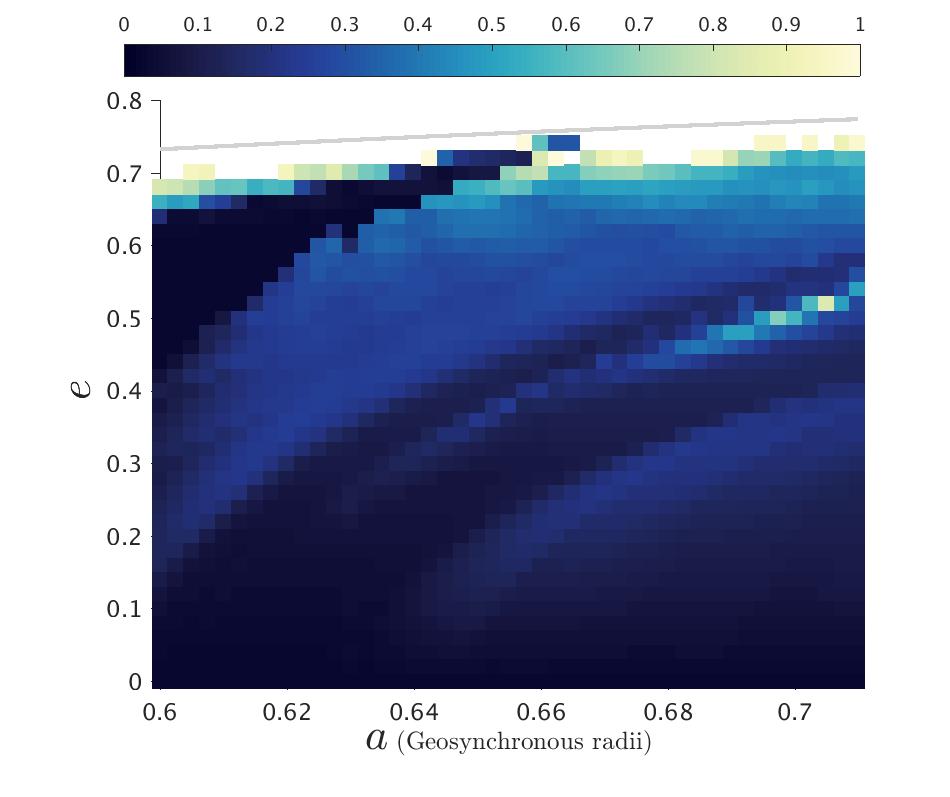} 
      \includegraphics[width=.49\textwidth]{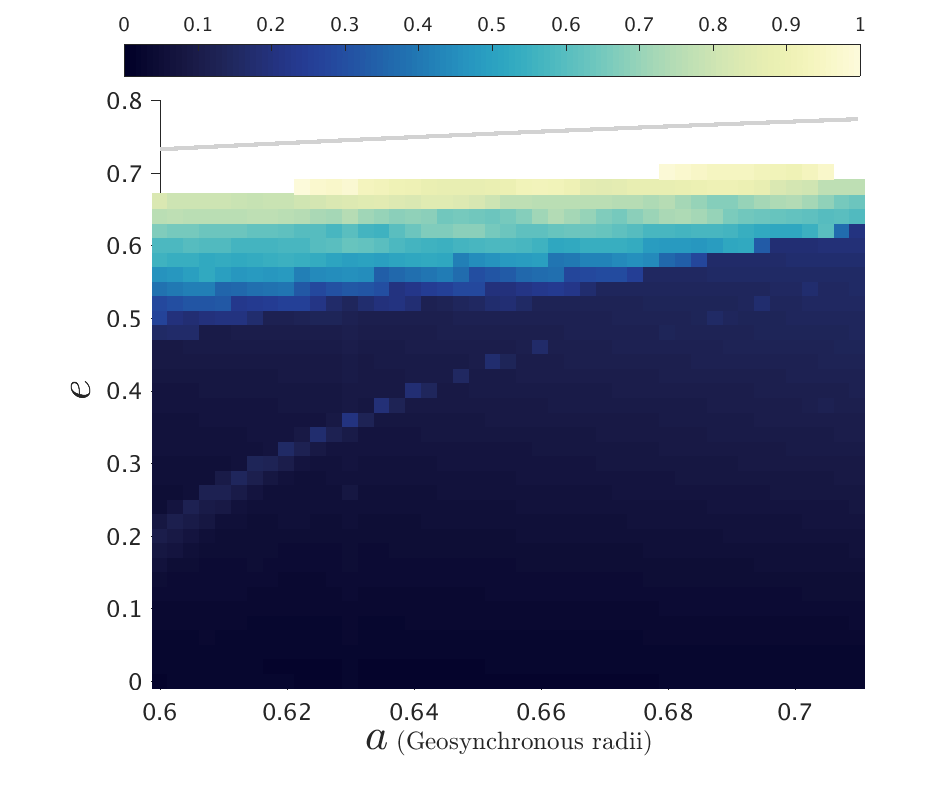}
    \end{subfigure}    
    \begin{subfigure}[b]{0.45\textwidth}
      \caption{$\bm{i}_{o}={\bf 54^{\circ}}$}
      \includegraphics[width=.49\textwidth]{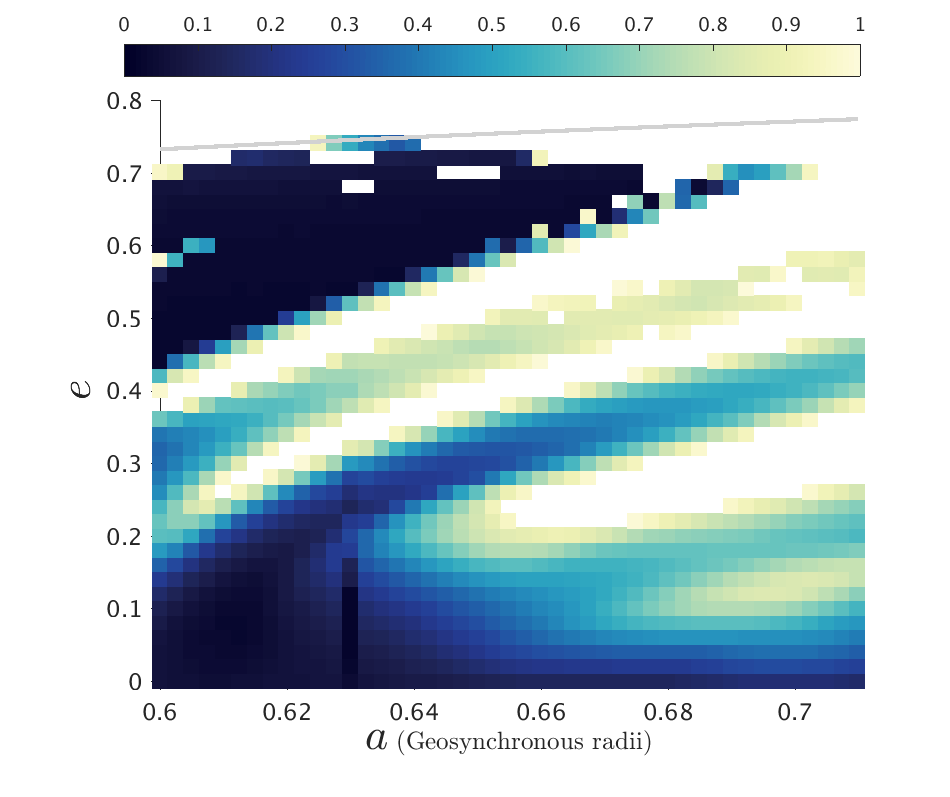} 
      \includegraphics[width=.49\textwidth]{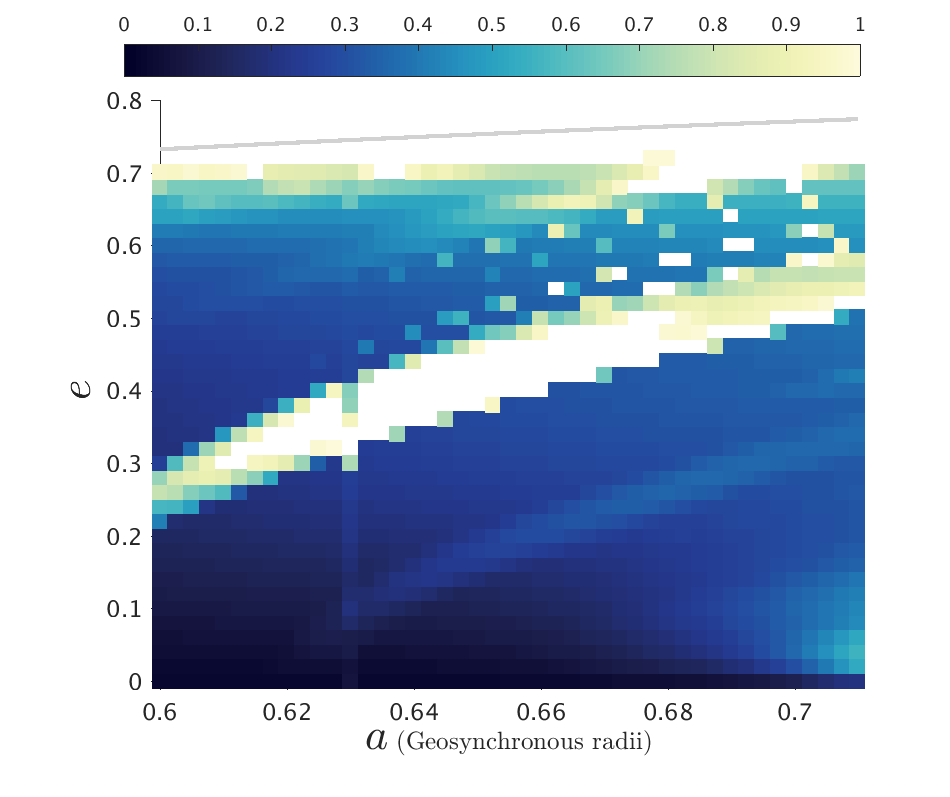}
    \end{subfigure}   
    \begin{subfigure}[b]{0.45\textwidth}
      \caption{$\bm{i}_{o}={\bf 58^{\circ}}$}
      \includegraphics[width=.49\textwidth]{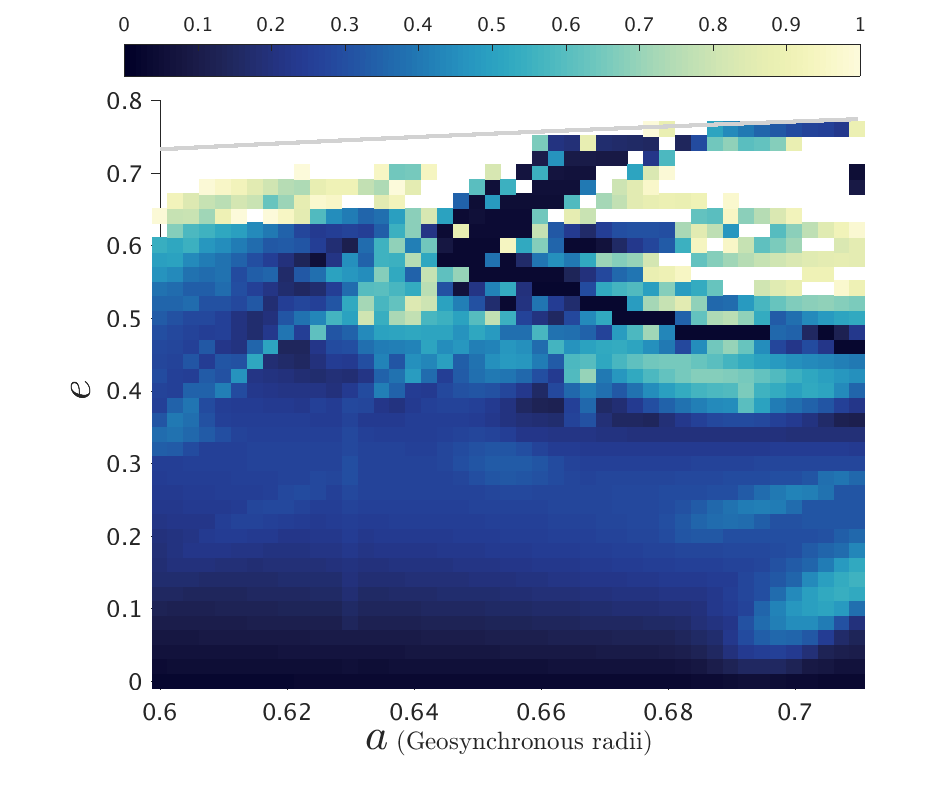} 
      \includegraphics[width=.49\textwidth]{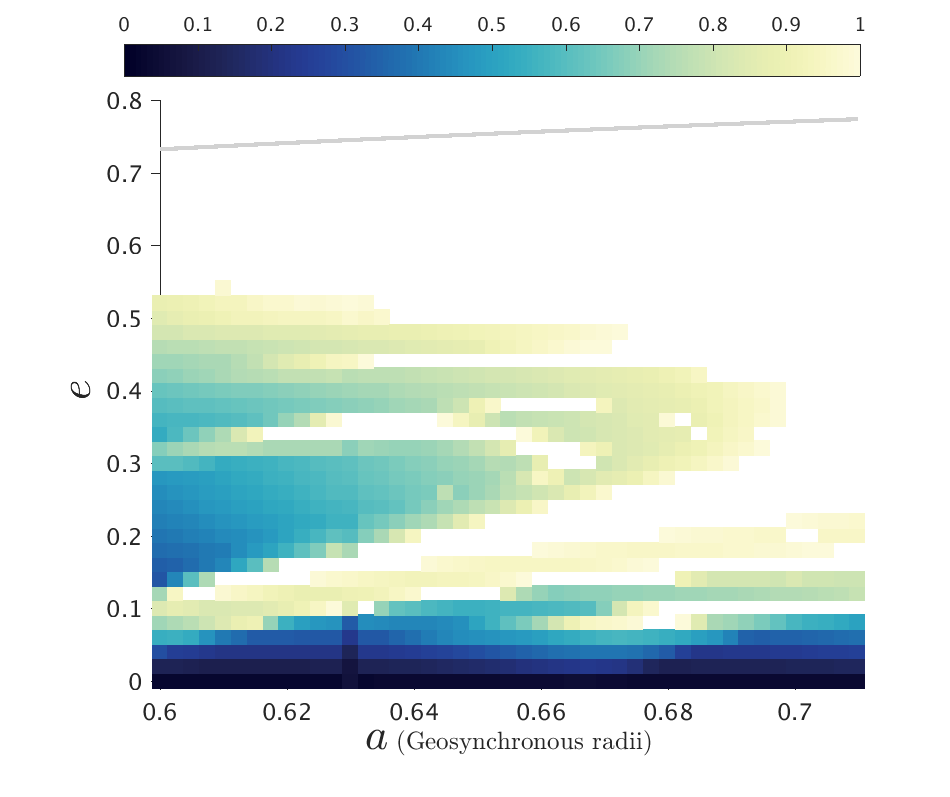}
    \end{subfigure}  
    \begin{subfigure}[b]{0.45\textwidth}
      \caption{$\bm{i}_{o}={\bf 66^{\circ}}$}
      \includegraphics[width=.49\textwidth]{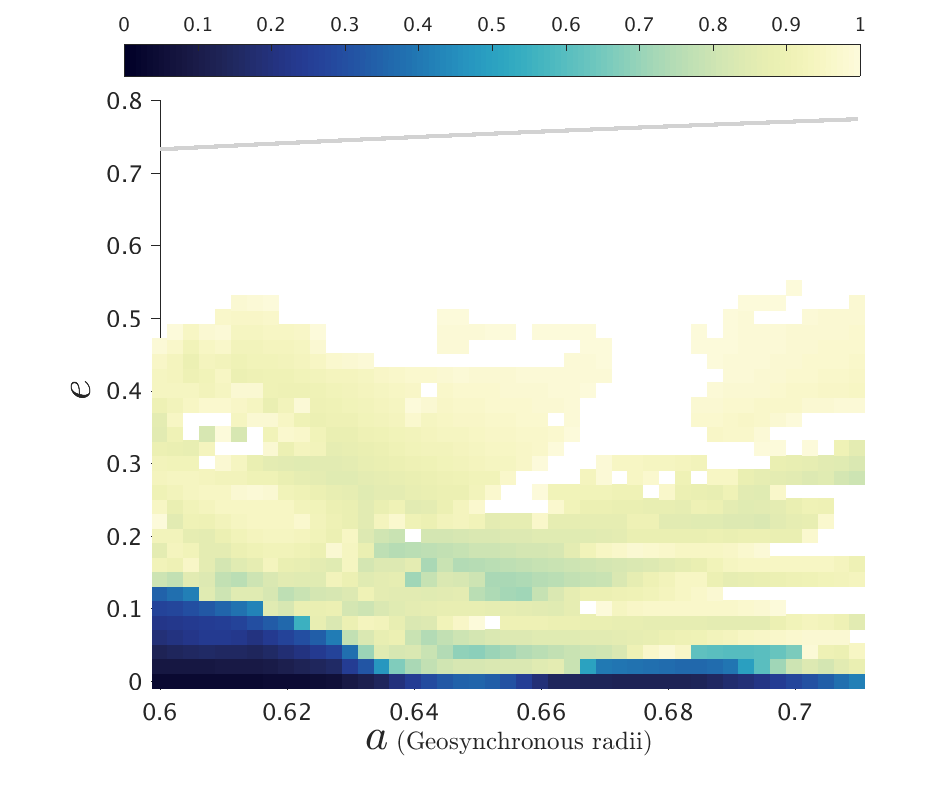} 
      \includegraphics[width=.49\textwidth]{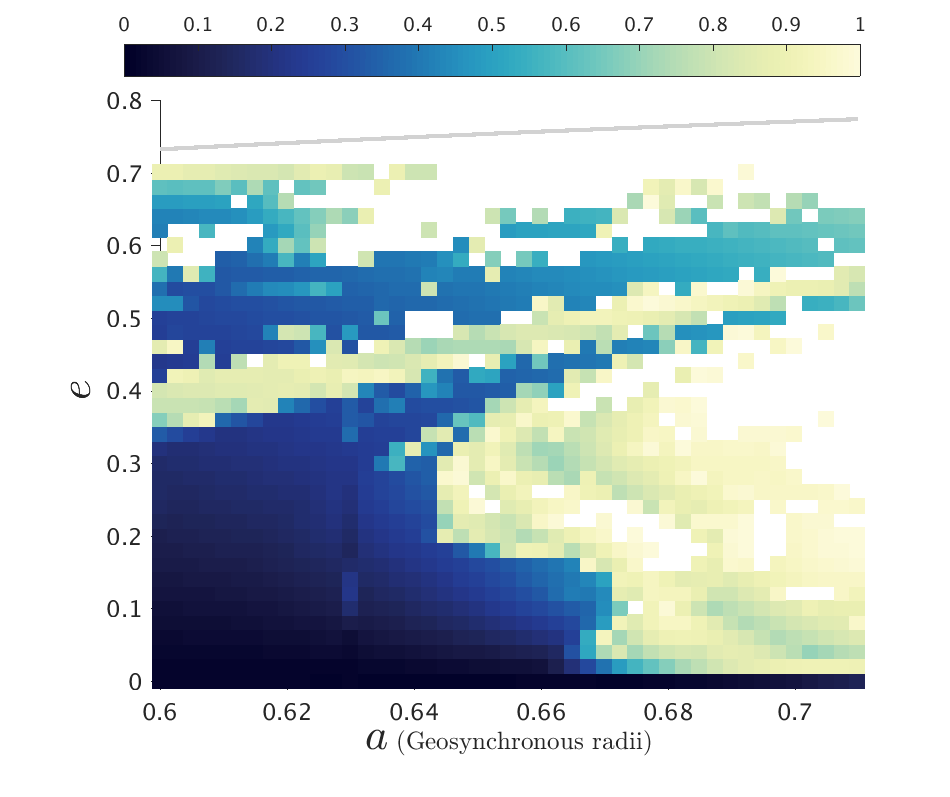}
    \end{subfigure}  
    \begin{subfigure}[b]{0.45\textwidth}
      \caption{$\bm{i}_{o}={\bf 68^{\circ}}$}
      \includegraphics[width=.49\textwidth]{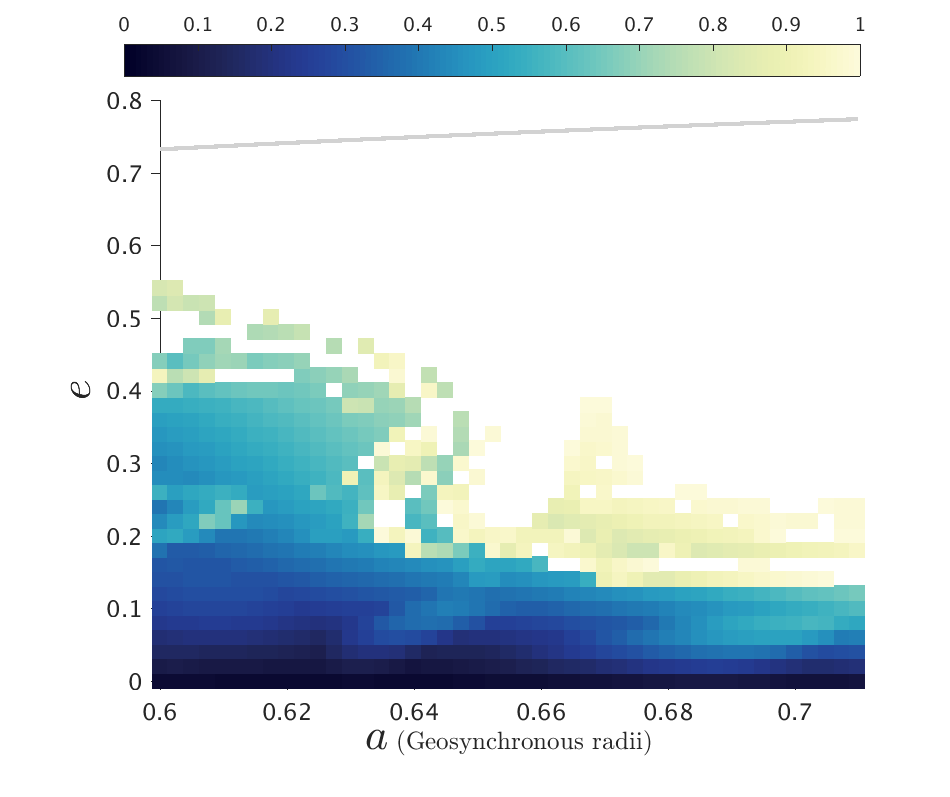} 
      \includegraphics[width=.49\textwidth]{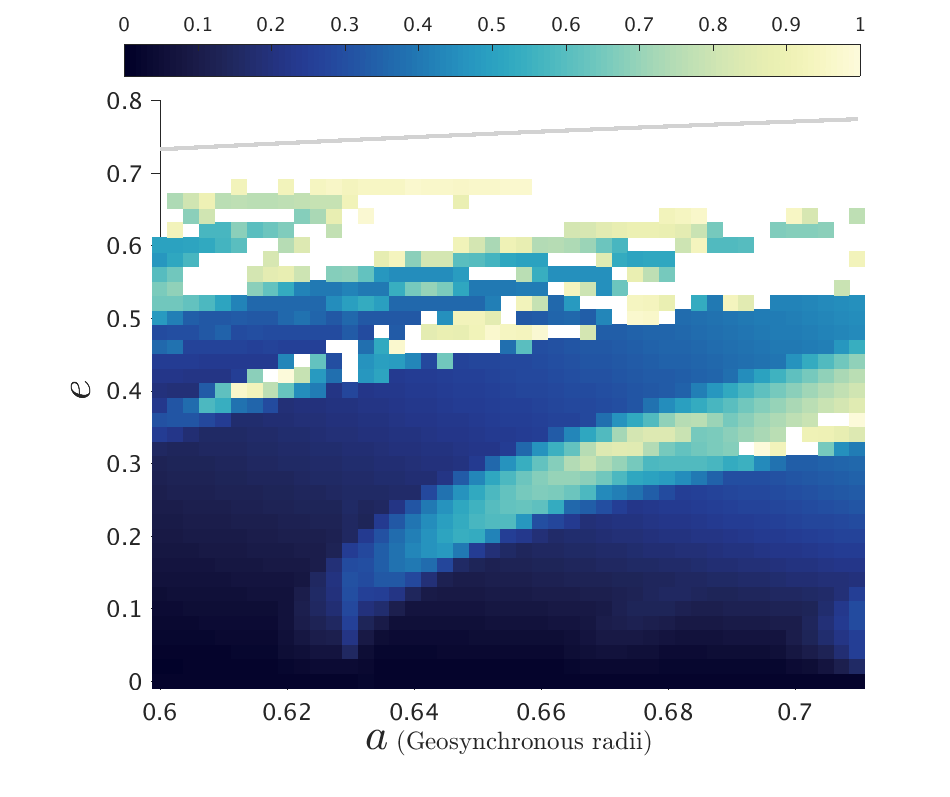}
    \end{subfigure}  
    \begin{subfigure}[b]{0.45\textwidth}
      \caption{$\bm{i}_{o}={\bf 70^{\circ}}$}
      \includegraphics[width=.49\textwidth]{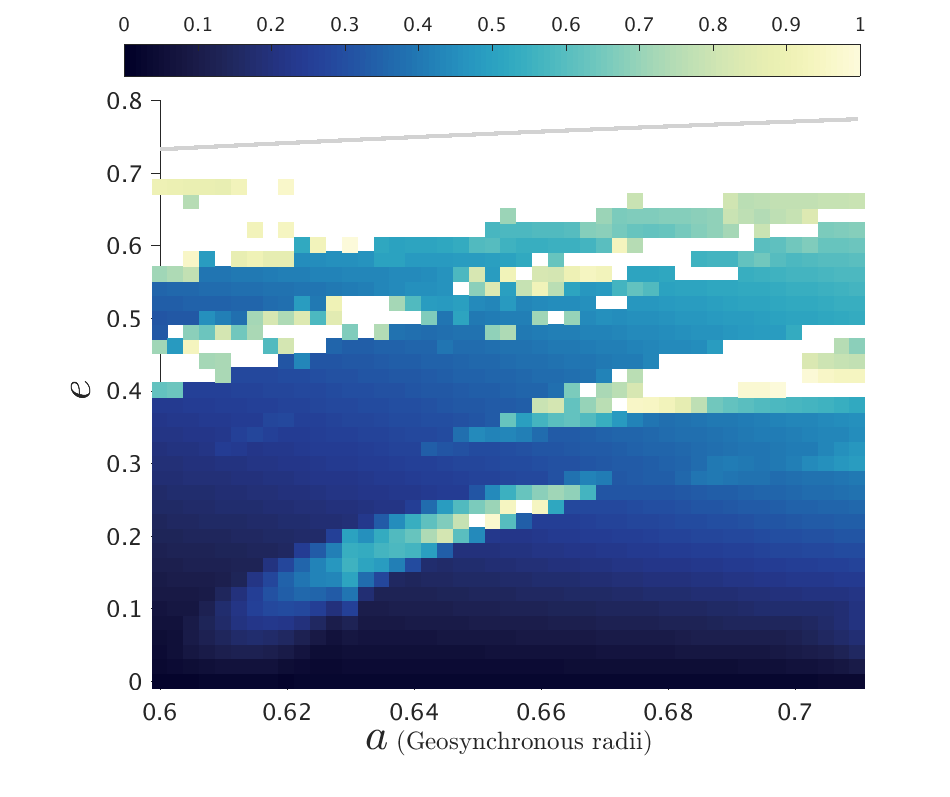} 
      \includegraphics[width=.49\textwidth]{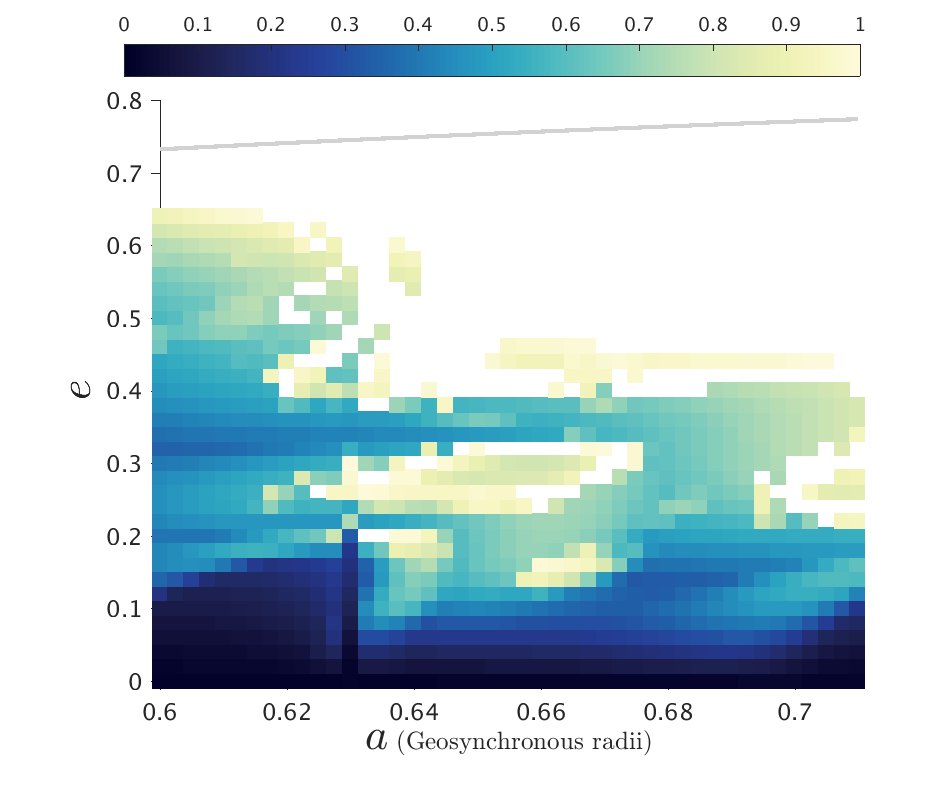}
    \end{subfigure}
    \begin{subfigure}[b]{0.45\textwidth}
      \caption{$\bm{i}_{o}={\bf 90^{\circ}}$}
      \includegraphics[width=.49\textwidth]{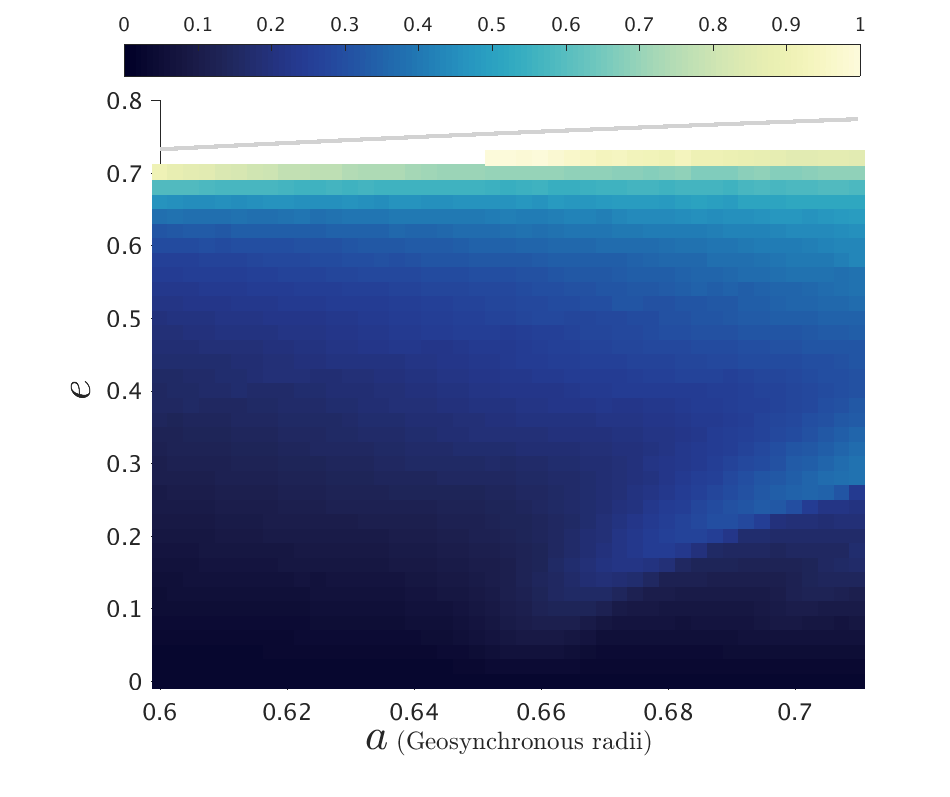} 
      \includegraphics[width=.49\textwidth]{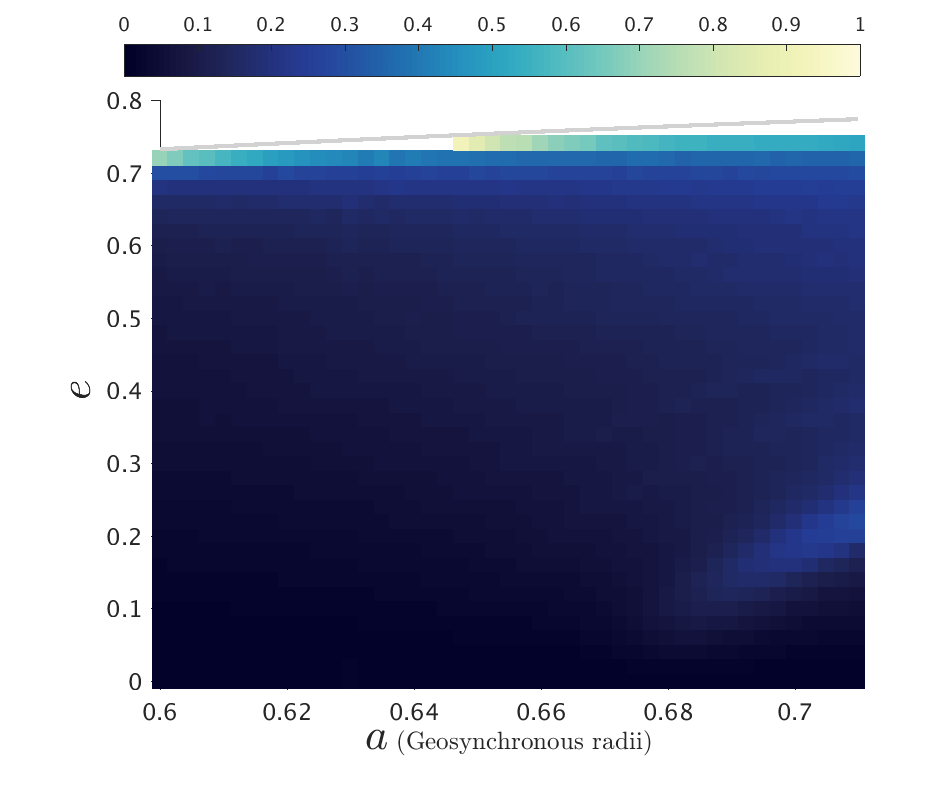}
    \end{subfigure}    
  \caption{$De$ maps of the \textit{MEO-general} phase space for various $\bm{i}_{o}$, 
  $\bm{\Delta}\bm{\Omega} = {\bf 180^\circ}$, $\bm{\Delta}\bm{\omega} = {\bf 270^\circ}$ (1st and 3rd columns) and 
  $\bm{\Delta}\bm{\Omega} = {\bf 270^\circ}$, $\bm{\Delta}\bm{\omega} = {\bf 90^\circ}$ (2nd and 4th columns),  
  for Epoch 2018, and for $C_{R}A/m=1$ m$^2$/kg.
  The colorbar for the $De$ maps is from 0 to 1, where the reentry particles were excluded (white).}
  \label{fig:MEO_gen_i3}
\end{figure}

\begin{figure}[htp!]
  \centering
    \begin{subfigure}[b]{0.45\textwidth}
      \caption{$\bm{i}_{o}={\bf 0}$}
      \includegraphics[width=.49\textwidth]{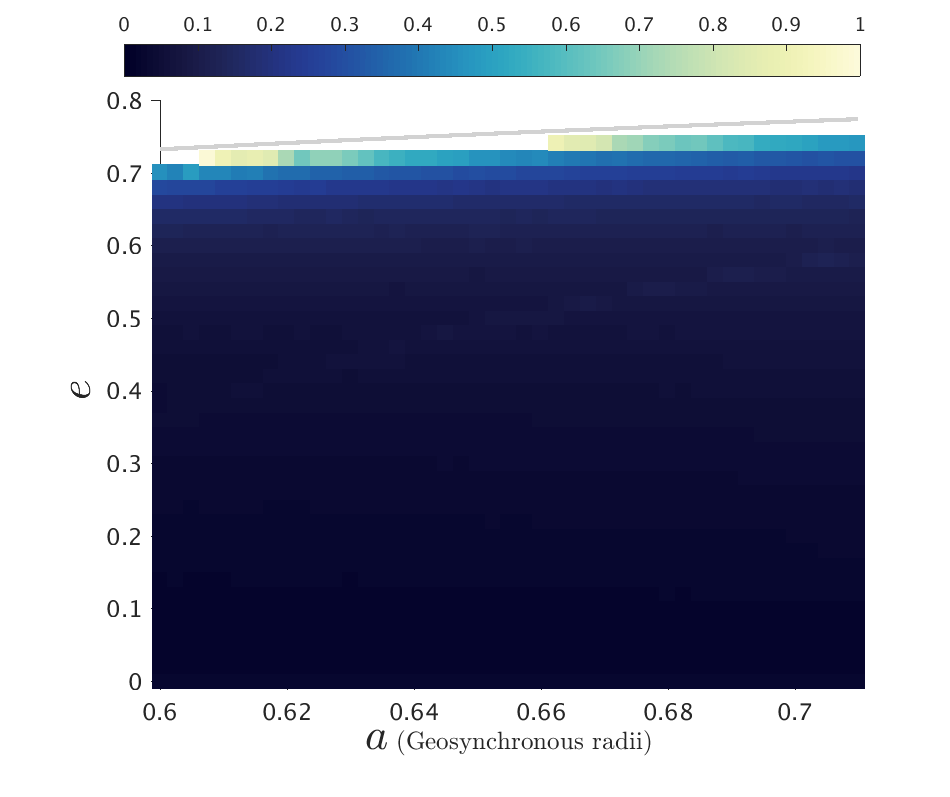} 
      \includegraphics[width=.49\textwidth]{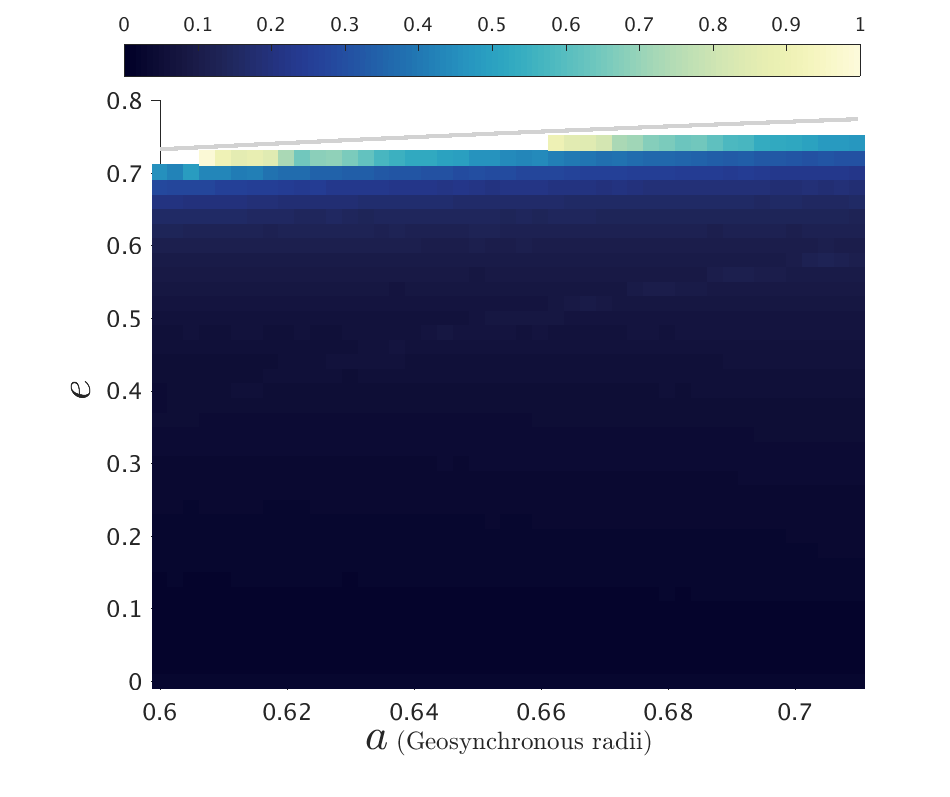}
    \end{subfigure} 
    \begin{subfigure}[b]{0.45\textwidth}
      \caption{$\bm{i}_{o}={\bf 28^{\circ}}$}
      \includegraphics[width=.49\textwidth]{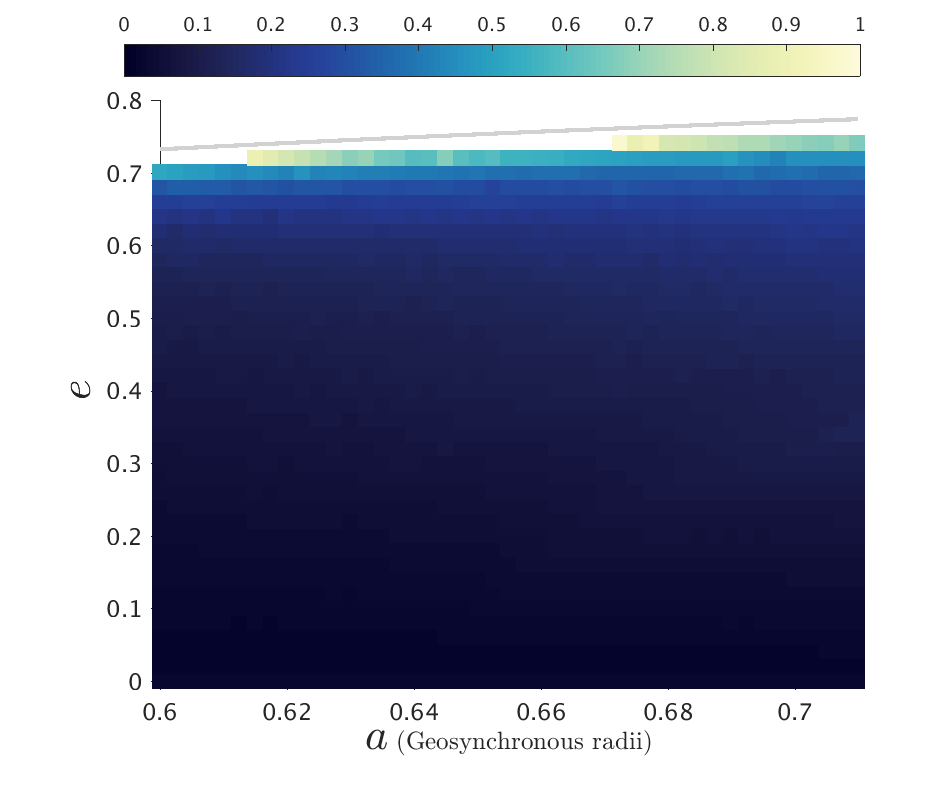} 
      \includegraphics[width=.49\textwidth]{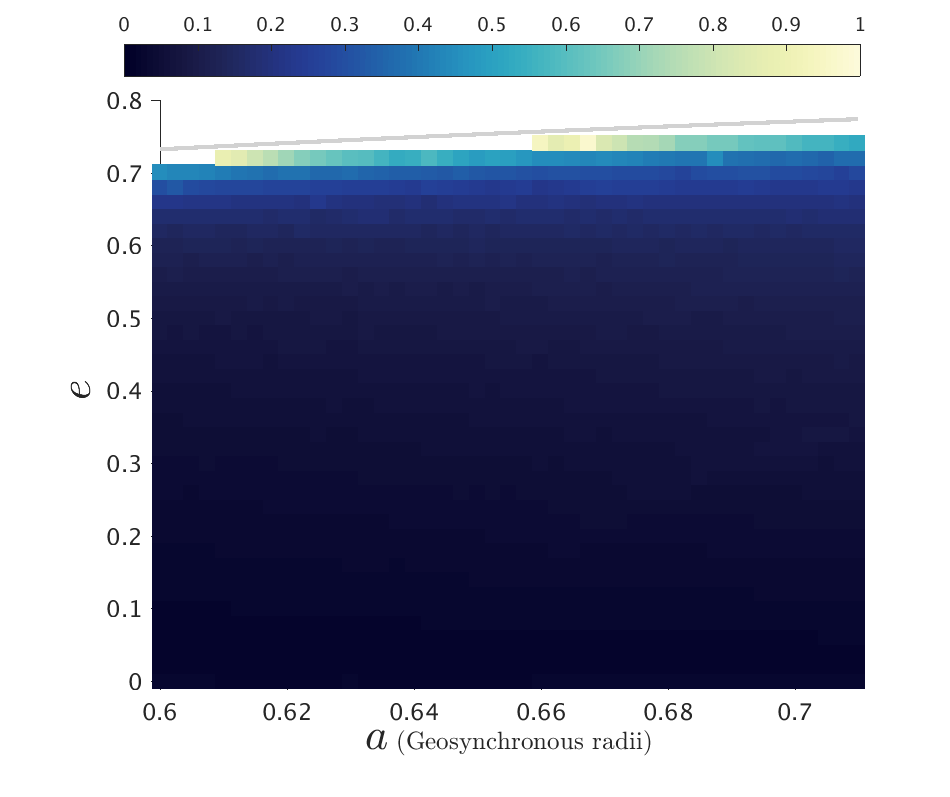}
    \end{subfigure}  
    \begin{subfigure}[b]{0.45\textwidth}
      \caption{$\bm{i}_{o}={\bf 44^{\circ}}$}
      \includegraphics[width=.49\textwidth]{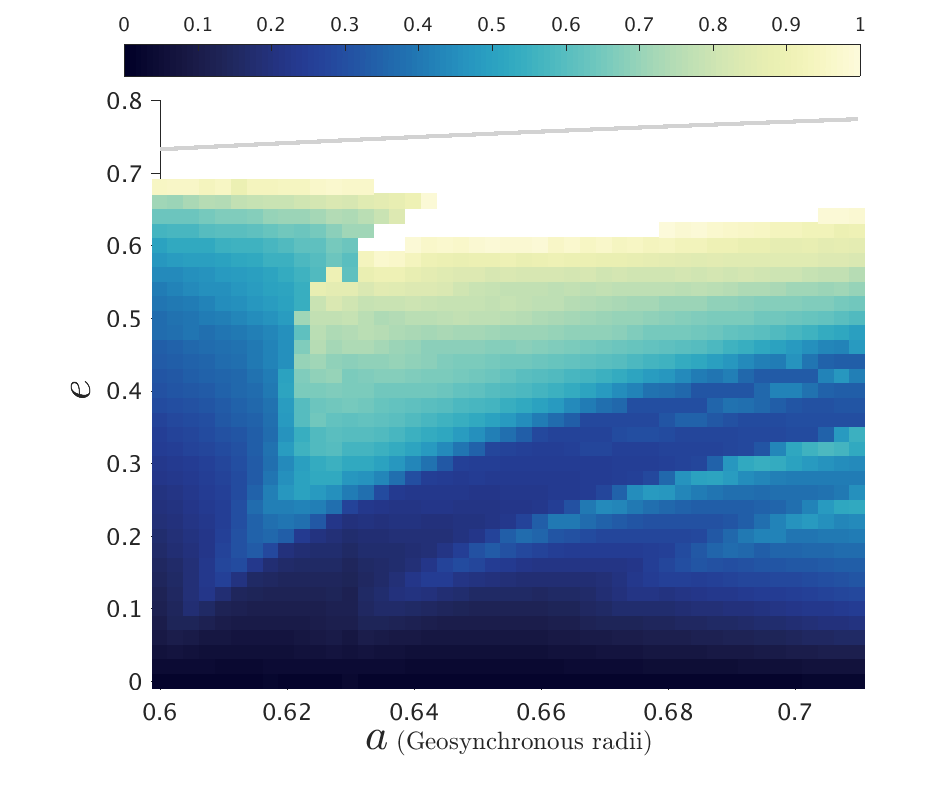} 
      \includegraphics[width=.49\textwidth]{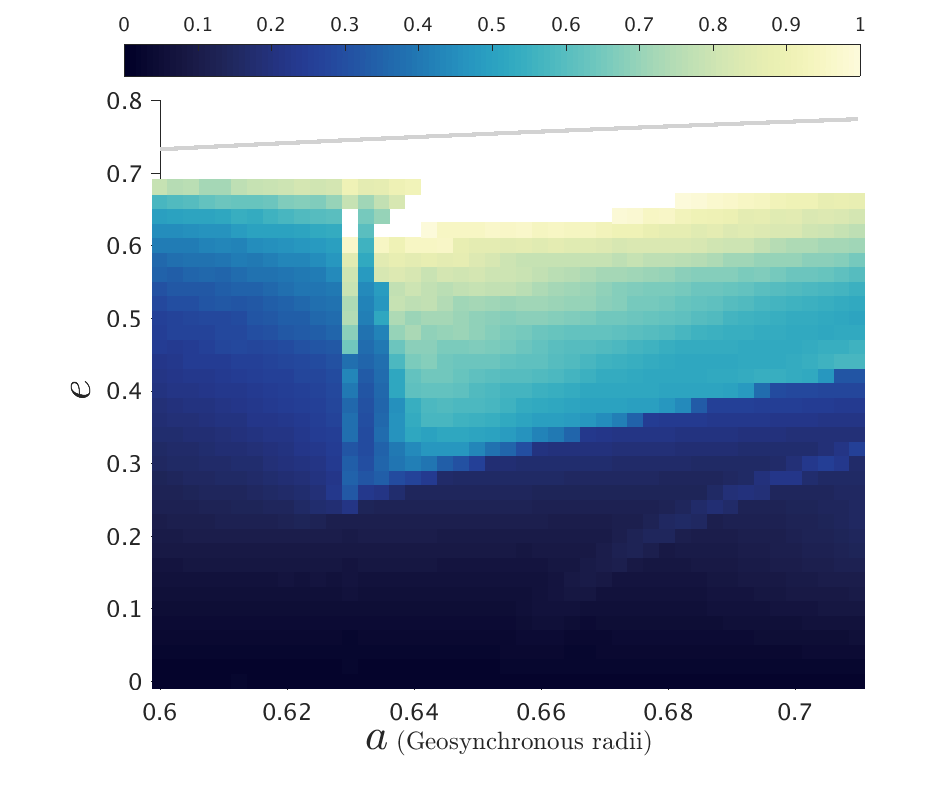}
    \end{subfigure}  
    \begin{subfigure}[b]{0.45\textwidth}
      \caption{$\bm{i}_{o}={\bf 46^{\circ}}$}
      \includegraphics[width=.49\textwidth]{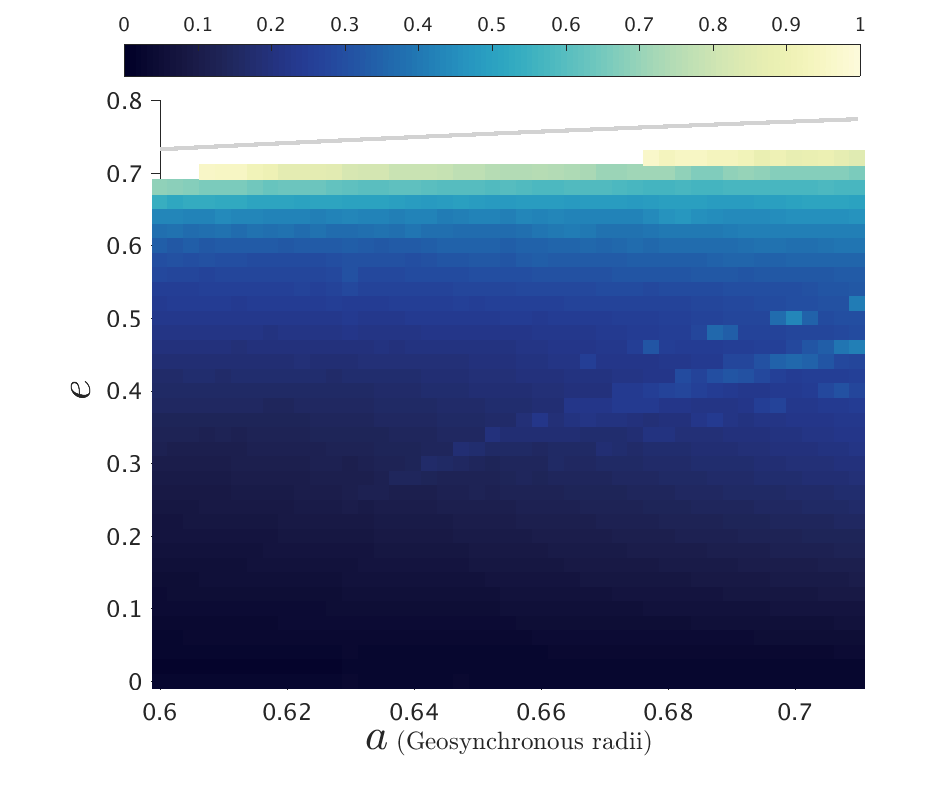} 
      \includegraphics[width=.49\textwidth]{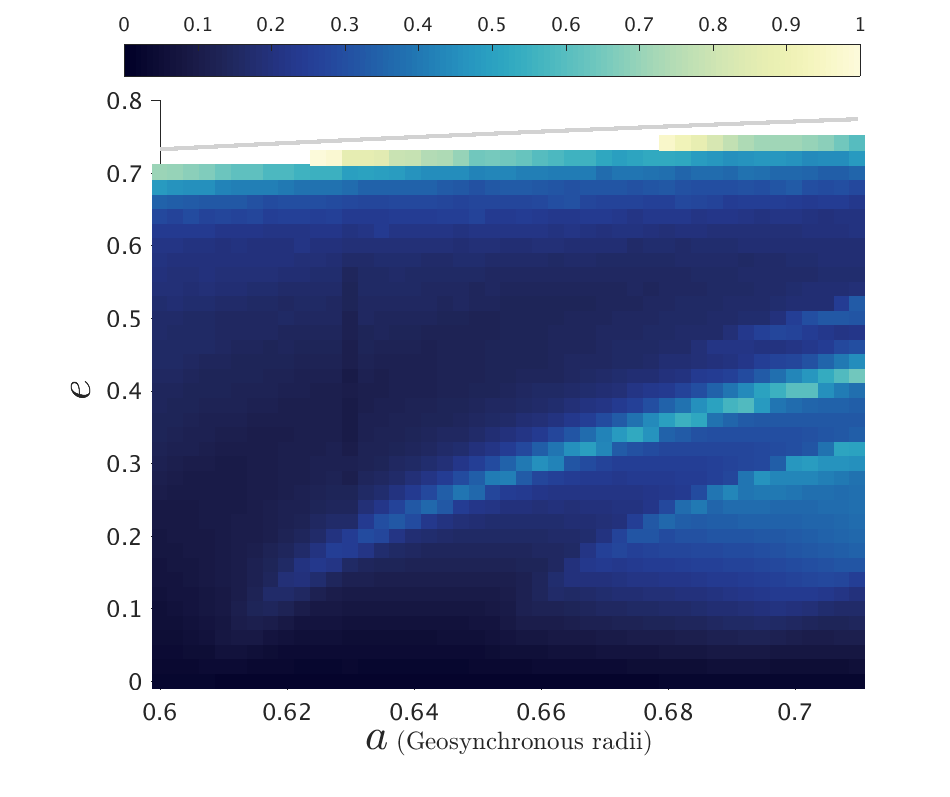}
    \end{subfigure}    
    \begin{subfigure}[b]{0.45\textwidth}
      \caption{$\bm{i}_{o}={\bf 54^{\circ}}$}
      \includegraphics[width=.49\textwidth]{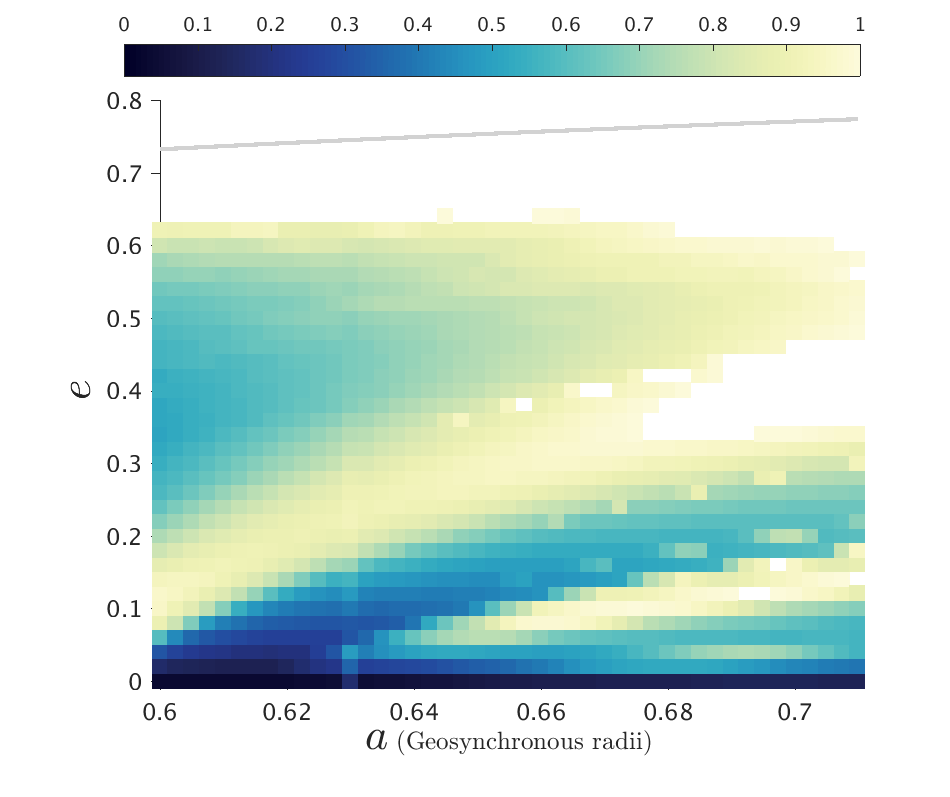} 
      \includegraphics[width=.49\textwidth]{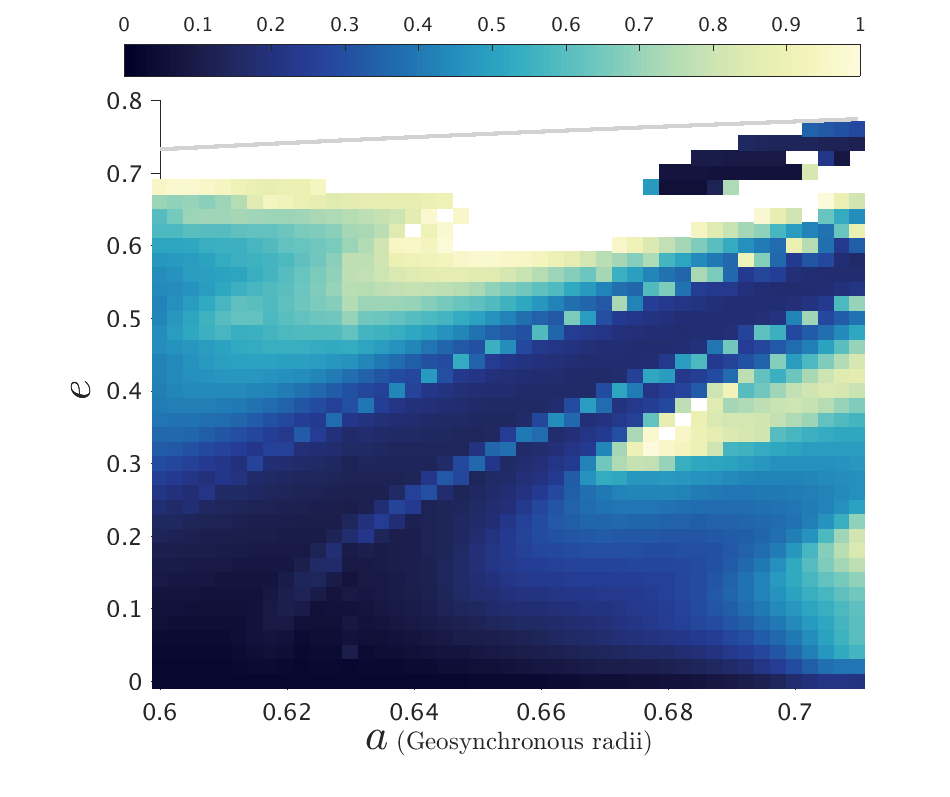}
    \end{subfigure}   
    \begin{subfigure}[b]{0.45\textwidth}
      \caption{$\bm{i}_{o}={\bf 58^{\circ}}$}
      \includegraphics[width=.49\textwidth]{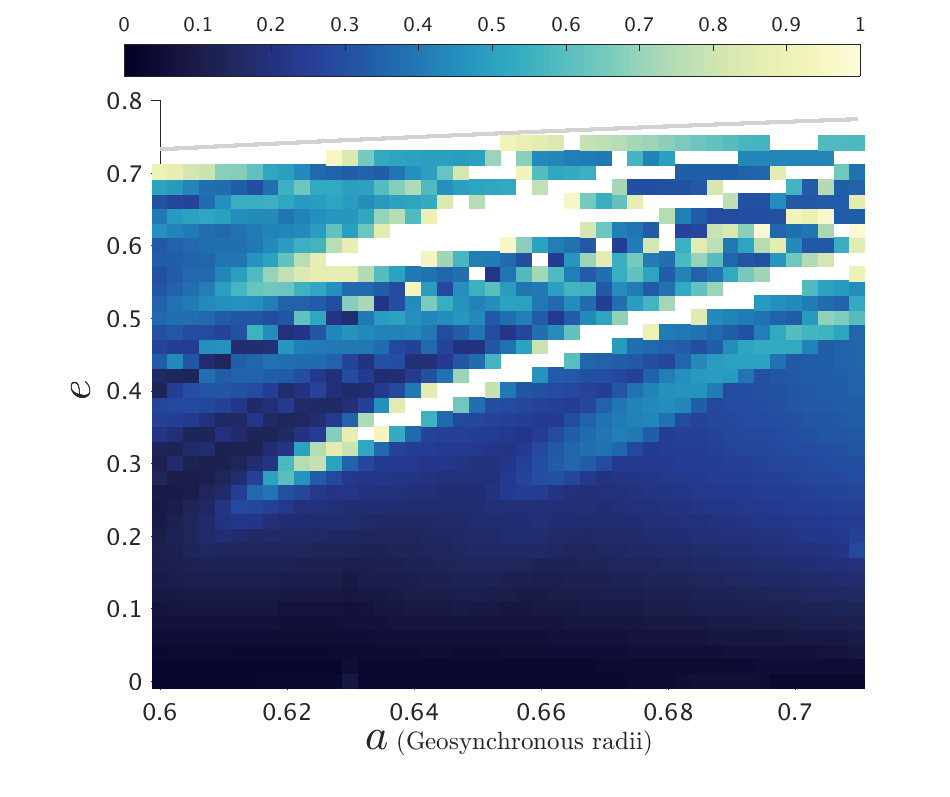} 
      \includegraphics[width=.49\textwidth]{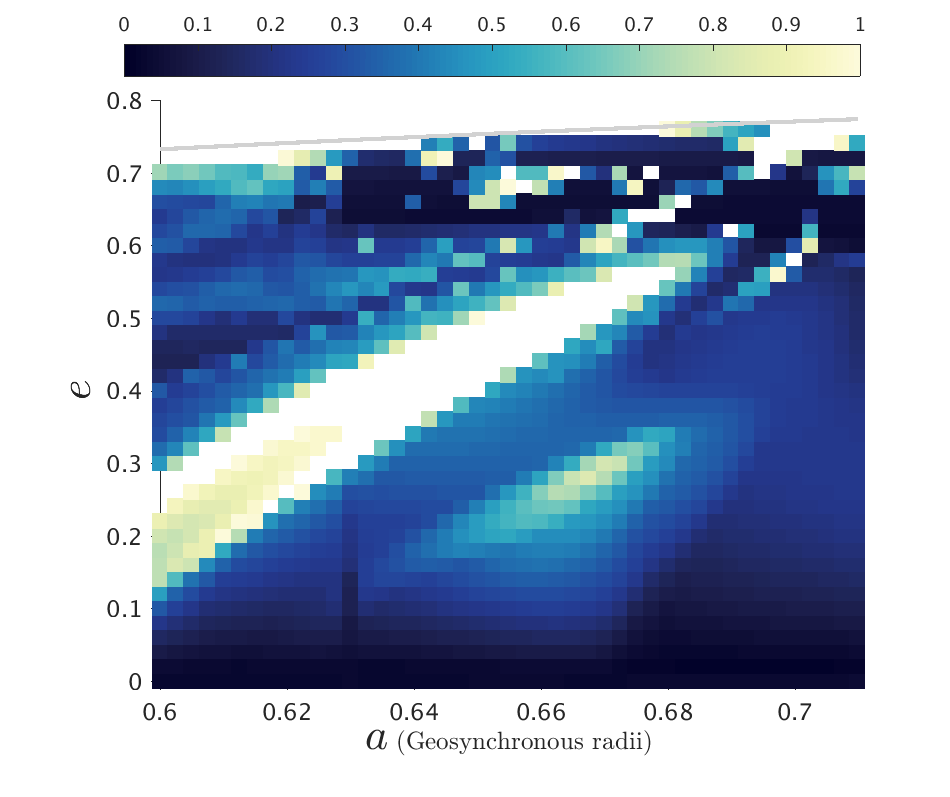}
    \end{subfigure}  
    \begin{subfigure}[b]{0.45\textwidth}
      \caption{$\bm{i}_{o}={\bf 66^{\circ}}$}
      \includegraphics[width=.49\textwidth]{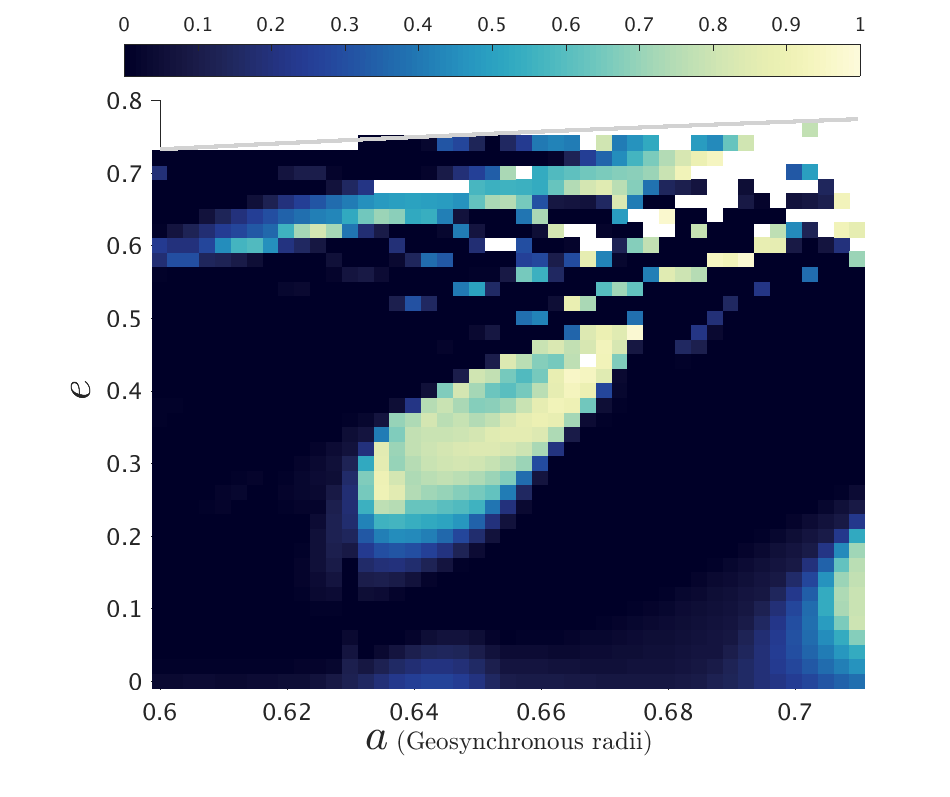} 
      \includegraphics[width=.49\textwidth]{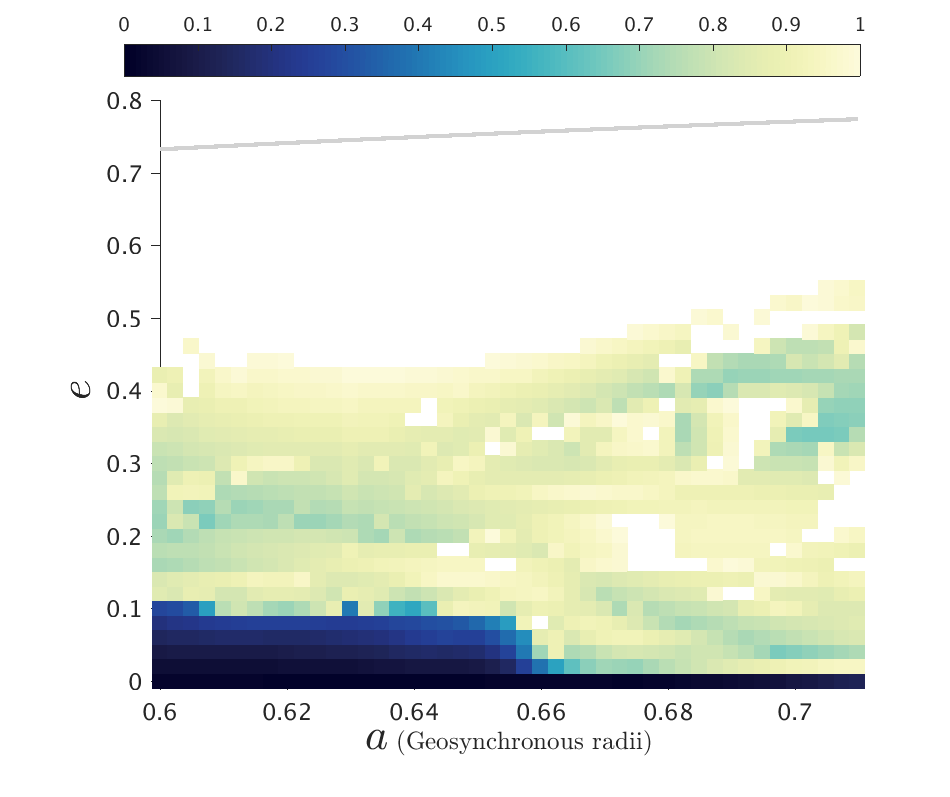}
    \end{subfigure}  
    \begin{subfigure}[b]{0.45\textwidth}
      \caption{$\bm{i}_{o}={\bf 68^{\circ}}$}
      \includegraphics[width=.49\textwidth]{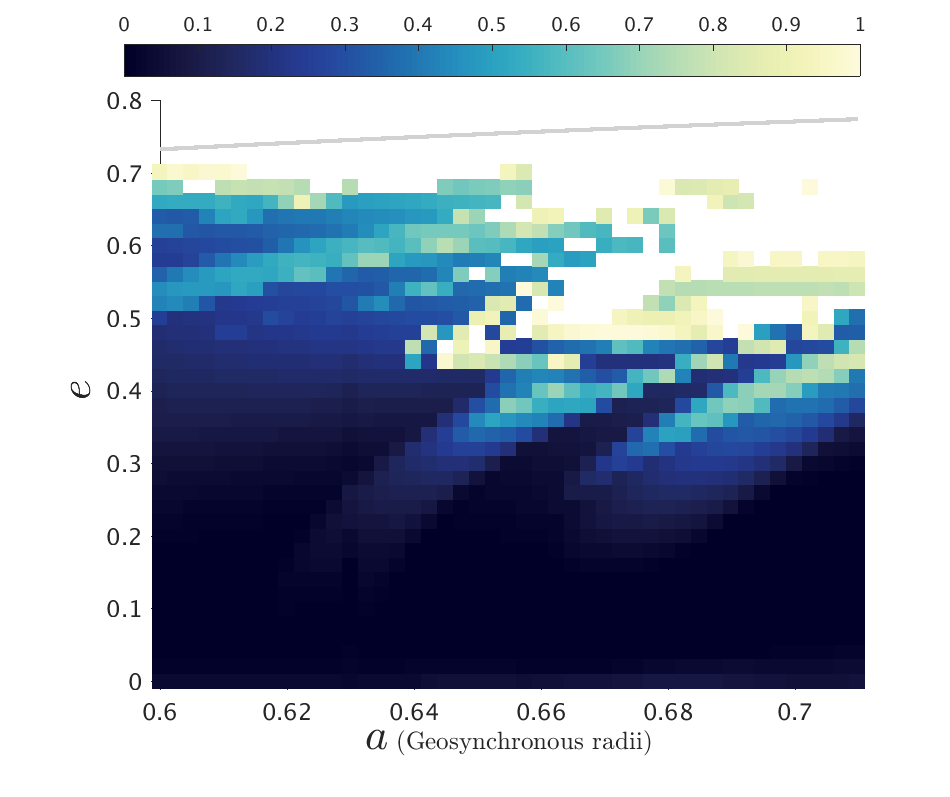} 
      \includegraphics[width=.49\textwidth]{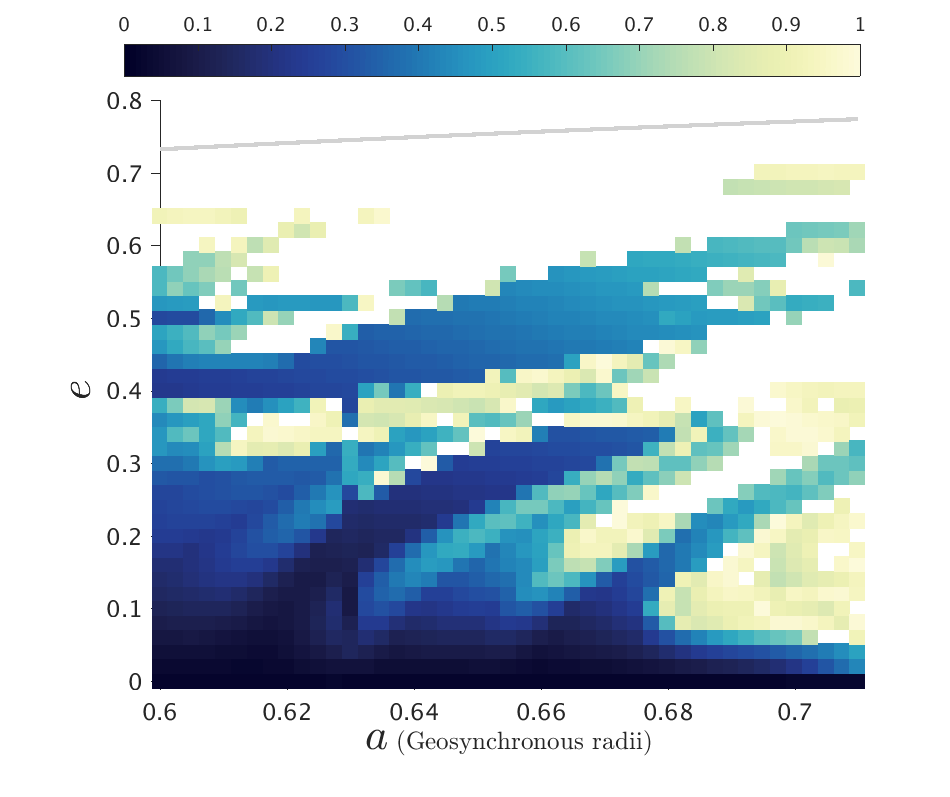}
    \end{subfigure}  
    \begin{subfigure}[b]{0.45\textwidth}
      \caption{$\bm{i}_{o}={\bf 70^{\circ}}$}
      \includegraphics[width=.49\textwidth]{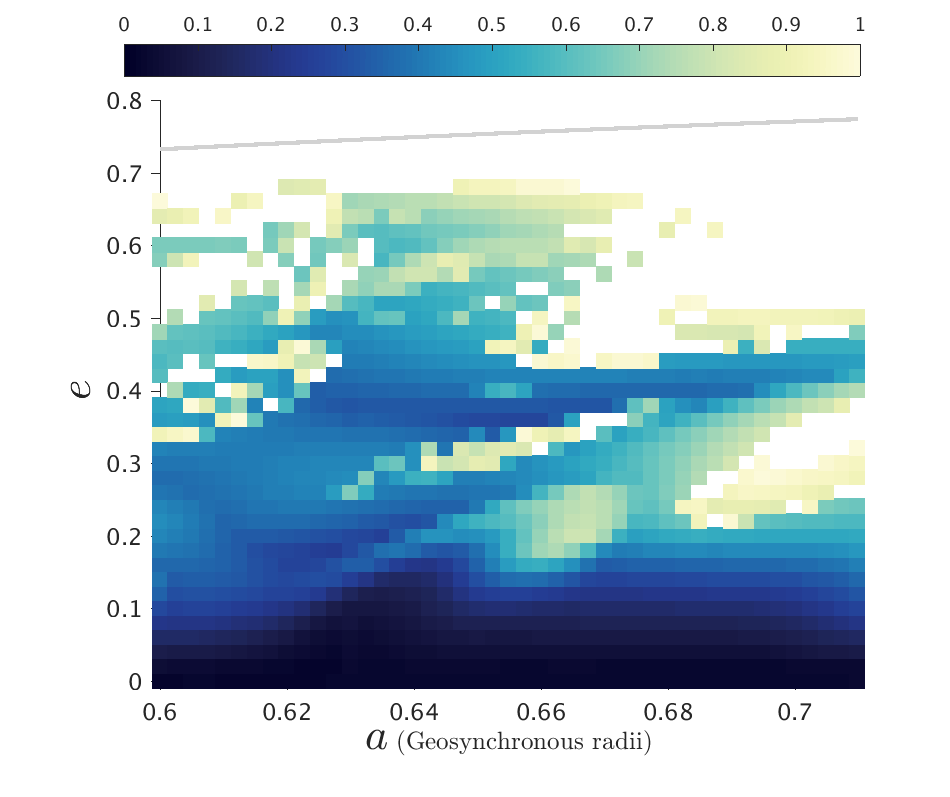} 
      \includegraphics[width=.49\textwidth]{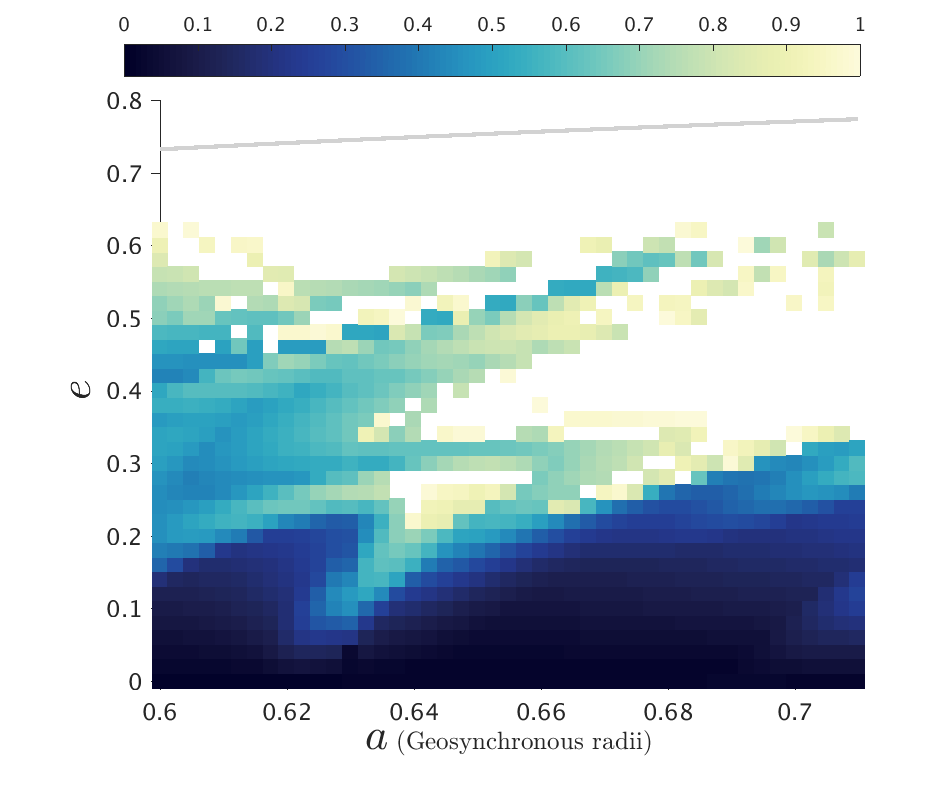}
    \end{subfigure}
    \begin{subfigure}[b]{0.45\textwidth}
      \caption{$\bm{i}_{o}={\bf 90^{\circ}}$}
      \includegraphics[width=.49\textwidth]{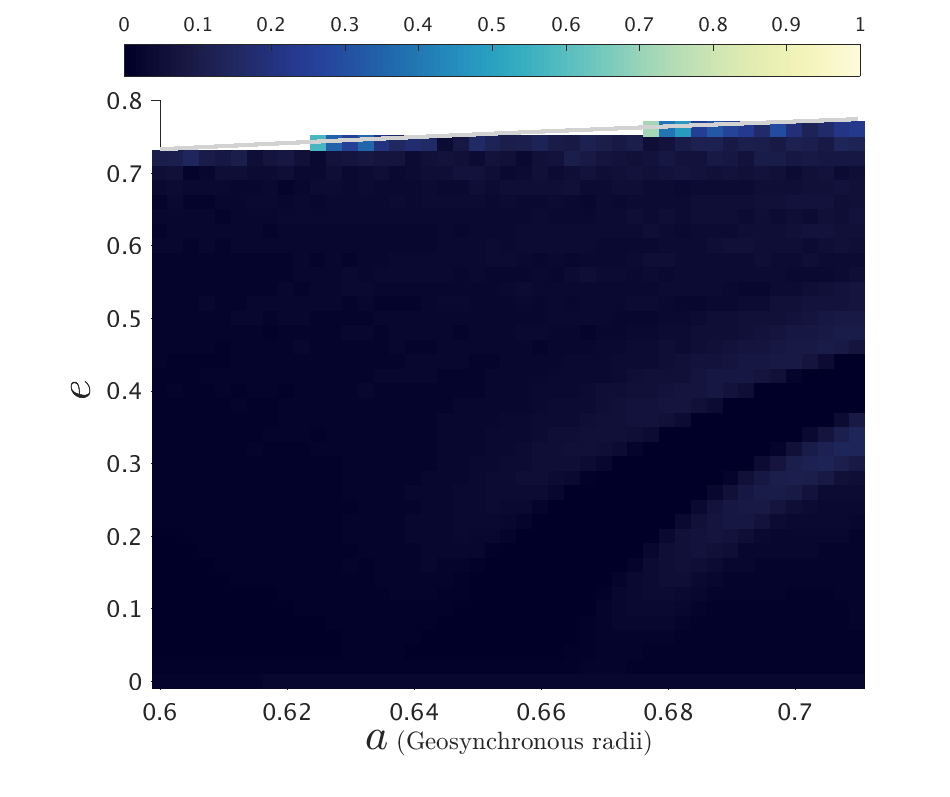} 
      \includegraphics[width=.49\textwidth]{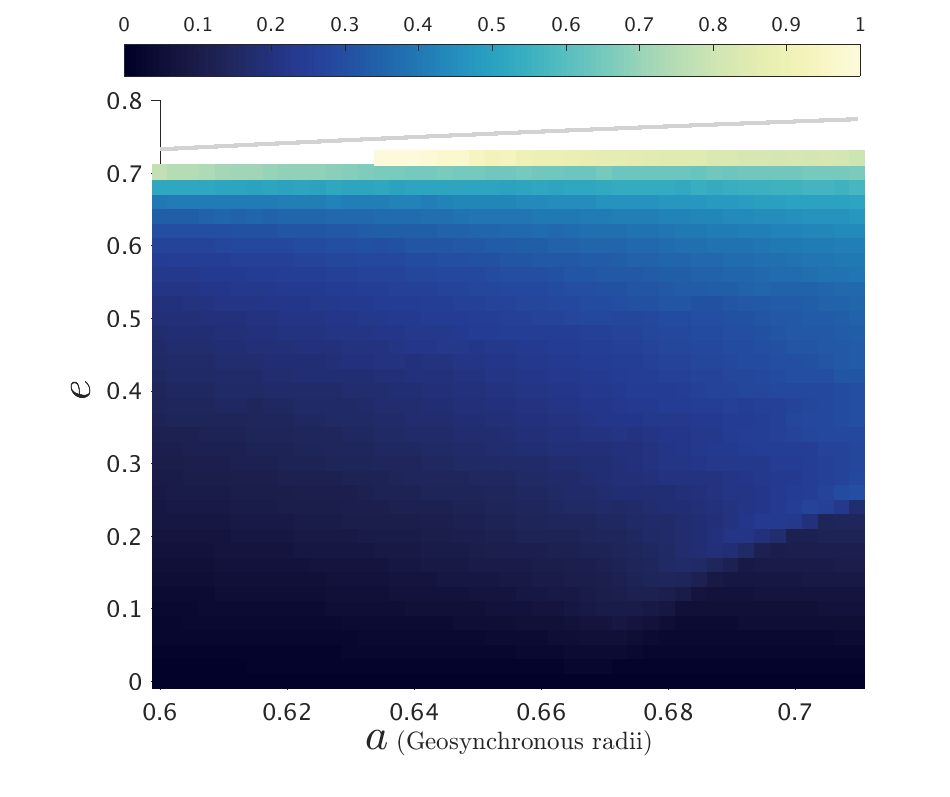}
    \end{subfigure}    
  \caption{$De$ maps of the \textit{MEO-general} phase space for various $\bm{i}_{o}$, 
  $\bm{\Delta}\bm{\Omega} = {\bf 180^\circ}$, $\bm{\Delta}\bm{\omega} = {\bf 270^\circ}$ (1st and 3rd columns) and 
  $\bm{\Delta}\bm{\Omega} = {\bf 270^\circ}$, $\bm{\Delta}\bm{\omega} = {\bf 90^\circ}$ (2nd and 4th columns),  
  for Epoch 2020, and for $C_{R}A/m=1$ m$^2$/kg.
  The colorbar for the $De$ maps is from 0 to 1, where the reentry particles were excluded (white).}
  \label{fig:MEO_gen_i4}
\end{figure}

\newpage
\section{Dynamical maps of GNSS-graveyard grid for $i_{nom}-0.5^{\circ}$ and $i_{nom}+0.5^{\circ}$}
\label{app:2}

\begin{figure}[htp!]
  \centering
    \begin{subfigure}[b]{0.45\textwidth}
      \caption{$\bm{\Delta}\bm{\Omega} = {\bf 0}$, $\bm{\Delta}\bm{\omega} = {\bf 0}$}
      \includegraphics[width=.49\textwidth]{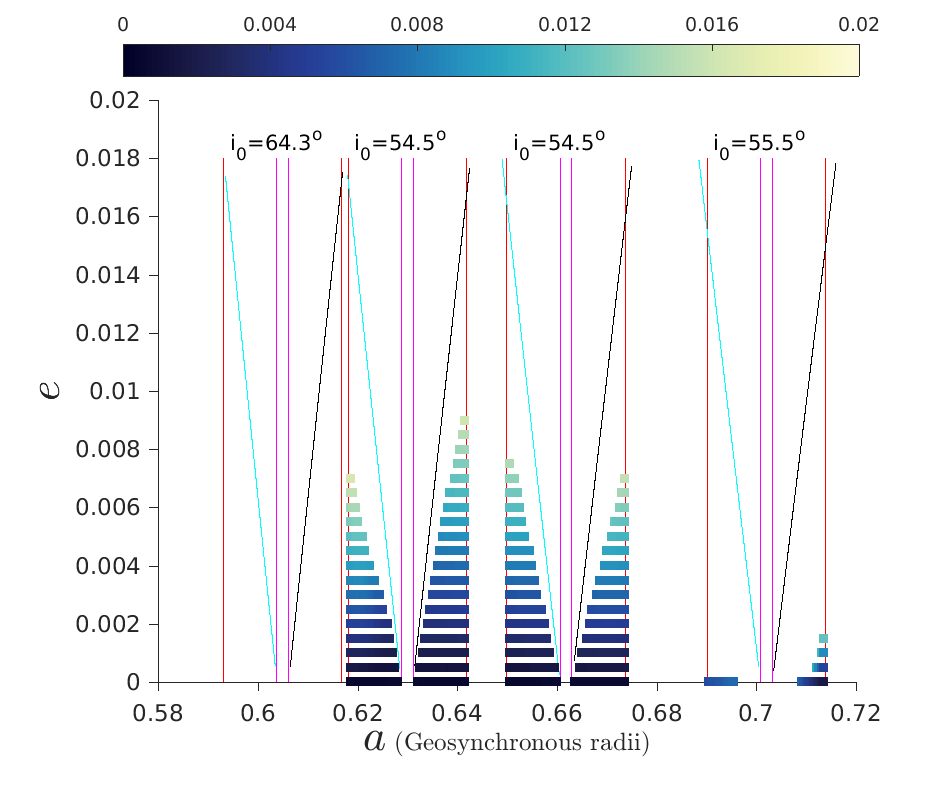} 
      \includegraphics[width=.49\textwidth]{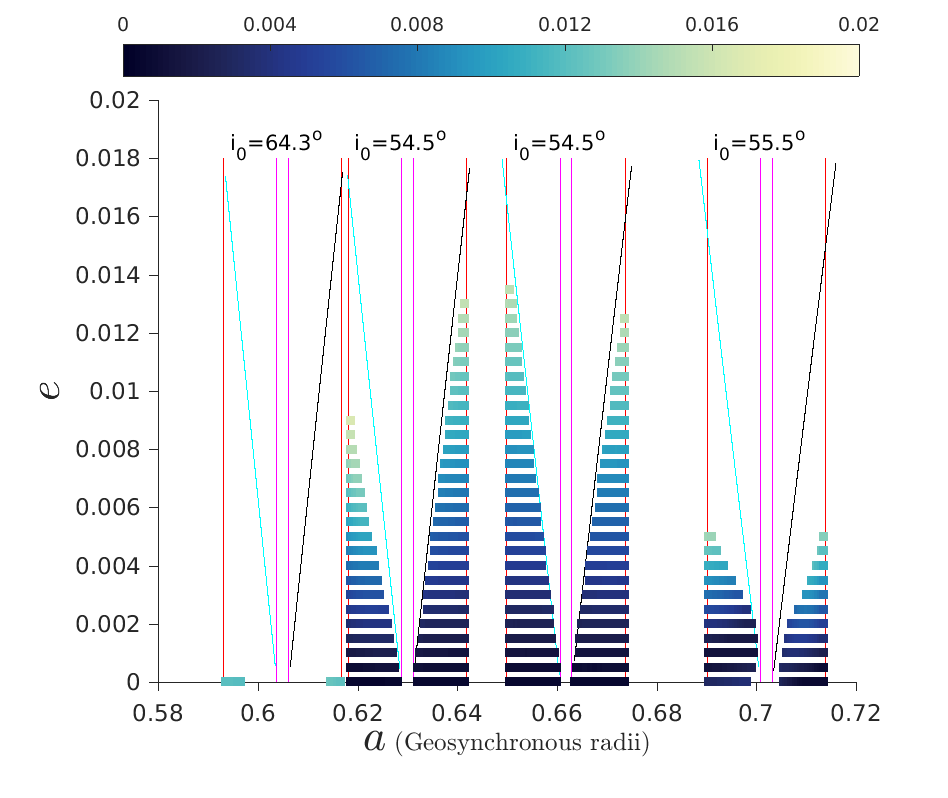}
    \end{subfigure}  
    \begin{subfigure}[b]{0.45\textwidth}
      \caption{$\bm{\Delta}\bm{\Omega} = {\bf 0}$, $\bm{\Delta}\bm{\omega} = {\bf 90^\circ}$}
      \includegraphics[width=.49\textwidth]{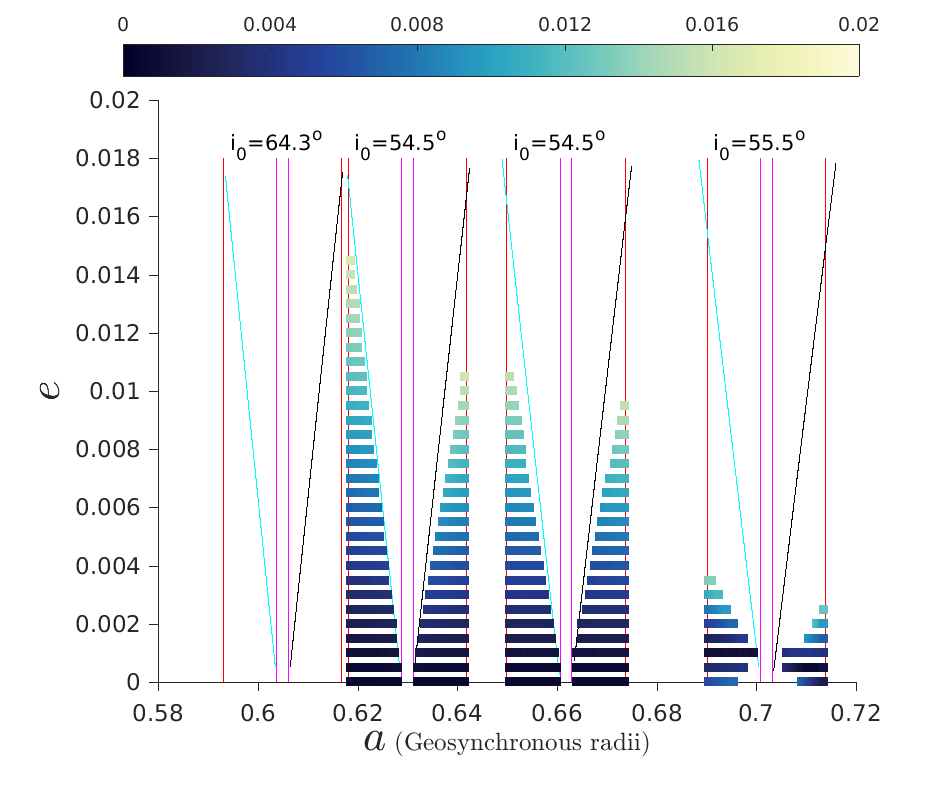} 
      \includegraphics[width=.49\textwidth]{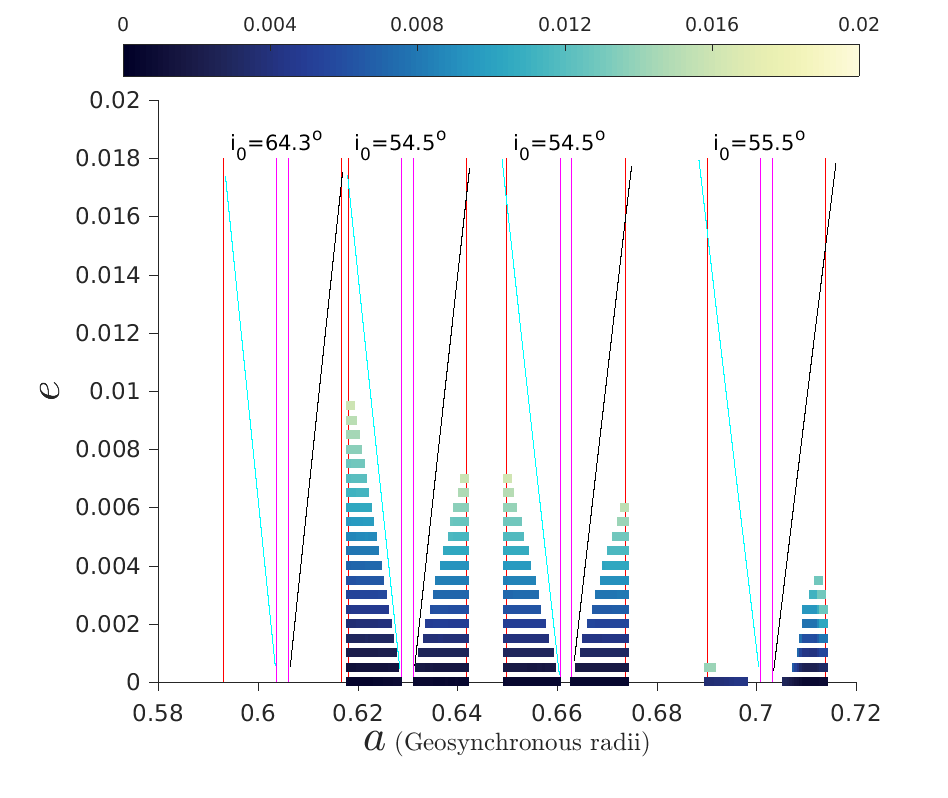}
    \end{subfigure} 
    \begin{subfigure}[b]{0.45\textwidth}
      \caption{$\bm{\Delta}\bm{\Omega} = {\bf 90^\circ}$, $\bm{\Delta}\bm{\omega} = {\bf 0}$}
      \includegraphics[width=.49\textwidth]{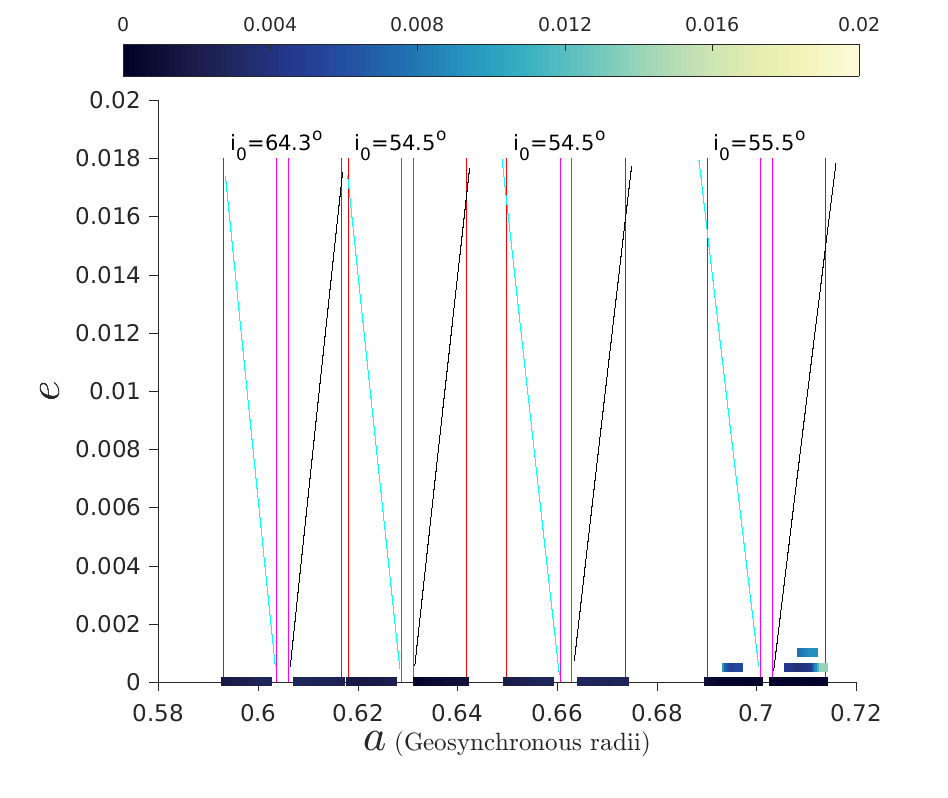} 
      \includegraphics[width=.49\textwidth]{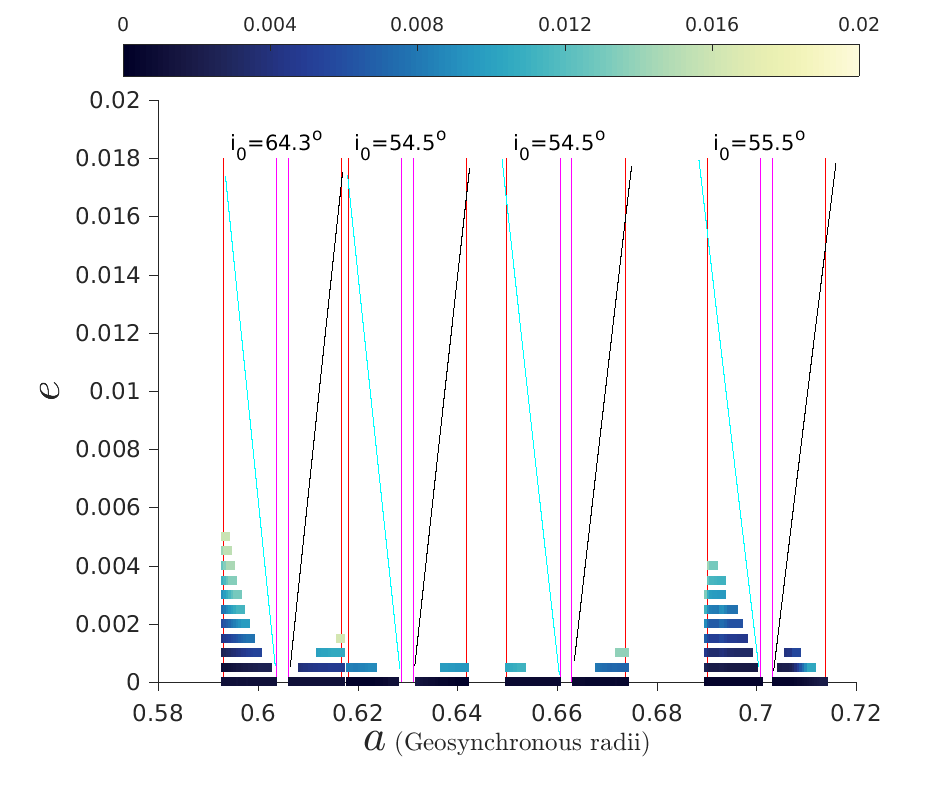}
    \end{subfigure}  
    \begin{subfigure}[b]{0.45\textwidth}
      \caption{$\bm{\Delta}\bm{\Omega} = {\bf 90^\circ}$, $\bm{\Delta}\bm{\omega} = {\bf 90^\circ}$}
      \includegraphics[width=.49\textwidth]{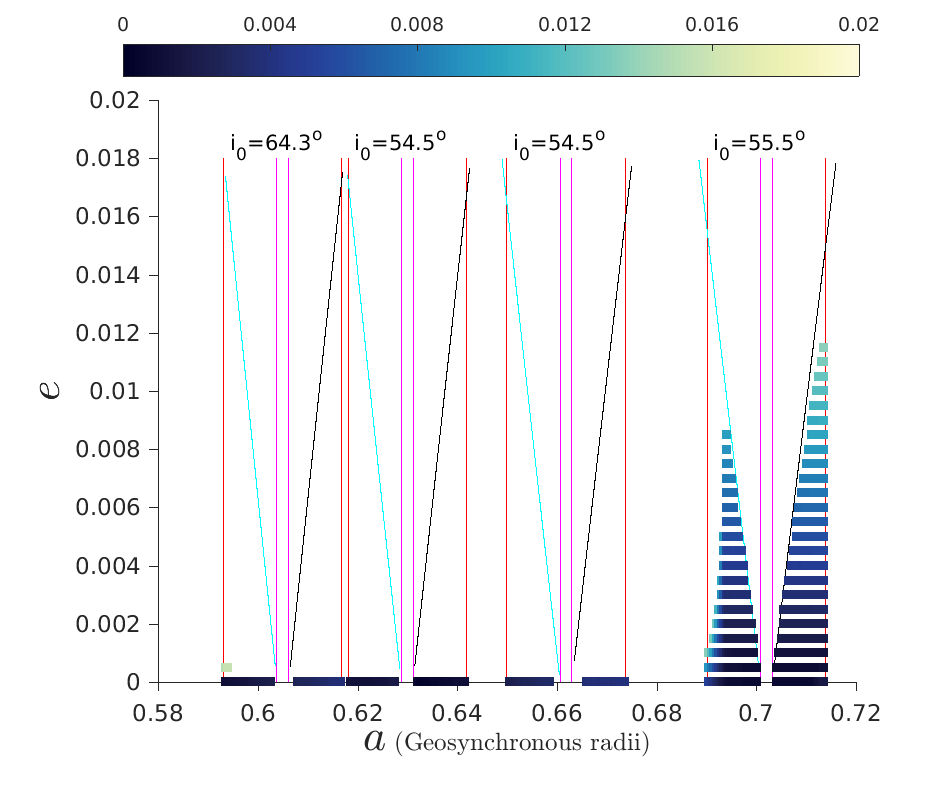} 
      \includegraphics[width=.49\textwidth]{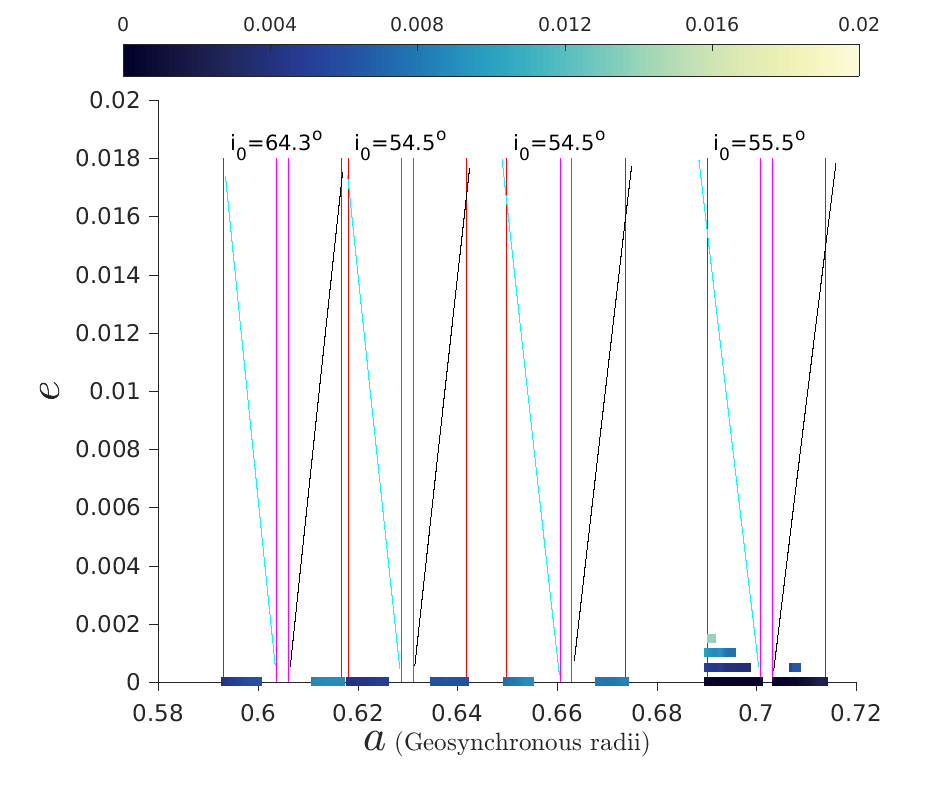}
    \end{subfigure}  
    \begin{subfigure}[b]{0.45\textwidth}
      \caption{$\bm{\Delta}\bm{\Omega} = {\bf 180^\circ}$, $\bm{\Delta}\bm{\omega} = {\bf 0}$}
      \includegraphics[width=.49\textwidth]{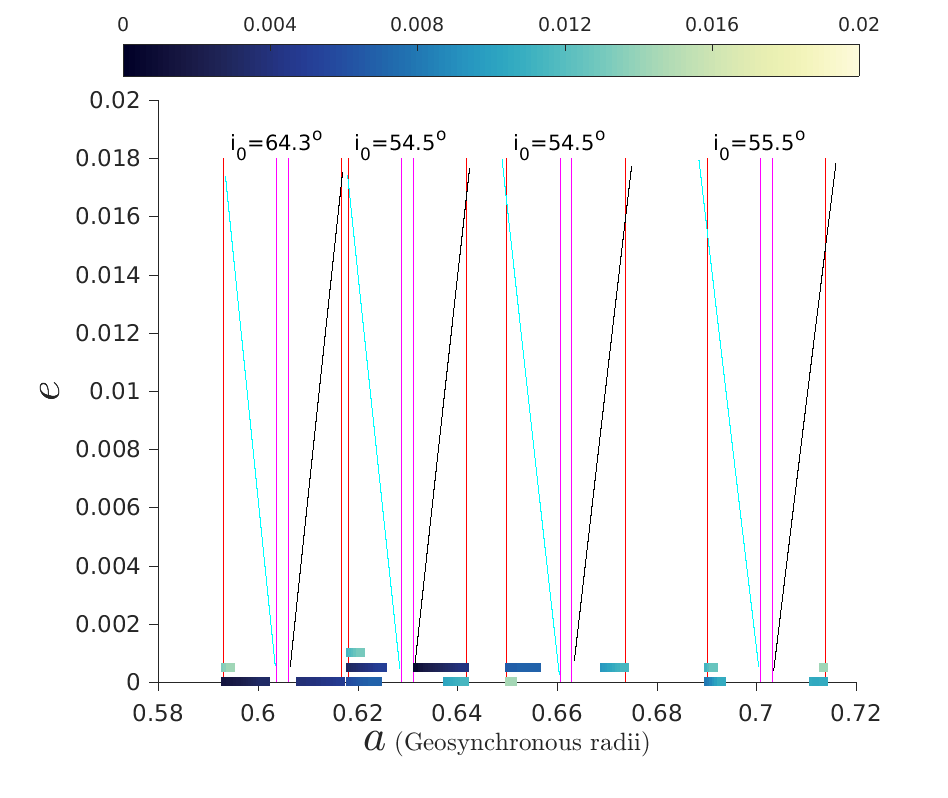}
      \includegraphics[width=.49\textwidth]{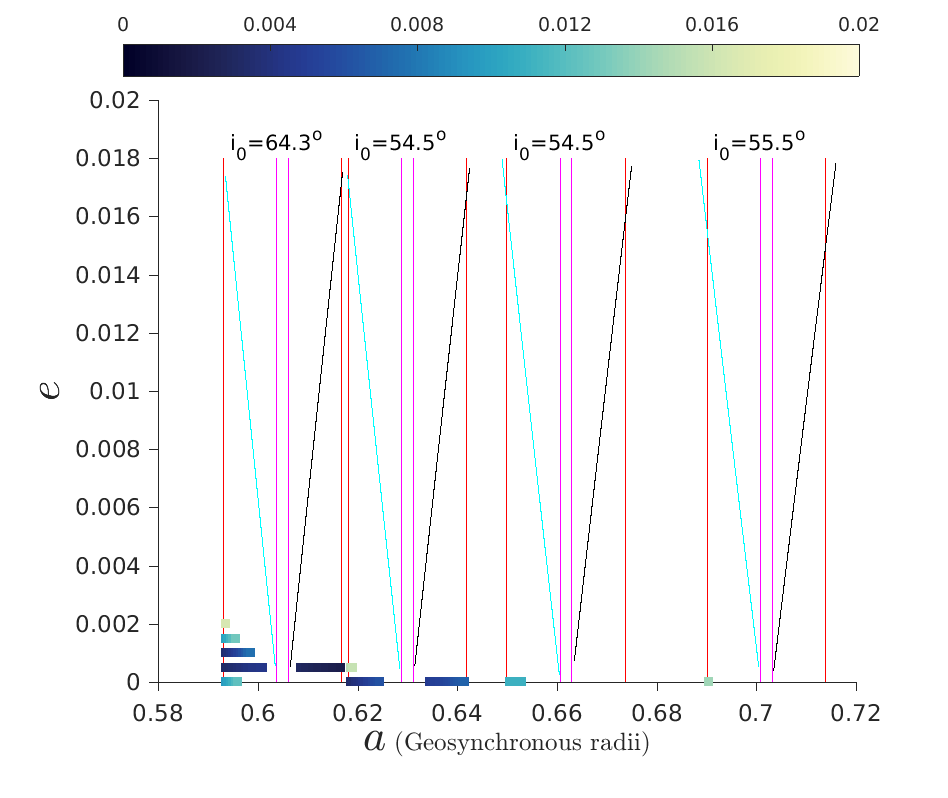}
    \end{subfigure}     
    \begin{subfigure}[b]{0.45\textwidth}
      \caption{$\bm{\Delta}\bm{\Omega} = {\bf 180^\circ}$, $\bm{\Delta}\bm{\omega} = {\bf 90^\circ}$}
      \includegraphics[width=.49\textwidth]{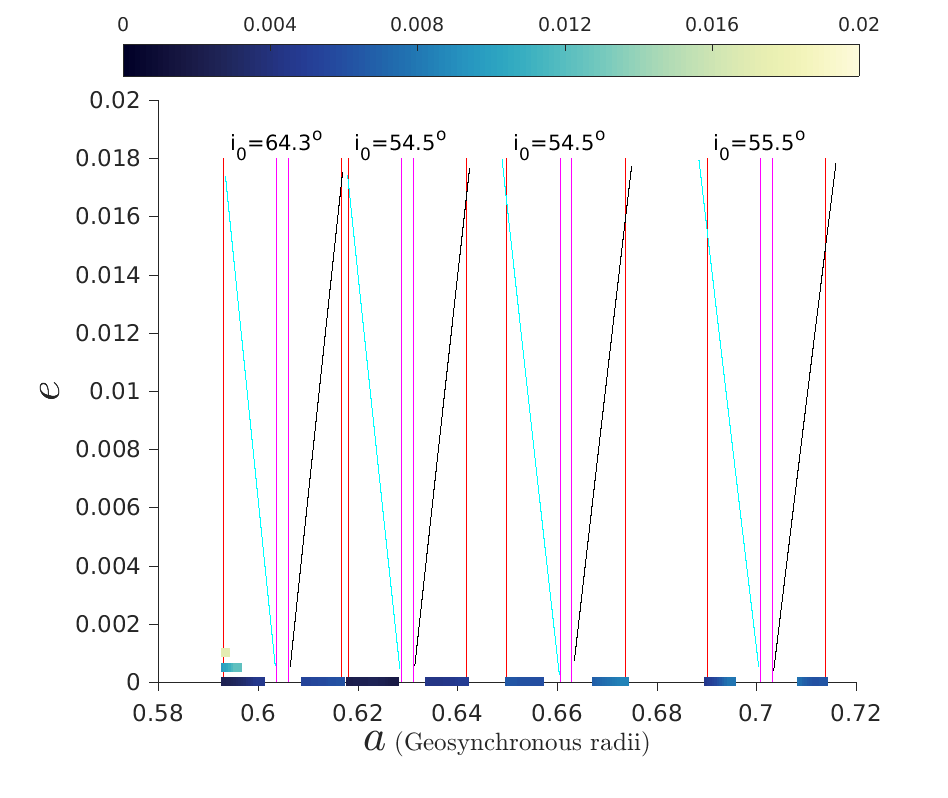}
      \includegraphics[width=.49\textwidth]{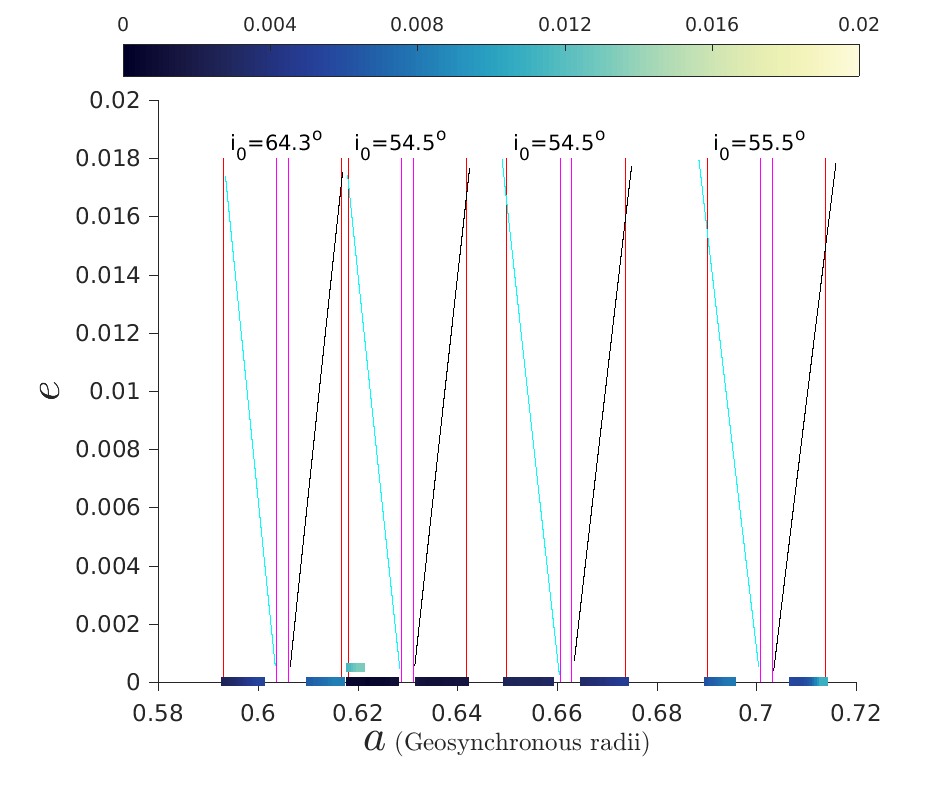}
    \end{subfigure}
    \begin{subfigure}[b]{0.45\textwidth}
      \caption{$\bm{\Delta}\bm{\Omega} = {\bf 270^\circ}$, $\bm{\Delta}\bm{\omega} = {\bf 0}$}
      \includegraphics[width=.49\textwidth]{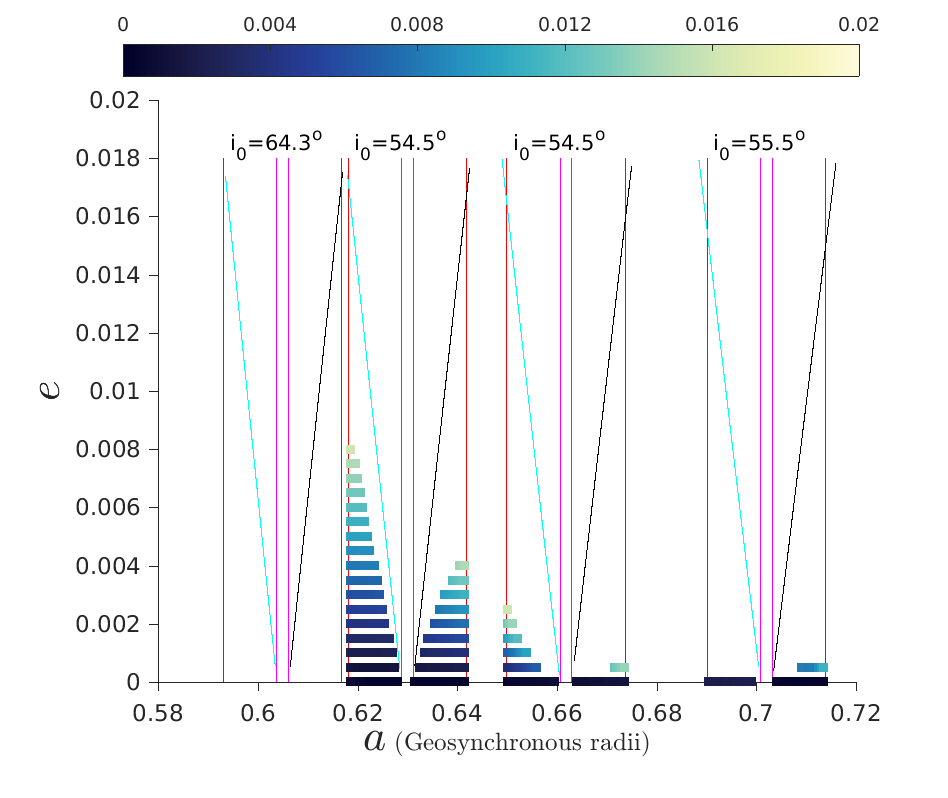}
      \includegraphics[width=.49\textwidth]{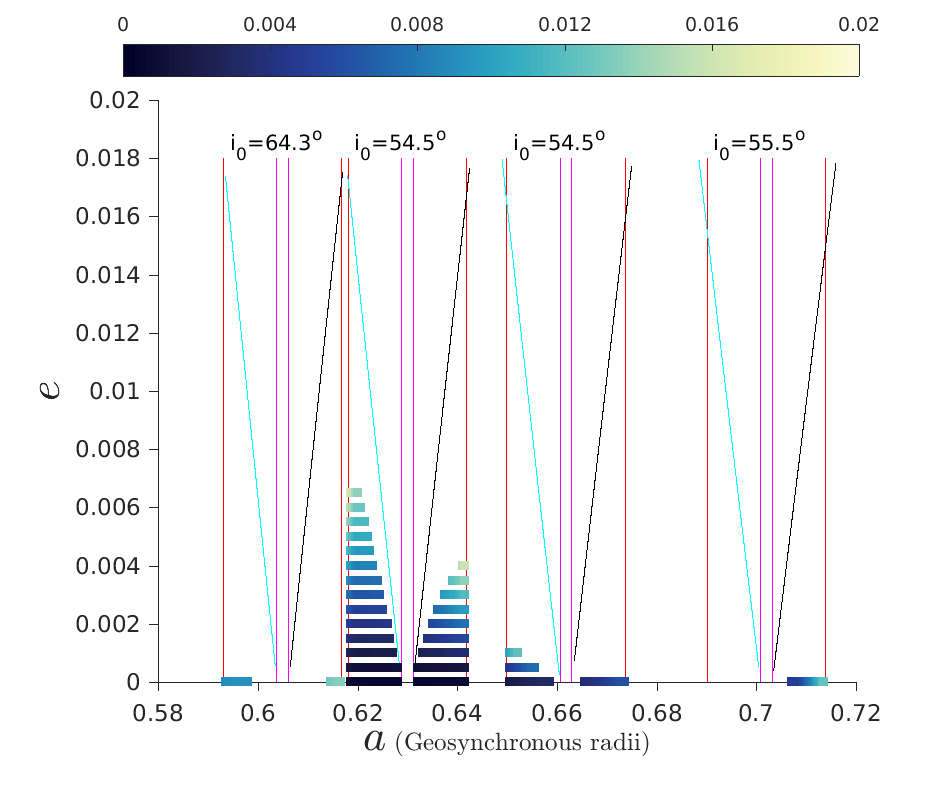}
    \end{subfigure} 
    \begin{subfigure}[b]{0.45\textwidth}
      \caption{$\bm{\Delta}\bm{\Omega} = {\bf 270^\circ}$, $\bm{\Delta}\bm{\omega} = {\bf 90^\circ}$}
      \includegraphics[width=.49\textwidth]{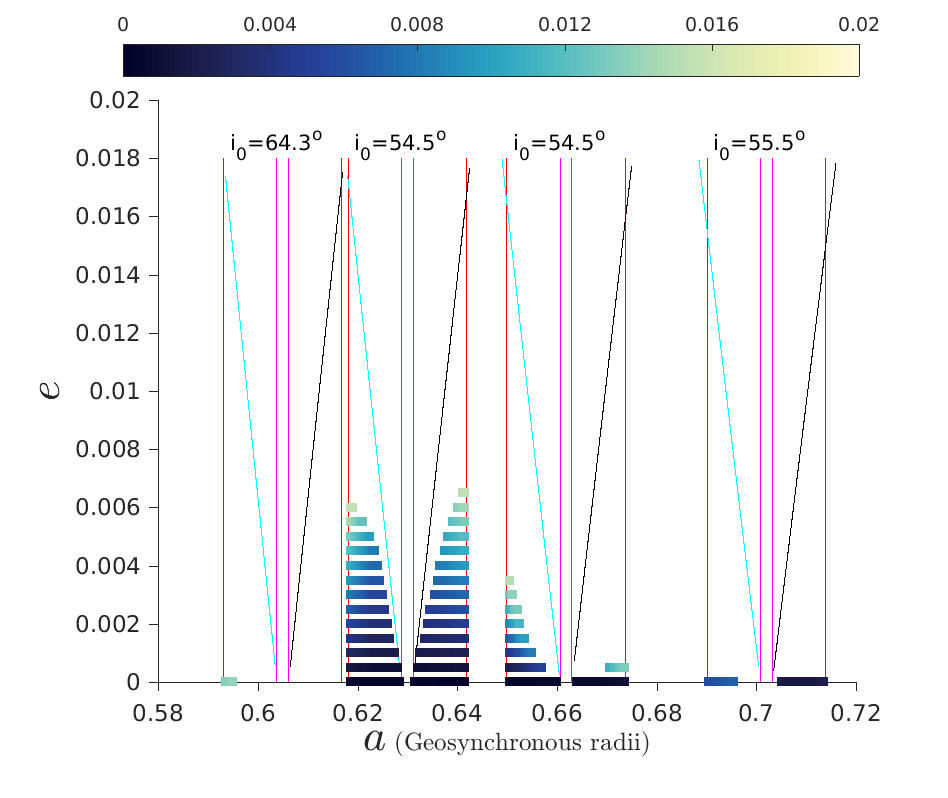}
      \includegraphics[width=.49\textwidth]{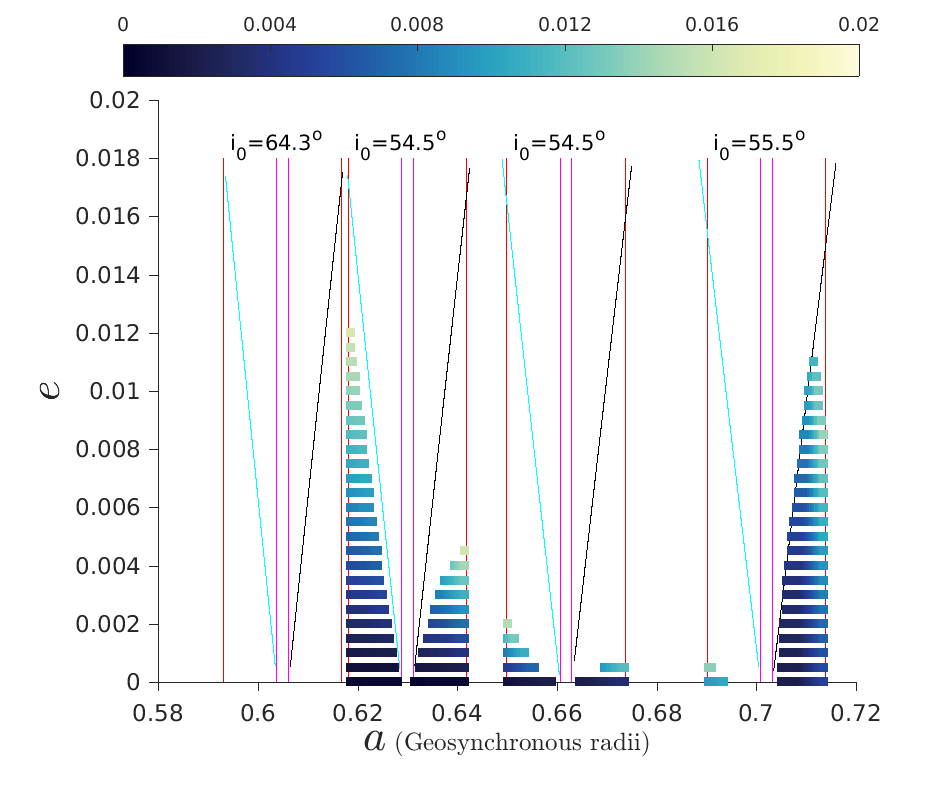}
    \end{subfigure}    
  \caption{Maximum eccentricity maps of the \textit{GNSS-graveyard} phase space for $\bm{i_{o}} = {\bf i_{nom}-0.5^{\circ}}$,  
  for Epoch 2018 (left) and Epoch 2020 (right), and for $C_{R}A/m=0.015$ m$^2$/kg. 
  $i=64.3^{\circ}$ for GLONASS, $54.5^{\circ}$ for GPS and BEIDOU, and $55.5^{\circ}$ for GALILEO. 
  The colorbar for maximum eccentricity maps is from 0 to 0.02.}
  \label{fig:GRAV_srp1_1}
\end{figure}

\begin{figure}[htp!]
  \centering
    \begin{subfigure}[b]{0.45\textwidth}
      \caption{$\bm{\Delta}\bm{\Omega} = {\bf 0}$, $\bm{\Delta}\bm{\omega} = {\bf 0}$}
      \includegraphics[width=.49\textwidth]{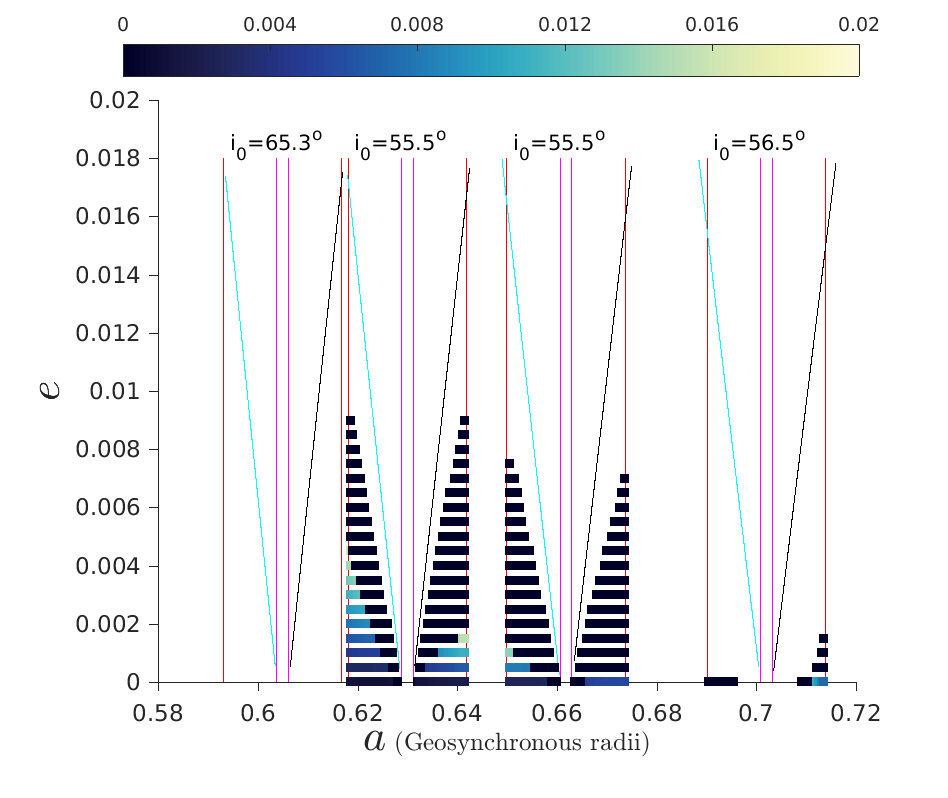} 
      \includegraphics[width=.49\textwidth]{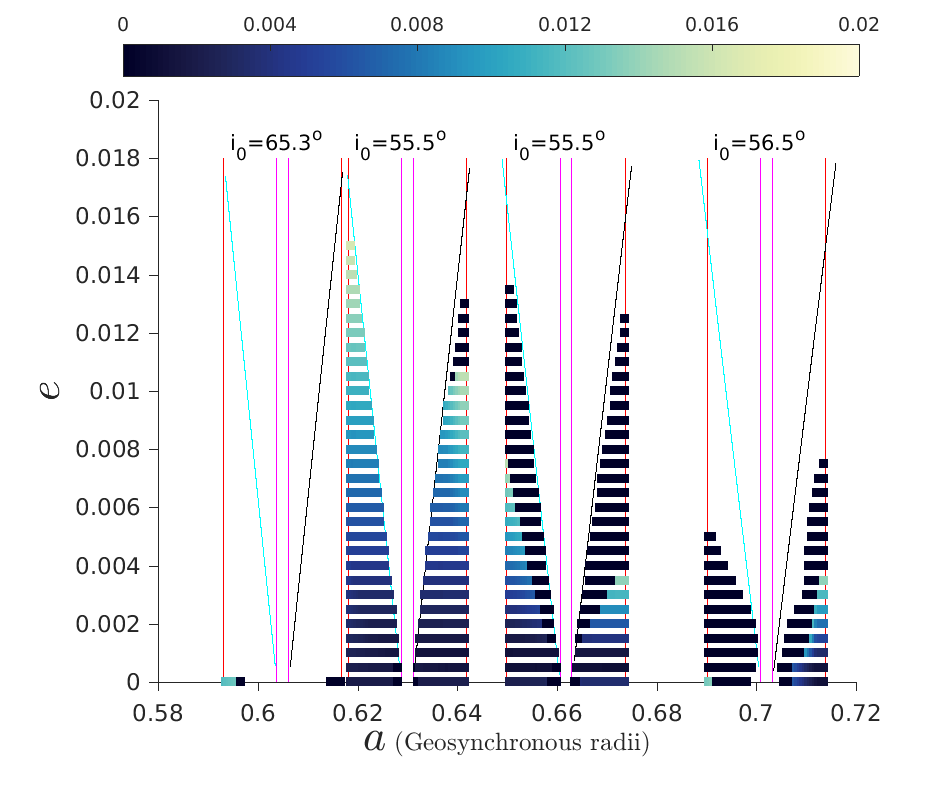}
    \end{subfigure}  
    \begin{subfigure}[b]{0.45\textwidth}
      \caption{$\bm{\Delta}\bm{\Omega} = {\bf 0}$, $\bm{\Delta}\bm{\omega} = {\bf 90^\circ}$}
      \includegraphics[width=.49\textwidth]{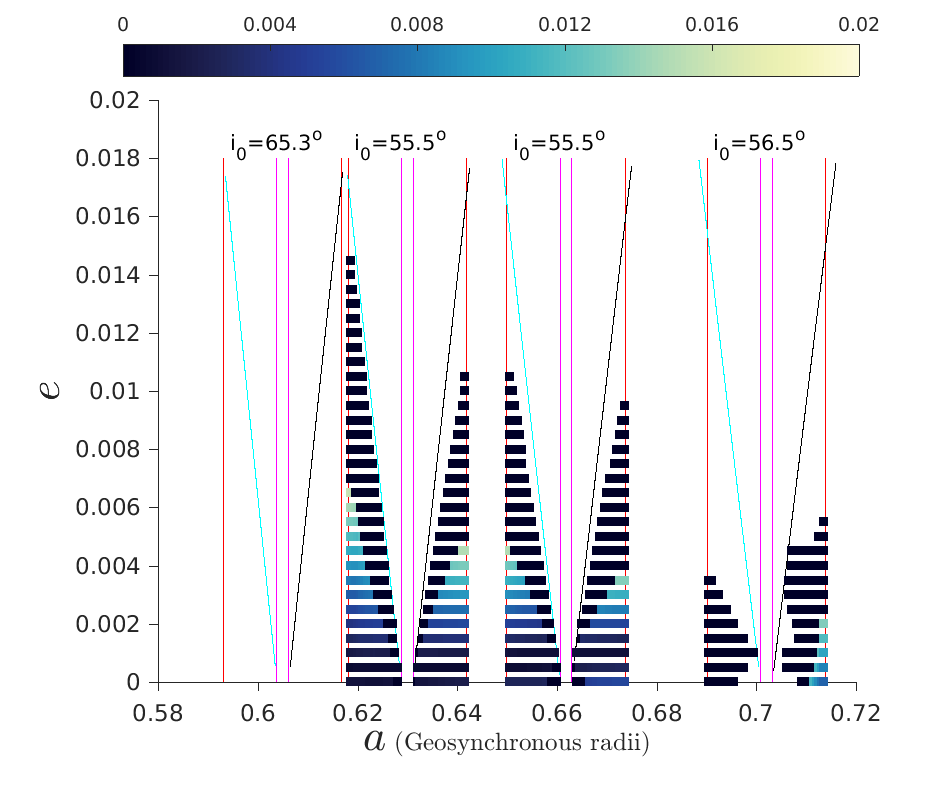} 
      \includegraphics[width=.49\textwidth]{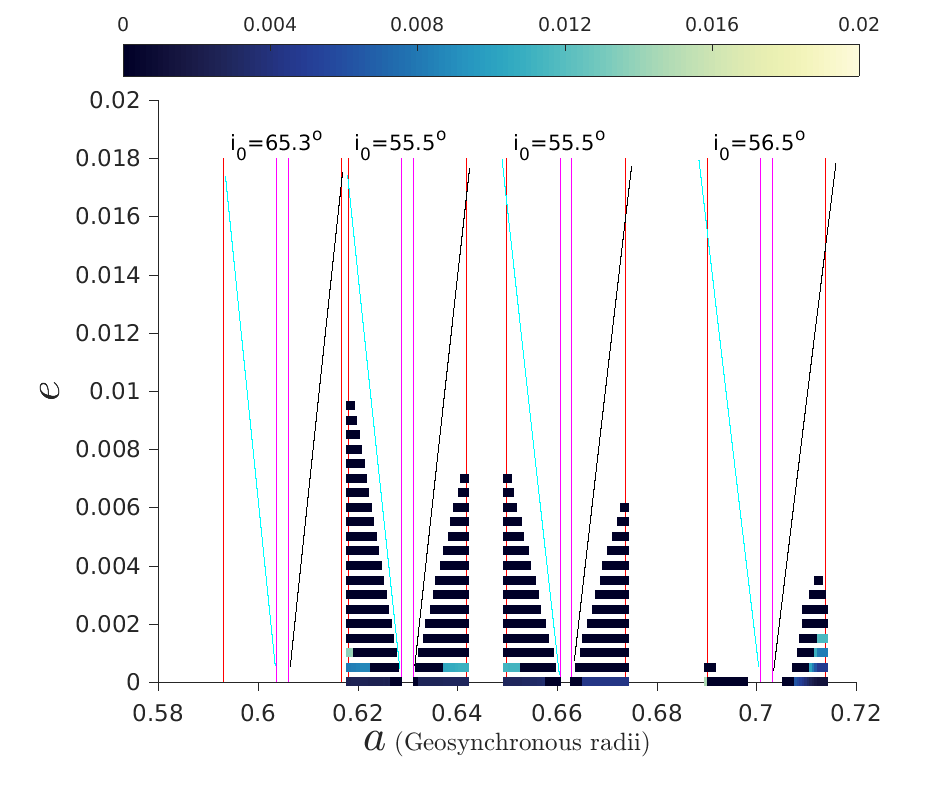}
    \end{subfigure} 
    \begin{subfigure}[b]{0.45\textwidth}
      \caption{$\bm{\Delta}\bm{\Omega} = {\bf 90^\circ}$, $\bm{\Delta}\bm{\omega} = {\bf 0}$}
      \includegraphics[width=.49\textwidth]{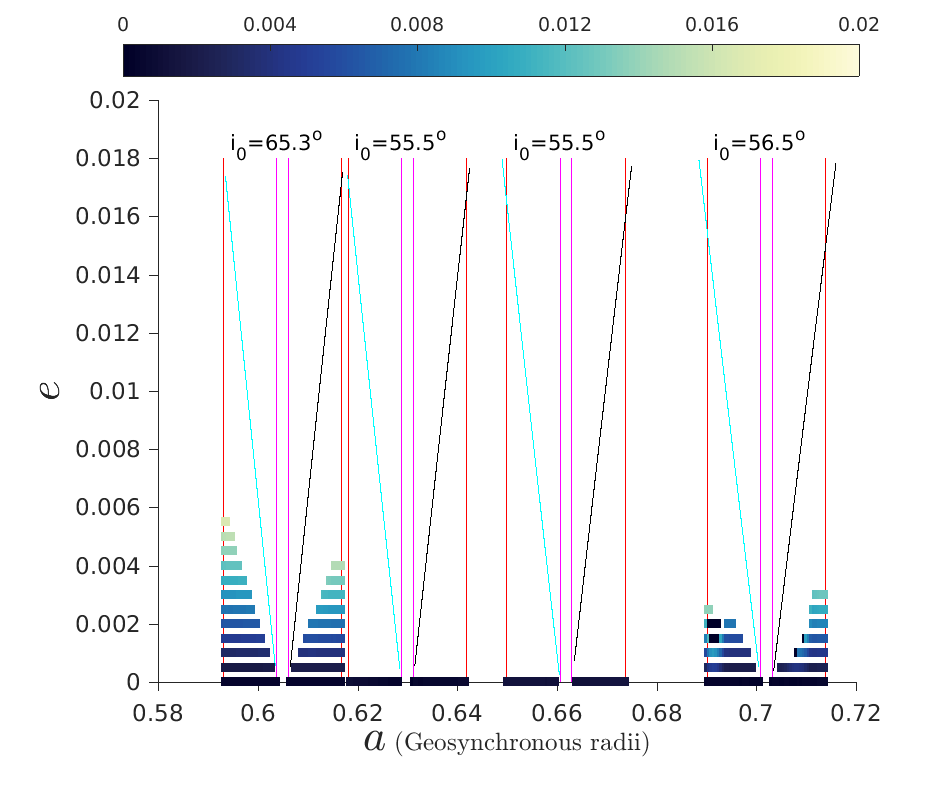} 
      \includegraphics[width=.49\textwidth]{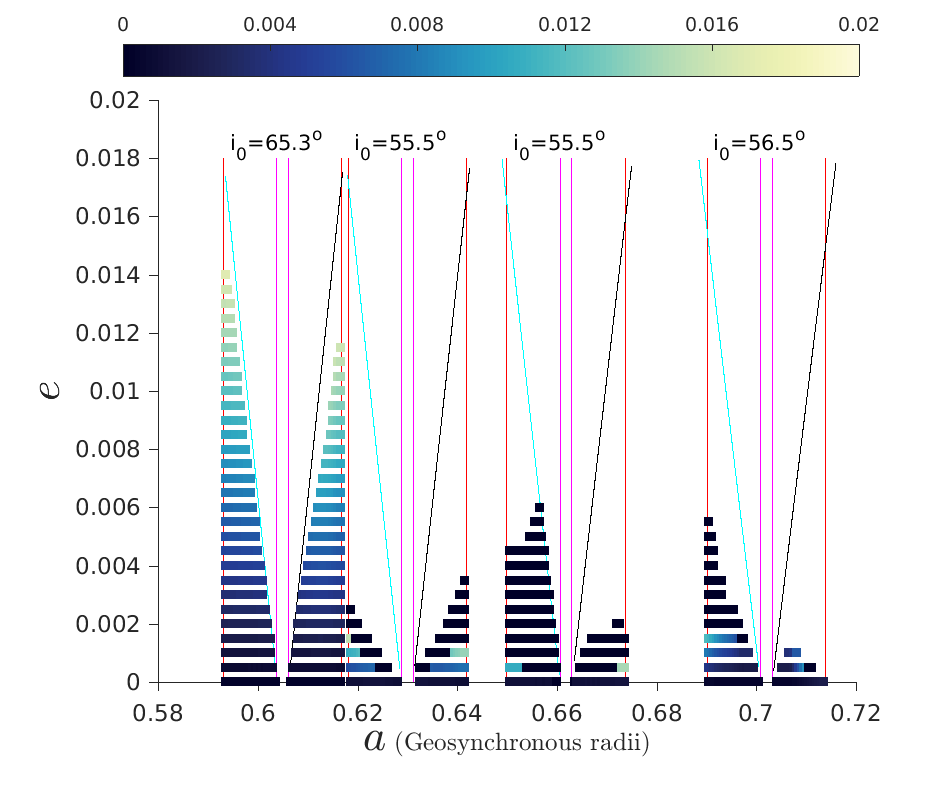}
    \end{subfigure}  
    \begin{subfigure}[b]{0.45\textwidth}
      \caption{$\bm{\Delta}\bm{\Omega} = {\bf 90^\circ}$, $\bm{\Delta}\bm{\omega} = {\bf 90^\circ}$}
      \includegraphics[width=.49\textwidth]{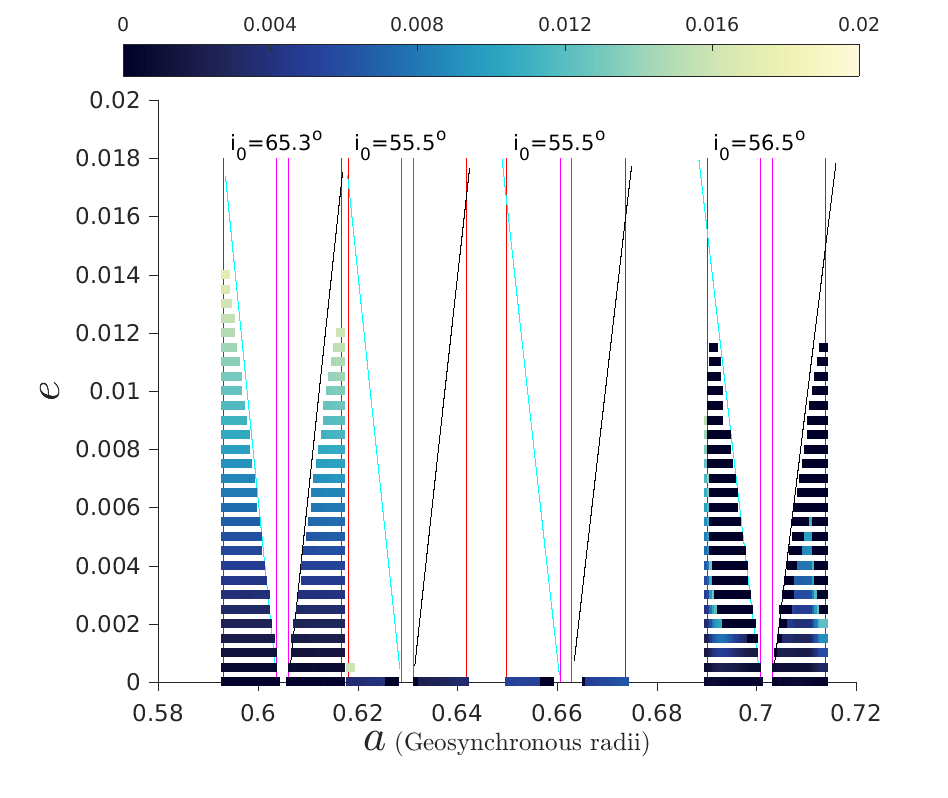} 
      \includegraphics[width=.49\textwidth]{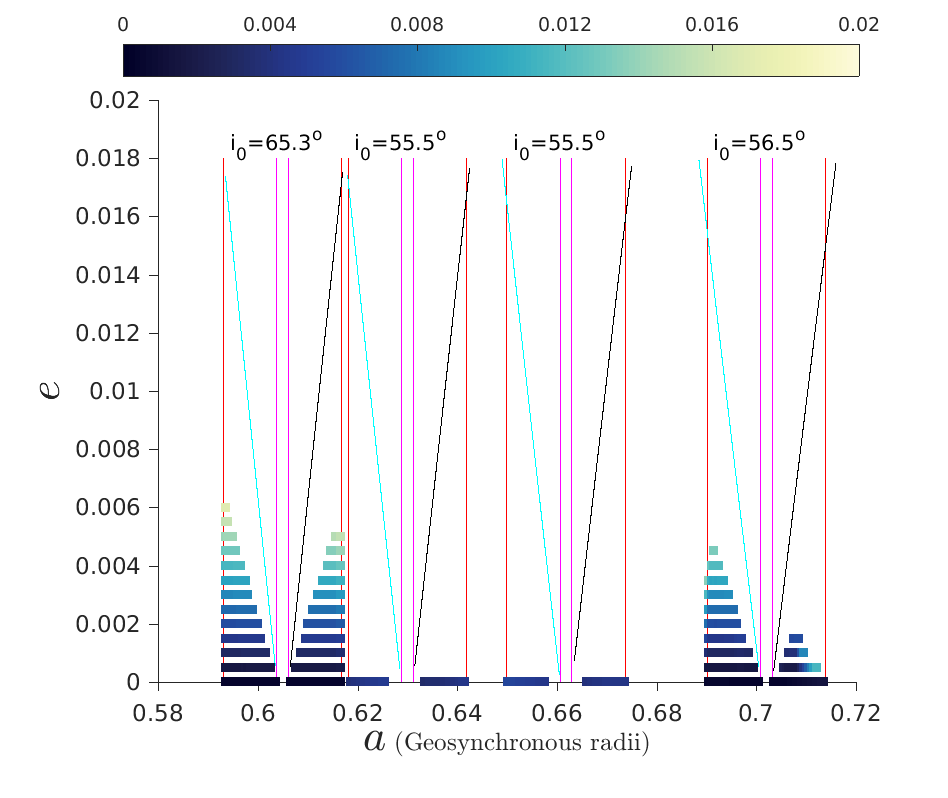}
    \end{subfigure}  
    \begin{subfigure}[b]{0.45\textwidth}
      \caption{$\bm{\Delta}\bm{\Omega} = {\bf 180^\circ}$, $\bm{\Delta}\bm{\omega} = {\bf 0}$}
      \includegraphics[width=.49\textwidth]{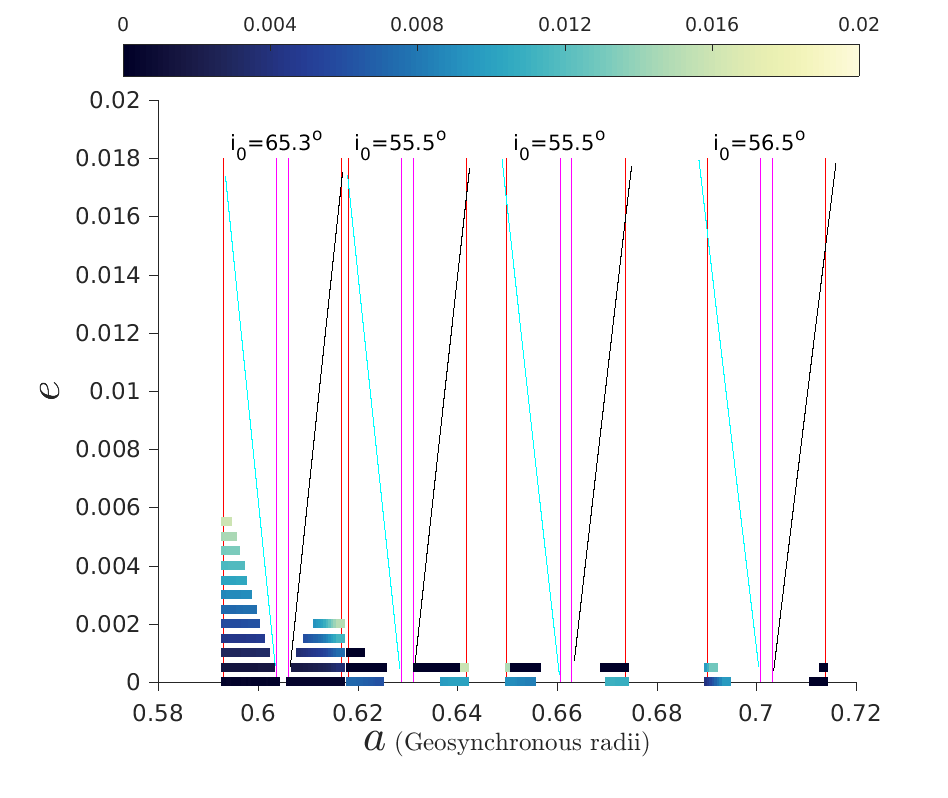}
      \includegraphics[width=.49\textwidth]{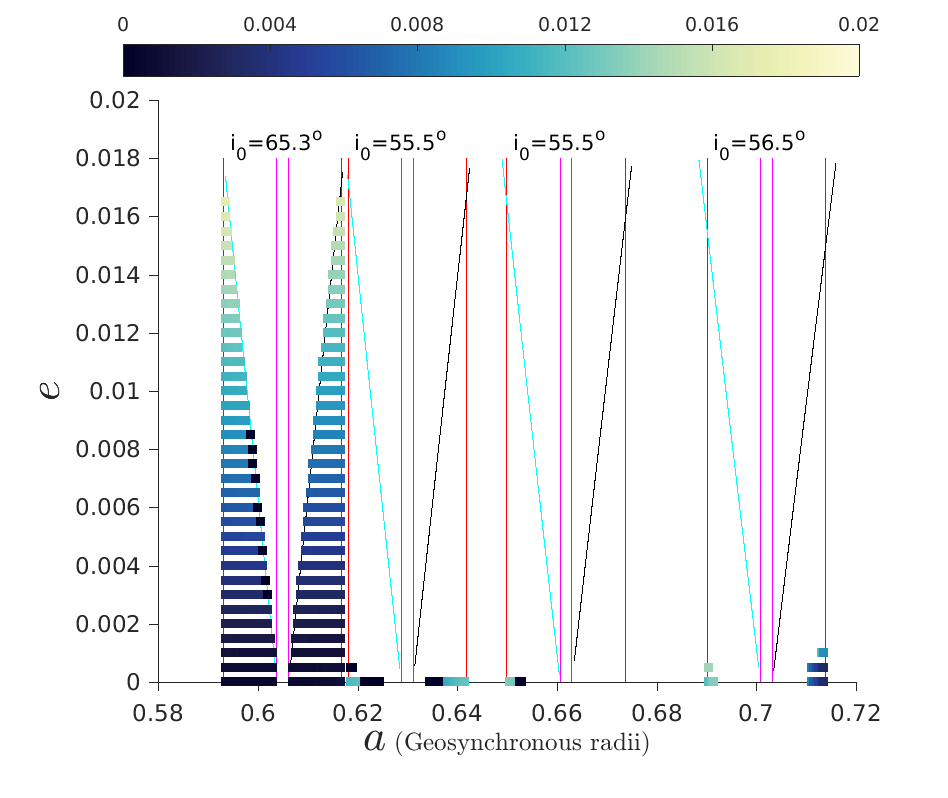}
    \end{subfigure}     
    \begin{subfigure}[b]{0.45\textwidth}
      \caption{$\bm{\Delta}\bm{\Omega} = {\bf 180^\circ}$, $\bm{\Delta}\bm{\omega} = {\bf 90^\circ}$}
      \includegraphics[width=.49\textwidth]{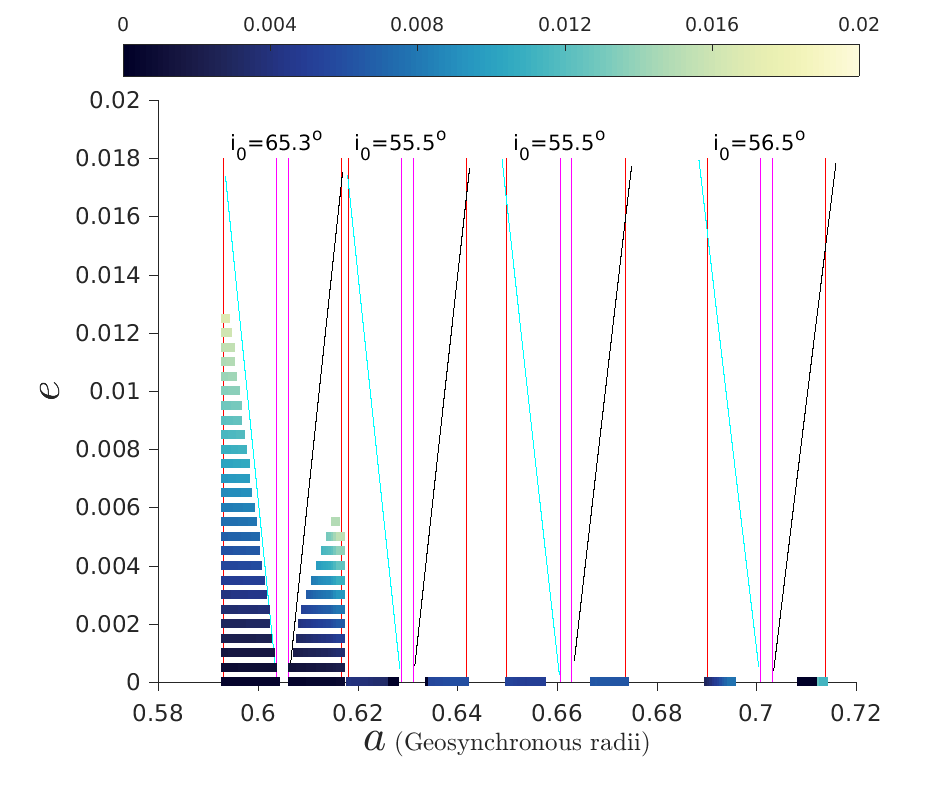}
      \includegraphics[width=.49\textwidth]{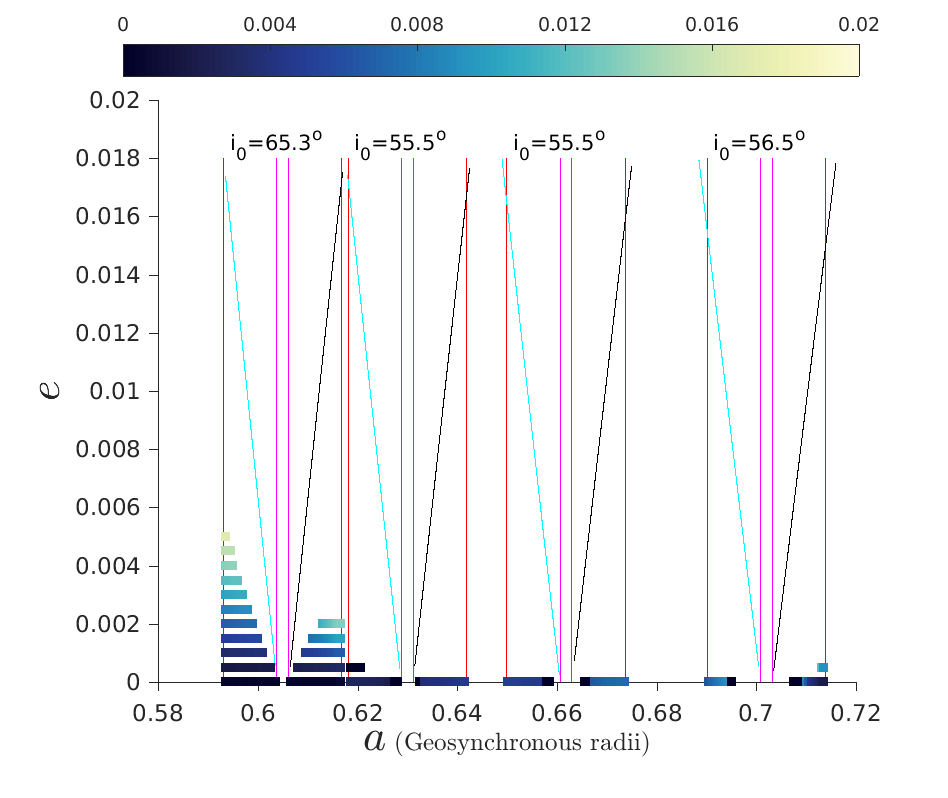}
    \end{subfigure}
    \begin{subfigure}[b]{0.45\textwidth}
      \caption{$\bm{\Delta}\bm{\Omega} = {\bf 270^\circ}$, $\bm{\Delta}\bm{\omega} = {\bf 0}$}
      \includegraphics[width=.49\textwidth]{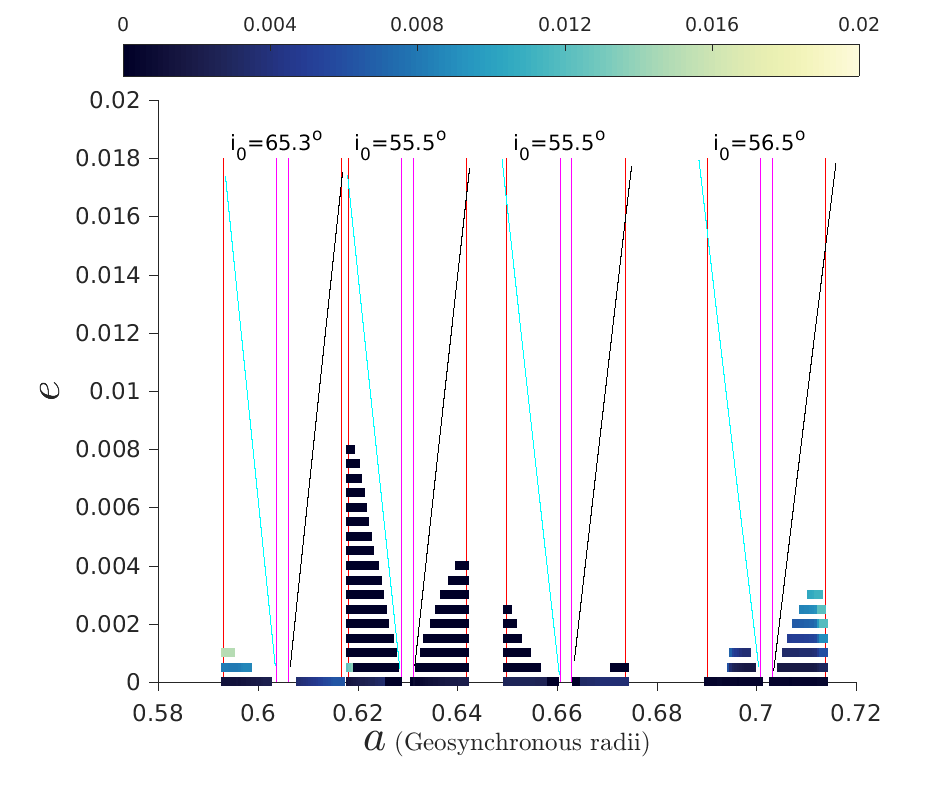}
      \includegraphics[width=.49\textwidth]{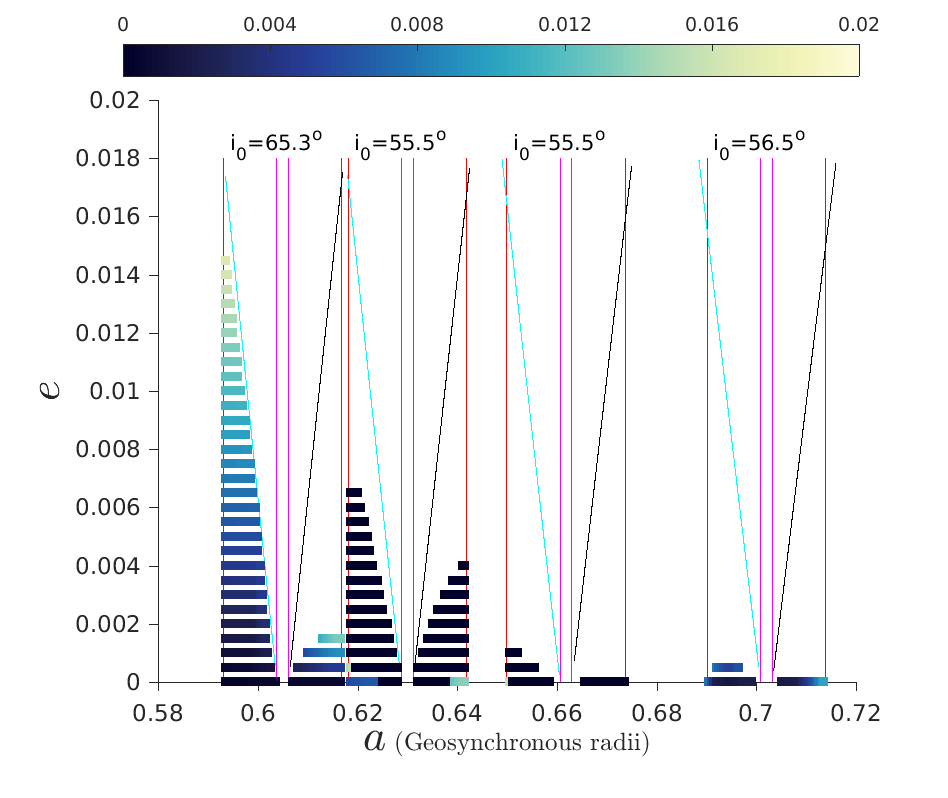}
    \end{subfigure} 
    \begin{subfigure}[b]{0.45\textwidth}
      \caption{$\bm{\Delta}\bm{\Omega} = {\bf 270^\circ}$, $\bm{\Delta}\bm{\omega} = {\bf 90^\circ}$}
      \includegraphics[width=.49\textwidth]{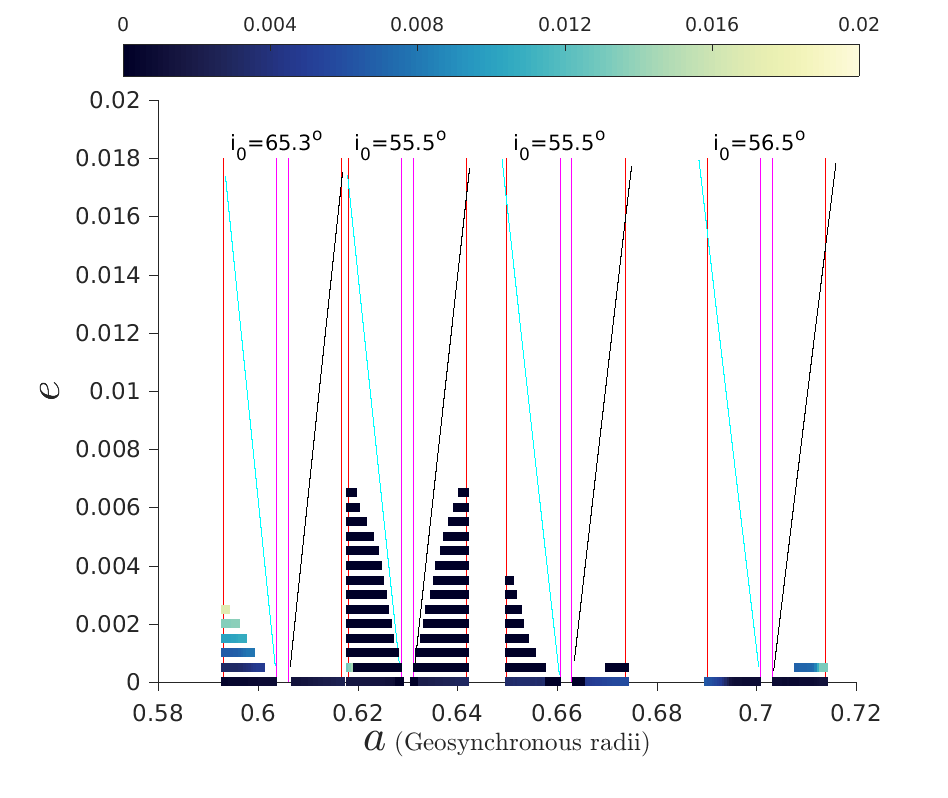}
      \includegraphics[width=.49\textwidth]{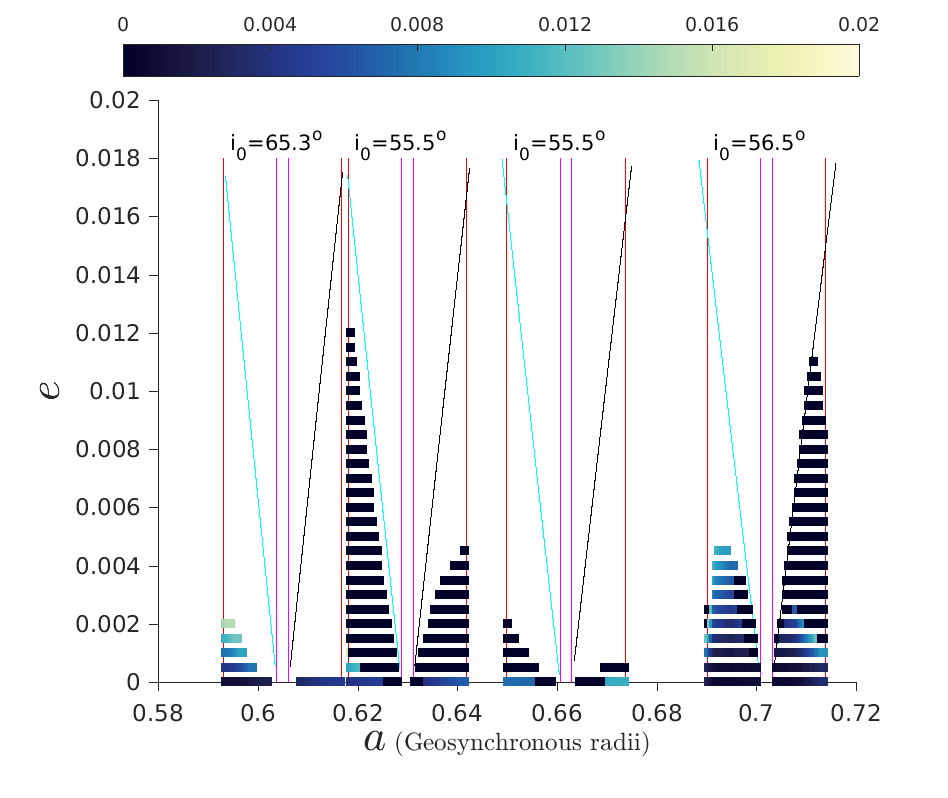}
    \end{subfigure}    
  \caption{Maximum eccentricity maps of the \textit{GNSS-graveyard} phase space for $\bm{i_{o}} = {\bf i_{nom}+0.5^{\circ}}$,  
  for Epoch 2018 (left) and Epoch 2020 (right), and for $C_{R}A/m=0.015$ m$^2$/kg. 
  $i=65.3^{\circ}$ for GLONASS, $55.5^{\circ}$ for GPS and BEIDOU, and $56.5^{\circ}$ for GALILEO. 
  The colorbar for maximum eccentricity maps is from 0 to 0.02.}
  \label{fig:GRAV_srp1_3}
\end{figure}

\end{appendix}

\clearpage

%-------------------------------------------------------------------------------------------------------------------------------------- 
%          REFERENCES
% -------------------------------------------------------------------------------------------------------------------------------------- 

\end{document}